\documentclass[fleqn,usenatbib]{mnras}

\usepackage{newtxtext,newtxmath}

\usepackage[T1]{fontenc}

\DeclareRobustCommand{\VAN}[3]{#2}
\let\VANthebibliography\thebibliography
\def\thebibliography{\DeclareRobustCommand{\VAN}[3]{##3}\VANthebibliography}

\usepackage{graphicx}	
\usepackage{amsmath}	
\usepackage{booktabs}
\usepackage{xspace}

\newcommand{\aesop}{{\sc aesopica}}
\newcommand{\fable}{{\sc fable}}
\newcommand{\ha}{H$\alpha$}
\newcommand{\Msun}{\mathrm{M_{\odot}}}
\newcommand{\jwst}{\textit{JWST}\xspace}

\title[The AESOPICA Simulations]{How to raise a supermassive black hole: interpreting early JWST AGN with the AESOPICA simulations}

\author[S. Koudmani et al.]{Sophie Koudmani,$^{1,2}$\thanks{E-mail: s.koudmani@herts.ac.uk}
Jan Scholtz,$^{2,3}$
Anthony J. Taylor,$^{4,5}$
Ignas Juod\v{z}balis,$^{2,3}$
Debora Sijacki,$^{2,6}$ \newauthor
Rachel S. Somerville,$^{7}$ 
Roberto Maiolino,$^{2,3,8}$
Emma Curtis-Lake,$^{1}$
Steven L. Finkelstein,$^{4,5}$ \newauthor
Martin A. Bourne,$^{1,2}$
Francesco D'Eugenio,$^{2,3}$
Sophia Geris,$^{2,3}$
Lucy R. Ivey,$^{2,3}$
and Hannah \"Ubler$^{9}$
\\
$^{1}$Centre for Astrophysics Research, Department of Physics, Astronomy and Mathematics, University of Hertfordshire, College Lane, Hatfield, AL10 9AB, UK\\
$^{2}$Kavli Institute for Cosmology, Cambridge, University of Cambridge, Madingley Road, Cambridge, CB3 0HA, UK\\
$^{3}$Cavendish Laboratory, University of Cambridge, 19 JJ Thomson Avenue, Cambridge CB3 0HE, UK\\
$^{4}$Department of Astronomy, The University of Texas at Austin, Austin, TX, USA\\
$^{5}$Cosmic Frontier Center, The University of Texas at Austin, Austin, TX, USA\\
$^{6}$Institute of Astronomy, University of Cambridge, Madingley Road, Cambridge, CB3 0HA, UK\\
$^{7}$Center for Computational Astrophysics, Flatiron Institute, 162 Fifth Avenue, New York, NY 10010, USA\\
$^{8}$Department of Physics and Astronomy, University College London, Gower Street, London WC1E 6BT, UK\\
$^{9}$Max-Planck-Institut f\"ur extraterrestrische Physik, Gie{\ss}enbachstra{\ss}e 1, 85748 Garching, Germany
}

\date{MNRAS, accepted}

\pubyear{\the\year{}}

\begin{document}
\label{firstpage}
\pagerange{\pageref{firstpage}--\pageref{lastpage}}
\maketitle

\begin{abstract}
The active galactic nuclei (AGN) uncovered by \jwst in the early Universe are highly abundant and seemingly overmassive with respect to local scaling relations, challenging standard models of black hole formation and growth. Yet it remains unclear whether they trace an efficient early growth channel, represent the observable tail of a broader population, or reflect systematic uncertainties in mass estimates. We introduce the \aesop~project, a suite of twelve mid-volume ($L = 60\,\mathrm{cMpc}$) cosmological simulations based on the \fable~galaxy formation model, varying the black hole seed mass across the theoretical formation channels ($M_\mathrm{seed} = 10^{2}$--$10^{5}\,\Msun$), the accretion efficiency, including super-Eddington bursts, and the supernova feedback strength. We forward-model observational selection with \textsc{balmersopica}, a mock \jwst broad-line survey pipeline that assesses the detectability of each simulated AGN for a given grating and exposure time. We find that the bulk of the \jwst AGN population can be assembled from any seed mass provided the accretion efficiency is high, although light seeds require the most favourable accretion conditions explored. Applying broad-line selection naturally recovers the apparently overmassive population, with the detected AGN lying furthest above the intrinsic $M_\mathrm{BH}$--$M_\mathrm{stellar}$ relation for inefficient accretion models. The selected AGN lie on the local, weakly evolving $M_\mathrm{BH}$--$\sigma_\mathrm{stellar}$ relation, supporting a scenario where black holes assemble before the stellar component is fully established. Since efficient accretion rapidly erases the imprint of the initial seed mass, the low-mass end of the black hole mass function and host gas-phase metallicities offer the most promising discriminants between seeding channels.
\end{abstract}




\section{Introduction}

The James Webb Space Telescope (\jwst) has revolutionised our view of black hole -- galaxy co-evolution in the early Universe by uncovering a large population of active galactic nuclei (AGN) at $4 \lesssim z \lesssim 11$ which appear to be overmassive with respect to the local scaling relations \citep[e.g.][]{ubler_ga-nifs_2023,harikane_jwstnirspec_2023,maiolino_jades_2024,taylor_broad-line_2025,gupta_rapid_2026}, including extremely compact objects with V-shaped spectra, the so-called Little Red Dots \citep[LRDs, see e.g.][]{matthee_little_2024,kokorev_census_2024,greene_uncover_2024,hviding_rubies_2025,kocevski_rise_2025}. The limited cosmic time available for growth of (over)massive black holes challenges standard models of black hole evolution. Several interpretations have been proposed to explain this emerging population. One possibility is that these systems represent a rapid black hole growth channel, potentially driven by dense gas inflows, compact host structures, mergers, or weak early feedback regulation  \citep[e.g.][]{trinca_episodic_2026,keitaanranta_rapid_2026,prole_seedz_2026}. Alternatively, the currently observed sources may constitute the extreme, observable tail of a much larger underlying population, with selection effects favouring the detection of the most massive and actively accreting black holes \citep[e.g.][]{habouzit_is_2025,li_tip_2025,geris_jades_2026}. A further possibility is that black hole masses are systematically overestimated, for example due to the impact of super-Eddington accretion on virial mass estimators \citep[e.g.][]{lupi_size_2024,lambrides_case_2026}, or radiative transfer and scattering effects \citep[e.g.][]{naidu_gas-enshrouded_2026,rusakov_little_2026}, in addition to significant uncertainties that remain for the stellar masses \citep[e.g.][]{narayanan_outshining_2024}. 

Discriminating between these scenarios is key to addressing the long-standing question of the origin of supermassive black holes. The main theoretical formation channels span several orders of magnitude in seed mass: light seeds ($\sim 10^{2}\,{\rm M_\odot}$) form as the remnants of Population~III stars \citep[e.g.][]{madau_massive_2001}, intermediate-mass seeds ($\sim 10^{3}$--$10^{4}\,{\rm M_\odot}$) arise from runaway stellar collisions in dense nuclear star clusters \citep[e.g.][]{portegies_zwart_runaway_2002,devecchi_formation_2009}, and heavy seeds ($\sim 10^{4}$--$10^{5}\,{\rm M_\odot}$) are produced by the direct collapse of pristine gas in atomic-cooling haloes \citep[e.g.][]{bromm_formation_2003,lodato_supermassive_2006}. These formation channels are expected to differ in number density: light seeds should be ubiquitous, whereas direct collapse likely requires a restrictive set of environmental conditions \citep[e.g.][]{omukai_can_2008,inayoshi_assembly_2020}. If the overmassive \jwst black holes reflect genuinely efficient early growth, heavy seeds provide a natural `head start' since they require fewer e-folding times to reach the supermassive regime and have a dynamical advantage for gas accretion. However, light seeds must sustain (super-)Eddington accretion for extended periods and may wander in their host galaxies impeding gas accretion. 

The question of whether efficient early black hole growth is \textit{physically plausible} can be directly addressed with cosmological simulations. Furthermore, observational selection functions can be applied directly to the simulation outputs, enabling synthetic survey catalogues that assess the role of selection effects in uncovering this seemingly unusual black hole population. The main challenge in modelling black hole evolution is the vast range of scales, spanning 14 orders of magnitude from the event horizon ($\sim 10^{-6}$~pc for Sgr A*) to the cosmic web ($\sim 10^{8}$~pc). This renders an ab-initio treatment computationally infeasible and hence cosmological simulations have to resort to including black hole physics via so-called subgrid models \citep[also see discussion in][]{koudmani_unified_2024}. The majority of large-volume cosmological simulations \citep[e.g.][]{dubois_dancing_2014,schaye_eagle_2015,sijacki_illustris_2015,henden_fable_2018,pillepich_simulating_2018} adopt a common set of assumptions for black hole evolution: (i) black hole growth follows Bondi--Hoyle--Lyttleton accretion, (ii) accretion is capped at the Eddington limit, and (iii) black holes are seeded with relatively large initial masses, typically $M_{\rm seed} \sim 10^{5}\,{\rm M_\odot}$. We note that the third assumption is necessitated by both resolution considerations and the first assumption since the strong dependency of the Bondi accretion rate on black hole mass ($\dot{M}_\mathrm{Bondi} \propto M_\mathrm{BH}^2$) stunts the growth of light black hole seeds.

While this framework has been successful in reproducing a range of low-redshift observables \citep[e.g.][]{habouzit_supermassive_2021}, it struggles to reproduce the observed AGN luminosity functions \citep[e.g.][]{habouzit_supermassive_2022} and its validity in the early Universe is increasingly uncertain \citep[e.g.][]{habouzit_is_2025}. The Bondi accretion prescription makes several simplifying assumptions, including spherical symmetry and neglecting angular momentum transfer in its standard implementation, though see \citet{krumholz_bondi_2005,rosas-guevara_impact_2015,curtis_resolving_2016}. In addition, most cosmological simulations do not resolve the characteristic Bondi radius, which typically ranges from $\sim 5$ to $5000$ pc for supermassive black holes \citep[also see][]{curtis_resolving_2015}. To compensate for unresolved multi-phase structure in the interstellar medium (ISM), simulations often introduce a multiplicative boost factor (commonly denoted $\alpha$), which effectively sets the halo mass above which black holes can grow efficiently \citep{booth_cosmological_2009} and can therefore equivalently be interpreted as a numerically inexpensive means of enabling an efficient accretion regime for low-mass black holes \citep{koudmani_two_2022}. An alternative class of models instead ties the accretion rate to the gravitational torques that drive gas inward on galactic scales \citep[e.g.][]{hopkins_analytic_2011,angles-alcazar_gravitational_2017}, circumventing the steep black hole mass dependence of the Bondi prescription and thereby permitting efficient growth of lighter seeds \citep[e.g.][]{wellons_exploring_2023}.

Motivated by the growing tension between observations and simulations in the high-redshift regime, recent theoretical work has begun to systematically relax the above assumptions. Semi-analytical models have explored alternative growth pathways and seeding prescriptions via systematic parameter studies \citep[e.g.][]{schneider_are_2023,spinoso_multiflavour_2023,trinca_seeking_2023,trinca_episodic_2026,dayal_exploring_2024,dayal_light_2026}. Meanwhile, hydrodynamical simulations have investigated efficient accretion models in the low-mass regime \citep[e.g.][]{dave_simba_2019,koudmani_two_2022,weinberger_accretion_2025,bhowmick_supermassive_2026,ortame_small_2026}, the impact of relaxing the Eddington limit \citep[e.g.][]{bennett_growth_2024,lupi_sustained_2024,rennehan_obsidian_2024,husko_effects_2025,quadri_super-eddington_2025}, and a broader range of black hole seed masses, including light seeds \citep[e.g.][]{taylor_seeding_2014,volonteri_black_2020,volonteri_exploring_2025,beckmann_population_2023,jeon_observability_2023,sanati_rapid_2025,bhowmick_dynamics_2025,mehta_growth_2026}. However, the higher numerical costs of hydrodynamical simulations have made it difficult to systematically vary all three aspects within one cosmological simulation suite.

In this context, we introduce the \aesop~project, a suite of mid-volume ($L = 60\,\mathrm{cMpc}$) cosmological simulations based on the \fable~model \citep{henden_fable_2018,henden_redshift_2019,henden_baryon_2020}. \aesop~is designed to systematically explore black hole growth in the early Universe by varying key modelling assumptions, including a wide range of seed masses ($10^{2}$--$10^{5}\,{\rm M_\odot}$) spanning the main theoretical seeding channels introduced above. Furthermore, we explore alternative accretion models that allow for efficient and super-Eddington growth. We aim to assess whether such models can reproduce the abundance and properties of massive black holes inferred at high redshift, and whether the emerging population can be understood as a natural outcome of early galaxy formation and/or as a consequence of observational selection.

The structure of the remainder of this paper is as follows. In Section~\ref{sec:methods} we provide an overview of the methodology including the galaxy formation model and parameter variations (see Section~\ref{subsec:methods-aesopica}), the galaxy identification and merger tree procedure (see Section~\ref{subsec:methods-galid}), and our tool for the synthetic AGN survey selection \textsc{balmersopica} (see Section~\ref{subsec:methods-mocks}). We present our results in Section~\ref{sec:results} including visualisations of the simulations and black hole demographics (see Section~\ref{subsec:results-visuals}), evolutionary tracks of the most massive halo at high redshift (see Section~\ref{subsec:results-infantgalsandbhs}), our simulated scaling relations (see Section~\ref{subsec:results-scalrel-raw}) as well as the scaling relations based on synthetic survey selection (see Section~\ref{subsec:results-scalrel-synth}), the bolometric AGN luminosity function (see Section~\ref{subsec:results-bollumfunc}), the black hole mass function (see Section~\ref{subsec:results-bhmf}) and the relationship between black hole masses and gas-phase metallicities (see Section~\ref{subsec:results-metals}). We discuss our results in the context of other theoretical work as well as observational caveats in Section~\ref{sec:discussion} and conclude in Section~\ref{sec:conclusions}.

\section{Methodology} \label{sec:methods}

In this section, we describe the methodology underlying the \aesop~project. Section~\ref{subsec:methods-aesopica} provides an overview of the simulation suite, including the galaxy formation model and the systematic variations of the black hole seeding and accretion prescriptions that define the twelve \aesop~runs. Section~\ref{subsec:methods-galid} summarises our galaxy identification and merger tree procedures. Finally, Section~\ref{subsec:methods-mocks} introduces \textsc{balmersopica}, our framework for constructing synthetic broad-line AGN surveys from the simulation outputs, which mimics the selection of \jwst observing programmes for different gratings and exposure times.

\subsection{Overview of the simulations} \label{subsec:methods-aesopica}

The \aesop~simulations were carried out with the massively parallel \textsc{arepo} code \citep{springel_e_2010}, where fluid dynamics is discretized on a moving mesh using a quasi-Lagrangian finite volume technique. The unstructured mesh is based on the Voronoi tessellation of a set of discrete points that cover the whole computational domain. These mesh-generating points are allowed to move with the local flow velocity, with minor corrections to avoid excessive distortion of the gas cells. Gravitational interactions are modelled using the TreePM approach, with stars and dark matter modelled as collisionless particles.

The \aesop~suite consists of twelve mid-volume simulations, systematically varying the implementation of stellar feedback and black hole models. In each case, the comoving 40~$h^{-1}$~Mpc box was evolved using initial conditions for a uniformly sampled cosmological volume based on a Planck cosmology \citep[see][]{planck_collaboration_planck_2016}. The volume contains $512^3$ dark matter particles with masses $m_\mathrm{DM} = 3.4 \times 10^7 ~ h^{-1}\, \Msun$ and initially $512^3$ gas resolution elements with target mass $\overline{m}_\mathrm{gas} = 6.4 \times 10^{6} ~ h^{-1}\, \Msun$. The gravitational softening length is set to $2.393 ~ h^{-1} \, \mathrm{kpc}$ in physical coordinates below $z=5$ and fixed in comoving coordinates at higher redshift, following the empirical recommendation by \citet{power_inner_2003} for optimal numerical convergence.

The \aesop~simulation suite is built upon the \fable~galaxy formation model \citep{henden_fable_2018,henden_redshift_2019,henden_baryon_2020}, which is in turn largely based on the Illustris galaxy formation model \citep{vogelsberger_introducing_2014}. Whilst the models for star formation \citep{springel_cosmological_2003}, radiative cooling \citep{katz_cosmological_1996,wiersma_effect_2009}, and chemical enrichment \citep{wiersma_chemical_2009} are unchanged from Illustris, the stellar feedback \citep{vogelsberger_model_2013} has been updated to include a thermal component; see Section~\ref{subsubsec:methods-galform} for a brief overview of the stellar feedback model. The Illustris AGN feedback model \citep{sijacki_illustris_2015} is further enhanced with a quasar duty cycle in \fable~to match the observed galaxy stellar mass function and gas fractions in groups and clusters \citep{henden_fable_2018}.  \aesop~introduces additional targeted updates for modelling the growth of infant black holes in the early Universe, exploring three key modifications to fiducial galaxy formation models: enabling efficient accretion in the low-mass regime \citep{koudmani_two_2022}, incorporating super-Eddington accretion, and examining a broad range of seed masses ($10^{2}~\mathrm{M_{\odot}}$ to $10^{5}~\mathrm{M_{\odot}}$) following seed evolution from early cosmic epochs ($z \sim 18$); see Section~\ref{subsubsec:methods-bhs} for further details on the black hole model.

\subsubsection{Stellar feedback model} \label{subsubsec:methods-galform}
Following the Illustris galactic wind model \citep{vogelsberger_model_2013}, wind particles are launched from star-forming regions driven by the available energy from core-collapse supernovae. The \fable~model, however, makes a few modifications to the parameters that govern the wind energetics compared to Illustris. Specifically, the wind energy factor $\epsilon_\mathrm{W,SN}$, which gives the fraction of energy available from each core-collapse supernova, is increased to $\epsilon_\mathrm{W,SN} = 1.5$ in \fable \ compared to the Illustris value of $\epsilon_\mathrm{W,SN} = 1.09$. Furthermore, one-third of the wind energy is injected as thermal energy in \fable, whilst in Illustris the stellar-feedback-driven winds are purely kinetic. Overall, this leads to more energetic stellar feedback, which more efficiently dissipates the released energy to the gas, and somewhat more effectively regulates star formation in low-mass haloes (see \citealt{henden_fable_2018} for details; the same method is used by \citealt{marinacci_formation_2014}). In our previous work, we found that whilst this more efficient feedback provides an improved fit to the observed galaxy stellar mass function in the local Universe, it also acts to suppress black hole activity in low-mass galaxies to the extent that the bright dwarf AGN detected by X-ray surveys are not reproduced by the \fable~simulations \citep[see][for details]{koudmani_little_2021}. We found that if a more moderate coupling efficiency for the stellar feedback and a more efficient black hole accretion model are assumed, then star formation can be regulated at the same level as the strong supernova feedback assumed in the original \fable~\citep[see][for details]{koudmani_two_2022}. Given the observational hints of highly efficient black hole growth in low-mass galaxies at high redshift, we therefore test both the standard \fable \ stellar feedback as well as a set-up with more moderate stellar feedback ($\epsilon_\mathrm{W,SN} = 0.5$) following \citet{koudmani_two_2022}.

\subsubsection{Black hole model} \label{subsubsec:methods-bhs}

Black holes are modelled as collisionless particles. For the \aesop~simulations, we focus on investigating black hole growth in the early Universe, and we therefore modify the black hole seeding model from the Illustris and \fable~implementations by lowering the halo mass threshold and varying the black hole seed mass. In \aesop, black holes are seeded into the smallest resolved dark matter haloes above a mass threshold of $2 \times 10^{9} \ h^{-1} \, \Msun$; this allows us to track black hole seeds from early epochs. As in \fable \ and Illustris, black holes are only seeded into haloes that do not already host a black hole. For the seed masses, we explore four different options: very light seeds ($M_\mathrm{seed} = 10^{2}~\Msun$), light seeds ($M_\mathrm{seed} = 10^{3}~\Msun$), intermediate seeds ($M_\mathrm{seed} = 10^{4}~\Msun$) and heavy seeds ($M_\mathrm{seed} = 10^{5}~\Msun$), also see Table~\ref{tab:aesopica}. These could represent a range of seeding channels including Pop III stars, runaway collapse of a dense star cluster and direct collapse of gas clouds. However, we caution that due to our limited resolution, we do not apply any seeding criteria beyond the halo mass threshold and therefore our seeding model likely represents an optimistic scenario for massive seeds. For any given simulation run, the seed mass is kept constant as `mixed' seeding scenarios are beyond the scope of this work. 

Further, we note that the black holes are advected to the potential minimum of their host halo to prevent spurious black hole movement due to numerical heating \citep[see][for details on the black hole seeding and growth models]{sijacki_unified_2007,vogelsberger_model_2013}. Specifically, at every time step, the black hole is moved to the position of the resolution element with the lowest gravitational potential within the black hole's smoothing length. However, for light seeds, wandering black holes may actually be physical and could result in suppressed accretion rates \citep[e.g.][]{bellovary_origins_2021,ma_seeds_2021,sharma_hidden_2022}. Hence, the repositioning scheme represents another optimistic choice for our early black hole evolution \citep[also see][]{bhowmick_dynamics_2025}.

\begin{table*} 
  \centering
  \caption{Accretion-model variants in the \aesop~suite as a function of black hole seed mass. For each run we list the accretion boost factor $\alpha$, the maximum Eddington ratio $f_{\rm Edd,max}$, and the supernova feedback efficiency $\epsilon_{\rm W,SN}$.}
  \label{tab:aesopica}
  \begin{tabular}{l@{\hskip 1.5em}cccc}
    \toprule
     & \multicolumn{4}{c}{\textbf{Seed mass}} \\
    \cmidrule(lr){2-5}
    \textbf{Accretion} & $10^{2}\,{\rm M_{\odot}}$ & $10^{3}\,{\rm M_{\odot}}$
              & $10^{4}\,{\rm M_{\odot}}$ & $10^{5}\,{\rm M_{\odot}}$ \\
     & (very light) & (light) & (intermediate) & (heavy) \\
    \midrule
    Fiducial & \multicolumn{4}{c}{$\alpha = 10^{2},\quad f_{\rm Edd,max} = 1,\quad \epsilon_{\rm W, SN} = 1.5$} \\
    \addlinespace
    \midrule
    \addlinespace
    SE-Boost & \multicolumn{4}{c}{$\alpha = 10^{3},\quad f_{\rm Edd,max} = 10      ,\quad \epsilon_{\rm W, SN} = 1.5$} \\
    \addlinespace
    \midrule
    \addlinespace
    SE-BoostMax
      & $\alpha = 10^{5}$       & $\alpha = 10^{4}$       & $\alpha = 10^{3}$       & $\alpha = 10^{3}$   \\
      & $f_{\rm Edd,max} = 10$  & $f_{\rm Edd,max} = 10$  & $f_{\rm Edd,max} = 10$  & $f_{\rm Edd,max} = 10$  \\
      & $\epsilon_{\rm W, SN} = 1.5$ & $\epsilon_{\rm W, SN} = 1.5$ & $\epsilon_{\rm W, SN} = 0.5$ & $\epsilon_{\rm W, SN} = 0.5$\\
    \bottomrule
  \end{tabular}
\end{table*}

The black holes may grow via black hole -- black hole mergers and via gas accretion following the Bondi-Hoyle-Lyttleton-like accretion rate. The black hole accretion rate, $\dot{M}_\mathrm{BH}$, is then given by:
\begin{equation}
    \dot{M}_\mathrm{BH} = \alpha \dot{M}_\mathrm{Bondi} = \alpha \frac{4 \pi \mathrm{G}^{2} M_\mathrm{BH}^{2} \rho}{c_\mathrm{s}^{3}},
    \label{eq:ZoomsBondiRate}
\end{equation}
where $\mathrm{G}$ is the gravitational constant, $M_\mathrm{BH}$ is the black hole mass, $\rho$ is the density and $c_\mathrm{s}$ is the sound speed in the vicinity of the black hole\footnote{The traditional Bondi-Hoyle-Lyttleton model also includes a relative velocity term in the denominator, though we neglect this here for simplicity (also note that our relative velocities may not always be accurate given the repositioning scheme employed).}. The radiative efficiency is set to a constant $\epsilon_\mathrm{r} = 0.1$ and $(1 - \epsilon_\mathrm{r})$ of the accreted mass is added to the black hole particle mass at each timestep. The boost factor $\alpha$ was introduced in the Bondi accretion prescription to account for the unresolved multi-phase nature of the ISM \citep[e.g.][]{springel_modelling_2005,booth_cosmological_2009,johansson_evolution_2009,sijacki_gravitational_2011} and for our simulation runs with the `\textit{Fiducial}' model, we employ the standard value $\alpha = 100$, following Illustris and \fable. However, it has been found by various groups that the fiducial Bondi implementation may underproduce accretion rates, in particular for low-mass black holes \citep[e.g.][]{hopkins_analytic_2011,gaspari_chaotic_2013,koudmani_little_2021,koudmani_two_2022,beckmann_population_2023,kho_signatures_2025}. Hence, we additionally perform simulations where we further enhance the boost factor to $\alpha = 10^{3}$, allowing for efficient black hole growth in low-mass galaxies \citep{koudmani_two_2022}. We also modify the maximum accretion rate limit, where instead of limiting the black hole accretion rates to the Eddington limit, as is standard for the vast majority of cosmological simulations, we increase the maximum accretion rate limit to ten times the Eddington rate motivated by the increasing observational evidence for super-Eddington accretion in the early Universe \citep[e.g.][]{maiolino_jades_2024,gravity_collaboration_spatially_2026,lambrides_case_2026}. These runs with $f_\mathrm{Edd,max}=10$, allowing for super-Eddington bursts and the enhanced boost factor $\alpha=10^{3}$, are labelled as \textit{`SE-Boost'}. However, we note that for the (very) light seed runs this boost factor is not sufficient to compensate for the low seed masses within the Bondi framework. Since $\dot{M}_\mathrm{Bondi} \propto \alpha M_\mathrm{BH}^{2}$ and $\dot{M}_\mathrm{Edd} \propto M_\mathrm{BH}$, we then have for the same ambient gas conditions that $f_\mathrm{Edd} \propto \alpha M_\mathrm{BH}$. Hence to obtain the same initial Eddington ratio as for the fiducial model with the heavy seeds ($M_\mathrm{seed}=10^{5}~\Msun$ and $\alpha=10^{2}$), we enhance the boost parameter in an inversely proportional manner for the lighter seeds, setting $\alpha=10^{4}$ for $M_\mathrm{seed}=10^{3}~\Msun$ and $\alpha=10^{5}$ for $M_\mathrm{seed}=10^{2}~\Msun$; this represents our most `optimistic' scenario for (very) light seed growth. For the most optimistic scenario for the intermediate and heavy seeds, we retain the \textit{`SE-Boost'} accretion parameters as $\alpha=10^{3}$ and $f_\mathrm{Edd,max}=10$, but additionally lower the supernova wind energy factor to $\epsilon_\mathrm{W,SN} = 0.5$ following \citet{koudmani_two_2022} to enhance the early seed growth. These respective most optimistic scenarios for our different seed masses are labelled as \textit{`SE-BoostMax'}. All three accretion variations (\textit{`Fiducial'}, \textit{`SE-Boost'}, \textit{`SE-BoostMax'}) together with the four seed masses then yield twelve simulation runs that form the \aesop~suite; see Table~\ref{tab:aesopica} for an overview. 

The AGN feedback in \aesop~is left the same as in \fable~and is based on a two-mode model where the quasar mode operates at high Eddington ratios \citep[see][]{di_matteo_energy_2005,springel_modelling_2005} and the radio mode is activated at low Eddington ratios \citep[see][]{sijacki_unified_2007}, with the switchover occurring at an Eddington ratio of $f_\mathrm{Edd,QM} = 0.01$.

In the quasar mode, a fraction $\epsilon_\mathrm{f}=0.1$ of the AGN luminosity is isotropically injected as thermal energy. In Illustris, this thermal energy injection happens continuously, which can lead to artificial overcooling due to the limited gas mass resolution. In \fable, this issue is alleviated by introducing a duty cycle with an approach similar to that of \citet{booth_cosmological_2009}, whereby thermal energy is accumulated over $\delta t_\mathrm{QM} = 25$~Myr before being released in a single event, allowing high feedback temperatures, and hence longer cooling times, to be reached. Such an intermittent feedback cycle is also, at least qualitatively, consistent with episodic accretion observed in high-resolution simulations \citep{ciotti_feedback_2010,torrey_instability_2017,costa_quenching_2018,angles-alcazar_cosmological_2021}. In the radio mode, the feedback energy is injected as buoyant bubbles mimicking those inflated by jets \citep{mcnamara_heating_2007,fabian_observational_2012,bourne_recent_2023}, with the injection duty cycle set by the fractional black hole growth $\delta_\mathrm{BH} = \delta M_\mathrm{BH} / M_\mathrm{BH}$. In \fable, this threshold is set to $\delta_\mathrm{BH} = 0.01$, much lower than the Illustris value of $0.15$. The bubble energy is $\epsilon_\mathrm{m} \epsilon_\mathrm{r} c^{2} \delta M_\mathrm{BH}$, with $\epsilon_\mathrm{m} = 0.8$, giving an effective efficiency $\epsilon_\mathrm{m}\epsilon_\mathrm{r} = 0.08$, close to Illustris' $0.07$. The smaller $\delta_\mathrm{BH}$ thus produces more frequent, less energetic bubbles in the \fable~model. The remainder of the feedback energy that is not coupled to the surrounding gas by the quasar or radio mode goes into radiative electromagnetic feedback, which is approximated as an additional radiation field around the black hole superposed with the redshift-dependent ultraviolet background \citep{vogelsberger_model_2013}.

\subsection{Galaxy identification} \label{subsec:methods-galid}
Haloes and subhaloes are identified via friends-of-friends (FoF) and \textsc{subfind} algorithms \citep{davis_evolution_1985,springel_gadget_2001,dolag_substructures_2009}. For the FoF search, we choose a linking length of $0.2$ multiplied by the mean inter-particle separation. We then use \textsc{subfind} to identify gravitationally self-bound subhaloes within each FoF halo. The `central subhalo' is the subhalo at the minimum gravitational potential of the FoF halo. All other subhaloes in the FoF halo are categorised as satellite subhaloes. We use \textsc{sublink} to create galaxy merger trees, allowing us to identify the progenitors and descendants of each subhalo across cosmic time \citep[see][for details]{rodriguez-gomez_merger_2015}. 

\subsection{Mock high-z survey selection: \textsc{balmersopica}} \label{subsec:methods-mocks}
The vast majority of black hole mass estimates in the early Universe are based on virial estimates using broad lines \citep[though see][for recent dynamical measurements at high redshift]{juodzbalis_direct_2026,gravity_collaboration_spatially_2026}. The selection bias affecting the identification of AGN with broad emission lines is significant since it does not just favour brighter AGN but also higher black hole masses due to the necessity of distinguishing the broad line \citep[also see][]{ding_testing_2020}. We have therefore devised a `mock \jwst survey' selection procedure that we apply to our simulation outputs to assess whether an AGN at a given redshift with a given luminosity and black hole mass would be detected for a given \jwst grating and exposure time. We caution that we only carry out this selection with respect to the final step of AGN selection (the fitting of the broad-line region) and hence this is likely to be an upper limit. We create simulated broad lines from $z=3$ to $z=11$ in increments of 0.25, black hole mass from $10^4~\Msun$ to $10^9~\Msun$ in increments of 0.2 dex, and bolometric luminosity from $10^{41}~\mathrm{erg\,s^{-1}}$ to $10^{46}~\mathrm{erg\,s^{-1}}$ in increments of 0.2 dex. Firstly, we assume a bolometric correction of $130$ \citep{stern_type_2012} to obtain the flux of the H$\alpha$ line, which is commonly used for high-z AGN \citep[e.g.][]{maiolino_jades_2024,taylor_broad-line_2025,geris_jades_2026,juodzbalis_jades_2026}. Next, we estimate the width of the simulated broad component based on the H$\alpha$ luminosities and black hole masses using virial black hole mass estimators from \citet{greene_active_2004,greene_new_2007,reines_dwarf_2013}. We then run this grid of fluxes and FWHMs through the \jwst exposure calculator \citep[][]{pontoppidan_pandeia_2016} for NIRSpec (including R1000 and R2700 gratings) to obtain mock observations of the broad line. Finally, we attempt to recover the properties of the line with Bayesian fitting. For a signal-to-noise ratio greater than three (computed from the median and 16th--84th percentiles of the fitted flux posterior) and provided the fitted line flux and FWHM are within a factor of two of their true values, the black hole is counted as `detectable'. We assume low-level continuum emission, noting that this is a sub-dominant effect as we are mainly noise-limited in this regime, which is taken into account with the \jwst exposure calculator. We also note that this assumes that all of the flux is from the broad-line component (no narrow component from star formation or narrow-line region; NLR). The actual procedure for identifying a broad-line AGN includes distinguishing a broad component from a narrow component, meaning signal-to-noise in the wings of the line is more important than peak/total line flux signal-to-noise. Hence our selection effects are optimistic, and strong star formation or NLRs might make the disentangling of the broad component harder \citep[also see][]{ziparo_selection-aware_2026}. 

We ignore objects with broad-line FWHMs greater than $10\,000~\mathrm{km\,s^{-1}}$ as these are generally not seen in observations (and are difficult to fit and confirm). We have checked that this criterion only has minimal impact on our high-z synthetic survey selection (up to one black hole discarded at $z=6$ and up to five black holes discarded at $z=4$). We also ignore objects with broad-line FWHMs smaller than $750~\mathrm{km\,s^{-1}}$, as these are challenging to distinguish from outflows\footnote{We note that in principle outflows could also be excluded observationally if the broad line is only in hydrogen but not seen in forbidden lines. Incorporating this type of selection would require full synthetic observations which are beyond the scope of this paper. Our broad line selection criteria may therefore be viewed as conservative \citep[e.g.][]{zamora_ga-nifs_2025,rodriguez_del_pino_ga-nifs_2026}.}, following the high-z AGN analysis for CEERS/RUBIES in \citet{taylor_broad-line_2025}. This range of FWHMs is also in close agreement with the broad-line priors used for the analysis of the JADES AGN \citep[approximately uniform priors from $800~\mathrm{km\,s^{-1}}$ to $10\,000~\mathrm{km\,s^{-1}}$, see][]{juodzbalis_jades_2026} as well as the resultant FWHMs measured for the high-z JADES AGN which range from $803~\mathrm{km\,s^{-1}}$ to $6564~\mathrm{km\,s^{-1}}$. We performed the line profile simulations assuming the grating spectrum error extensions for PRISM, G395M, G395H and F444W-slitless spectroscopy, testing exposure times between 1 hour and 100 hours. In practice, to compare our results with two major \jwst surveys and also provide a prediction for future observations, we focus on three configurations of the G395M grating for exposure times $1$~hr, $2.6$~hr and $7$~hr, corresponding to the CEERS/RUBIES and JADES (Medium Tier and Deep Tier) configurations, respectively, as well as G395H for exposure times $30$~h and $100$~h to assess the theoretical insights we could gain from an even deeper notional survey that could be carried out in the future. We generally focus on fitting the H$\alpha$ line, however, for redshifts $z>7$, we fit H$\beta$ profiles instead due to the H$\alpha$ line being redshifted out of the observable wavelength range of NIRSpec and NIRCam slitless spectroscopy. Similarly, above $z\sim9.6$, we fit the H$\gamma$ line. We note that the usage of these single-epoch virial relations in the high-redshift Universe is hotly debated in the literature, in particular for super-Eddington black holes \citep[e.g.][]{lupi_size_2024,juodzbalis_direct_2026,lambrides_case_2026,rusakov_little_2026}. Fully correcting for these effects is beyond the scope of this paper; however, we specifically highlight black holes that are `observed' as super-Eddington in our mock survey analysis. We note that super-Eddington black holes generally form a minority of the simulated AGN population due to relatively short burst phases, within our modelling assumptions. We also highlight that in the context of our simulation-based analysis the traditional bolometric corrections and virial relations assumed are \textit{conservative} choices. Lower bolometric corrections \citep[e.g.][]{greene_what_2026} and larger line widths due to scattering \citep[e.g.][]{rusakov_little_2026} would lead to even higher detectability of the simulated AGN than what we have assumed with \textsc{balmersopica}, also see Section~\ref{subsec:discuss_mbh_problem} for a discussion of the black hole mass measurement uncertainties at high redshift.

We publicly release the code and full tables for the \textsc{balmersopica} tool on GitHub at \url{https://github.com/skoudmani/balmersopica}.

\section{Results} \label{sec:results}

\subsection{Visual inspection} \label{subsec:results-visuals}

\begin{figure*}
    \centering
    \includegraphics[width=\textwidth]{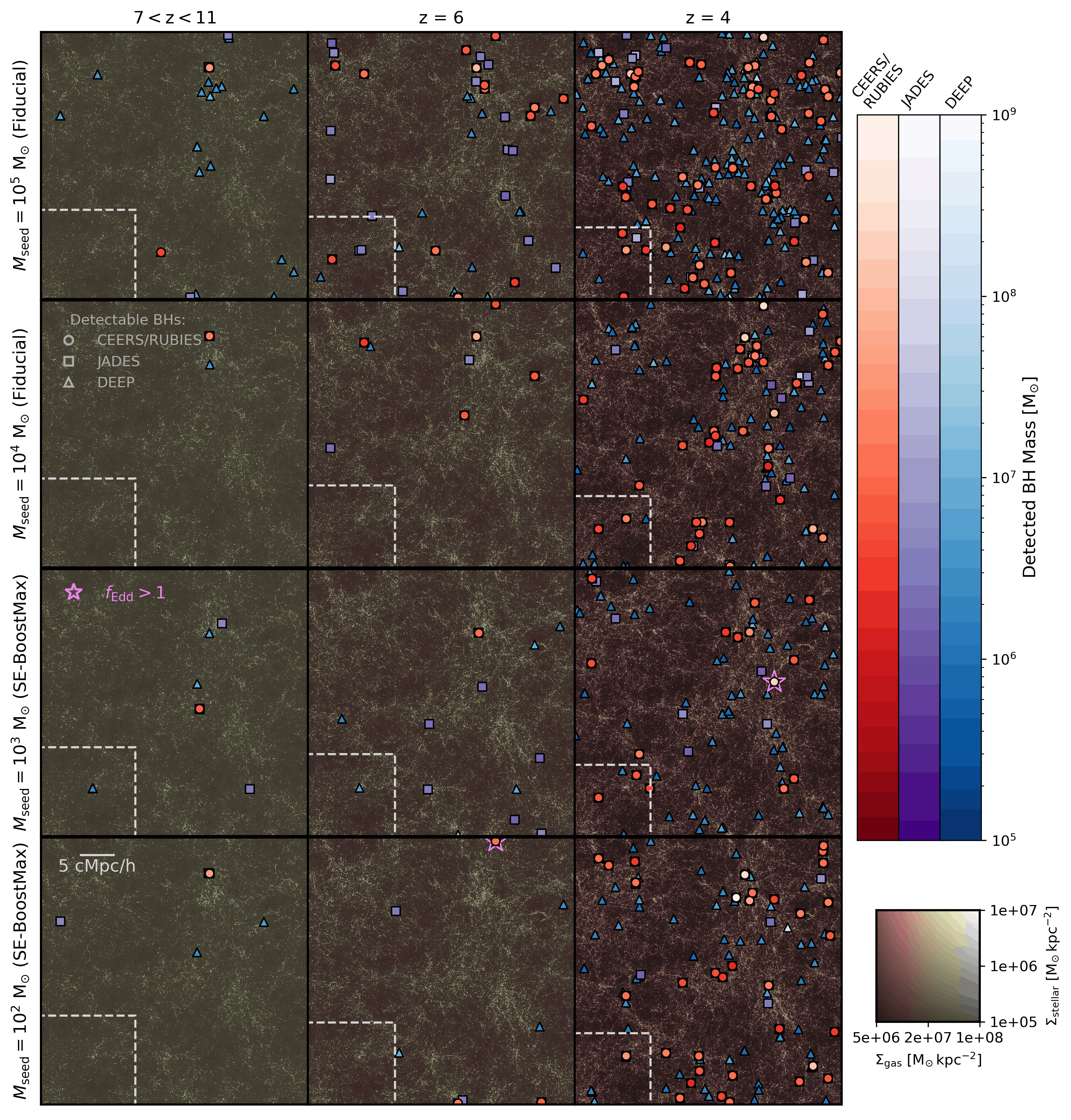}
    \caption{Gas and stellar projection of the whole \aesop~simulation box at very high redshift ($7 < z < 11$) and at $z=6$ and $z=4$. The markers show the locations of black holes detectable by CEERS/RUBIES (red circles), JADES (medium tier, purple squares) and a hypothetical future survey with a finer grating and deeper exposure (blue triangles), based on the broad line selection criteria for \jwst outlined in Section~\ref{subsec:methods-mocks}. The four rows correspond to the four seed masses explored. The top rows show the \aesop~models with heavy and intermediate seeds ($M_\mathrm{seed}=10^{5} \ \Msun$ and $M_\mathrm{seed}=10^{4} \ \Msun$) and the fiducial accretion model. The bottom rows show the \aesop~models based on light seeds ($M_\mathrm{seed}=10^{3} \ \Msun$ and $M_\mathrm{seed}=10^{2} \ \Msun$) with the `\textit{SE-BoostMax}' accretion model. For these models, super-Eddington bursts (up to $f_\mathrm{Edd}=10$) are permitted and black holes that would be observed as super-Eddington at the redshifts shown are highlighted with pink stars. To demonstrate the impact of cosmological selection effects, the field of view of the NIRSpec MSA ($3.6\ \mathrm{arcmin} \times 3.4\ \mathrm{arcmin}$) is indicated by dashed lines in the bottom-left corner of each panel. Light seeds with super-Eddington accretion may lead to broadly similar observed AGN distributions to heavy seeds; however, deeper surveys may increasingly be able to distinguish the number densities at high redshift.}
    \label{fig:star_bh_proj}
\end{figure*}

We begin our analysis with a visual inspection of a selection of our simulation runs. Fig.~\ref{fig:star_bh_proj} shows projections of the stellar and gas density distributions as well as synthetically selected AGN distributions. The high-z AGN uncovered by \jwst broad-line selection criteria mostly fall in the redshift range $11 > z > 4$. For redshifts $z>7$, it becomes significantly harder to identify broad-line AGN due to the  H$\alpha$ line being redshifted out of the observable wavelength range of NIRSpec and NIRCam slitless spectroscopy, so that the significantly fainter H$\beta$ or H$\gamma$ have to be employed instead. We therefore divide our analysis of the detectable AGN distributions into three redshift bins including very high redshift objects ($7 < z < 11$, left column) and AGN at $z=6$ (middle column) and $z=4$ (right column). The four rows correspond to the four different seed masses explored as indicated by the row titles. For the light seeds, we show the detectable AGN distributions assuming the \textit{`SE-BoostMax'} model and for the heavy seeds we show the detectable AGN distribution for the fiducial accretion model.

To identify the detectable AGN, we follow the procedure described in Section~\ref{subsec:methods-mocks} based on the exposure times for the CEERS/RUBIES survey (red circles, G395M grating with 1~hr exposure) and JADES Medium Tier (purple squares, G395M grating with 2.6~hr exposure). Additionally, we indicate the AGN detectable by a hypothetical even deeper survey (blue triangles). For this `DEEP' survey, we assume 30~hr exposure for $z<7$ and 100~hr exposure for sources at $z>7$, mimicking the time that may be allocated for a notional future \jwst Large Program. 

The applicability of the virial relations for black hole mass estimates at high redshifts remains hotly debated, especially for super-Eddington sources \citep[e.g.][]{lupi_size_2024,lambrides_case_2026}. We therefore indicate sources that are accreting at super-Eddington rates at the redshift shown by a pink star. We note that, within our theoretical framework, due to the short nature of the super-Eddington bursts, only a small number of sources are observed actually accreting above the Eddington limit (also see discussion of super-Eddington fractions in Section~\ref{subsubsec:results_scal_rel_lum}), and this fraction rapidly decreases with survey depth.

Overall, light seeds with `\textit{SE-BoostMax}' accretion and heavy seeds with fiducial accretion result in similar number densities of detectable AGN, yielding detected fractions of approximately one to six per cent across the different models considered, which is in a similar range to the estimated broad-line AGN fractions for JADES, which range from two to nine per cent \citep{juodzbalis_jades_2026}. We, however, caution that the broad-line selection criteria we apply here do not account for Type~2 AGN or for the potential of lensing to uncover fainter and less massive black holes \citep[e.g.][]{fei_glimpse_2026,golubchik_venus_2026}. Our `detectable' broad-line AGN population should therefore be taken as an upper limit since we assume that all AGN have observable broad lines. Indeed, the ratio of Type~1 and Type~2 AGN is challenging to estimate due to various selection effects of different surveys at high redshifts. Accounting for these is beyond the scope of this work. Hence, our fractions should not be interpreted as active fractions.

The distribution of detectable AGN clearly traces large-scale structure, with broad-line AGN preferentially found in nodes and filaments. This has implications for deeper surveys with only a small number of pointings. To illustrate the impact of cosmological selection effects, we show the field of view of the NIRSpec MSA ($3.6 \ \mathrm{arcmin} \times 3.4 \ \mathrm{arcmin}$) as dashed lines in the lower left-hand corner. Taking the JADES survey as an illustrative case, the two deep JADES pointings may remain subject to environmental selection effects and cosmic variance. By contrast, the larger area covered by the medium tier \citep[$75 \ \mathrm{arcmin^{2}}$, see][]{curtis-lake_jades_2026} is broadly comparable to our simulation volume (corresponding to a cosmic volume of side length $L\sim 60$~cMpc at $z=4$), and is therefore more likely to sample a region representative of the wider AGN population. CEERS probes a similar volume to our simulations. The RUBIES spectroscopic volume, however, is roughly twice the \aesop~volume \citep[see also discussion in][]{ivey_cliff_2026}, so direct comparisons with this much wider survey should be made with care. In terms of cosmic variance within our simulations and medium tier surveys, we expect host galaxies with stellar masses above $\log(M_\mathrm{stellar}/\mathrm{M_\odot})=10.35$ at $z=4$ to be affected (see discussion in Appendix~\ref{appsec:gsmf}). This is again similar to the stellar mass range probed by JADES, but is an important caveat when comparing with the RUBIES AGN, which have more massive hosts \citep{gupta_rapid_2026}.

\subsection{Infant galaxies and black holes} \label{subsec:results-infantgalsandbhs}
\subsubsection{Early black hole accretion}

Having established an overview of the detectable AGN population, we now focus on an individual system to develop intuition for how the initial black hole seed mass and accretion parametrization shape massive black hole evolution in the early Universe. For this purpose, we analyse the earliest-forming structure in the \aesop~box, which we define as the most massive halo at $9<z<20$ that also hosts the most massive black hole over this redshift interval for all seed masses and accretion parametrizations explored. We construct the merger tree of this halo in post-processing using \textsc{sublink} \citep{rodriguez-gomez_merger_2015}, and track its primary black hole, defined as the black hole closest to the potential minimum of the main subhalo\footnote{We note that the repositioning advects the black hole towards the potential minimum of the subhalo by moving it to the position of the resolution element with the lowest potential within the black hole's smoothing length. The black hole is therefore not necessarily at the potential minimum of its host subhalo at any given time.}.

\begin{figure*}
    \centering
    \includegraphics[width=\textwidth]{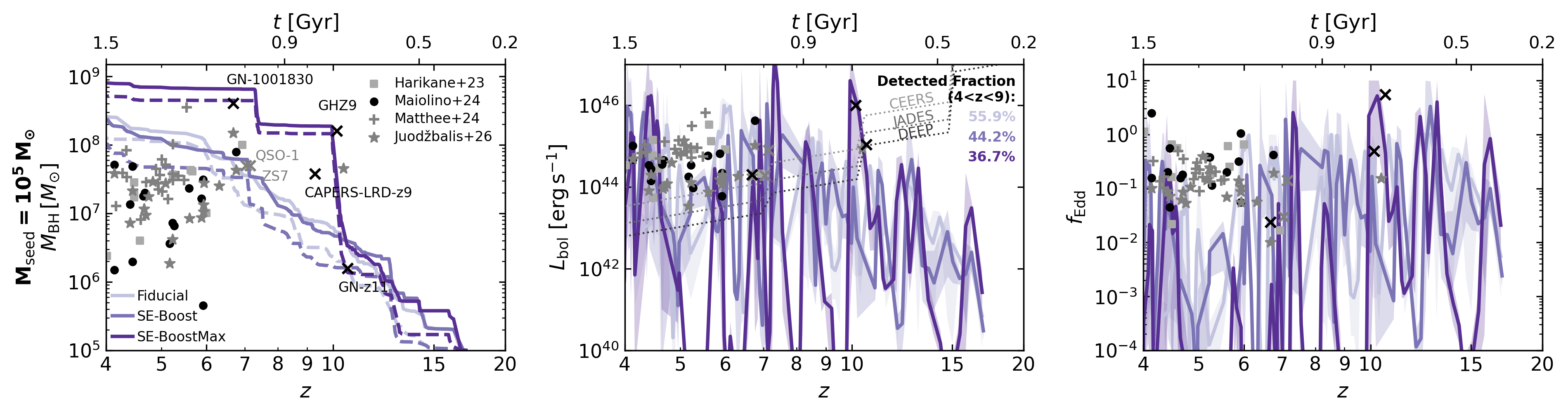}\\
    \includegraphics[width=\textwidth]{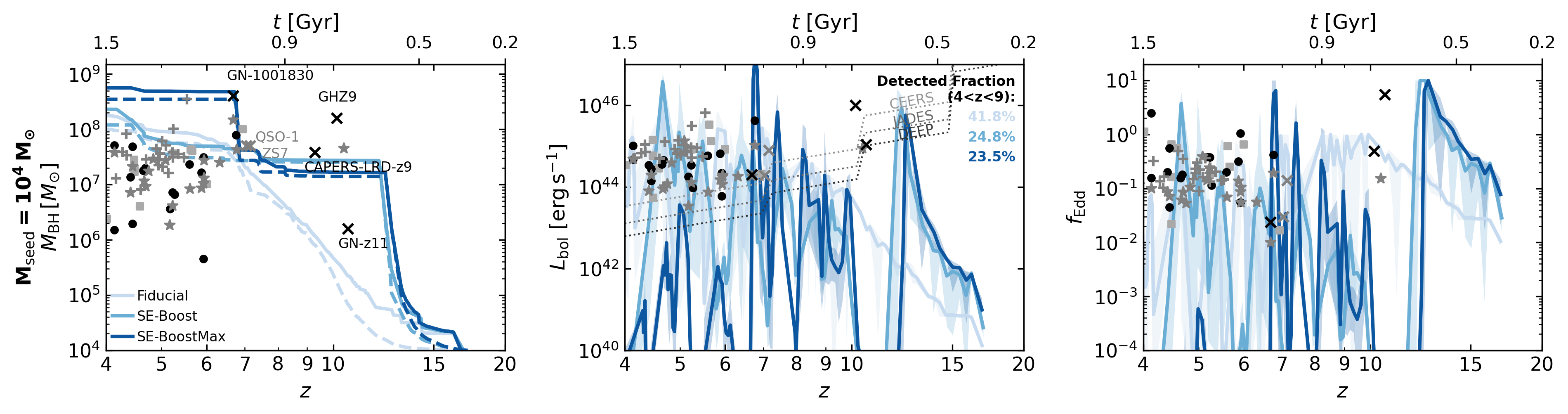}\\
    \includegraphics[width=\textwidth]{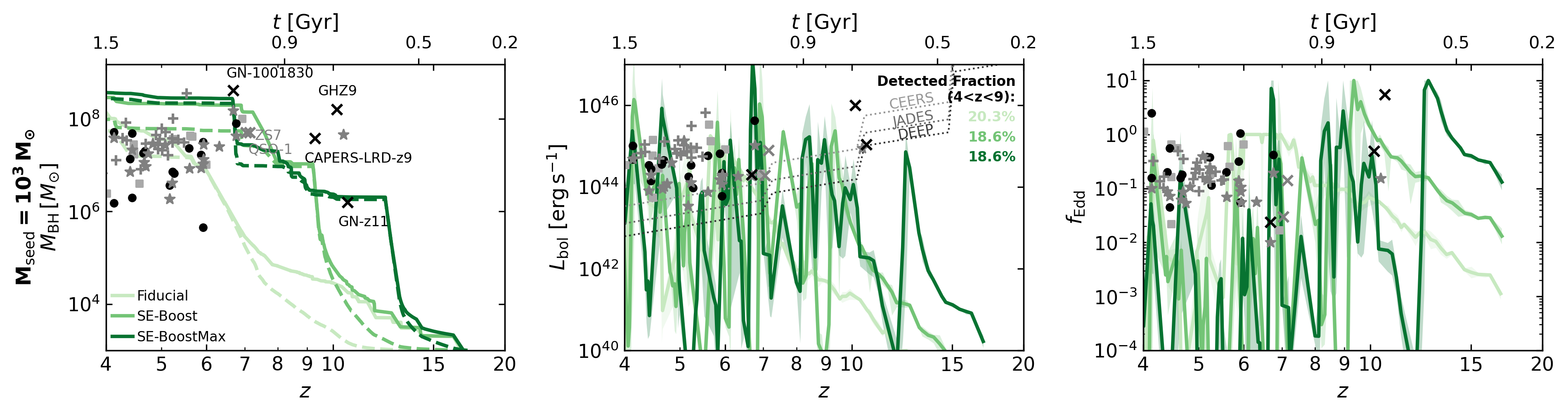}\\
    \includegraphics[width=\textwidth]{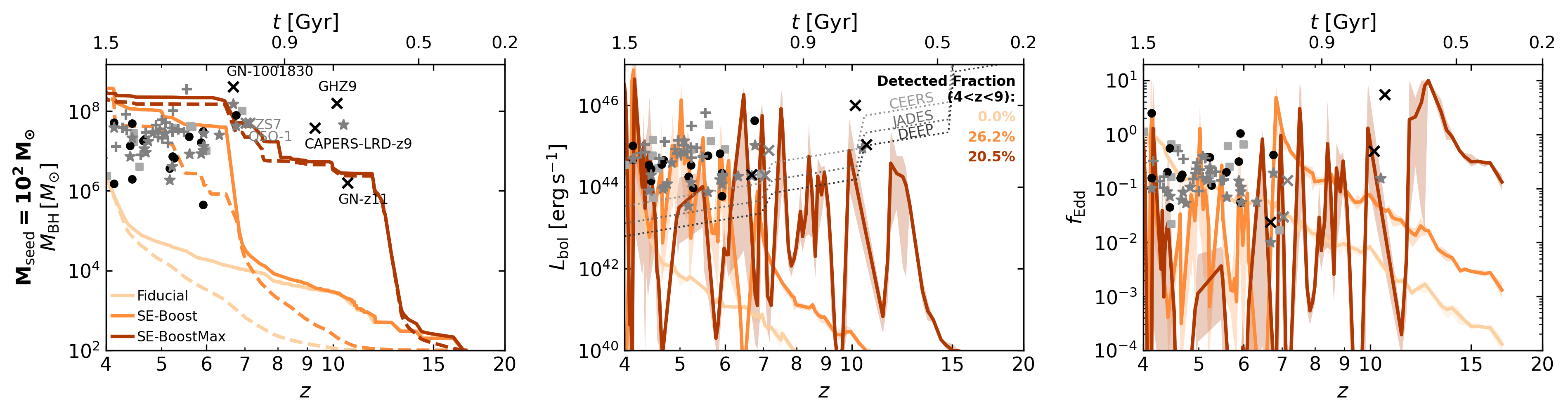}
    \caption{Redshift evolution of the first infant massive black hole hosted by the most massive halo in the \aesop~simulations. The four rows represent the four different seed masses we explore ($M_\mathrm{seed}=10^{5} \ \Msun,10^{4} \ \Msun,10^{3} \ \Msun,10^{2} \ \Msun$). The line-shading indicates the different accretion models, with light shading representing the fiducial accretion parametrization, whilst medium shading and dark shading correspond to our `\textit{SE-Boost}' and `\textit{SE-BoostMax}' models with super-Eddington bursts permitted. The first column shows black hole mass evolution (dashed lines correspond to accreted mass), the second column shows the black hole luminosity evolution, and the third column shows the Eddington fraction -- the latter two are binned over $\Delta t = 10$~Myr with the instantaneous values indicated by the shaded regions. We note that our simulation volume may limit the direct comparison to rarer bright AGN at $z \gtrsim 9$, which would be hosted by large overdensities not contained within our box. These most extreme \jwst AGN can only be reproduced by heavy seeds with super-Eddington bursts and lowered stellar feedback within our small volume. The vast majority of \jwst AGN can be grown from any seed mass provided that the accretion efficiency is boosted to compensate for the strong mass dependency in the Bondi accretion model.}
    \label{fig:bh_z_evol}
\end{figure*}

The evolution of this first `infant' supermassive black hole in the \aesop~simulations is displayed in Fig.~\ref{fig:bh_z_evol}. The left-hand panel shows the evolution of the black hole mass (total black hole mass is shown as solid lines and black hole mass acquired through gas accretion is shown as dashed lines), while the middle and right-hand panels show the evolution of the black hole luminosity and Eddington fraction, respectively. 

We note that for a robust statistical comparison with the \jwst AGN it is crucial to contrast the observed AGN samples with the \textit{whole} simulated AGN population, rather than individual objects, with detailed analysis in Sections~\ref{subsec:results-scalrel-raw}, \ref{subsec:results-scalrel-synth}, \ref{subsec:results-bollumfunc}, \ref{subsec:results-bhmf} and \ref{subsec:results-metals}. Nevertheless, it is instructive to analyse the evolution of the first massive black hole in \aesop~within the observational context of the \jwst AGN as reference points. Hence we also show the virial black hole mass estimates for broad-line AGN uncovered in several \jwst surveys \citep[][see legend]{harikane_jwstnirspec_2023,maiolino_jades_2024,matthee_little_2024,juodzbalis_jades_2026} as well as their bolometric luminosity estimates and resultant Eddington fractions. In addition, we plot individual massive black holes at $z\gtrsim7$ detected by \jwst, including the dormant overmassive black hole \textit{GN-1001830} \citep{juodzbalis_dormant_2024}, \textit{QSO1} \citep{furtak_high_2024}, \textit{ZS7} \citep{ubler_ga-nifs_2024}, \textit{CAPERS-LRD-z9} \citep{taylor_capers-lrd-z9_2025}, \textit{GHZ9} \citep{kovacs_candidate_2024} and \textit{GN-z11} \citep{maiolino_small_2024}, as listed in the plot annotations (only shown in first column, for clarity). We note that only a fraction of the selected objects have exponential wings or are LRDs. The fact that these are a minority implies that our comparison will be less affected by uncertainties in the black hole mass estimates\footnote{Although it is not confirmed that these properties affect the black hole mass estimates \citep[e.g.][]{juodzbalis_jades_2026,madau_wings_2026,scholtz_little_2026}.}, also see discussion in Section~\ref{subsec:discuss_mbh_problem}. 

First, we focus on the simulations initialised with heavy seeds ($M_{\rm seed}=10^{5}\,{\rm M_\odot}$, top row, purple lines). All accretion parametrizations result in rapid growth reaching the supermassive regime before $z=10$. In all cases, the accretion rates are bursty and strongly feedback-regulated. Keeping in mind the up to one order of magnitude uncertainties that may affect both the mass estimates (see Section~\ref{subsec:discuss_mbh_problem}) and luminosity estimates \citep[e.g.][]{greene_what_2026}, in particular for LRDs (though again we highlight that these form a minority of our comparison sample), we also compare the growth tracks to our observational reference points. With heavy seeding, we can reproduce the bulk of the massive black hole population uncovered by \jwst at $z\lesssim7$, both in terms of the relatively large black hole masses and high inferred bolometric luminosities. For the comparison with the most massive and highest redshift objects, we also need to take into account the limited box size of our simulations. The most extreme AGN such as \textit{GHZ9} or \textit{CAPERS-LRD-z9} were identified in large fields and are likely hosted in rare overdensities that our limited box size does not sample. Hence, it is unsurprising that such objects are difficult to reproduce for environmental reasons alone. However, we note that with our most optimistic model, the `\textit{SE-BoostMax}' configuration with abundant heavy seeding, we can obtain similar black hole masses to these objects. In particular, lowering the supernova strength to $\epsilon_\mathrm{W,SN}=0.5$ has a significant impact on the growth rates as it allows for a much stronger initial accretion peak. The `\textit{SE-BoostMax}' run cycles through several powerful accretion bursts corresponding to luminosities of $L_\mathrm{bol} \sim 10^{46} \ \mathrm{erg\, s^{-1}}$, in agreement with the luminosities of the most extreme \jwst AGN observed at $z\sim 10$. With this parametrization, the simulated black hole is able to reach masses comparable to those inferred for the brightest \jwst AGN by $z\sim10$, though we highlight that these extreme AGN observations should not be seen as a target given the limited box size. 

A similar level of agreement with the general \jwst AGN population can be achieved for intermediate-mass seeds with $M_{\rm seed}=10^{4}\,{\rm M_\odot}$. Interestingly, here enhancing the boost factor to $\alpha=10^{3}$ (\textit{SE-Boost} set-up) has a notable impact since the black hole does not reach the feedback-regulated phase until $z\sim 12$. Lowering the supernova strength therefore only leads to a weak additional enhancement; however, at later times ($z \lesssim 8$), when the black hole growth is strongly feedback-regulated, the \textit{`SE-BoostMax'} configuration has markedly enhanced growth rates and high-luminosity bursts. We also note that the large systematic uncertainties in virial mass estimates in the super-Eddington regime mean that caution must be applied for those objects that may be observed at the time of a burst.

Remarkably, even seed masses as low as $10^{3}\ {\rm M_\odot}$ can grow rapidly enough to match the majority of the inferred black hole masses at high redshift. At early times ($z > 12$), the Eddington fraction is almost directly proportional to the boost factor, since mass growth by accretion is negligible and feedback is inefficient, leading to similar gas conditions, until the \textit{`SE-BoostMax'} run with $\alpha=10^{4}$ enters the feedback-regulated phase. This run then experiences more rapid early growth, however, from $z\sim 10$ onwards it has a similar black hole growth history to the \textit{SE-Boost} run with $\alpha=10^{3}$ and both of these match most of the observed data apart from the most extreme objects at $z\sim 9$ which are likely not represented by our simulation volume.

In contrast, models seeded with $10^{2}\,{\rm M_\odot}$ black holes fail to reach the observed mass scale unless extreme boost factors are employed (\textit{SE-BoostMax} with $\alpha=10^{5}$). In all other set-ups explored, black hole growth remains largely merger-dominated until $z\sim8$, after which there is insufficient time to achieve the required mass through gas accretion. It is nevertheless notable that there is a parameter space that allows the light seeds to reach the regime of overmassive black holes as well as the high luminosities observed by \jwst.

Overall, we find that with each accretion model, the black hole evolution undergoes three main phases: early inefficient growth that is only weakly self-regulated, rapid growth resulting in a strong feedback burst and, finally, strongly self-regulated, moderate black hole growth. With the heavy seeds, the second and third phases are reached almost immediately after the black hole is seeded, resulting in bursty accretion histories. We note that the duration of these phases is likely also impacted by the limited resolution of our simulations. Indeed, the growth of light seeds may be much more efficient in high-resolution simulations that resolve the multi-phase ISM \citep[also see][]{mehta_growth_2026}.

From the right-hand panel, we can see that even if the maximum accretion limit is increased to ten times the Eddington rate, super-Eddington accretion phases are only very short-lived ($\sim10$~Myr) due to the self-regulated nature of the set-up. Notably, by $z\sim 7$, all four seed mass configurations explored here have experienced at least one super-Eddington burst. We highlight that whilst this work is focused on high redshifts, we performed all simulations until $z=0$ and found that by $z=2$, the black hole masses for this object have all converged, emphasising that whilst our model variations lead to markedly different early growth histories, they may not be distinguishable at cosmic noon and in the local Universe.

For the redshift intervals in which H$\alpha$ and H$\beta$ fall within the \jwst wavelength coverage ($4 \lesssim z \lesssim 9$), we additionally estimate the fraction of time for which this black hole would be observationally detectable by \jwst assuming the configuration of the JADES survey (medium tier), based on its mass and binned bolometric luminosity, allowing us to infer an effective duty cycle for this infant supermassive black hole. For the fiducial accretion model (faint lines), the detected fraction increases dramatically with the seed mass. Whilst the lightest seeding set-up ($M_\mathrm{seed}=10^{2} \ \Msun$) would \textit{not} be detectable assuming the fiducial accretion model throughout the redshift range considered here, the most massive seed ($M_\mathrm{seed}=10^{5} \ \Msun$) would result in this black hole being detectable based on broad-line selection for 56 per cent of the same redshift interval (not including obscuration effects). With constant boosted accretion ($\alpha=10^{3}$) and super-Eddington bursts permitted, the detectability of the $10^{5}~\Msun$ black hole is reduced to 44 per cent due to the higher accretion efficiency also resulting in stronger feedback self-regulation. With the supernova feedback efficiency lowered, the detectability fraction decreases further to 37 per cent as a strong super-Eddington burst at $z\sim 7.5$ suppresses black hole activity until $z \sim 5$. Similarly, for the $M_\mathrm{seed}=10^{4} \ \Msun$ runs, boosting the accretion results in lower detected fractions. The trade-offs between more efficient accretion and the resultant stronger feedback result in very similar detected fractions for the `\textit{SE-BoostMax}' runs with lighter seeds, leading to detected fractions around 20 per cent.

Finally, the simulated infant black hole evolution provides a natural context for interpreting dormant, overmassive black holes such as \textit{GN-1001830} \citep{juodzbalis_dormant_2024}. In our models, such systems can arise when a black hole is observed during the declining phase of a rapid accretion episode, after having already assembled the bulk of its mass. In all cases, this is associated with super-Eddington accretion. We note that the fiducial \fable~model is unable to reproduce such a dormant overmassive black hole, even in zoom-in simulations of overdense regions; also see discussion in \citet{juodzbalis_dormant_2024}. This suggests that apparently quiescent yet overmassive black holes may represent short-lived transitional phases, following intense black hole accretion (and potentially star formation bursts) in their host galaxies, which lead to strong feedback events, thereby shutting down both star formation and black hole activity \citep[also see][where AGN quenching is observed at very high redshift]{prole_black_2026}. This highlights the importance of bursty accretion histories when connecting black hole growth to galaxy evolution at the highest redshifts as well as interpreting this in the context of galactic star formation histories.

\subsubsection{Early star formation} \label{subsubsec:early_SF}

\begin{figure*}
    \centering
    \includegraphics[width=\textwidth]{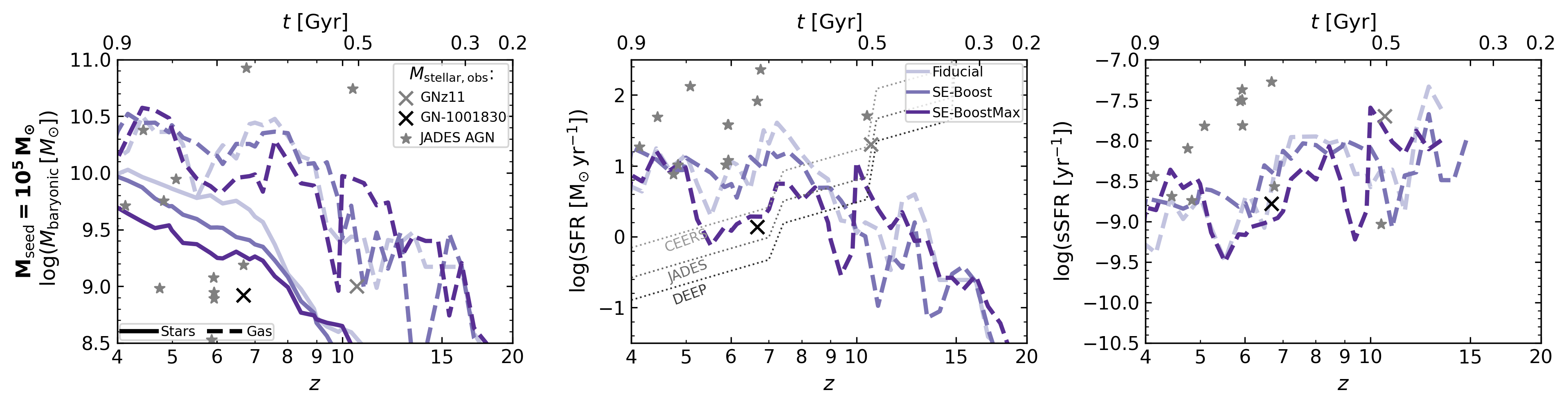}\\
    \includegraphics[width=\textwidth]{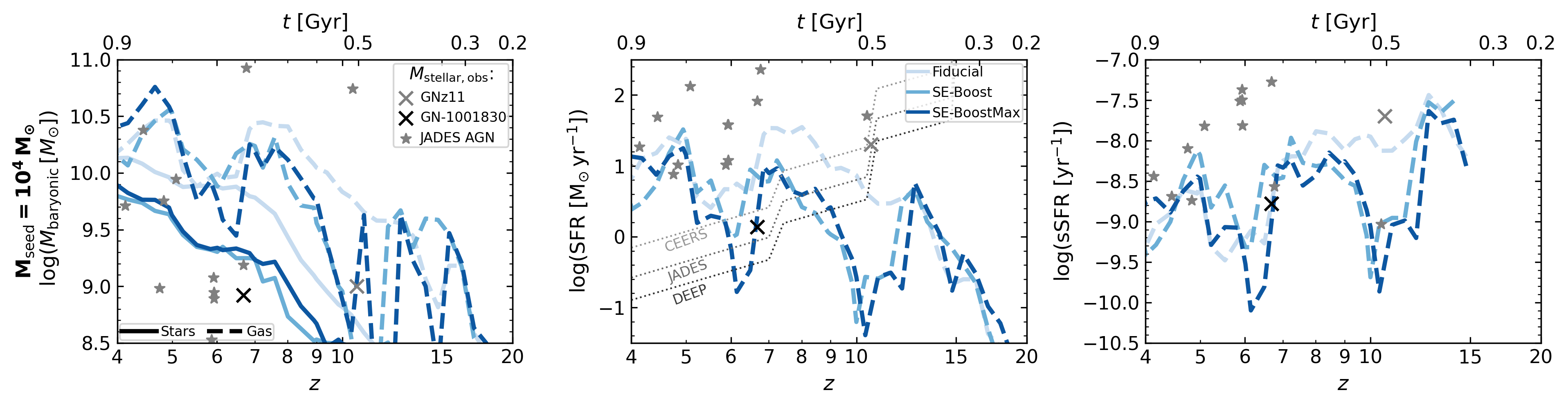}\\
    \includegraphics[width=\textwidth]{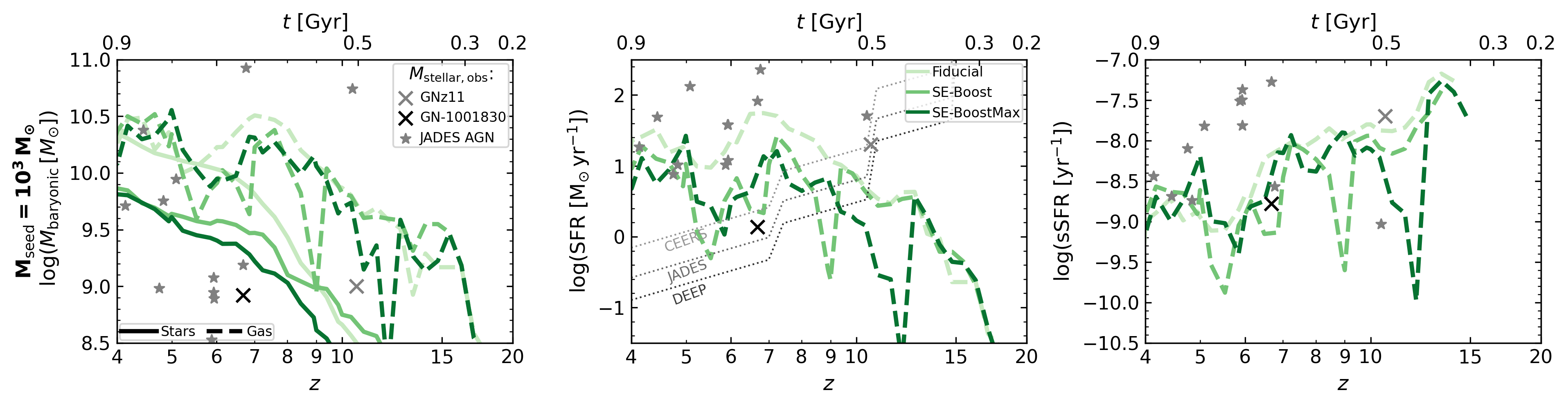}\\
    \includegraphics[width=\textwidth]{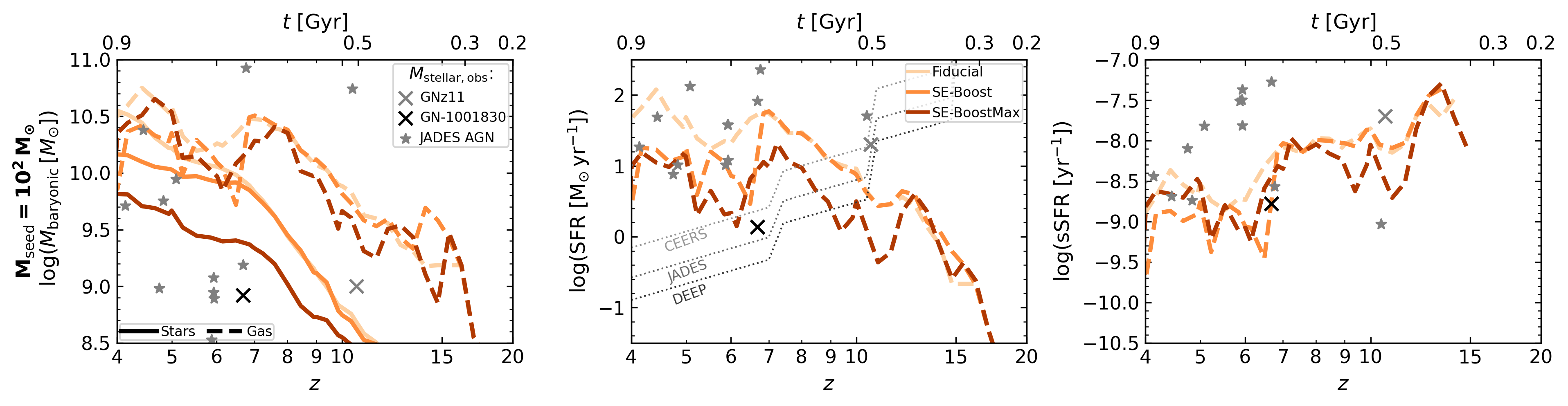}
    \caption{Redshift evolution of the stellar mass and gas mass (first column), the star formation rate (second column) and the specific star formation rate (third column) of the most massive halo in the \aesop~simulations. The line-shading indicates the different accretion models, with light shading representing the fiducial accretion parametrization, medium shading corresponding to the `\textit{SE-Boost}' accretion model and dark shading showing the `\textit{SE-BoostMax}' model. For reference, we also plot the inferred stellar masses and (specific) star formation rates of the host galaxies of the JADES AGN presented in \citet{juodzbalis_jades_2026} as well as \textit{GN-1001830} \citep{juodzbalis_dormant_2024} and \textit{GN-z11} \citep{maiolino_small_2024}. Furthermore, we indicate the detectability of star formation rates based on the Balmer lines for three different survey configurations with the grey dotted lines. The star formation rate levels are broadly similar to the observed range, keeping in mind selection effects. With super-Eddington accretion enabled, strong AGN activity bursts can lead to quiescent episodes similar to star formation and accretion levels observed for the dormant overmassive black hole \textit{GN-1001830}.}
    \label{fig:sfgas_z_evol}
\end{figure*}

Fig.~\ref{fig:sfgas_z_evol} places the growth of the central black hole in the broader context of early galaxy formation, showing the redshift evolution of the stellar mass and gas mass (left-most column), the star formation rate (middle column) and the specific star formation rate (right-most column) of the most massive halo in the \aesop~simulations. The four rows correspond to the four different seed masses explored here and, in each row, the star formation evolution is shown for all three accretion prescriptions.

Before a significant stellar component has formed ($M_{\star} > 5 \times 10^{8}\,\Msun$), we compute the gas mass and star formation rate using all gas cells gravitationally bound to the main subhalo. After this point, both quantities are measured within twice the stellar half-mass radius. For reference, we overplot the inferred stellar masses and star formation rates of the host galaxies of the JADES AGN presented in \citet{juodzbalis_jades_2026}, as well as the systems \textit{GN-1001830} \citep{juodzbalis_dormant_2024} and \textit{GN-z11} \citep{maiolino_small_2024}. Grey dotted lines indicate the detectability limits for star formation rates inferred from Balmer lines \citep{kennicutt_global_1998} for three representative \jwst survey configurations (CEERS/RUBIES, JADES and a notional DEEP survey; see Section~\ref{subsec:methods-mocks}). We base these estimates on the flux limit of the JADES Medium tier ($2.6$~h exposure), $F_\mathrm{min} \sim 5 \times 10^{-19}~\mathrm{erg \, s^{-1} \, cm^{-2}}$ \citep{scholtz_jades_2026}, and rescale this according to the exposure times for CEERS/RUBIES ($1$~h) and the notional DEEP survey ($30$~h). We note that this is a conservative estimate since we assume no contribution to the narrow H$\alpha$ flux from the NLR, which would boost the line flux. The exact contribution of AGN and star formation to the narrow line is currently debated \citep[also see][]{scholtz_jades_2025}, so we do not attempt to correct for this.

Overall, the observational comparison is biased towards actively star-forming systems, with all JADES AGN host galaxies that have reliably measured star formation rates inferred to be highly star forming with SFRs $\gtrsim 10~\mathrm{M_{\odot} \, yr^{-1}}$. \textit{GN-z11} lies close to the observational detection threshold due to its high redshift but still experiences relatively high star formation levels. Across the runs with the fiducial black hole accretion model, the simulated star formation rates broadly span the observed range between $6 > z > 4$, indicating that early black hole growth does not generically suppress star formation in massive haloes at high redshift. However, we note that the specific star formation rates of our halo are systematically lower than most of the observations. This could reflect that we are missing high star formation bursts due to the effective equation of state employed to model the ISM. Potentially, with a multi-phase ISM implementation, we may capture these high star formation bursts that are inferred by JADES \citep[also see][]{keller_can_2023,martin-alvarez_pandora_2023,mccaffrey_no_2023,mcclymont_thesan-zoom_2025}.

In our simulations, notable star formation suppression only occurs when (episodic) super-Eddington accretion is permitted \citep[also see][]{chaikin_importance_2026}. This produces (temporarily) quiescent hosts in accordance with the star formation levels observed for the dormant, overmassive black hole host \textit{GN-1001830} with SFRs $\lesssim 1~\mathrm{M_{\odot} \, yr^{-1}}$ for all seed masses explored (see dark-shaded lines for the `\textit{SE-BoostMax}' configuration). The inferred luminosities and star formation rates for \textit{GN-1001830} are consistent with observing such a galaxy during the downturn of an accretion burst, when strong AGN feedback temporarily regulates both star formation and black hole growth. We find that these quenching episodes typically last $\sim$100--200\,Myr, consistent with the `mini-quenching' scenario proposed for bursty star formation histories \citep[e.g.][]{dome_mini-quenching_2024}. In our case, this mini-quenching is primarily driven by AGN feedback, as our simulations with inefficient AGN accretion predominantly exhibit smooth star formation histories \citep[though we note this would likely be different with a multi-phase ISM model, see e.g.][]{sun_bursty_2023,bhagwat_spice_2024,mcclymont_thesan-zoom_2025}. 

The stellar mass build-up of \textit{GN-z11} provides an additional comparison point. In our models, reproducing the high stellar mass of this system requires a relatively light black hole seed. Heavier seeds regulate star formation too strongly at early times, preventing sufficient stellar mass assembly by $z \sim 11$. Hence, only the set-ups with $M_\mathrm{seed}=10^{3} \ \Msun$ and boosted accretion with super-Eddington bursts are in agreement with the star formation rate, luminosity, stellar mass and black hole mass inferred for \textit{GN-z11}. Whilst our most massive halo does match the inferred halo mass for \textit{GN-z11}, we caution that \textit{GN-z11} may be hosted by a large overdensity, possibly a cluster progenitor \citep[][]{scholtz_gn-z11_2024}. Hence, both cosmic variance and selection effects will also impact our ability to match the observed properties of \textit{GN-z11} with the \aesop~simulations, in addition to constraints on our black hole modelling due to the limited resolution (also see discussion in Section~\ref{subsec:discuss_acc_feed}).

More generally, we find that quenched galaxies at high redshift only arise in simulations that allow for super-Eddington accretion, and are always associated with high black hole masses and powerful AGN activity. We also computed the quiescent fraction for massive galaxies in our simulations and found that we can only reproduce the observational constraints with the `\textit{SE-Boost}' and `\textit{SE-BoostMax}' models. Simulations that do not allow for super-Eddington bursts underpredict the quenched fraction of massive galaxies compared to observational constraints \citep{carnall_surprising_2023,gould_cosmos2020_2023,valentino_atlas_2023,long_efficient_2024,de_graaff_efficient_2025,stevenson_primer_2026}. While our simulation volume ($L = 60$\,cMpc) is too small to make robust statistical statements about the abundance of such rare systems, this result is qualitatively consistent with recent findings that cosmological simulations such as ASTRID and IllustrisTNG underpredict the fraction of massive quenched galaxies compared to \jwst observations \citep[e.g.][]{weller_discrepancies_2025}. Interestingly, the super-Eddington runs also lead to powerful AGN-driven outflows in agreement with high-redshift outflows with AGN signatures observed by \jwst \citep{ivey_exploring_2026}, hinting at a possible link between AGN-driven outflows and star formation regulation in the early Universe \citep[also see][]{farcy_mistral_2025}. We stress that the comparisons presented in this Section are meant to build intuition for how the black holes and their host galaxies evolve under the different model variants. To constrain the models, population-level comparisons are required which we present in the following Sections. Whilst the focus of this work is on black hole evolution, we also present the galaxy stellar mass functions for all twelve models explored in Appendix~\ref{appsec:gsmf}.

\subsection{Simulated scaling relations} \label{subsec:results-scalrel-raw}

\subsubsection{Black hole mass -- stellar mass scaling relations}

\begin{figure*}
    \centering
    \includegraphics[width=\textwidth]{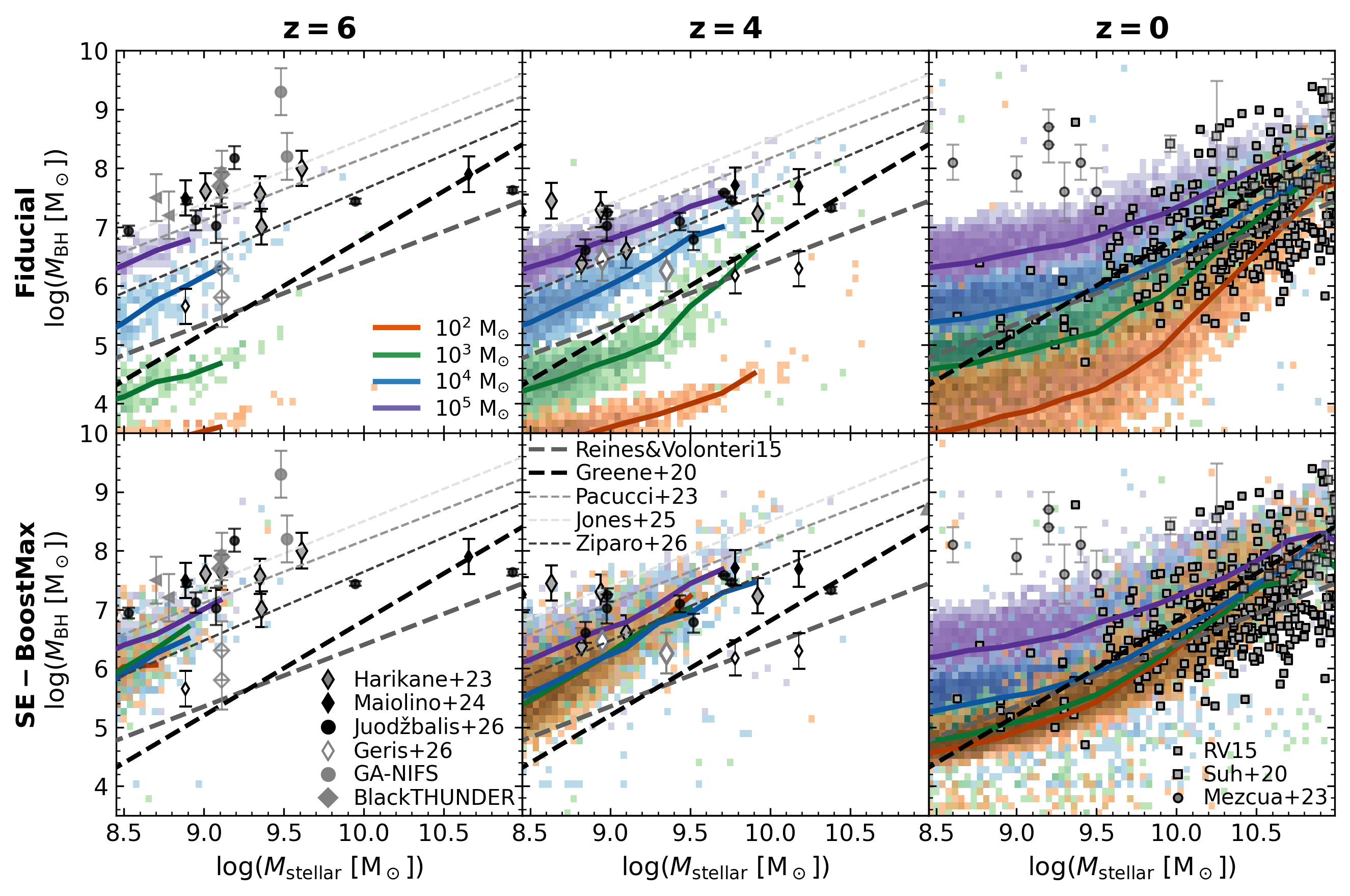}
    \caption{Black hole -- stellar mass scaling relations in \aesop. We show the distributions and mean relations for all four black hole seed masses (see legend) and two accretion prescriptions: the fiducial model (top row) and `\textit{SE-BoostMax}' accretion model (bottom row). Columns correspond to three redshifts ($z = 6, 4, 0$). Observational constraints are overplotted, including high-redshift measurements from \citet{harikane_jwstnirspec_2023,perna_ga-nifs_2023,ubler_ga-nifs_2023,ubler_blackthunder_2026,maiolino_jades_2024,parlanti_ga-nifs_2024,marshall_ga-nifs_2025,deugenio_blackthunder_2026,deugenio_jades_2026-1,jones_blackthunder_2026,juodzbalis_jades_2026,geris_jades_2026,maiolino_black_2026}. The JADES stacks \citep{geris_jades_2026} as well as the binary and triple companion black holes from \citet{maiolino_jades_2024} and \citet{ubler_blackthunder_2026} are highlighted by empty diamonds. In the $z=0$ panel, we plot low-redshift observational data for comparison \citep{reines_relations_2015,suh_no_2020,mezcua_overmassive_2023}. We also show scaling relations inferred from high-redshift AGN \citep{pacucci_jwst_2023,jones_m_rm_2025,ziparo_selection-aware_2026} and in the local Universe \citep{reines_relations_2015,greene_intermediate-mass_2020}. With the fiducial accretion model, the high-z overmassive black hole population can only be reproduced by heavy seeds, whereas with the `\textit{SE-BoostMax}' accretion model, light seeds may also result in overmassive black holes at high redshift. At low redshift, the overmassive population identified by \citet{mezcua_overmassive_2023} may have originated as outliers grown from any seed mass for all of the accretion configurations explored. All scaling relations converge at the high-mass end, demonstrating that the seeding signatures for individual black holes may have been erased for massive galaxies in the local Universe.}
    \label{fig:scal_rel}
\end{figure*}

Having gained intuition for the impact of the different seeding and accretion models on the evolution of individual galaxies and black holes, we now turn to analyse the whole massive black hole population in \aesop. Fig.~\ref{fig:scal_rel} shows the black hole mass--stellar mass scaling relations in the \aesop~simulations for the fiducial accretion model (top row) and the `\textit{SE-BoostMax}' model (bottom row) at three different redshifts ($z = 6, 4, 0$). For each seed mass, the full galaxy population is shown as a colour-coded two-dimensional histogram. As in the previous section, the stellar mass is measured within twice the stellar half-mass radius. Mean scaling relations are overplotted as solid lines, computed in bins of width 0.2\,dex in stellar mass and requiring a minimum of 10 objects per bin. Observational constraints are included for comparison, with high-redshift AGN from \citet{harikane_jwstnirspec_2023, maiolino_jades_2024,juodzbalis_jades_2026}, the JADES stack measurement \citep{geris_jades_2026} as well as the high-redshift GA-NIFS AGN \citep{perna_ga-nifs_2023,ubler_ga-nifs_2023,parlanti_ga-nifs_2024,marshall_ga-nifs_2025} and BlackTHUNDER AGN \citep{deugenio_blackthunder_2026,deugenio_jades_2026-1,jones_blackthunder_2026,maiolino_black_2026,ubler_blackthunder_2026} plotted in the $z=6$ and $z=4$ panels alongside scaling relations inferred for high-redshift AGN samples \citep{pacucci_jwst_2023,jones_m_rm_2025,ziparo_selection-aware_2026}. In the $z=0$ panels, we also include low-redshift data from \citet{reines_relations_2015}, \citet{suh_no_2020} and \citet{mezcua_overmassive_2023} as well as the local scaling relations derived by \citet{reines_relations_2015} and \citet{greene_intermediate-mass_2020}.

At high redshift, the fiducial accretion model produces a strong dependence on seed mass. In this case, only intermediate and heavy seeds are able to reach the overmassive regime by $z \gtrsim 6$, broadly matching the locus of \jwst-detected AGN in low-mass galaxies. Black holes grown from (very) light seeds remain systematically undermassive relative to the observed high-$z$ population, indicating that standard Bondi accretion is insufficient to erase the initial seeding conditions on these timescales.

The `\textit{SE-BoostMax}' accretion model substantially relaxes this constraint. Very light and light seeds are able to grow rapidly and populate the overmassive region of parameter space at early times. In these models, the distinction between seed masses becomes insignificant as early as $z \sim 6$, demonstrating that rapid early growth can compensate for low initial black hole masses and reproduce the observed high-redshift scaling relations without requiring exclusively heavy seeds.

By contrast, at low redshift (see $z=0$ panel), the simulated black hole populations show a high degree of convergence across all seed masses and accretion prescriptions, with the mean relations for all models converging towards the massive galaxy regime ($M_\mathrm{stellar} \sim 3 \times 10^{10} \ \Msun$) onto the local relation \citep[also see][]{taylor_time_2016}. Overmassive black holes in dwarf galaxies, such as those reported by \citet{mezcua_overmassive_2023}, may originate from any of the seed masses considered here. This behaviour is observed in all accretion models, including the fiducial model, and reflects the progressive loss of seeding signatures once black holes enter a feedback-regulated growth phase. At this stage, black hole growth becomes largely self-regulated and only weakly sensitive to both the initial seed mass and the details of the accretion prescription. Consequently, the overall distributions at $z=0$ are remarkably similar in the massive galaxy regime, regardless of accretion efficiency, though other diagnostics such as metallicities may still be able to distinguish between these scenarios \citep[e.g.][]{ortame_small_2026}. The low-mass end may still be sensitive to the accretion modelling, providing strong motivation for future observations to further push dynamical black hole mass measurements into the dwarf galaxy regime.

We also inspected the full black hole mass -- stellar velocity dispersion scaling relations (see Fig.~\ref{fig:scalrel_veldisp} in Appendix~\ref{appsec:stellarveldisprel}). Here the massive galaxies converge onto the local relations already by $z \sim 4$ for all heavy seed models and the light seed models with `\textit{SE-BoostMax}' accretion. This reflects that there is only insignificant evolution between the average BH mass and the host potential across cosmic time in \aesop, similarly to the Illustris simulation which our galaxy formation model is derived from \citep[see][for detailed discussion of the evolution of the scaling relations in Illustris]{sijacki_illustris_2015,habouzit_supermassive_2021}. This effect is also seen in observations which find that the high-redshift \jwst AGN lie on the local black hole mass -- stellar velocity dispersion relation \citep[e.g.][]{maiolino_jades_2024,juodzbalis_jades_2026}. In this sense, black holes appear to `lock in' to the host potential before the galaxy assembles the bulk of its stellar mass. This reflects a strong linkage of the black hole growth to the gas content in the inner region of the galaxy, which decreases with cosmic time in the Illustris and \fable~models, whilst star formation proceeds with a constant efficiency so that the stellar mass is able to catch up \citep[][]{habouzit_supermassive_2021,koudmani_little_2021}. Indeed, we find that the high-redshift overmassive black holes are in agreement with the local scaling relations if we add the gas mass to the stellar mass. Therefore the evolution of the black hole mass -- stellar mass scaling relations represents the shifting conversion efficiencies of the gas reservoir into black hole growth and star formation \citep[also see][]{wu_redshift_2026}. The evolution of the black hole mass -- stellar velocity dispersion relation, on the other hand, tracks the total potential and therefore also the baryonic mass budget and is hence unaffected by these shifting efficiencies, resulting in only a mild evolution with redshift. This is also in line with observational results \citep[e.g.][]{maiolino_jades_2024,ma_undermassive_2026} which find that the black holes which appear to be overmassive with respect to their stellar mass do lie on the black hole mass -- stellar velocity dispersion scaling relations.

We also compare our simulated mean relations to the observationally inferred scaling relations from the literature. The relation with the highest normalisation from \citet{jones_m_rm_2025} (thin light-grey dashed line) lies above all our simulated mean relations, even for our most optimistic accretion configuration `\textit{SE-BoostMax}' with heavy seeds. The relation reported by \citet{pacucci_jwst_2023} (thin medium-grey dashed line) has a slightly lower normalisation ($\sim 0.2$~dex) and is in moderate agreement with the heavy-seed mean relations at $z=6$, although even in this case the simulated relation tends to lie below the observed one by $0.1$--$0.2$ dex. Indeed, we find that our simulations are generally in closest agreement with the high-redshift relation inferred by \citet{ziparo_selection-aware_2026} (thin black dashed line) which matches the mean scaling relations for the \textit{`SE-BoostMax'} simulations with (very) light and intermediate seeds, whilst the heavy seed simulations lie in between the \citet{ziparo_selection-aware_2026} and \citet{pacucci_jwst_2023} relations. Interestingly, \citet{ziparo_selection-aware_2026} apply a very similar algorithm to the observed AGN for determining completeness limits as our \textsc{balmersopica} tool using the broad-line selection criteria. 

Whether the overmassive black holes uncovered by \jwst represent an intrinsically different black hole--galaxy relation at high redshift remains hotly debated. Whilst all of the three relations that we have reported above lie above the local relations (with varying degrees of redshift evolution), other groups find the high-redshift population to be in agreement with the local relations once selection effects are taken into account \citep[e.g.][]{li_tip_2025}. Our results highlight that disentangling early black hole evolution from observational selection requires a careful treatment of accretion physics at high redshift as the early growth phase is particularly sensitive to accretion model parameters. It will also be crucial to push the broad-line measurements to lower black hole masses by analysing the forbidden lines to rule out or confirm outflow components \citep[e.g.][]{maiolino_jades_2024,ubler_blackthunder_2026}.

\subsubsection{Black hole mass -- luminosity scaling relations} \label{subsubsec:results_scal_rel_lum}

\begin{figure*}
    \centering
    \includegraphics[width=\columnwidth]{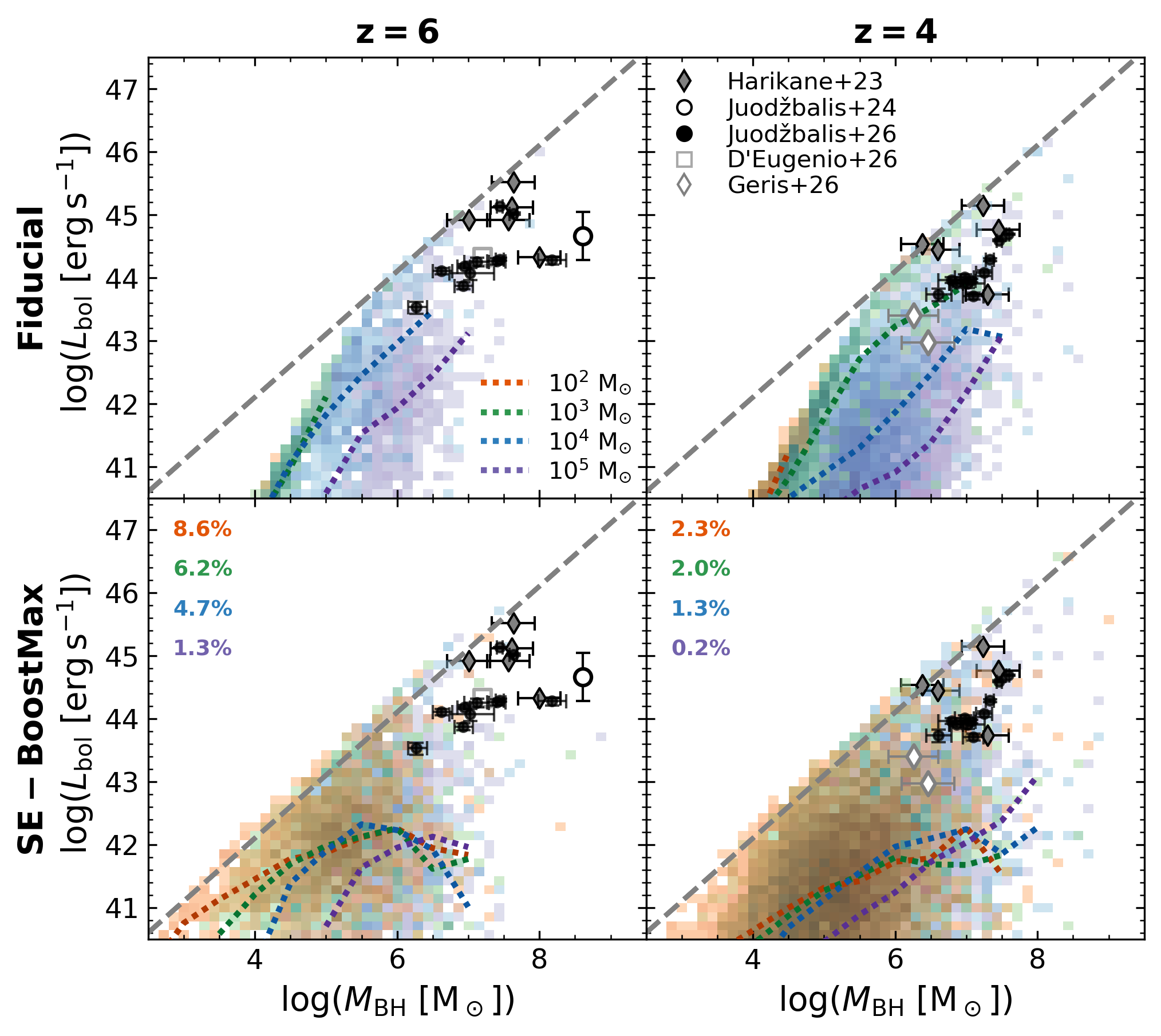}
    \includegraphics[width=\columnwidth]{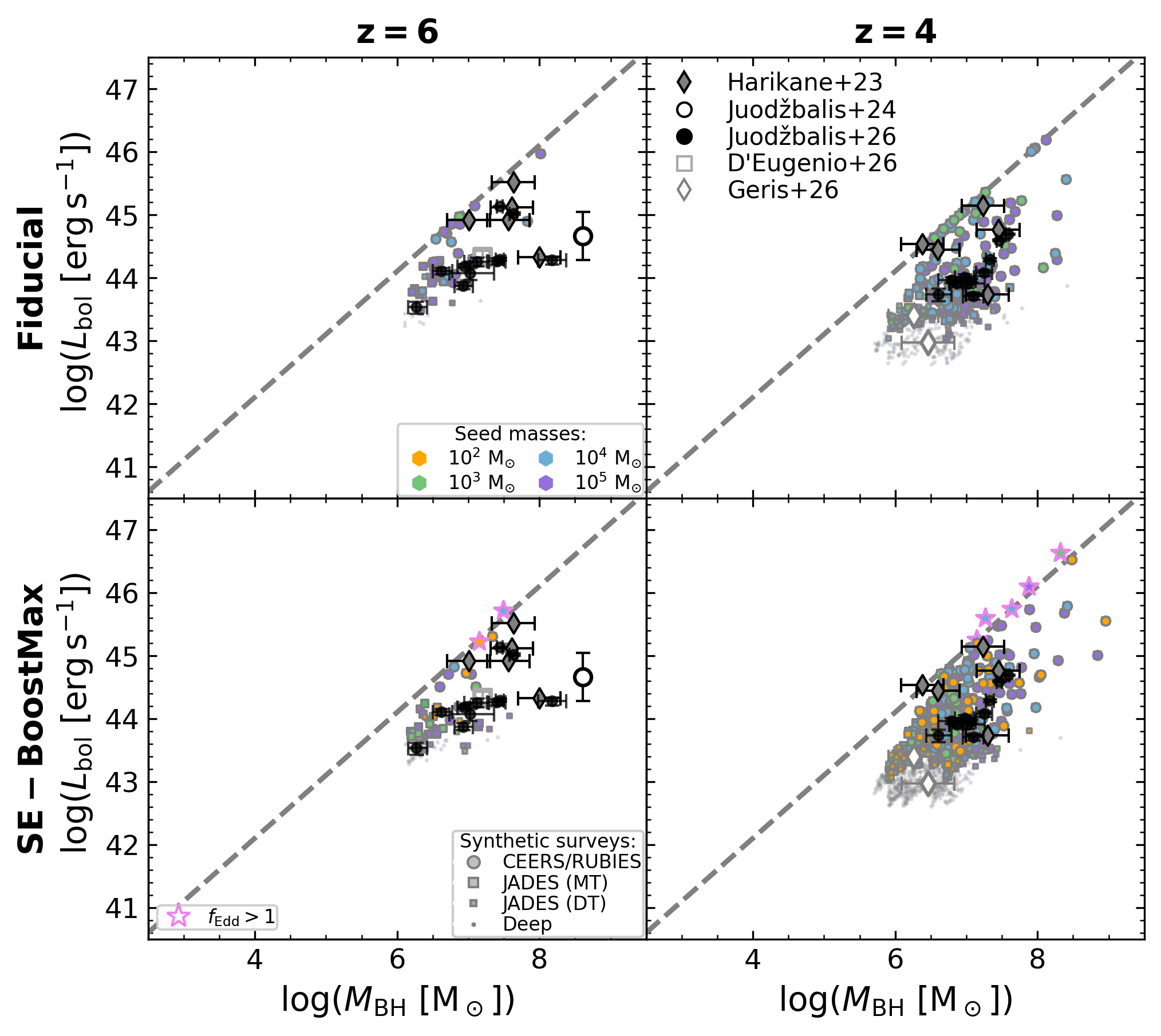}
    \caption{Black hole luminosity versus black hole mass in the \aesop~simulations at redshifts $z=6$ and $z=4$. We plot these relations for our fiducial (top row) and `\textit{SE-BoostMax}' accretion prescriptions (bottom row). Observational constraints are overplotted, including high-redshift AGN observations \citep{harikane_jwstnirspec_2023,juodzbalis_dormant_2024,juodzbalis_jades_2026,deugenio_blackthunder_2026} and the JADES AGN stack \citep{geris_jades_2026}. We also show the luminosities corresponding to $f_\mathrm{Edd}=1.0$ as a dashed grey line. \textit{Left panel:} We plot colour-coded 2D histograms and mean relations for the different seed masses in each row, as indicated by the legend. We list the fraction of black holes accreting at super-Eddington rates for the `\textit{SE-BoostMax}' models in order of seed mass. \textit{Right panel}: We only include \jwst-detectable black holes based on broad-line selection criteria for CEERS/RUBIES, JADES (medium and deep tiers) as well as the notional DEEP survey as scatter points using the \textsc{balmersopica} tool (see Section~\ref{subsec:methods-mocks}). This illustrates that the high-z AGN uncovered by \jwst are generally very bright sources that lie above the mean relations for all models explored here. Hence, it is crucial to include observational selection in the comparison between simulated and observed high-z AGN populations.}
    \label{fig:scal_rel_lum}
\end{figure*}

We further explore the connection between black hole growth and observable AGN properties in Fig.~\ref{fig:scal_rel_lum}, which shows the black hole bolometric luminosity as a function of black hole mass at $z=6$ and $z=4$. We include the runs with the fiducial (top row) and `\textit{SE-BoostMax}' accretion models (bottom row) for all four seed masses explored. In the left panel, we plot the full black hole population as colour-coded two-dimensional histograms together with the corresponding mean relations for each seed mass, shown as colour-coded dotted lines. Mean relations are computed in bins of width 0.5\,dex in black hole mass and require a minimum of 10 objects per bin. Observational constraints from high-redshift AGN uncovered by \jwst are overplotted, including the samples presented by \citet{harikane_jwstnirspec_2023} and \citet{juodzbalis_jades_2026}. We only show the error bars for the black hole masses for clarity, however, the luminosities also come with significant uncertainties in the bolometric correction, which may change the inferred bolometric luminosities by up to an order of magnitude \citep[e.g.][]{greene_what_2026}.

The grey dashed line indicates the luminosity corresponding to an Eddington ratio of $f_{\mathrm{Edd}}=1$, providing a reference for the characteristic accretion states of the simulated and observed systems. We note that, as in \fable, we assume a constant radiative efficiency when converting accretion rates to bolometric luminosities. In the super-Eddington regime, however, the effective radiative efficiency is expected to decrease due to photon trapping and large optical depths. In \aesop, we do not explicitly model this decrease in the radiative efficiency. However, for the luminosities presented in Fig.~\ref{fig:scal_rel_lum}, we have applied the slim disc formula from \citet{madau_super-critical_2014} based on the GRRMHD simulations from \citet{sadowski_numerical_2014}. Following \citet{prole_seedz_2026} and \citet{quadri_super-eddington_2025}, we assume a constant black hole spin of $a=0.7$. In practice, since we are only considering mildly super-Eddington cases, all of these corrections only make a difference of up to $\sim 0.5$~dex. We reiterate that these super-Eddington bursts in the simulations are merely brief phases spanning only tens of Myr (see e.g. Fig.~\ref{fig:bh_z_evol}). To illustrate the length of the duty cycle, we indicate in Fig.~\ref{fig:scal_rel_lum} the percentage of black holes hosted by well-resolved galaxies (defined as a baryonic mass of at least $2 \times 10^{9}~\Msun$) that are accreting at super-Eddington rates. These range from $\sim 1$ to $9$ per cent at $z=6$ and $\sim 0.2$ to $2$ per cent at $z=4$. For both redshifts, the lighter seeds have a higher fraction of super-Eddington black holes since they are more weakly feedback-regulated and have their efficient growth phase shifted to later redshifts (also see Fig.~\ref{fig:bh_z_evol}).

Across all accretion models, the black hole luminosity generally increases with black hole mass, but the detailed shape of the relation depends on the accretion prescription. In the fiducial model, the luminosity rises steadily with mass, reflecting relatively inefficient feedback regulation. In contrast, most boosted-accretion models show a flattening, and in some cases turnover, of the relation at high black hole masses, signalling the onset of self-regulated growth. 

This effect is especially pronounced for our `\textit{SE-BoostMax}' accretion models, as shown in Fig.~\ref{fig:scal_rel_lum}, where episodic super-Eddington bursts are allowed, and powerful feedback events can temporarily quench both accretion and star formation, leading to a suppressed mean luminosity at the high-mass end. At early times ($z=6$), we find that for a fixed black hole mass the mean luminosity tends to increase systematically with decreasing seed mass. Black holes originating from lighter seeds will have undergone more sustained and moderated accretion histories to reach a given black hole mass, avoiding strong early feedback episodes. This, in turn, allows for more efficient accretion, resulting in higher luminosities at fixed mass. This behaviour is also evident in the individual growth histories shown in Fig.~\ref{fig:bh_z_evol}. By $z=6$, luminous AGN comparable to the observed \jwst population can be produced by black holes originating from any of the seed masses considered here with our `\textit{SE-BoostMax}' accretion model. This underlines that the degeneracy between seed masses that we found for the individual black hole growth tracks in Fig.~\ref{fig:bh_z_evol} also persists on a population level.

Importantly, Fig.~\ref{fig:scal_rel_lum} highlights that the current population of high-redshift AGN uncovered by \jwst is \textit{not} representative of the broader AGN population predicted by \aesop. Even in our most optimistic configuration, the `\textit{SE-BoostMax}' model, the observed sources mostly lie well above the mean relations of the simulated population\footnote{On this note, it is also worth emphasising that when comparing simulated bolometric luminosities to observed samples, additional uncertainties arise from bolometric corrections and obscuration. Observational luminosities are typically inferred from rest-frame UV or optical emission using bolometric corrections that may evolve with redshift, Eddington ratio, and black hole mass, and which remain poorly constrained at $z \gtrsim 6$ \citep[see discussion in][]{greene_what_2026}.}. 

We therefore also show an equivalent black hole luminosity -- mass plot where we only include the black holes that would be detectable via their broad lines in the CEERS/RUBIES, JADES and our notional DEEP survey in the right panel of Fig.~\ref{fig:scal_rel_lum} based on our \textsc{balmersopica} selection tool (see Section~\ref{subsec:methods-mocks}). This leads to very good agreement with the observed distribution apart from a few of the super-Eddington sources that are moving towards very high luminosities, more typical of high-z quasars. These super-Eddington sources may form up to 33 per cent of sources for shallow surveys at $z=6$ and up to 7 per cent at $z=4$. With the DEEP survey (30 hours exposure), it should be feasible to observe broad-line AGN in a similar parameter space as the JADES stacks from \citet{geris_jades_2026}, allowing us to potentially probe the low-mass end of the high-z black hole mass function and black holes that are more in line with the local scaling relations.

Our analysis highlights that, in addition to invoking efficient early black hole growth to explain the most extreme high-redshift AGN (see Fig.~\ref{fig:bh_z_evol}), it is essential to account for the observational selection effects when comparing simulations to the observed AGN population at high redshift. An `apples-to-apples' comparison approach that explicitly accounts for survey selection is required to ensure that we are comparing simulated and observed populations consistently.

\subsection{Scaling relations from synthetic surveys} \label{subsec:results-scalrel-synth}

\begin{figure*}
    \centering
    \includegraphics[width=\columnwidth]{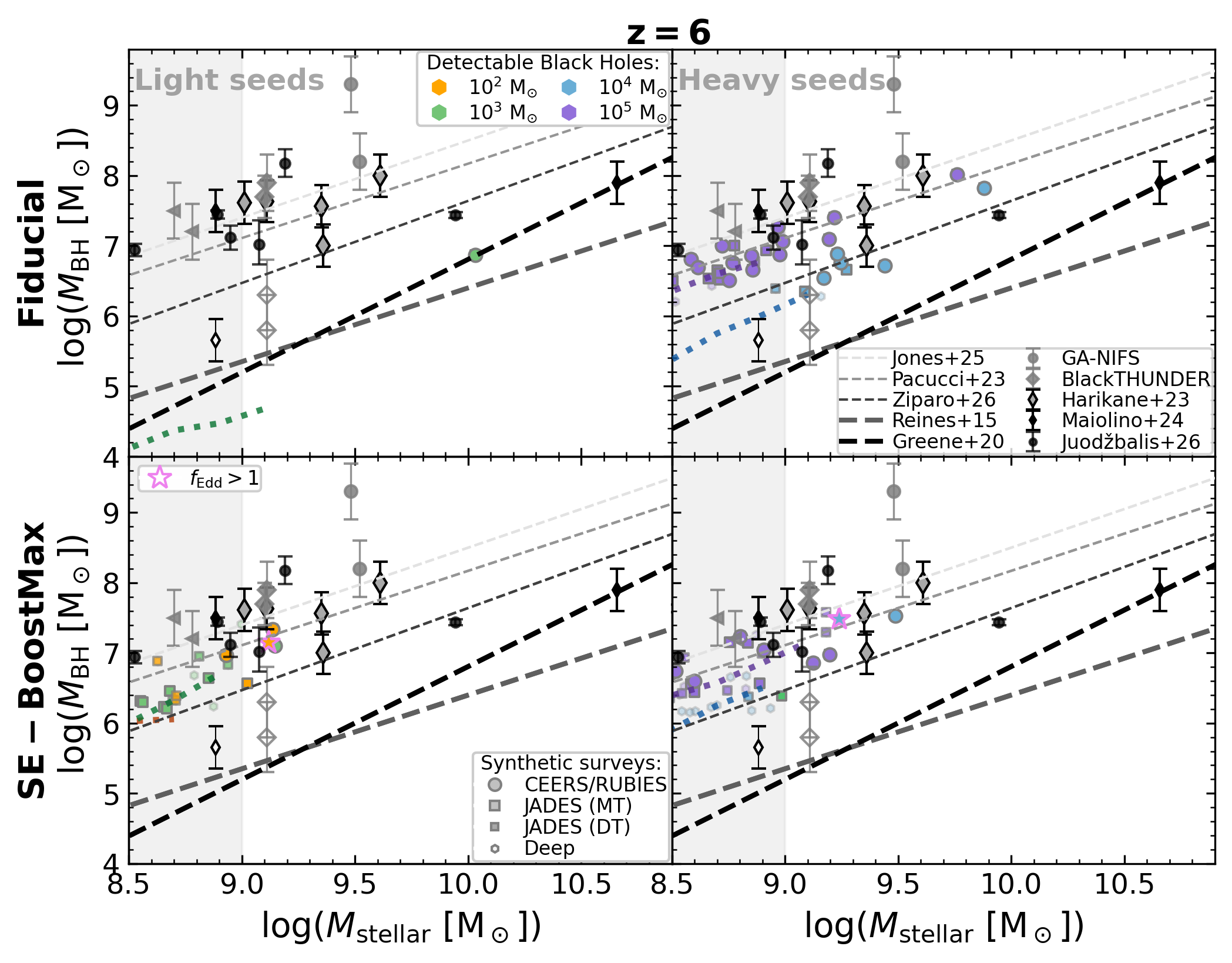}
    \includegraphics[width=\columnwidth]{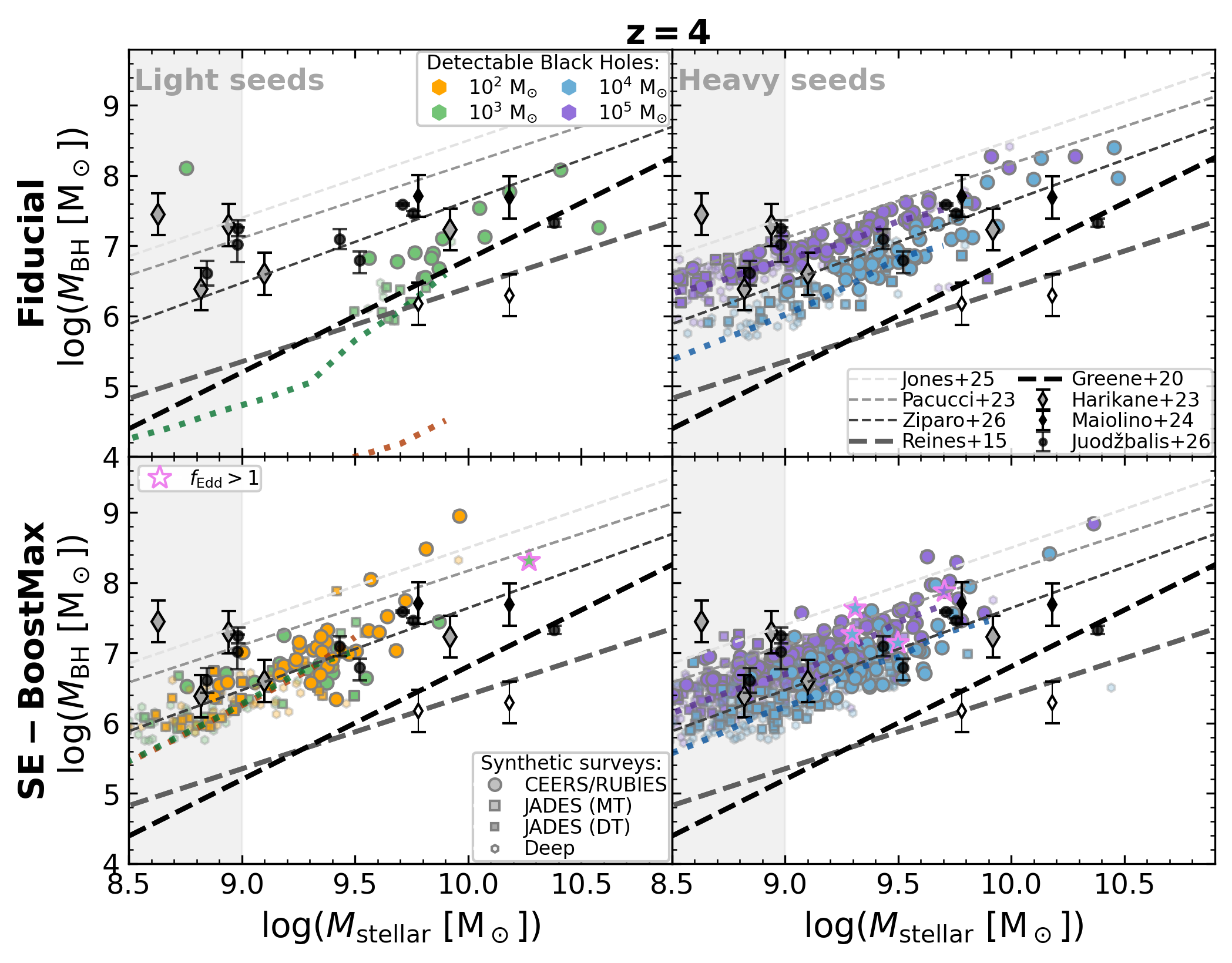}
    \caption{Black hole mass -- stellar mass scaling relations from the \aesop~simulations for the fiducial (top row) and `\textit{SE-BoostMax}' accretion models (bottom row). Left and right panels show results at $z=6$ and $z=4$, respectively. The columns separate light seeds ($10^{2}$ and $10^{3}\,\Msun$, left) from heavy seeds ($10^{4}$ and $10^{5}\,\Msun$, right). We only include simulated active black holes that meet broad-line AGN selection criteria mimicking \jwst survey detection limits. The scatter symbols indicate detectability for different \jwst survey configurations: CEERS/RUBIES (circles), JADES Medium Tier (opaque squares), JADES Deep Tier (transparent squares) and our notional DEEP survey (transparent hexagons). Observational constraints are overplotted, including high-redshift measurements from \citet{harikane_jwstnirspec_2023,perna_ga-nifs_2023,ubler_ga-nifs_2023,ubler_blackthunder_2026,maiolino_jades_2024,parlanti_ga-nifs_2024,marshall_ga-nifs_2025,deugenio_blackthunder_2026,deugenio_jades_2026-1,jones_blackthunder_2026,juodzbalis_jades_2026,maiolino_black_2026}. The binary and triple companion black holes from \citet{maiolino_jades_2024} and \citet{ubler_blackthunder_2026} are highlighted by empty diamonds. The grey shaded region marks the effective stellar mass resolution limit of the simulations ($\sim 10^{9}\,\Msun$). Pink stars denote black holes observed during super-Eddington accretion bursts. Heavy seeds always result in an overmassive AGN population once selection cuts are applied, whilst for lighter seeds, the `\textit{SE-BoostMax}' model with enhanced accretion including super-Eddington bursts is required.}
    \label{fig:scal_rel_filter}
\end{figure*}

\begin{figure*}
    \centering
    \includegraphics[width=\columnwidth]{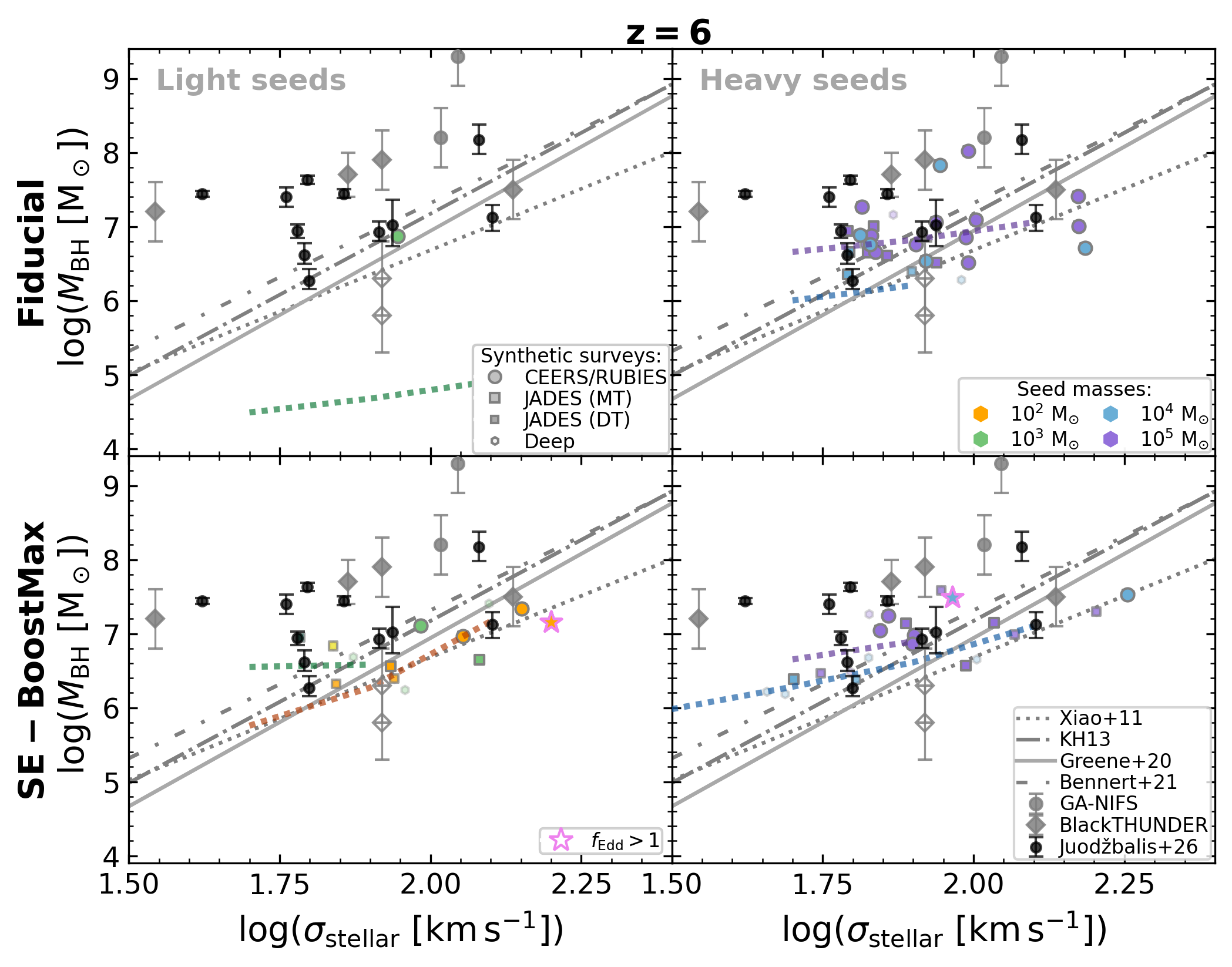}
        \includegraphics[width=\columnwidth]{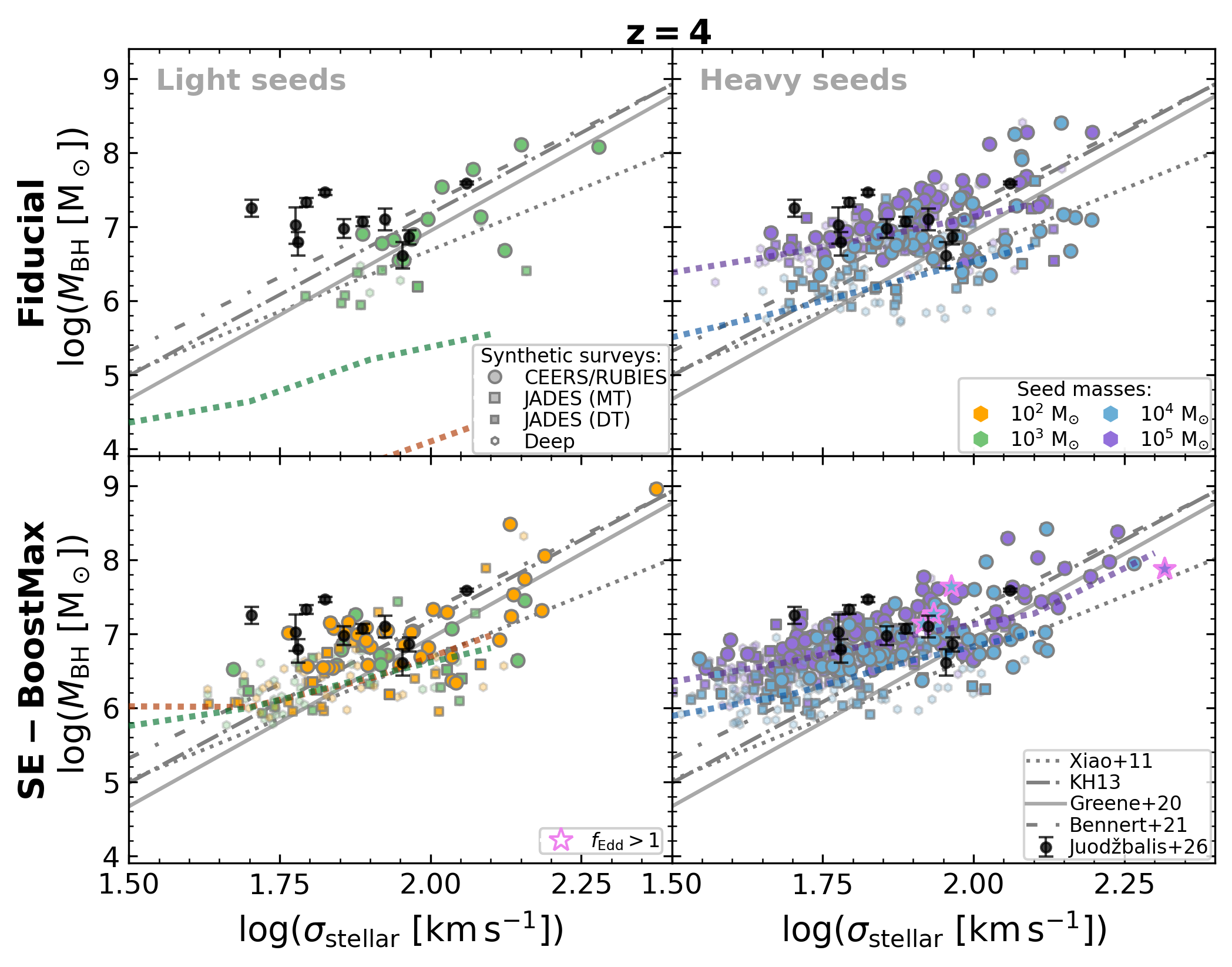}
    \caption{Black hole mass -- stellar velocity dispersion scaling relations from the \aesop~simulations for the fiducial (top row) and `\textit{SE-BoostMax}' accretion models (bottom row). Left and right panels show results at $z=6$ and $z=4$, respectively. The columns separate light seeds ($10^{2}$ and $10^{3}\,\Msun$, left) from heavy seeds ($10^{4}$ and $10^{5}\,\Msun$, right). We only include simulated active black holes that meet broad-line AGN selection criteria mimicking \jwst survey detection limits. The scatter symbols indicate detectability for different \jwst survey configurations: CEERS/RUBIES (circles), JADES Medium Tier (opaque squares), JADES Deep Tier (transparent squares) and our notional DEEP survey (transparent hexagons). Observed high-redshift AGN from \citet{juodzbalis_jades_2026} as well as the high-redshift GA-NIFS AGN \citep{perna_ga-nifs_2023,ubler_ga-nifs_2023,parlanti_ga-nifs_2024,marshall_ga-nifs_2025} and BlackTHUNDER AGN \citep{deugenio_blackthunder_2026,deugenio_jades_2026-1,jones_blackthunder_2026,maiolino_black_2026,ubler_blackthunder_2026} are overplotted for comparison. Pink stars highlight simulated black holes `observed' during super-Eddington accretion bursts. In agreement with the observations, the broad-line selected \aesop~AGN lie on the local black hole mass -- stellar velocity dispersion scaling relations.}
    \label{fig:scal_rel_filter_veldisp}
\end{figure*}

To assess the impact of observational selection effects on the black hole -- galaxy scaling relations, we apply our broad-line AGN selection criteria (see Section~\ref{subsec:methods-mocks} for details) to the simulation outputs of the fiducial and `\textit{SE-BoostMax}' accretion models. We consider the black hole mass--stellar mass relation (Fig.~\ref{fig:scal_rel_filter}) and the black hole mass--stellar velocity dispersion relation (Fig.~\ref{fig:scal_rel_filter_veldisp}). In both cases, we plot the `detectable' black holes as colour-coded scatter points.

To mimic the \jwst survey selection function, we assume an R1000 grating and exposure times of 1\,hr (CEERS/RUBIES; circles), 2.6\,hr (JADES Medium Tier; opaque squares), and 7\,hr (JADES Deep Tier; transparent squares). We also include AGN that would be detectable with our notional DEEP survey (R2700 grating and exposure time of 30\,hr). Results are shown at $z=6$ (left panels) and $z=4$ (right panels). We further separate the simulations into two seed-mass categories: `light seeds' ($10^{2}$ and $10^{3}\,\Msun$; left columns in each panel) and `heavy seeds' ($10^{4}$ and $10^{5}\,\Msun$; right columns in each panel).

As discussed in Section~\ref{subsec:methods-mocks}, applying virial mass estimators at high redshift is subject to several uncertainties. For the purposes of our filtering analysis, however, this implies that we are adopting a \textit{conservative} detectability threshold. In practice, a larger fraction of simulated AGN may be observable, as the broad-line FWHM may be enhanced by additional scattering processes \citep[e.g.][]{rusakov_little_2026}, and the H$\alpha$ flux may be higher than assumed if bolometric corrections decrease at high redshift \citep[e.g.][]{greene_what_2026}. We also highlight black holes that would be observed during a super-Eddington burst phase with pink star symbols, to caution that these would be affected by perhaps the most significant uncertainties in the virial relations \citep[e.g.][]{lupi_size_2024,lambrides_case_2026}.

We first focus on the black hole mass--stellar mass relation (Fig.~\ref{fig:scal_rel_filter}). For comparison, we also show the observed apparently overmassive AGN reported by \citet{harikane_jwstnirspec_2023}, \citet{maiolino_jades_2024}, \citet{juodzbalis_jades_2026} as well as GA-NIFS and BlackTHUNDER, together with local scaling relations from \citet{reines_relations_2015} and \citet{greene_intermediate-mass_2020}. We additionally plot the high-redshift relations reported by \citet{jones_m_rm_2025}, \citet{pacucci_jwst_2023} and \citet{ziparo_selection-aware_2026}. The effective stellar-mass resolution limit of \aesop\ ($\sim 10^{9}\,\Msun$) is indicated by the grey-shaded region.

At $z=4$, the apparently overmassive black hole population is naturally reproduced once observational selection effects are applied, for \textit{all} intermediate-to-heavy seed models, largely independent of the accretion prescription. For the (very) light seeds, `\textit{SE-BoostMax}' accretion is required, but even in these cases the observed locus can be mostly recovered. At $z=6$, all accretion models remain broadly consistent with the observed data when assuming intermediate or heavy seeds. For lighter seeds, reproducing the observed population becomes increasingly challenging. However, the filtered data points for the `\textit{SE-BoostMax}' accretion parametrization still occupy a similar locus to the observations, albeit at somewhat lower density and shifted to lower stellar masses. In all cases, the notional DEEP survey would enable additional detections at lower black hole and stellar masses due to improved resolution and higher sensitivity.

For reference, we also show the mean scaling relations based on the whole simulated AGN sample from \aesop, as shown in Fig.~\ref{fig:scal_rel}. We note that for the `\textit{SE-BoostMax}' model, the detectable AGN generally scatter around the intrinsic mean relation with a small offset towards higher BH masses (in particular at lower stellar masses). Due to efficient accretion and self-regulation in these models, there is only an extremely weak correlation between the black hole mass offsets and luminosities (also see Fig.~\ref{fig:bh_z_evol}). Therefore, once the efficient, self-regulated state is reached within our modelling framework, the black hole mass -- stellar mass relation is merely undersampled by the detectable AGN. Hence, whilst the scatter would be underestimated by the detectable sample, the intrinsic normalisation could be inferred with relatively high accuracy. In the inefficient growth model, however, the detectable AGN lie significantly above the intrinsic mean relation (see e.g. the fiducial model with light seeds) so that the normalisation of the inferred scaling relations may be significantly overestimated. We note that in this work, we consider mostly models that lead to either mostly very inefficient or mostly very efficient black hole growth. In our Universe, it could be a much broader mixture of seed masses, accretion efficiencies and stellar feedback regulation; and we emphasise that the observable `tip of the iceberg' will not necessarily represent the broader black hole population with potentially a large population of undetected inefficient black holes present.

Considering the ability of our models to match the observed overmassive black holes, we find that either relatively ubiquitous heavy black hole seeding is required, or, if supermassive black holes predominantly originate from light seeds, highly efficient accretion, including super-Eddington episodes, must be invoked. If the apparently overmassive black hole population has a genuine physical origin, rather than being driven purely by uncertainties in the virial estimators, this raises the question of what such systems imply for early black hole -- galaxy co-evolution. We also checked the distribution of our synthetically selected black holes in black hole mass -- baryonic mass space (see Fig.~\ref{fig:scal_rel_filter_dyn} and discussion in Appendix~\ref{appsec:scal_rel}) and found that the synthetically detected AGN move onto the local relations once the gas mass is included in the host mass budget. This demonstrates that sufficient baryonic mass is already present in these systems to support a `black hole first' scenario, in which the black hole assembles its mass early and the stellar component subsequently catches up as black hole growth slows at late times due to depletion of the inner gas reservoir. However, as discussed in Appendix~\ref{appsec:scal_rel} \citep[also see][]{ortame_small_2026}, there are various caveats when comparing dynamical masses inferred from observations based on gas velocity dispersions to those computed directly from the physical mass distribution in simulations. 

To assess the evolution of the early black holes within the overall potential of their host galaxies, we therefore focus on examining the black hole mass--stellar velocity dispersion relation, which requires fewer post-processing assumptions than dynamical mass estimates in observations. Fig.~\ref{fig:scal_rel_filter_veldisp} shows the `detectable' \aesop~black hole population as a function of the 1D stellar velocity dispersion within twice the stellar half-mass radius (see Section~\ref{appsec:stellarveldisprel} for details on how the velocity dispersion is calculated in the simulations). For comparison, we include several local relations \citep{xiao_exploring_2011,kormendy_coevolution_2013,greene_intermediate-mass_2020,bennert_local_2021}. We also plot the JADES AGN with reliably measured stellar velocity dispersions for their host galaxies \citep{juodzbalis_jades_2026} as well as the high-redshift GA-NIFS AGN \citep{perna_ga-nifs_2023,ubler_ga-nifs_2023,parlanti_ga-nifs_2024,marshall_ga-nifs_2025} and BlackTHUNDER AGN \citep{deugenio_blackthunder_2026,deugenio_jades_2026-1,jones_blackthunder_2026,maiolino_black_2026,ubler_blackthunder_2026}.

At $z=4$, we find good agreement between the simulations and observations for all accretion models explored when assuming heavy seeds, and also for the `\textit{SE-BoostMax}' accretion scenarios with (very) light seeds. At $z=6$, heavy seeds remain in good agreement across all accretion prescriptions, while light seeds are only consistent with the observed population for `\textit{SE-BoostMax}' accretion. Both the simulated and observed populations are broadly consistent with the local relation reported by \citet{bennert_local_2021}, albeit with an overall offset towards slightly higher black hole masses at fixed velocity dispersion, in particular for low-mass systems. This behaviour further supports the picture that black hole co-evolution with the host potential, as traced by the stellar velocity dispersion, is established early, at least for massive galaxies. In our simulations, we find that with all heavy seed models as well as with the \textit{`SE-BoostMax'} accretion models, black hole activity with respect to the halo mass remains at similar levels throughout cosmic history. The coupling to the stellar mass of the galaxy, however, changes throughout cosmic history as the balance between black hole growth efficiency and star formation efficiency with respect to the total gas supply shifts at late cosmic times. This enables star formation to `catch up' at low redshift following a `black hole first' evolution scenario  \citep[also see discussion in][]{maiolino_jades_2024,ubler_blackthunder_2026}. It is important to caveat that explicitly modelling the multi-phase ISM in our simulations may lead to higher densities, increased star formation and therefore potentially higher stellar masses at early times. However, star formation in our galaxies is already relatively efficient compared to observational constraints (see also Fig.~\ref{fig:gsmf_z40}) so we would not expect even higher amounts of stellar mass build-up in these early phases.

\begin{figure*}
    \centering
    \includegraphics[width=\textwidth]{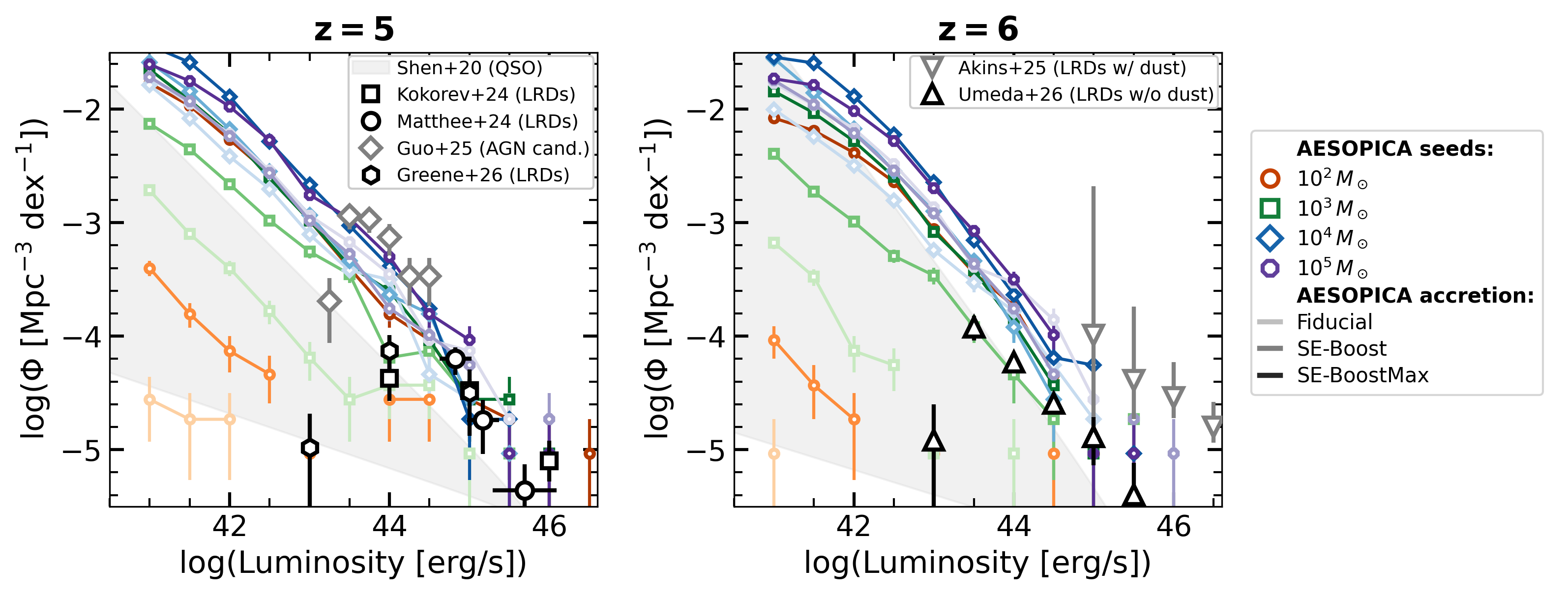}
    \caption{Bolometric AGN luminosity functions from the \aesop~simulations at $z=5$ (left) and $z=6$ (right). Simulated luminosity functions are shown for all seed masses and three different accretion prescriptions. Line shading denotes the accretion model, with light shading corresponding to the fiducial model, medium shading to boosted accretion with episodic super-Eddington bursts (`\textit{SE-Boost}'), whilst the dark shading is the `\textit{SE-BoostMax}' model. Black holes are grouped in luminosity bins of width 0.5\,dex, with symbols indicating the seed mass: $M_{\mathrm{seed}}=10^{2}\,\Msun$ (circles), $10^{3}\,\Msun$ (squares), $10^{4}\,\Msun$ (diamonds), and $10^{5}\,\Msun$ (hexagons). Vertical error bars indicate the Poisson error in the simulations. Only resolved galaxies with total baryonic masses $\geq 2 \times 10^{9}\,\Msun$ are included. Observational constraints are compiled from the literature, including the bolometric quasar luminosity function from \citet{shen_bolometric_2020} (grey shaded region), luminosity functions inferred from \jwst-selected LRD samples at comparable redshifts \citep{kokorev_census_2024,matthee_little_2024,greene_what_2026,akins_cosmos-web_2025,umeda_black_2026} and AGN candidates identified as dual-line emitters by \citet{guo_search_2025}. These observational constraints span a wide range of assumptions regarding bolometric corrections, dust reddening, and obscuration, and the scatter in the observational data should therefore be interpreted as indicative of the associated uncertainties in inferring bolometric luminosities in the observations. Most of our models are in agreement with the observational constraints, however, for (very) light seeds early efficient accretion with super-Eddington bursts (`\textit{SE-BoostMax}') is required to match the observational constraints.}
    \label{fig:lum_func}
\end{figure*}

\subsection{Bolometric AGN luminosity function} \label{subsec:results-bollumfunc}

We have demonstrated that several of the \aesop~models, when combined with observational selection effects, can occupy a parameter space similar to that probed by current observations and thereby explain the presence of unexpectedly early AGN activity in the high-redshift Universe. However, in addition to explaining the most extreme objects as well as the scaling relations of the observed population, we also need to account for their high inferred abundance. To this end, we compare the bolometric AGN luminosity function predicted by our simulations to observational constraints from the literature in Fig.~\ref{fig:lum_func}.

There are substantial uncertainties associated with such a comparison, including bolometric corrections, dust content, obscuration, and selection effects. A fully self-consistent comparison would require a forward-modelling approach, which is beyond the scope of this paper. We therefore do not attempt to correct for these effects, but instead compile a range of observational constraints derived under different assumptions to illustrate the associated uncertainties. We then compare these to the raw bolometric luminosity functions from the \aesop~simulations, without explicitly accounting for observational selection\footnote{We did assess the impact on the AGN luminosity function of excluding all radiatively inefficient AGN, with Eddington fractions smaller than 0.01 or larger than 1.0. The former only has a minimal impact at the low-luminosity end which is most significant for heavy seeds. The latter marginally lowers the high-luminosity end. However, these adjustments are much smaller ($<0.5$~dex) than the uncertainties in the observations.}.

We consider two redshifts, $z=5$ (left panel) and $z=6$ (right panel), and show the bolometric luminosity functions for various \aesop~runs at these epochs. We restrict our analysis to `resolved' galaxies with a total baryonic mass of at least $2 \times 10^{9}\,\Msun$, i.e. twice the mass we require for our nominal resolution limit in terms of galaxy stellar mass only. In practice, this cut has little impact on the bright end of the luminosity function, as luminous AGN are typically hosted by well-resolved systems (see also Fig.~\ref{fig:scal_rel_filter_dyn}). Black holes are grouped in luminosity bins of width 0.5\,dex, with bin means indicated by different symbols corresponding to the seed mass, as listed in the figure legend.

For reference, we include the bolometric quasar luminosity function from \citet{shen_bolometric_2020}, shown as a grey-shaded region reflecting uncertainties in the evolution of the faint-end slope. In addition, we plot several luminosity functions derived from \jwst-selected LRD samples at comparable redshifts. We emphasise that LRDs only form a minority of the population (10 to 40 per cent) and therefore the LRD luminosity function represents a lower limit to be corrected by a factor around two to ten to account for the full population \citep[e.g.][]{hainline_investigation_2025,taylor_broad-line_2025,madau_little_2026}.

In the left-hand panel ($z=5$), we show AGN candidates identified as dual-line emitters by \citet{guo_search_2025}, constraints on the LRD luminosity function at $z=4.5$--$6.5$ from \citet{kokorev_census_2024}, and measurements at $z=4$--$6$ from \citet{matthee_little_2024} and \citet{greene_what_2026}. For \citet{matthee_little_2024}, the luminosity function is presented in terms of \ha\ luminosities; following \citet{habouzit_is_2025}, we apply the bolometric correction from \citet{richards_spectral_2006} to convert these to bolometric luminosities. In the right-hand panel ($z=6$), we show bolometric luminosity functions for LRDs at $z=5$--$7$ from \citet{akins_cosmos-web_2025} and \citet{umeda_black_2026}. These studies are based on the same COSMOS-Web sample but adopt different physical interpretations: \citet{akins_cosmos-web_2025} assume dust-reddened emission, while \citet{umeda_black_2026} reanalyse the data assuming a black hole envelope model without dust reddening, resulting in significantly lower inferred luminosities. Broad \ha\ selection effects and the focus on LRDs imply that these inferred high-redshift luminosity functions should generally be regarded as lower limits, with an additional uncertainty of at least a factor of $\sim$2. Nevertheless, the high-redshift data points from \jwst generally lie above even the most generous extrapolations of the quasar bolometric luminosity function from \citet{shen_bolometric_2020}.

Keeping these caveats in mind, we find that the inferred high-redshift AGN abundance is consistent with all accretion prescriptions for intermediate and heavy seeds ($M_{\mathrm{seed}} = 10^{4}$--$10^{5}\,\Msun$). For light seeds, models with $M_{\mathrm{seed}} = 10^{3}\,\Msun$ are only able to reproduce the observational constraints when the accretion activity is boosted. In contrast, for the very light seed models ($M_{\mathrm{seed}} = 10^{2}\,\Msun$), the most efficient accretion prescription `\textit{SE-BoostMax}' is required to match the observations. The other two very light seed models lead to inefficient accretion and underpredict the observed luminosity function by a substantial margin. Taken together with the scaling-relation results, this strengthens the conclusion that assembling the high-redshift AGN population uncovered by \jwst from exclusively stellar-remnant seeds may be challenging and would require extremely favourable accretion conditions. 

\subsection{Black hole mass function} \label{subsec:results-bhmf}

\begin{figure*}
    \centering
    \includegraphics[width=\textwidth]{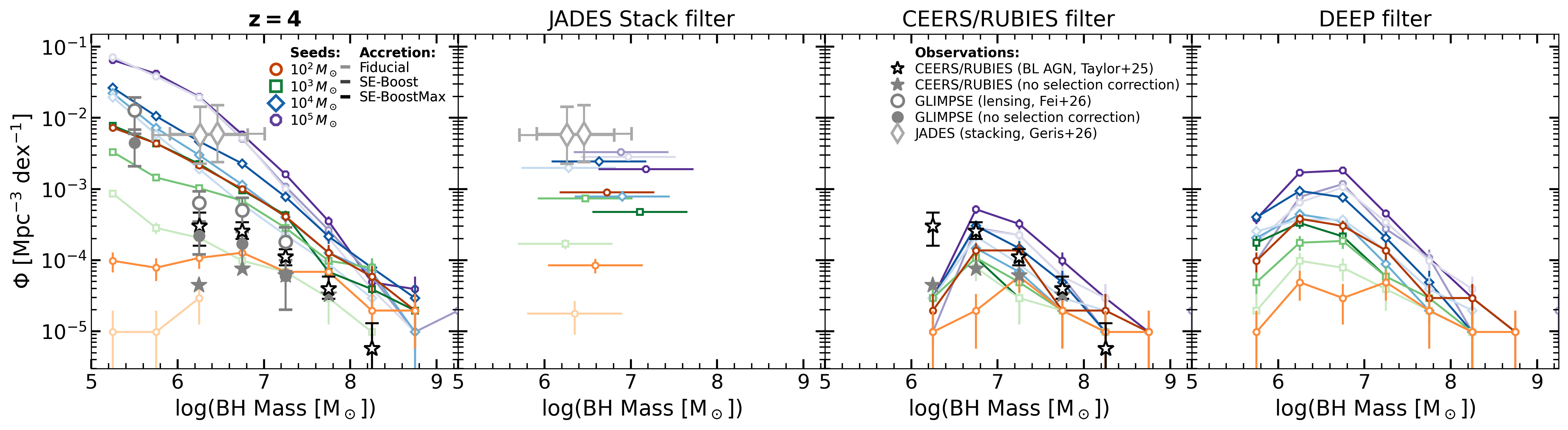}
    \caption{Black hole mass functions from the \aesop~simulations and \jwst surveys at $z=4$. Simulated black hole mass functions are shown for all seed masses and three different accretion prescriptions. Line shading denotes the accretion model, with light shading corresponding to the fiducial model, medium shading to `\textit{SE-Boost}', whilst the dark shading represents the `\textit{SE-BoostMax}' model. The symbols show the different seed masses: $M_{\mathrm{seed}}=10^{2}\,\Msun$ (circles), $10^{3}\,\Msun$ (squares), $10^{4}\,\Msun$ (diamonds), and $10^{5}\,\Msun$ (hexagons). Vertical error bars indicate the Poisson error in the simulations. The first panel shows the `raw' black hole mass function from the simulations, the second panel shows the results from our synthetic stacking procedure mimicking the selection of the JADES stack in \citet{geris_jades_2026}, and the third panel has a broad-line selection criterion applied mimicking the grating and exposure for CEERS/RUBIES following \citet{taylor_broad-line_2025}. Lastly, the fourth panel shows the black hole mass function based on the broad-line selection for a notional DEEP survey (R2700 grating and 30 hr exposure). The low-mass end of the black hole mass function retains significant sensitivity to the seed mass at $z=4$ with the heavy seeds exceeding the efficient light seed model number densities by an order of magnitude. Stacking approaches as well as future deep surveys may be starting to unveil distinguishing features in the black hole mass distribution.}
    \label{fig:bh_mass_f}
\end{figure*}

\begin{figure*}
    \centering
    \includegraphics[width=\textwidth]{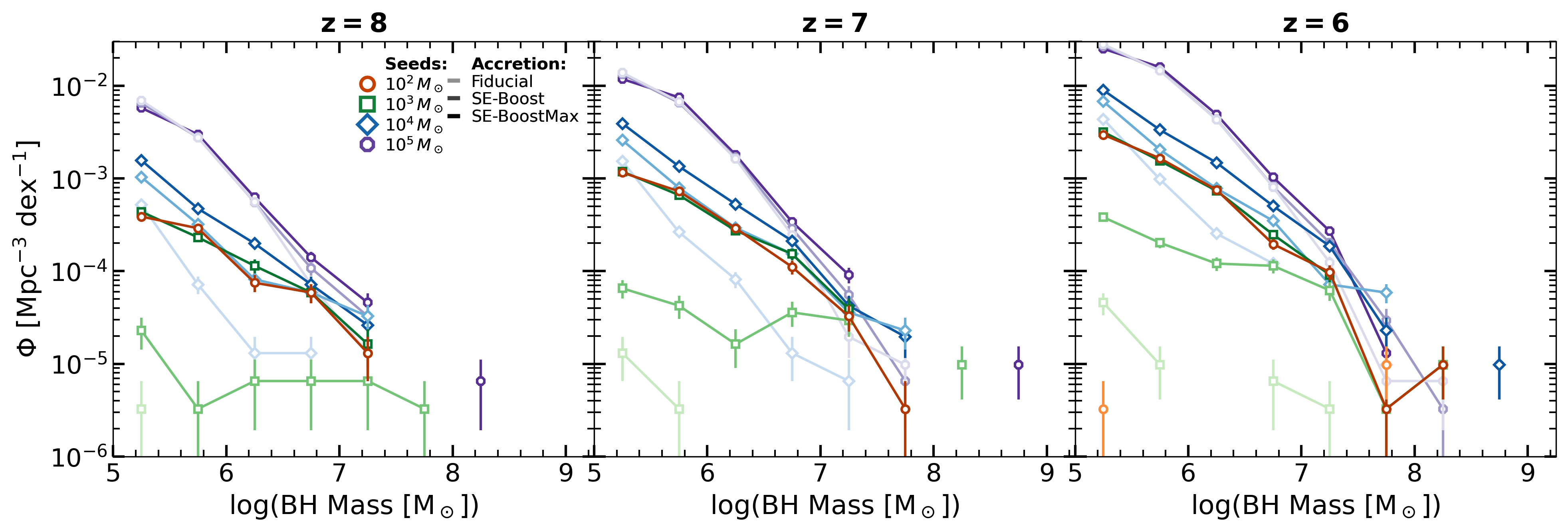} \\
    \includegraphics[width=\textwidth]{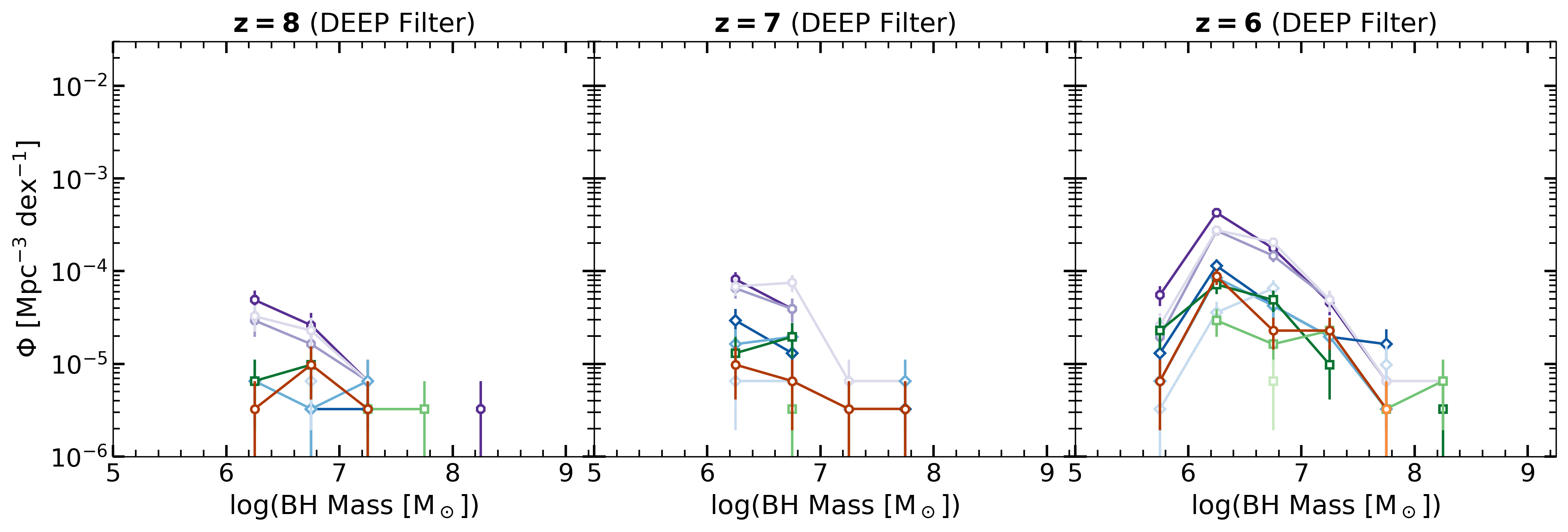}
    \caption{Black hole mass functions from the \aesop~simulations at $z=8$ (first column), $z=7$ (second column) and $z=6$ (third column). The top row shows the `raw' black hole mass function and the bottom row shows the black hole mass function that would be observed by a notional DEEP survey (R2700 grating, 30 hr exposure for $z<7$ and 100 hr exposure for $z>7$). Simulated black hole mass functions are shown for all seed masses and three different accretion prescriptions. Line shading denotes the accretion model, with light shading corresponding to the fiducial model, medium shading to `\textit{SE-Boost}', whilst the dark shading represents the `\textit{SE-BoostMax}' model. The symbols show the different seed masses: $M_{\mathrm{seed}}=10^{2}\,\Msun$ (circles), $10^{3}\,\Msun$ (squares), $10^{4}\,\Msun$ (diamonds), and $10^{5}\,\Msun$ (hexagons). Vertical error bars indicate the Poisson error in the simulations. At $z=6$, a DEEP broad-line AGN survey would retain sufficient completeness to distinguish heavy black hole seeds from lighter ones: their mass function remains offset by $\sim 0.3$~dex across all accretion models. However, at higher redshifts, low-number statistics and strong selection effects erode this discriminating power outside the low-mass end of the supermassive black hole population ($6.0 < \log(M_\mathrm{BH}\,[\Msun]) < 6.5$).}
    \label{fig:bh_mass_f_predict}
\end{figure*}

In this section, we analyse whether the different accretion and seeding models explored lead to realistic black hole mass distributions compared with observational constraints. Fig.~\ref{fig:bh_mass_f} shows the black hole mass functions from the \aesop~simulations and \jwst surveys at $z=4$. Simulated black hole mass functions are shown for all seed masses and three different accretion prescriptions. Line shading denotes the accretion model, with light shading corresponding to the fiducial model, medium shading to constant boosted accretion with episodic super-Eddington bursts (`\textit{SE-Boost}'), whilst the dark shading is the `\textit{SE-BoostMax}' model with (very) light seeds inversely proportionally boosted and the supernova coupling strength reduced for heavy seeds. The symbols show the different seed masses: $M_{\mathrm{seed}}=10^{2}\,\Msun$ (circles), $10^{3}\,\Msun$ (squares), $10^{4}\,\Msun$ (diamonds), and $10^{5}\,\Msun$ (hexagons). 

For comparison, we also show various observed black hole mass functions from the literature. We caution that our simulation provides the mass function of all black holes, while observations only provide the mass function of black holes that are in the active phase. The active phase is not well-defined and set by the sensitivity of the spectroscopic observations as well as their target selection. These comparisons should therefore be treated with caution. The optimal approach would be to forward model our simulation to take into account the observational biases and sensitivity limits and indeed we also show the simulated black hole mass functions resulting from our \textsc{balmersopica} selection tool in the second, third and fourth panels.

We plot the black hole mass function inferred by \citet{taylor_broad-line_2025} from 62 broad-line AGN in the CEERS/RUBIES surveys for $3.5 < z < 6$ (open star symbols). For the fully processed black hole mass functions, incompleteness corrections both due to detection limits and observational limits have been applied. The former accounts for incompleteness due to the limited resolution and sensitivity of the instrument for detecting the broad lines. The latter is defined as incompleteness due to candidate sources that fall within the MSA pointing areas but were not assigned open shutters \citep[see][for details]{taylor_broad-line_2025}. Estimating the broad line detection completeness for observational samples is very complex since the underlying black hole and luminosity distributions are not known. We also plot the sample that has been only `observationally' corrected as grey stars. 

We show the black hole mass function inferred by \citet{geris_jades_2026} from JADES stacking for $3<z<5$. We include two of their stack estimates here. The first is for the most luminous bin (by $M_\mathrm{UV}$) which shows evidence for a broad component corresponding to a black hole mass of $1.7 \times 10^{6}~\Msun$ and inferred Eddington fraction of $f_\mathrm{Edd} \sim 0.1$. The most luminous bin therefore still corresponds to relatively active black holes and may not be representative of the overall population. They also repeat the black hole mass function estimation for their stacking bin with the lowest Eddington ratio, which has $f_\mathrm{Edd} \sim 0.025$. Because of detectability limits, this lower Eddington ratio bin then corresponds to a higher average mass of $3\times 10^{6}~\Msun$. Both analyses lead to relatively similar estimations of the black hole mass function (see Fig.~\ref{fig:bh_mass_f}, grey diamonds).

Finally, we include the black hole mass function derived by \citet{fei_glimpse_2026} based on 10 lensed broad-line AGN at $4.5<z<7.0$, allowing them to push to lower luminosities and black hole masses (open grey circles). Similarly to \citet{taylor_broad-line_2025}, this black hole mass function has been post-processed to account for observational and detection completeness limits. We show the `raw' black hole mass function directly assembled from the lensed observations without the completeness corrections as filled grey circles, for reference.

For the simulated black hole mass function, we analyse the simulation outputs at $z=4.0$ in black hole mass bins of 0.5 dex. This matches the bin width of \citet{taylor_broad-line_2025} and roughly corresponds to the median redshift of their sample ($z=4.2$). We note this is also close to the median redshift of \citet{geris_jades_2026}, $z\sim3.8$, and we checked that the simulated black hole mass function only experiences negligible evolution between $z=3.8$--$4.2$. As with the AGN luminosity function, we only include black holes hosted by resolved galaxies with a total baryonic mass of at least $2\times10^{9}~\Msun$. This only has a minimal impact on the black hole mass function of the heavy seed runs in the lowest mass bin considered (difference is less than 0.3 dex). For all other simulation set-ups, significant mass growth needs to have taken place even to reach the lowest black hole mass shown here, so these black holes are then generally hosted by well-resolved galaxies.

In the first panel of Fig.~\ref{fig:bh_mass_f}, we show the `complete' black hole mass function from our \aesop~simulations without applying any observational corrections. Vertical error bars indicate the Poisson error. For the intermediate ($M_\mathrm{seed}=10^{4}~\Msun$) and heavy seed ($M_\mathrm{seed}=10^{5}~\Msun$) set-ups, the full black hole mass function at $z\sim 4$ is relatively insensitive to the black hole accretion model, leading to comparatively high black hole abundances across the black hole mass range in agreement with the observational stacking constraints. For the very light ($M_\mathrm{seed}=10^{2}~\Msun$) and light ($M_\mathrm{seed}=10^{3}~\Msun$) seeds, on the other hand, the results for the different accretion prescriptions vary by orders of magnitude. With these low seed masses, the simulation runs based on the fiducial accretion models lie well below the observational constraints and can therefore likely be excluded. The `\textit{SE-BoostMax}' runs with (very) light seeds converge towards the black hole mass function of the heavier seeds at high masses, but at lower black hole masses ($M_\mathrm{BH} < 10^{7}~\Msun$), these still fall below our set-ups with more massive seeds. The `\textit{SE-BoostMax}' runs with (very) light seeds and all runs with more massive seeds appear to be consistent with the constraints from the JADES stacks within the uncertainties. We note, however, that the majority of the simulation runs lie above the constraints inferred by \citet{taylor_broad-line_2025}. This is not too surprising since these are based on relatively active broad-line AGN for which completeness corrections are extremely challenging. The median Eddington fractions in our simulations span two orders of magnitude, from $3\times10^{-4}$ to 0.06, and generally lie below the typical Eddington fractions sampled in observations.

For comparison with the lensed sample, the `corrected' black hole mass function (grey open circles) from \citet{fei_glimpse_2026} appears consistent with the JADES stacking at low masses and the CEERS/RUBIES broad-line AGN at high masses. This leads to a departure from the expected Schechter shape for the corrected sample, which the authors hypothesise may be explained by an injection mass scale feature at the seed mass scale. We note, however, that in our model we do not observe such features despite seeding relatively late in cosmic time due to the limited resolution of our simulations. We instead hypothesise that the difference between the observed low-mass end and high-mass end constraints may be driven by selection. The lensed galaxy sample has a very complex selection function, so we do not attempt to mimic this for our simulations. Instead, we separately apply a JADES stack selection (second panel) as well as CEERS/RUBIES broad-line selection (third panel) to see whether we can identify simulation set-ups that are consistent with both the JADES stack at the low-mass end and the CEERS/RUBIES results at the high-mass end.

For the JADES stack selection in the second panel, we only retain the simulated galaxies that would be included in the JADES stack, i.e. galaxies that would be observable via their emission lines based on the JADES Medium tier sensitivity. As in Section~\ref{subsubsec:early_SF}, we base this on the typical flux limits for narrow lines \citep{scholtz_jades_2026}, $F_\mathrm{min}=5 \times 10^{-19} \ \mathrm{erg \, s^{-1} \, cm^{-2}}$, and calculate the corresponding star formation rate limit assuming the \citet{kennicutt_global_1998} relation between H$\alpha$ luminosity and star formation rate. We also discard all black holes from the sample that would be detectable by JADES as broad-line AGN, based on their masses and luminosities using \textsc{balmersopica} (see Section~\ref{subsec:methods-mocks}), since all broad-line-detected AGN are excluded from the \citet{geris_jades_2026} sample for the purposes of the stacking. The remainder of this simulated sample then forms the `synthetic stacks'. For each stack, we calculate the luminosity-weighted mean black hole mass and then compute the black hole mass function for a bin centred at this mean black hole mass with bin width 1.1 dex, matching the bin width of the stacks from \citet{geris_jades_2026}, see second panel of Fig.~\ref{fig:bh_mass_f}. For the heavy seed runs (purple hexagons), there is a clear impact since the star formation criterion dramatically suppresses the low-mass end of the black hole mass function due to the efficient AGN feedback in these runs. Overall, the mean black hole masses from the synthetic stacks appear to be consistent with the observed stack black hole mass given the uncertainties in the black hole mass measurements and the selection function. In particular, we note that our median Eddington ratios ($0.002$--$0.05$) in the synthetic stacks are still generally lower than what is found for the observed stacks (0.025 and 0.1), hinting that there are additional important selection criteria (such as the exclusion of known Type 2 AGN) which we have not accounted for here. To fully match the stack selection function, we would need to compute mock SEDs and line ratios requiring full synthetic observations, which is beyond the scope of this paper. Keeping these caveats in mind, we find that there is good agreement with the black hole mass function constraints inferred from the stacks and all heavy and intermediate-mass seed simulations. For the (very) light seeds, even the `\textit{SE-BoostMax}' synthetic stacks fall below the constraints from the observed stacks \citep[also see][]{dayal_light_2026} though the raw black hole mass functions for the `\textit{SE-BoostMax}' runs for the (very) light seeds are just within the constraints (see left panel), in agreement with results from semi-analytical models \citep[e.g.][]{trinca_seeking_2023,trinca_episodic_2026}.

We move on to consider the shape of the black hole mass function when we only retain simulated black holes which would be detectable via their broad lines based on the exposure time and grating of CEERS/RUBIES using \textsc{balmersopica} (see Section~\ref{subsec:methods-mocks}). The resulting simulated black hole mass function alongside the observational results from \citet{taylor_broad-line_2025} is plotted in the third panel of Fig.~\ref{fig:bh_mass_f}. The median Eddington fractions for the filtered sample range from $0.07$ to $0.6$ which is in good agreement with the range of Eddington ratios reported for the \citet{taylor_broad-line_2025} sample, where the two highest mass bins probe Eddington ratios in the range $0.06$--$0.15$  (10th --90th percentiles), whilst the lower mass bins ($M_\mathrm{BH} < 3 \times 10^{7}~\Msun$) probe Eddington ratios of $0.07$--$0.6$. This demonstrates that our synthetic observational selection is reasonable. Nevertheless, we note that with additional line broadening effects, lower-mass and/or lower-luminosity black holes may make it into this selection (also see Section~\ref{subsec:discuss_mbh_problem}). Overall, our simulations reproduce the black hole mass function estimates from CEERS/RUBIES that have not been corrected for detection incompleteness, which show a turnover at $\log(M_\mathrm{BH})=6.75$. In terms of the overall normalisation, the most appropriate sample to compare to would be the one that has been corrected for observational incompleteness due to the shutters but not for line detection incompleteness (shown by the grey symbols), as we account for the latter with our synthetic \textsc{balmersopica} selection. This would suggest that the heavy seeds ($M_\mathrm{seed}=10^{5}~\Msun$) with `\textit{SE-BoostMax}' accretion result in an overabundance of black holes at $z=4$, whilst most of the intermediate and light seed runs as well as the very light seeds with `\textit{SE-BoostMax}' accretion are in good agreement with the \citet{taylor_broad-line_2025} results.

In the fourth panel, we also show a prediction for the black hole mass function constraints that could be obtained by our notional DEEP survey (R2700 grating, 30 hr exposure for $z<7$ and 100 hr exposure for $z>7$). This demonstrates that with higher resolution and longer exposure time, we can move close to the intrinsic population at high masses ($M_\mathrm{BH} > 10^{7} \ \Msun$). In particular, this may offer constraints on seed masses as high mass function values ($\Phi \gtrsim 10^{-3}~\mathrm{Mpc^{-3}\,dex^{-1}}$) measured in such a deep survey would be indicative of heavy seeds (or very efficiently growing intermediate seeds). It would be difficult to obtain such high densities with (very) light seeds, even with our most optimistic `\textit{SE-BoostMax}' configuration.

Whilst the current observational constraints on the black hole mass function are mostly restricted to $z \sim 4$, we also make some predictions for the black hole mass function at higher redshifts from \aesop. Fig.~\ref{fig:bh_mass_f_predict} shows the intrinsic black hole mass function (top row) as well as the black hole mass function that would be inferred by the notional DEEP survey (bottom row) at redshifts $z=8$, $z=7$ and $z=6$. We note that due to the low number statistics at very high redshifts (also see Fig.~\ref{fig:star_bh_proj}), we consider black holes in redshift bins of $\delta z \sim 0.5$ (rather than just at the snapshot corresponding to the target redshift). 

As we move to higher redshifts, the distinction between the heavy seed runs and simulations initialised with less massive seeds becomes more pronounced due to the seed mass leaving a stronger imprint at early times. However, detecting broad-line AGN becomes increasingly challenging. At $z=6$, the black hole mass function that would be recovered with our notional DEEP survey set-up still has relatively high completeness and therefore future broad-line AGN surveys at this redshift may deliver significant constraints on the nature and origin of the high-z AGN population. In particular the black hole mass function based on heavy seed runs is significantly offset ($\sim 0.3$ dex) from the other simulation runs for all three accretion models. As we move to even higher redshifts, however, the low-number statistics in our 60 cMpc box combined with the strong selection make it increasingly difficult to distinguish between the different scenarios. Notably all seed masses remain detectable up to $z=8$ with the `\textit{SE-BoostMax}' model. At intermediate masses ($6.0 < \log(M_\mathrm{BH} \ [\Msun]) < 6.5$), the heavy seed mass function is still clearly offset across all redshifts, though for the other mass bins, the Poisson errors become comparable to the differences between the models. 

Overall, whilst the bright end of the luminosity function does not yet constrain most of the models we have explored, the mass function is starting to allow us to distinguish between the different scenarios. However, the uncertainties both on the observational and the theoretical side are still significant, in particular, with regard to modelling the selection function. This demonstrates that we do not just need to push to lower black hole masses to constrain our models but also to higher completeness levels with deep spectroscopic surveys to enable direct comparisons with simulations and constrain black hole formation and growth in the early Universe.

\subsection{Constraints from metallicities} \label{subsec:results-metals}

\begin{figure}
    \centering
    \includegraphics[width=\columnwidth]{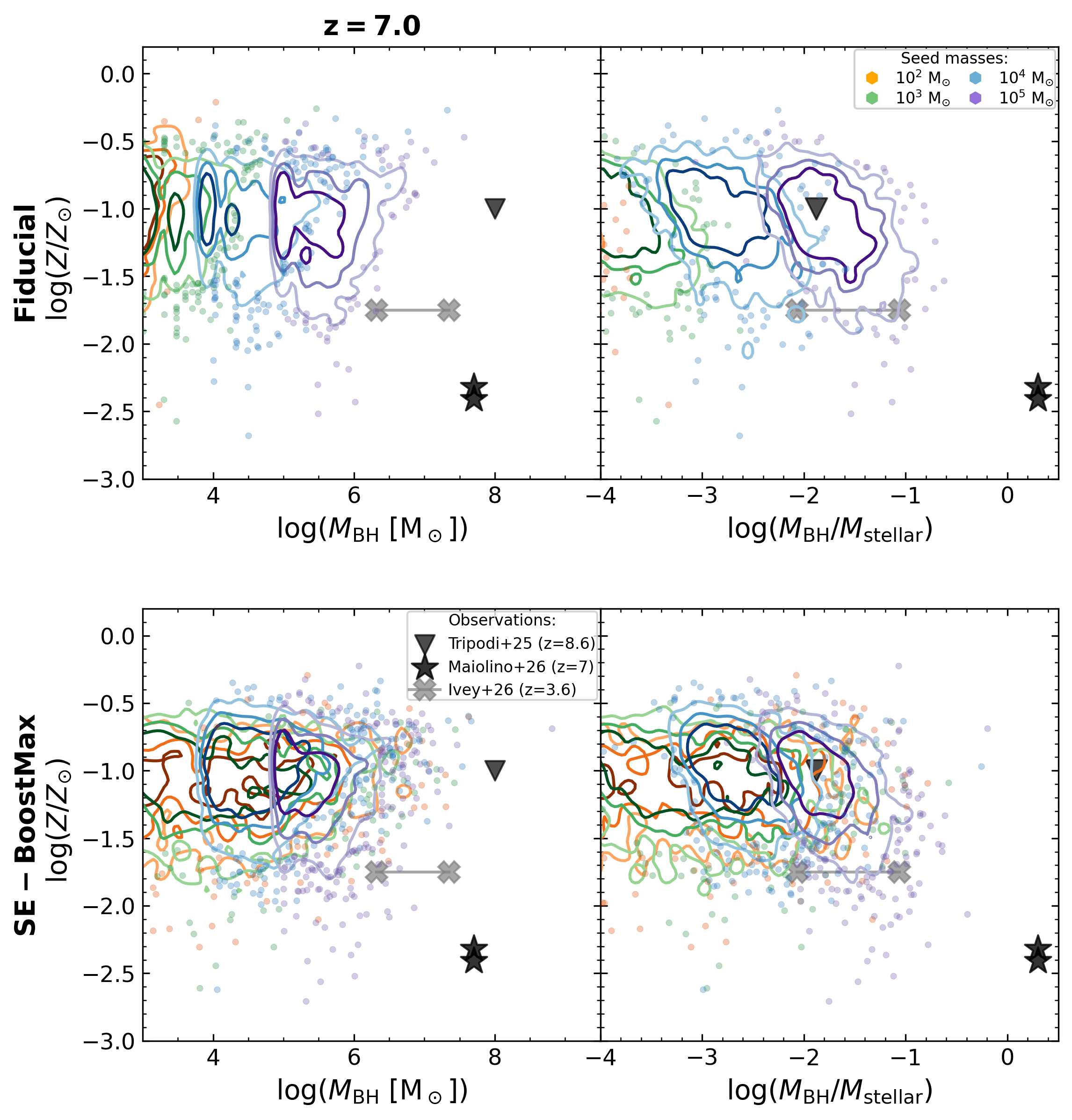}
    \caption{Gas-phase metallicity as a function of black hole mass (left column) and black hole-to-stellar mass ratio (right column), for our simulation runs with fiducial (top row) and `\textit{SE-BoostMax}' (bottom row) accretion at $z=7$. Each panel shows all four seed masses, colour-coded as indicated in the legend. For each run, the population is binned into a 2D histogram on a fixed grid, with the bulk of the distribution shown as density contours drawn at fixed fractions of the peak bin count (5 per cent, 15 per cent, and 30 per cent), colour-coded by seed mass. Objects whose (smoothed) bin density falls below the outermost contour level are instead plotted as individual scattered points. For comparison, we also show recent \jwst observations of massive AGN at high redshift hosted by seemingly pristine galaxies from CANUCS-LRD-z8.6 at $z=8.6$ \citep{tripodi_extreme_2025}, QSO1 at $z=7$ \citep{maiolino_black_2026} and the Cliff at $z=3.6$ \citep{ivey_cliff_2026}. For QSO1, we show the metallicity measured in both the inner ($R<150$~pc) and outer ($150~\mathrm{pc}<R<300$~pc) regions, and for the Cliff we show the black hole mass from both the standard virial relations and the estimate assuming additional scattering processes. With the exception of the scattering-corrected Cliff, these extreme low-metallicity AGN lie outside the region spanned by our models, even for our most optimistic accretion parametrizations.}
    \label{fig:metallicities}
\end{figure}

Recently, there have been observational indications of a few high-redshift AGN hosted by galaxies that are not just lower in stellar mass than expected, but also have very low gas-phase metallicities (mostly measured from oxygen). Such pristine hosts are difficult to reconcile with high levels of AGN activity, since the conditions favourable to black hole accretion are difficult to achieve without also creating conditions favourable to star formation, which would enrich the gas. We investigate this apparent tension in Fig.~\ref{fig:metallicities}, where we plot the gas-phase metallicity as a function of black hole mass (left column) and of black hole to stellar mass ratio (right column). We include simulation runs with fiducial (top row) and `\textit{SE-BoostMax}' (bottom row) accretion for all four seed masses. For each run, the population is binned into a 2D histogram on a fixed grid, with the bulk of the distribution shown as density contours drawn at fixed fractions of the peak bin count (5 per cent, 15 per cent, and 30 per cent), colour-coded by seed mass. Objects whose (smoothed) bin density falls below the outermost contour level are instead plotted as individual scattered points. For comparison, we also show recent \jwst observations of massive AGN at high redshift hosted by seemingly pristine galaxies from CANUCS-LRD-z8.6 at $z=8.6$ \citep{tripodi_extreme_2025}, QSO1 at $z=7$ \citep{maiolino_black_2026} and the Cliff at $z=3.6$ \citep{ivey_cliff_2026}. We present the data from our simulations at $z=7$, matching the redshift of QSO1, which is the most extreme example of an overmassive black hole in a pristine galaxy. 

For QSO1, we plot the metallicity measured in both the inner ($R<150$~pc) and outer ($150~\mathrm{pc}<R<300$~pc) regions, and for the Cliff of \citet{ivey_cliff_2026} we include the black hole mass derived from the standard virial relations as well as the value obtained when additional scattering processes are assumed \citep{rusakov_little_2026}. With the exception of this scattering-corrected `Cliff' estimate, none of these extreme low-metallicity AGN overlap with our simulation models in black hole mass--metallicity space, despite the very optimistic parametrizations we employ. We note an important caveat, however: our limited box size of $L=60$~cMpc \citep[see also the discussion in][]{ivey_cliff_2026}. The rarity of such `pristine AGN' may simply mean that larger simulation volumes are needed to capture these extreme environments. Conversely, if \jwst were to identify larger samples of overmassive `pristine' black holes, this would place strong constraints on the models. Whilst almost all seed masses can match the overmassive black hole population under our `\textit{SE-BoostMax}' accretion parametrization, galaxies with $M_\mathrm{BH}/M_\mathrm{stellar}>0.1$ and $Z/\mathrm{Z_{\odot}} < 0.01$ are produced only by our heavy seed set-ups. Hence, metallicity provides a valuable additional axis along which to discriminate between the models. Indeed, for the most extreme objects such as QSO1, even our most efficient heavy seed models cannot reproduce this region of parameter space. Given that QSO1 is a single object, we need to be careful when interpreting this mismatch within our theoretical framework. In particular, since we do not resolve the multi-phase structure of the ISM in \aesop, it may be that this discrepancy could be alleviated by explicitly modelling the dense gas phase and better resolving the metallicity distribution. Higher resolution would also enable us to seed our black holes at higher redshifts which may be necessary to enable such rapid growth by $z=7$. Other works have also highlighted the importance of exploring alternative origins for these overmassive black holes in low-metallicity environments, such as primordial black holes \citep[see also][]{dayal_exploring_2024,dayal_light_2026,dayal_properties_2026,prole_primordial_2025,zhang_primordial_2026}. Intriguingly, further high-$z$ AGN that may reside in pristine environments have since been discovered \citep[e.g.][]{trefoloni_missing_2025,caputi_pseudo_2026}, and the crucial next step will be to investigate their black hole, stellar and gas properties in detail and to link these to hydrodynamical simulations to learn what accretion and feedback processes might be the driving mechanisms behind this observed extreme population.

\section{Discussion} \label{sec:discussion}

\subsection{Black hole accretion and feedback} \label{subsec:discuss_acc_feed}

Several other works have begun to relax the standard assumptions governing black hole growth in cosmological simulations, and it is instructive to place our parameter variations for black hole accretion in \aesop~within this wider context.

The recently introduced COLIBRE simulations \citep{schaye_colibre_2026,chaikin_importance_2026} permit super-Eddington accretion by default, capping the accretion rate at $100\,\dot{M}_\mathrm{Edd}$, whilst retaining a Bondi-based accretion estimate with turbulence and vorticity corrections \citep{krumholz_bondi_2005}. Notably, COLIBRE aims to account for the cold, multi-phase structure of the ISM, so that very high gas densities in the vicinity of the black hole are (partially) resolved rather than approximated via a boost factor. In this regime, the Bondi rate can become very large and effectively needs to be calibrated \textit{down} (rather than boosted up) in the super-Eddington regime, with the accretion efficiency with respect to the Bondi rate set to 1 per cent \citep[see][]{schaye_colibre_2026}. A similar behaviour is seen in high-resolution simulations that employ Bondi-like estimators within a resolved multi-phase ISM, where the high densities reached in dense clouds allow even light seeds to grow efficiently via super-Eddington accretion \citep[e.g.][]{gordon_conditions_2025,zana_super-eddington_2026}. These results suggest that our `\textit{SE-BoostMax}' parametrization can be viewed as an optimistic limit, in which the inversely-proportional boost factor acts as a computationally inexpensive proxy for the high-density accretion environments that may emerge once the cold ISM phase is resolved \citep[also see][]{koudmani_two_2022,ortame_small_2026}. We caution, however, that simulations explicitly following Pop~III remnants find that only a small fraction of light seeds experience such rapid growth \citep[e.g.][]{mehta_growth_2026,prole_seedz_2026}. Since we place a single seed in each halo above $\sim2\times10^{9}\,h^{-1}\,\Msun$, our light seed is best interpreted as the most `successful' remnant in its host, and the `\textit{SE-BoostMax}' parametrization as an upper limit on light-seed growth.

Other cosmological simulations have also found that powerful super-Eddington bursts are required to reproduce the abundance of massive quiescent galaxies at high redshift \citep[e.g.][]{rennehan_obsidian_2024,chaikin_importance_2026}, in agreement with the behaviour of our `\textit{SE-Boost}' and `\textit{SE-BoostMax}' runs (see Section~\ref{subsubsec:early_SF}). We caution, however, that this does not mean that super-Eddington accretion is a necessary condition for obtaining massive quiescent galaxies. More generically, it appears that powerful feedback bursts in the high accretion-rate regime are required to regulate star formation in massive galaxies throughout cosmic history. Indeed, the stochastic MISTRAL implementation achieves early quenching at sub-Eddington accretion rates via powerful kinetic winds \citep{farcy_mistral_2025}, and jet-based feedback models offer yet another route to efficient self-regulation \citep[e.g.][]{sullivan_arkenstone_2026}. In \aesop, we deliberately kept the AGN feedback model fixed to the \fable~implementation in order to isolate the impact of the seeding and accretion prescriptions. Disentangling the relative roles of the accretion regime and the feedback channel will require simulation suites that vary both simultaneously, which we defer to future work.

Nevertheless, our simulations indicate that light seeds can only account for the \jwst AGN population if efficient super-Eddington bursts are invoked, a growth regime that the traditional Bondi model, with its steep black hole mass dependence, is ill-suited to explore \citep[also see][]{bhowmick_growth_2024,bhowmick_supermassive_2026}. Alternative accretion estimators with weaker black hole mass dependencies, such as the free-fall-based model of \citet{weinberger_accretion_2025} or the sink-based accretion model from \citet{ortame_small_2026}, similarly find that the choice of accretion model predominantly affects the intermediate-mass black hole population at high redshift whilst leaving the local massive black hole population largely unchanged \citep[also see][]{kho_learning_2026}, mirroring the convergence of our models at low redshifts.

Ultimately, it will be imperative to move from the phenomenological black hole growth models towards more comprehensive models that are tied to the state of the black hole accretion disc. This will allow us to self-consistently include all of the radiative regimes and determine how much gas reaches the black hole on small scales \citep[e.g.][]{power_accretion_2011,dubois_black_2014,fiacconi_galactic_2018,beckmann_dense_2019,bustamante_spin_2019,husko_spin-driven_2022,husko_winds_2024,massonneau_how_2023,koudmani_unified_2024,kao_novel_2026,piana_blackholeweather_2026,piana_blackholeweather_2026-1}. This approach has already been implemented by several groups in idealised galaxy simulations \citep[e.g.][]{talbot_simulations_2024,partmann_importance_2025,petersson_noctua_2025,shin_mandelzoom_2025,shin_mandelzoom_2026}. These high-resolution simulations have also emphasised the complex interplay between the multi-phase ISM and the black hole modelling, including feedback, which is an aspect that we do not capture in \aesop. Indeed, it is possible that the efficacy of our feedback may be reduced once the cold dense gas component is explicitly included \citep[e.g.][]{bourne_resolution_2015,valentini_impact_2020,sivasankaran_agn_2025} -- though this may also boost accretion rates as stellar feedback is also weakened at higher ISM densities in the early Universe \citep[e.g.][]{somerville_density-modulated_2025}. Dedicated zoom-in follow-up simulations would be required to explore this \citep[also see][]{chon_overmassive_2026}.

Crucially, the aforementioned high-resolution simulations have demonstrated that there is not a straightforward `correction factor' that could be used to adjust the Bondi rate to account for small-scale accretion disc and ISM physics \citep[see][for a direct comparison]{shin_mandelzoom_2025}. Instead, `accretion disc particles' must be implemented directly to track the black hole growth within a multi-phase ISM. However, resolving the outer edge of the accretion disc, as these prescriptions require, demands a resolution of $\Delta x \lesssim 10^{-2}$~pc, which is many orders of magnitude below the typical resolution of cosmological simulations ($\Delta x \sim 100$ -- $1000$~pc). Bridging this gap will therefore require analytical scaling relations for the inflow rates onto the subgrid accretion disc \citep[e.g.][]{cho_bridging_2023,guo_toward_2023} and/or machine-learning-accelerated approaches building on surrogate models pioneered for supernova feedback \citep[e.g.][]{hirashima_asura-fdps-ml_2025}.

\subsection{Seeding mechanisms} \label{subsec:discuss_seeding_constraints}

Most large-volume cosmological simulations do not have the required resolution to directly capture black hole formation processes, and instead employ a halo mass threshold, placing a heavy seed in all massive haloes \citep[e.g.][]{sijacki_unified_2007,schaye_eagle_2015,sijacki_illustris_2015,henden_fable_2018,pillepich_simulating_2018,dave_simba_2019}. We follow the same approach, but significantly lower this threshold to the smallest resolved haloes and explore a wide range of seed masses \citep[also see][who explore light seeds with a hybrid black hole particle]{taylor_seeding_2014}. Our simple prescription allows us to efficiently explore the growth of different seeds within a cosmological context, however, there are several caveats to highlight.

The timing of our seeding is likely conservative. Our resolution limits seeding to haloes above $2\times10^{9}\,h^{-1}\,\Msun$, corresponding to seeding black holes from $z\sim18$. Physically, Pop~III remnants and direct collapse black holes may form in atomic-cooling haloes at significantly earlier epochs, so our seeds are deprived of $\sim100$~Myr of potential growth precisely when the competition between seed growth and host enrichment is evolving most rapidly. Higher-resolution simulations with earlier seeding will be needed to assess how much of this early window matters in practice for the metallicity constraints discussed in Section~\ref{subsec:results-metals}, in addition to including a resolved ISM and better modelling of the thermochemistry. The unresolved Pop III and early Pop II star formation, and the metal enrichment that accompanies it, are a further limitation of our approach. Because this enrichment is not captured, host metallicities at the earliest epochs are likely underestimated.

In terms of number densities, however, and in particular for the heavy seeds, our halo-mass-based seeding is very likely \textit{over}-optimistic. Direct collapse requires a restrictive set of environmental conditions (dense, metal-poor gas, turbulent accretion and/or suppression of $\mathrm{H_2}$ cooling by a sufficiently strong Lyman--Werner flux, and low gas angular momentum) and simulations that impose such criteria find that the abundance of viable seeding sites is reduced dramatically. This has been demonstrated both with on-the-fly environmental seeding criteria, e.g. in the BRAHMA \citep{bhowmick_growth_2024,bhowmick_supermassive_2026}, MELIORA \citep{cenci_little_2025} and SEEDZ \citep{prole_seedz_2026,prole_black_2026,prole_seedz_2026-1} simulations, and with post-processing approaches applied to existing simulations \citep{degraf_cosmological_2020}, with the assumed seeding criteria making a substantial difference to the predicted black hole abundances. On the other hand, our intermediate seeds with $10^{4}~\Msun$ from runaway star clusters may be relatively common. For the (very) light seeds from Pop III remnants, we may even be underestimating the seed densities as we are restricting black hole formation to one seed per halo \citep[also see][]{prole_seedz_2026,prole_black_2026,prole_seedz_2026-1}.

We further caution that all seeding channels explored in \aesop~are astrophysical in origin. Primordial black holes have (re-)emerged as an alternative seeding mechanism that naturally produces high $M_\mathrm{BH}/M_\mathrm{stellar}$ ratios in the early Universe \citep[e.g.][]{zhang_primordial_2026,prole_primordial_2025,dayal_properties_2026,maiolino_black_2026}. Exploring this channel is, however, beyond the scope of this work, and we aim to explore this in future simulations.

Finally, we have treated the seed mass as a single, fixed parameter within each simulation run, whereas recent theoretical work suggests that the seed mass spectrum is better described as a continuum \citep[e.g.][]{regan_massive_2024}. Exploring `mixed' seeding scenarios with physically motivated relative abundances, as has been done in semi-analytic frameworks \citep[e.g.][]{spinoso_multiflavour_2023,trinca_seeking_2023,dayal_light_2026}, is a natural next step for the \aesop~programme. Indeed since we find that multiple seed and growth scenarios are largely degenerate by $z\sim 4$, the current observations would allow for such a mixed seeding scenario.

\subsection{Black hole dynamics} \label{subsec:discuss_bh_dynamics}

In \aesop, black holes are advected to the potential minimum of their host halo to prevent spurious scattering by numerical heating following e.g. \fable~\citep[][]{henden_fable_2018} and Illustris \citep{sijacki_illustris_2015}. Whilst this is a common and numerically robust choice, it represents another optimistic assumption for early black hole growth. In particular, for low-mass galaxies in the early Universe (with velocity dispersions as small as tens of $\mathrm{km\,s^{-1}}$), unless the black hole is very massive (at least $10^{5}~\Msun$) or surrounded by a nuclear star cluster effectively boosting the dynamical mass \citep[e.g.][]{partmann_importance_2025,shin_mandelzoom_2025,shin_mandelzoom_2026}, then the sinking time is longer than Gyrs if the distance is larger than $\sim100$~pc \citep[see][]{pfister_erratic_2019}. Therefore, in the case of light seeds, `wandering' is likely to be physical rather than numerical and such seeds may never settle to the galactic centre, or `sink', within a Hubble time \citep[e.g.][]{ma_seeds_2021}. Off-centre black holes tend to sample lower ambient gas densities, suppressing their accretion rates and luminosities \citep[also see][]{sharma_hidden_2022}.

A number of subgrid schemes have been developed to model unresolved black hole dynamics more faithfully, including calibrated and analytical dynamical friction models for large-volume simulations \citep[e.g.][]{tremmel_off_2015,pfister_erratic_2019,chen_dynamical_2022,genina_calibrated_2024,bhowmick_dynamics_2025,li_ramcoal_2025} as well as regularised integration techniques incorporating post-Newtonian corrections \citep[e.g.][]{rantala_ketju_2017,mannerkoski_signatures_2022,partmann_importance_2025}. These studies consistently find delayed sinking, suppressed merger rates and substantial offset populations for low-mass black holes \citep[also see][]{buttigieg_premature_2025}. We note that our simulated overmassive black holes at high redshift mostly stem from accretion-dominated growth (see Fig.~\ref{fig:bh_z_evol}) with only an insignificant portion of the growth stemming from mergers; hence, our key findings will be less sensitive to the modelling of the dynamics. However, inevitably, these accretion rates would be lowered if the seeds were wandering. Quantifying this effect will require dedicated follow-up zoom-in simulations with explicit black hole dynamics.

\subsection{Observational selection} \label{subsec_discuss_obs_selection}

Our analysis highlights that any comparison between simulated and observed high-redshift black hole populations has to take into account the selection effects. Indeed, there is a growing body of `selection-aware' interpretations of the \jwst AGN population, which find that the finite detection limits and broad-line selection requirements of current surveys preferentially uncover the most massive and most actively accreting black holes \citep[e.g.][]{li_tip_2025,geris_jades_2026,juodzbalis_jades_2026,ziparo_selection-aware_2026}. Our \textsc{balmersopica} framework represents a step towards a forward-modelling approach, but it only targets the final stage of the selection chain, i.e. the detectability of the broad-line component. A fully self-consistent treatment would require synthetic spectral energy distributions and emission-line ratios for each simulated system, allowing for photometric pre-selection, Type~1/Type~2 classification, obscuration and dust reddening to be modelled explicitly. Developing such synthetic catalogues from the \aesop~outputs, e.g. by coupling our simulations to emission-line and radiative transfer post-processing \citep[e.g.][]{smith_lyman_2015,hirschmann_emission-line_2023,wilkins_first_2025}, is a priority for future work.

A closely related uncertainty concerns the intrinsic SEDs of the high-redshift objects themselves, which remain hotly debated, in particular for the LRD population \citep[e.g.][]{matthee_little_2024}. Proposed interpretations range from dust-reddened AGN \citep[e.g.][]{madau_little_2026} to `black hole stars', in which a dense gas envelope reprocesses the accretion luminosity \citep[e.g.][]{ji_blackthunder_2025,de_graaff_little_2026,naidu_gas-enshrouded_2026}, whilst statistical studies suggest that LRDs may represent a continuous extension of the broader AGN population, evolving as the host stellar component assembles, rather than a physically distinct class \citep{billand_little_2026}. Since the mapping from black hole mass and accretion rate to observed fluxes, colours and line profiles differs substantially between these scenarios, they directly affect any synthetic selection function. First steps in this direction are being taken \citep[e.g.][]{marszewski_little_2026,roy_little_2026}, but a self-consistent end-to-end model will require synthetic observations built on simulations that resolve the relevant physical scales spanning the accretion disc, the gas envelope or torus, and the host galaxy.

\subsection{Accuracy of observational black hole mass estimates}\label{subsec:discuss_mbh_problem}

Recent works on the high-z AGN with \jwst have discussed the accuracy of the black hole mass estimates \citep[e.g.][]{rusakov_little_2026, brazzini_ruling_2025,brazzini_little_2026, deugenio_irony_2025,deugenio_blackthunder_2026,juodzbalis_direct_2026,scholtz_little_2026}. The discovery of exponential line profiles in the broad components of Balmer lines has suggested that the broadening of the line emission could be due to electron scattering \citep{laor_evidence_2006, rusakov_little_2026} in addition to the motion of broad-line region clouds around the black hole. This electron scattering model implies that the intrinsic widths of the broad lines could be 5--10 times narrower than currently measured, potentially leading to systematic overestimates of black hole masses by up to two orders of magnitude when using standard virial relations (since $M_{\rm BH;virial}\propto {\rm FWHM_{broad-line}^2}$).

However, \citet{scholtz_little_2026, brazzini_ruling_2025, brazzini_little_2026,madau_wings_2026} find that exponential line profiles can also be explained by kinematics from a stratified broad line region. Indeed, the direct measurement of black hole mass by \citet{juodzbalis_direct_2026} in an LRD at $z=7.04$ (QSO1; \citealt{furtak_high_2024}) is in agreement with the virial estimation excluding electron scattering. Furthermore, \citet{parlanti_doubling_2026} find black hole masses from virial relations consistent with masses from reverberation mapping for black holes with low Eddington ratios. At high Eddington ratios, however, the virial relations may no longer be applicable \citep[e.g.][]{gravity_collaboration_spatially_2026}, requiring new black hole mass calibrations in this regime \citep[e.g.][]{woo_new_2026}. Beyond additional scattering, there are also hints that the bolometric correction assumed for these AGN may be overestimated \citep{greene_what_2026}, though see \citet{geris_little_2026}. However, we note that since we follow the reverse procedure (i.e. mapping from black hole masses and bolometric luminosities to Balmer line widths and fluxes), both of these factors mean that our observability estimates are conservative, i.e. a larger fraction of simulated black holes may be observable if additional electron scattering or lower bolometric corrections lead to larger line widths and fluxes. Therefore, at this time we use classical virial calibration of black hole masses in this work. Future direct measurements of the black hole mass at high-z will further test the validity of these calibrations and improve the comparison between simulations and observations as we do in this work.

\subsection{How to break the degeneracies?} \label{subsec:discuss_degeneracies}

The \aesop~simulations demonstrate that once efficient accretion channels are permitted, the imprint of the initial seed mass is rapidly erased: by $z\sim4$ the scaling relations, luminosity functions and black hole mass functions of most of our models occupy overlapping regions of parameter space, and at low redshifts they are mostly indistinguishable. Breaking these degeneracies will require combining several complementary approaches.

Firstly, whilst the black hole masses of the \jwst~AGN can be reproduced from any seed mass once the standard accretion assumptions are relaxed, the chemical state of the host galaxy may retain some imprint of the seeding channel. Within our framework, even the most optimistic light-seed runs do not grow their black holes fast enough to leave the host pristine, such that only heavy seeds populate the most extreme region of parameter space with $M_\mathrm{BH}/M_\mathrm{stellar}>0.1$ and $Z/\mathrm{Z_{\odot}} < 0.01$ (see Section~\ref{subsec:results-metals}). The most pristine observed systems are not reached by any of our models. We stress, however, that this result is strongly model-dependent. The IMF, yields and final fates of Pop~III stars, the existence of supermassive stars, and the rates of core-collapse supernovae and hypernovae all remain poorly constrained \citep[e.g.][]{karlsson_pregalactic_2013,nomoto_nucleosynthesis_2013,maiolino_re_2019}. Moreover, \aesop~does not follow the early enrichment from unresolved Pop~III/II star formation, which would raise the host metallicities and hence make our estimates for light seeds conservative. Conversely, our effective equation of state for the ISM yields relatively smooth star formation histories and does not capture bursty, metal-loaded outflows or the dilution by pristine cold-stream inflows \citep[also see discussion in][]{maiolino_black_2026}, both of which could lower the metallicity in the vicinity of the black hole. The net impact of a more sophisticated treatment of the ISM and early enrichment is therefore uncertain. Establishing whether the tension we identify for light seeds is robust will require simulations that resolve Pop~III/II star formation and the multi-phase ISM, and that explore the range of plausible stellar yields. On the observational side, growing samples of high-redshift AGN with (low) metallicity measurements will be essential to test these predictions.

Secondly, on the theoretical side, we need to make self-consistent predictions of the black hole growth rates based on AGN accretion disc physics. It is also crucial to model these next-generation prescriptions in tandem with better resolved AGN feedback models which may be more directional/collimated and thereby allow for higher and more sustained accretion rates \citep[e.g.][]{curtis_resolving_2015,talbot_simulations_2024,su_self-regulation_2025}. We stress that accretion-disc-based models are not a straightforward multiplicative correction to the Bondi rate. On one hand, their weaker dependence on black hole mass permits substantially higher accretion rates for low-mass black holes. On the other hand, the angular momentum barrier can strongly suppress the rates at which gas is actually circularised onto the disc and delivered to the black hole \citep[see][]{fiacconi_galactic_2018,koudmani_unified_2024,shin_mandelzoom_2025}. High-resolution simulations that resolve the multi-phase ISM around low-mass black holes and follow the disc mass and angular momentum evolution self-consistently \citep[e.g.][]{shin_mandelzoom_2025} will be crucial for calibrating the next generation of subgrid accretion models, replacing the effective boost factors employed here with physically grounded prescriptions.

Finally, on the observational side, it will be crucial to push to higher redshifts, lower black hole masses and higher completeness. As shown in Section~\ref{subsec:results-bhmf}, the low-mass end of the black hole mass function retains sensitivity to the seeding and accretion models even when the high-mass end has converged. This makes deep surveys, e.g. via stacking \citep{geris_jades_2026} or lensing \citep{fei_glimpse_2026}, particularly valuable. In particular, the new NIRSpec `slitless' mode \citep[also known as DarkHorse,][]{deugenio_jades_2026} will allow for deep, unbiased and large-scale samples necessary to push towards breaking the degeneracy in the seeding and black hole growth models. Furthermore, we are currently relying on the low redshift virial calibration for black hole masses, poorly tested at $z>4$. It is necessary to perform more direct black hole measurements similar to those of \citet{juodzbalis_direct_2026} with NIRSpec/IFS or with interferometry with GRAVITY+ \citep[][]{gravity_collaboration_size-luminosity_2024,gravity_collaboration_spatially_2026} to verify the virial calibrations for black hole masses. 

Gravitational-wave signatures offer complementary constraints on black hole growth histories: in particular, the gravitational-wave background measured by pulsar timing arrays may be better matched by models with efficient black hole growth in the early Universe \citep[e.g.][]{buttigieg_comparing_2026}. In the longer term, LISA will detect black hole mergers with $10^{4}~\Msun \lesssim M_\mathrm{BH} \lesssim 10^{7}~\Msun$ out to $z\sim20$ \citep[see][and references therein]{amaro-seoane_astrophysics_2023}, probing early black hole assembly independently of electromagnetic selection effects and virial mass calibrations. However, we note that there are important localisation and redshift degeneracies at play when identifying and analysing LISA sources. Firstly, gravitational waves constrain the redshifted mass $(1+z)M_\mathrm{BH}$ and the luminosity distance, so that black hole masses and redshifts must be inferred jointly. At high redshift, lower signal-to-noise and weak lensing increase the uncertainties for both quantities \citep[e.g.][]{hirata_reducing_2010}. Secondly, sky localisation degrades rapidly with redshift \citep[e.g.][]{mangiagli_observing_2020} and identifying individual hosts will therefore be very challenging. LISA constraints will instead be predominantly statistical, through merger rates, mass and redshift distributions, and spins, which trace the accretion mode \citep[e.g.][]{berti_cosmological_2008,sesana_reconstructing_2011,klein_science_2016}. Fully exploiting LISA will thus require joint analyses of gravitational-wave and electromagnetic populations, combined with simulations that follow black hole dynamics, accretion and feedback self-consistently.

\section{Conclusions} \label{sec:conclusions}

The abundant, seemingly overmassive black holes uncovered by \jwst in the early Universe challenge standard models of black hole formation and growth. However, whether they trace genuinely efficient early growth, observational selection, or systematic uncertainties in the mass estimates has remained unclear. To address this question, we have introduced the \aesop~project, a suite of twelve mid-volume ($L = 60\,\mathrm{cMpc}$) cosmological simulations based on the \fable~galaxy formation model, with updated seeding and accretion modelling. We systematically vary the black hole seed mass across the main black hole seeding channels ($M_\mathrm{seed} = 10^{2}$--$10^{5}\,\Msun$) together with the accretion efficiency, spanning the fiducial Bondi model, boosted accretion with super-Eddington bursts permitted (`\textit{SE-Boost}'), and the most optimistic configuration for each seed mass (`\textit{SE-BoostMax}') where for light seeds the boost factor is inversely proportionally enhanced with the seed mass and for heavy seeds the supernova feedback is moderated (see Table~\ref{tab:aesopica}). Crucially, we post-process our simulation outputs with \textsc{balmersopica}, a mock \jwst broad-line survey selection procedure that assesses the detectability of each simulated AGN for a given grating and exposure time, enabling more direct comparisons with observed samples. Our main results are as follows:

\begin{itemize}
    \item The bulk of the massive black hole population uncovered by \jwst in the early Universe can be reproduced from any seed mass explored here ($M_\mathrm{seed} = 10^{2}$--$10^{5}\,\Msun$), provided the accretion efficiency is high, allowing for brief episodes of super-Eddington accretion.
    \item With the fiducial accretion model, only intermediate-to-heavy seeds reach the observed overmassive locus in the $M_\mathrm{BH}$--$M_\mathrm{stellar}$ plane by $z \gtrsim 4$, whereas under our most optimistic accretion model with super-Eddington bursts permitted (`\textit{SE-BoostMax}'), the seeding signatures are erased as early as $z \sim 6$. At $z=0$, all seed masses and accretion prescriptions converge onto the local relations for massive galaxies. At the low-mass end ($M_\mathrm{stellar} \lesssim 3 \times 10^{9}~\Msun$), however, the normalisation of the scaling relations is significantly different by two to four orders of magnitude for all accretion models explored. Hence the high-redshift and low-mass regimes are very promising for constraining black hole formation and growth models with future surveys.
    \item Observational selection plays a decisive role: the \jwst-detected AGN are not representative of the underlying simulated population, lying well above the mean luminosity relations of all our models. Once our broad-line selection criteria are applied, the apparently overmassive population is naturally recovered for all intermediate-to-heavy seed models largely independently of the accretion prescription, whilst (very) light seeds require efficient accretion that allows for super-Eddington bursts.
    \item For efficient AGN models, the observable AGN population is largely representative of the intrinsic mean relation. For inefficient AGN models, however, the observable AGN population lies an order of magnitude above the intrinsic relation. In reality, a broader mixture of seed masses, accretion efficiencies and stellar feedback regimes is likely at play, and we emphasise that the observable `tip of the iceberg' need not be representative, potentially concealing a large population of undetected, inefficiently accreting black holes.
    \item Whilst the efficient \aesop~AGN appear overmassive with respect to the local $M_\mathrm{BH}$--$M_\mathrm{stellar}$ relation, they are consistent with the local relations once the gas mass is included in the host mass budget, and they lie on the local $M_\mathrm{BH}$--$\sigma_\mathrm{stellar}$ relation. This supports a `black hole first' scenario in which black holes lock in to the host potential before the bulk of the stellar component is assembled.
    \item The inferred abundances of high-redshift AGN, as traced by the bolometric AGN luminosity function and the black hole mass function, are consistent with all accretion prescriptions for intermediate and heavy seeds. Our results, however, disfavour (very) light seeds unless their growth is highly efficient with (mildly) super-Eddington bursts. We predict that future deep surveys will be able to strengthen these constraints at $z=4$ and constrain the black hole mass function out to redshift $z=8$, which would inform theoretical black hole formation modelling, in particular with regard to the need for heavy seeds (or lack thereof).
    \item Strongly quenched massive galaxies at high redshift only arise in our simulations when super-Eddington bursts are permitted, with AGN-driven `mini-quenching' episodes lasting $\sim$100--200~Myr. Dormant, overmassive systems such as \textit{GN-1001830} are naturally explained as black holes observed during the declining phase of such a burst.
    \item Gas-phase metallicities provide a valuable additional axis for discriminating between models: only our heavy seed set-ups produce systems with $M_\mathrm{BH}/M_\mathrm{stellar} > 0.1$ in hosts with $Z < 0.01\,\mathrm{Z_\odot}$, and the most extreme pristine AGN observed, such as QSO1, lie outside the parameter space of even our most optimistic models, pointing towards the need to incorporate more sophisticated, multi-phase ISM modelling to capture these systems. In addition, this also motivates exploring alternative origins such as primordial black holes.
\end{itemize}

Looking ahead, breaking the degeneracies between seeding and accretion scenarios will require physically grounded accretion prescriptions, explicit black hole dynamics and full synthetic observations on the theoretical side, together with deeper observational censuses that probe the low-mass end of the black hole mass function, where sensitivity to the models persists even after the high-mass end has converged. Ultimately, gravitational-wave observations with LISA will probe black hole mergers in precisely the mass range explored here, offering a constraint on early black hole assembly that is fully independent of electromagnetic selection effects and, in combination with the approach developed in this work, promising to finally disentangle the origin of the first supermassive black holes.

\section*{Acknowledgements}
The authors are grateful to the anonymous referee whose helpful comments improved the quality of this paper. The authors would like to thank Brian Bichang’a, Johannes Buchner, Pratika Dayal, Jenny Greene, Silvia Guida, Sugata Kaviraj, Chiaki Kobayashi, Erini Lambrides and Víctor Rodríguez Morales for useful comments and suggestions. Part of this work was completed during the 2026 Sesto Workshop on `The growth of galaxies in the early Universe' and benefitted from stimulating discussions with participants. SK acknowledges support from the Royal Society under grant number URF\textbackslash R1\textbackslash 251867. JS, RM, FDE and LRI acknowledge support by the Science and Technology Facilities Council (STFC), ERC Advanced Grant 695671 ``QUENCH'' and the UKRI Frontier Research grant RISEandFALL. RM also acknowledges funding from a research professorship from the Royal Society. SLF and AJT acknowledge support from STScI via JWST-GO-5893 and JWST-GO-6368. DS acknowledges support from the Science and Technology Facilities Council (STFC) under grant ST/W000997/1. ECL acknowledges support of an STFC Webb Fellowship (ST/W001438/1). MAB is supported by a UKRI Stephen Hawking Fellowship (EP/X04257X/1). SG is supported by a Woolf Fisher Scholarship from the Woolf Fisher Trust of New Zealand and Cambridge Commonwealth and European Trust. H\"U acknowledges support by the Max Planck Society through the Lise Meitner Excellence Program. H\"U acknowledges funding by the European Union (ERC APEX, 101164796). Views and opinions expressed are however those of the authors only and do not necessarily reflect those of the European Union or the European Research Council Executive Agency. Neither the European Union nor the granting authority can be held responsible for them. The Flatiron Institute is supported by the Simons Foundation.

This work used the DiRAC Data Intensive service (CSD3) at the University of Cambridge, managed by the University of Cambridge University Information Services on behalf of the STFC DiRAC HPC Facility (\url{www.dirac.ac.uk}). The DiRAC component of CSD3 at Cambridge was funded by BEIS, UKRI and STFC capital funding and STFC operations grants. This work used the DiRAC Memory Intensive service (Cosma8) at Durham University, managed by the Institute for Computational Cosmology on behalf of the STFC DiRAC HPC Facility (\url{www.dirac.ac.uk}). The DiRAC service at Durham was funded by BEIS, UKRI and STFC capital funding, Durham University and STFC operations grants. DiRAC is part of the UKRI Digital Research Infrastructure.

\section*{Data Availability}

The data underlying this article will be shared on reasonable request to the corresponding author. We publicly release the code and full tables for the \textsc{balmersopica} tool on GitHub at \url{https://github.com/skoudmani/balmersopica}.



\bibliographystyle{mnras}
\bibliography{references} 

@article{beckmann_population_2023,
	title = {Population statistics of intermediate-mass black holes in dwarf galaxies using the {NEWHORIZON} simulation},
	volume = {523},
	issn = {0035-8711},
	url = {https://ui.adsabs.harvard.edu/abs/2023MNRAS.523.5610B},
	doi = {10.1093/mnras/stad1544},
	urldate = {2024-09-02},
	journal = {MNRAS},
	publisher = {OUP},
	author = {Beckmann, R. S. and Dubois, Y. and Volonteri, M. and Dong-Páez, C. A. and Trebitsch, M. and Devriendt, J. and Kaviraj, S. and Kimm, T. and Peirani, S.},
	month = aug,
	year = {2023},
	note = {ADS Bibcode: 2023MNRAS.523.5610B},
	pages = {5610--5623},
}

@article{rennehan_obsidian_2024,
	title = {The {OBSIDIAN} model: three regimes of black hole feedback},
	volume = {532},
	issn = {0035-8711},
	shorttitle = {The {OBSIDIAN} model},
	url = {https://ui.adsabs.harvard.edu/abs/2024MNRAS.532.4793R},
	doi = {10.1093/mnras/stae1785},
	urldate = {2024-08-15},
	journal = {MNRAS},
	publisher = {OUP},
	author = {Rennehan, Douglas and Babul, Arif and Moa, Belaid and Davé, Romeel},
	month = aug,
	year = {2024},
	note = {ADS Bibcode: 2024MNRAS.532.4793R},
	pages = {4793--4809},
}

@article{koudmani_unified_2024,
	title = {A unified accretion disc model for supermassive black holes in galaxy formation simulations: method and implementation},
	volume = {532},
	issn = {0035-8711},
	shorttitle = {A unified accretion disc model for supermassive black holes in galaxy formation simulations},
	url = {https://ui.adsabs.harvard.edu/abs/2024MNRAS.532...60K},
	doi = {10.1093/mnras/stae1422},
	urldate = {2024-07-30},
	journal = {MNRAS},
	publisher = {OUP},
	author = {Koudmani, Sophie and Somerville, Rachel S. and Sijacki, Debora and Bourne, Martin A. and Jiang, Yan-Fei and Profit, Kasar},
	month = jul,
	year = {2024},
	note = {ADS Bibcode: 2024MNRAS.532...60K},
	pages = {60--88},
}

@article{sharma_hidden_2022,
	title = {A {Hidden} {Population} of {Massive} {Black} {Holes} in {Simulated} {Dwarf} {Galaxies}},
	volume = {936},
	issn = {0004-637X},
	url = {https://ui.adsabs.harvard.edu/abs/2022ApJ...936...82S},
	doi = {10.3847/1538-4357/ac8664},
	urldate = {2024-07-28},
	journal = {ApJ},
	publisher = {IOP},
	author = {Sharma, Ray S. and Brooks, Alyson M. and Tremmel, Michael and Bellovary, Jillian and Ricarte, Angelo and Quinn, Thomas R.},
	month = sep,
	year = {2022},
	note = {ADS Bibcode: 2022ApJ...936...82S},
	pages = {82},
}

@article{vogelsberger_introducing_2014,
	title = {Introducing the {Illustris} {Project}: simulating the coevolution of dark and visible matter in the {Universe}},
	volume = {444},
	issn = {0035-8711},
	shorttitle = {Introducing the {Illustris} {Project}},
	url = {https://doi.org/10.1093/mnras/stu1536},
	doi = {10.1093/mnras/stu1536},
	number = {2},
	urldate = {2024-07-23},
	journal = {MNRAS},
	author = {Vogelsberger, Mark and Genel, Shy and Springel, Volker and Torrey, Paul and Sijacki, Debora and Xu, Dandan and Snyder, Greg and Nelson, Dylan and Hernquist, Lars},
	month = oct,
	year = {2014},
	pages = {1518--1547},
}

@article{martin-alvarez_pandora_2023,
	title = {The {Pandora} project - {I}. {The} impact of radiation, magnetic fields, and cosmic rays on the baryonic and dark matter properties of dwarf galaxies},
	volume = {525},
	issn = {0035-8711},
	url = {https://ui.adsabs.harvard.edu/abs/2023MNRAS.525.3806M},
	doi = {10.1093/mnras/stad2559},
	urldate = {2024-07-17},
	journal = {MNRAS},
	publisher = {OUP},
	author = {Martin-Alvarez, Sergio and Sijacki, Debora and Haehnelt, Martin G. and Farcy, Marion and Dubois, Yohan and Belokurov, Vasily and Rosdahl, Joakim and Lopez-Rodriguez, Enrique},
	month = nov,
	year = {2023},
	note = {ADS Bibcode: 2023MNRAS.525.3806M},
	pages = {3806--3830},
}

@article{curtis_resolving_2016,
	title = {Resolving flows around black holes: the impact of gas angular momentum},
	volume = {463},
	issn = {0035-8711},
	shorttitle = {Resolving flows around black holes},
	url = {https://ui.adsabs.harvard.edu/abs/2016MNRAS.463...63C},
	doi = {10.1093/mnras/stw1944},
	urldate = {2024-06-04},
	journal = {MNRAS},
	publisher = {OUP},
	author = {Curtis, Michael and Sijacki, Debora},
	month = nov,
	year = {2016},
	note = {ADS Bibcode: 2016MNRAS.463...63C},
	pages = {63--77},
}

@article{ubler_ga-nifs_2024,
	title = {{GA}-{NIFS}: {JWST} discovers an offset {AGN} 740 million years after the big bang},
	volume = {531},
	issn = {0035-8711},
	shorttitle = {{GA}-{NIFS}},
	url = {https://ui.adsabs.harvard.edu/abs/2024MNRAS.531..355U},
	doi = {10.1093/mnras/stae943},
	urldate = {2024-06-04},
	journal = {MNRAS},
	publisher = {OUP},
	author = {Übler, Hannah and Maiolino, Roberto and Pérez-González, Pablo G. and D'Eugenio, Francesco and Perna, Michele and Curti, Mirko and Arribas, Santiago and Bunker, Andrew and Carniani, Stefano and Charlot, Stéphane and Rodríguez Del Pino, Bruno and Baker, William and Böker, Torsten and Cresci, Giovanni and Dunlop, James and Grogin, Norman A. and Jones, Gareth C. and Kumari, Nimisha and Lamperti, Isabella and Laporte, Nicolas and Marshall, Madeline A. and Mazzolari, Giovanni and Parlanti, Eleonora and Rawle, Tim and Scholtz, Jan and Venturi, Giacomo and Witstok, Joris},
	month = jun,
	year = {2024},
	note = {ADS Bibcode: 2024MNRAS.531..355U},
	pages = {355--365},
}

@article{talbot_simulations_2024,
	title = {Simulations of spin-driven {AGN} jets in gas-rich galaxy mergers},
	volume = {528},
	issn = {0035-8711},
	url = {https://ui.adsabs.harvard.edu/abs/2024MNRAS.528.5432T},
	doi = {10.1093/mnras/stae392},
	urldate = {2024-06-04},
	journal = {MNRAS},
	publisher = {OUP},
	author = {Talbot, Rosie Y. and Sijacki, Debora and Bourne, Martin A.},
	month = mar,
	year = {2024},
	note = {ADS Bibcode: 2024MNRAS.528.5432T},
	pages = {5432--5451},
}

@article{maiolino_small_2024,
	title = {A small and vigorous black hole in the early {Universe}},
	volume = {627},
	issn = {0028-0836},
	url = {https://ui.adsabs.harvard.edu/abs/2024Natur.627...59M},
	doi = {10.1038/s41586-024-07052-5},
	urldate = {2024-06-04},
	journal = {Nature},
	author = {Maiolino, Roberto and Scholtz, Jan and Witstok, Joris and Carniani, Stefano and D'Eugenio, Francesco and de Graaff, Anna and Übler, Hannah and Tacchella, Sandro and Curtis-Lake, Emma and Arribas, Santiago and Bunker, Andrew and Charlot, Stéphane and Chevallard, Jacopo and Curti, Mirko and Looser, Tobias J. and Maseda, Michael V. and Rawle, Timothy D. and Rodríguez del Pino, Bruno and Willott, Chris J. and Egami, Eiichi and Eisenstein, Daniel J. and Hainline, Kevin N. and Robertson, Brant and Williams, Christina C. and Willmer, Christopher N. A. and Baker, William M. and Boyett, Kristan and DeCoursey, Christa and Fabian, Andrew C. and Helton, Jakob M. and Ji, Zhiyuan and Jones, Gareth C. and Kumari, Nimisha and Laporte, Nicolas and Nelson, Erica J. and Perna, Michele and Sandles, Lester and Shivaei, Irene and Sun, Fengwu},
	month = mar,
	year = {2024},
	note = {ADS Bibcode: 2024Natur.627...59M},
	pages = {59--63},
}

@article{cho_bridging_2023,
	title = {Bridging {Scales} in {Black} {Hole} {Accretion} and {Feedback}: {Magnetized} {Bondi} {Accretion} in {3D} {GRMHD}},
	volume = {959},
	issn = {0004-637X},
	shorttitle = {Bridging {Scales} in {Black} {Hole} {Accretion} and {Feedback}},
	url = {https://ui.adsabs.harvard.edu/abs/2023ApJ...959L..22C},
	doi = {10.3847/2041-8213/ad1048},
	urldate = {2024-06-04},
	journal = {ApJ},
	publisher = {IOP},
	author = {Cho, Hyerin and Prather, Ben S. and Narayan, Ramesh and Natarajan, Priyamvada and Su, Kung-Yi and Ricarte, Angelo and Chatterjee, Koushik},
	month = dec,
	year = {2023},
	note = {ADS Bibcode: 2023ApJ...959L..22C},
	pages = {L22},
}

@article{bennett_growth_2024,
	title = {The growth of the gargantuan black holes powering high-redshift quasars and their impact on the formation of early galaxies and protoclusters},
	volume = {527},
	issn = {0035-8711},
	url = {https://ui.adsabs.harvard.edu/abs/2024MNRAS.527.1033B},
	doi = {10.1093/mnras/stad3179},
	urldate = {2024-04-22},
	journal = {MNRAS},
	publisher = {OUP},
	author = {Bennett, Jake S. and Sijacki, Debora and Costa, Tiago and Laporte, Nicolas and Witten, Callum},
	month = jan,
	year = {2024},
	note = {ADS Bibcode: 2024MNRAS.527.1033B},
	pages = {1033--1054},
}

@article{wellons_exploring_2023,
	title = {Exploring supermassive black hole physics and galaxy quenching across halo mass in {FIRE} cosmological zoom simulations},
	volume = {520},
	issn = {0035-8711},
	url = {https://ui.adsabs.harvard.edu/abs/2023MNRAS.520.5394W},
	doi = {10.1093/mnras/stad511},
	urldate = {2024-04-22},
	journal = {MNRAS},
	publisher = {OUP},
	author = {Wellons, Sarah and Faucher-Giguère, Claude-André and Hopkins, Philip F. and Quataert, Eliot and Anglés-Alcázar, Daniel and Feldmann, Robert and Hayward, Christopher C. and Kereš, Dušan and Su, Kung-Yi and Wetzel, Andrew},
	month = apr,
	year = {2023},
	note = {ADS Bibcode: 2023MNRAS.520.5394W},
	pages = {5394--5412},
}

@article{krumholz_bondi_2005,
	title = {Bondi {Accretion} in the {Presence} of {Vorticity}},
	volume = {618},
	issn = {0004-637X},
	url = {https://ui.adsabs.harvard.edu/abs/2005ApJ...618..757K},
	doi = {10.1086/426051},
	urldate = {2024-02-20},
	journal = {ApJ},
	author = {Krumholz, Mark R. and McKee, Christopher F. and Klein, Richard I.},
	month = jan,
	year = {2005},
	note = {ADS Bibcode: 2005ApJ...618..757K},
	pages = {757--768},
}

@article{husko_winds_2024,
	title = {Winds versus jets: a comparison between black hole feedback modes in simulations of idealized galaxy groups and clusters},
	volume = {527},
	issn = {0035-8711},
	shorttitle = {Winds versus jets},
	url = {https://ui.adsabs.harvard.edu/abs/2024MNRAS.527.5988H},
	doi = {10.1093/mnras/stad3548},
	urldate = {2024-01-04},
	journal = {MNRAS},
	author = {Huško, Filip and Lacey, Cedric G. and Schaye, Joop and Nobels, Folkert S. J. and Schaller, Matthieu},
	month = jan,
	year = {2024},
	note = {ADS Bibcode: 2024MNRAS.527.5988H},
	pages = {5988--6020},
}

@article{ubler_ga-nifs_2023,
	title = {{GA}-{NIFS}: {A} massive black hole in a low-metallicity {AGN} at z ∼ 5.55 revealed by {JWST}/{NIRSpec} {IFS}},
	volume = {677},
	issn = {0004-6361},
	shorttitle = {{GA}-{NIFS}},
	url = {https://ui.adsabs.harvard.edu/abs/2023A&A...677A.145U},
	doi = {10.1051/0004-6361/202346137},
	urldate = {2023-12-11},
	journal = {A\&A},
	author = {Übler, Hannah and Maiolino, Roberto and Curtis-Lake, Emma and Pérez-González, Pablo G. and Curti, Mirko and Perna, Michele and Arribas, Santiago and Charlot, Stéphane and Marshall, Madeline A. and D'Eugenio, Francesco and Scholtz, Jan and Bunker, Andrew and Carniani, Stefano and Ferruit, Pierre and Jakobsen, Peter and Rix, Hans-Walter and Rodríguez Del Pino, Bruno and Willott, Chris J. and Boeker, Torsten and Cresci, Giovanni and Jones, Gareth C. and Kumari, Nimisha and Rawle, Tim},
	month = sep,
	year = {2023},
	note = {ADS Bibcode: 2023A\&A...677A.145U},
	pages = {A145},
}

@article{husko_spin-driven_2022,
	title = {Spin-driven jet feedback in idealized simulations of galaxy groups and clusters},
	volume = {516},
	issn = {0035-8711},
	url = {https://ui.adsabs.harvard.edu/abs/2022MNRAS.516.3750H},
	doi = {10.1093/mnras/stac2278},
	urldate = {2023-12-03},
	journal = {MNRAS},
	author = {Huško, Filip and Lacey, Cedric G. and Schaye, Joop and Schaller, Matthieu and Nobels, Folkert S. J.},
	month = nov,
	year = {2022},
	note = {ADS Bibcode: 2022MNRAS.516.3750H},
	pages = {3750--3772},
}

@article{harikane_jwstnirspec_2023,
	title = {A {JWST}/{NIRSpec} {First} {Census} of {Broad}-line {AGNs} at z = 4-7: {Detection} of 10 {Faint} {AGNs} with {M} {BH} 106-108 {M} ⊙ and {Their} {Host} {Galaxy} {Properties}},
	volume = {959},
	issn = {0004-637X},
	shorttitle = {A {JWST}/{NIRSpec} {First} {Census} of {Broad}-line {AGNs} at z = 4-7},
	url = {https://ui.adsabs.harvard.edu/abs/2023ApJ...959...39H},
	doi = {10.3847/1538-4357/ad029e},
	urldate = {2023-12-11},
	journal = {ApJ},
	author = {Harikane, Yuichi and Zhang, Yechi and Nakajima, Kimihiko and Ouchi, Masami and Isobe, Yuki and Ono, Yoshiaki and Hatano, Shun and Xu, Yi and Umeda, Hiroya},
	month = dec,
	year = {2023},
	note = {ADS Bibcode: 2023ApJ...959...39H},
	pages = {39},
}

@article{greene_intermediate-mass_2020,
	title = {Intermediate-{Mass} {Black} {Holes}},
	volume = {58},
	issn = {0066-4146},
	url = {https://ui.adsabs.harvard.edu/abs/2020ARA%26A..58..257G/abstract},
	doi = {10.1146/annurev-astro-032620-021835},
	language = {en},
	urldate = {2022-03-23},
	journal = {ARA\&A},
	author = {Greene, Jenny E. and Strader, Jay and Ho, Luis C.},
	month = aug,
	year = {2020},
	pages = {257},
}

@article{pillepich_simulating_2018,
	title = {Simulating galaxy formation with the {IllustrisTNG} model},
	volume = {473},
	issn = {0035-8711},
	url = {https://ui.adsabs.harvard.edu/abs/2018MNRAS.473.4077P},
	doi = {10.1093/mnras/stx2656},
	urldate = {2022-03-02},
	journal = {MNRAS},
	author = {Pillepich, Annalisa and Springel, Volker and Nelson, Dylan and Genel, Shy and Naiman, Jill and Pakmor, Rüdiger and Hernquist, Lars and Torrey, Paul and Vogelsberger, Mark and Weinberger, Rainer and Marinacci, Federico},
	month = jan,
	year = {2018},
	note = {ADS Bibcode: 2018MNRAS.473.4077P},
	pages = {4077--4106},
}

@article{schaye_eagle_2015,
	title = {The {EAGLE} project: simulating the evolution and assembly of galaxies and their environments},
	volume = {446},
	issn = {0035-8711},
	shorttitle = {The {EAGLE} project},
	url = {https://ui.adsabs.harvard.edu/abs/2015MNRAS.446..521S},
	doi = {10.1093/mnras/stu2058},
	urldate = {2022-03-01},
	journal = {MNRAS},
	author = {Schaye, Joop and Crain, Robert A. and Bower, Richard G. and Furlong, Michelle and Schaller, Matthieu and Theuns, Tom and Dalla Vecchia, Claudio and Frenk, Carlos S. and McCarthy, I. G. and Helly, John C. and Jenkins, Adrian and Rosas-Guevara, Y. M. and White, Simon D. M. and Baes, Maarten and Booth, C. M. and Camps, Peter and Navarro, Julio F. and Qu, Yan and Rahmati, Alireza and Sawala, Till and Thomas, Peter A. and Trayford, James},
	month = jan,
	year = {2015},
	note = {ADS Bibcode: 2015MNRAS.446..521S},
	pages = {521--554},
}

@article{booth_cosmological_2009,
	title = {Cosmological simulations of the growth of supermassive black holes and feedback from active galactic nuclei: method and tests},
	volume = {398},
	issn = {00358711, 13652966},
	shorttitle = {Cosmological simulations of the growth of supermassive black holes and feedback from active galactic nuclei},
	url = {https://academic.oup.com/mnras/article-lookup/doi/10.1111/j.1365-2966.2009.15043.x},
	doi = {10.1111/j.1365-2966.2009.15043.x},
	language = {en},
	number = {1},
	urldate = {2022-01-25},
	journal = {MNRAS},
	author = {Booth, C. M. and Schaye, Joop},
	month = sep,
	year = {2009},
	pages = {53--74},
}

@article{henden_fable_2018,
	title = {The {FABLE} simulations: a feedback model for galaxies, groups, and clusters},
	volume = {479},
	issn = {0035-8711, 1365-2966},
	shorttitle = {The {FABLE} simulations},
	url = {https://academic.oup.com/mnras/article/479/4/5385/5051752},
	doi = {10.1093/mnras/sty1780},
	language = {en},
	number = {4},
	urldate = {2022-01-25},
	journal = {MNRAS},
	author = {Henden, Nicholas A and Puchwein, Ewald and Shen, Sijing and Sijacki, Debora},
	month = oct,
	year = {2018},
	pages = {5385--5412},
}

@article{inayoshi_assembly_2020,
	title = {The {Assembly} of the {First} {Massive} {Black} {Holes}},
	volume = {58},
	url = {https://doi.org/10.1146/annurev-astro-120419-014455},
	doi = {10.1146/annurev-astro-120419-014455},
	number = {1},
	urldate = {2022-01-25},
	journal = {ARA\&A},
	author = {Inayoshi, Kohei and Visbal, Eli and Haiman, Zoltán},
	year = {2020},
	note = {\_eprint: https://doi.org/10.1146/annurev-astro-120419-014455},
	pages = {27--97},
}

@article{sijacki_gravitational_2011,
	title = {Gravitational recoils of supermassive black holes in hydrodynamical simulations of gas-rich galaxies},
	volume = {414},
	issn = {0035-8711},
	url = {https://ui.adsabs.harvard.edu/abs/2011MNRAS.414.3656S},
	doi = {10.1111/j.1365-2966.2011.18666.x},
	urldate = {2022-01-25},
	journal = {MNRAS},
	author = {Sijacki, Debora and Springel, Volker and Haehnelt, Martin G.},
	month = jul,
	year = {2011},
	note = {ADS Bibcode: 2011MNRAS.414.3656S},
	pages = {3656--3670},
}

@article{vogelsberger_model_2013,
	title = {A model for cosmological simulations of galaxy formation physics},
	volume = {436},
	issn = {1365-2966, 0035-8711},
	url = {http://academic.oup.com/mnras/article/436/4/3031/984888/A-model-for-cosmological-simulations-of-galaxy},
	doi = {10.1093/mnras/stt1789},
	language = {en},
	number = {4},
	urldate = {2022-01-25},
	journal = {MNRAS},
	author = {Vogelsberger, Mark and Genel, Shy and Sijacki, Debora and Torrey, Paul and Springel, Volker and Hernquist, Lars},
	month = dec,
	year = {2013},
	pages = {3031--3067},
}

@article{springel_modelling_2005,
	title = {Modelling feedback from stars and black holes in galaxy mergers},
	volume = {361},
	issn = {0035-8711},
	url = {https://ui.adsabs.harvard.edu/abs/2005MNRAS.361..776S},
	doi = {10.1111/j.1365-2966.2005.09238.x},
	urldate = {2022-01-25},
	journal = {MNRAS},
	author = {Springel, Volker and Di Matteo, Tiziana and Hernquist, Lars},
	month = aug,
	year = {2005},
	note = {ADS Bibcode: 2005MNRAS.361..776S},
	pages = {776--794},
}

@article{springel_e_2010,
	title = {E pur si muove: {Galilean}-invariant cosmological hydrodynamical simulations on a moving mesh},
	volume = {401},
	issn = {0035-8711},
	shorttitle = {E pur si muove},
	url = {https://ui.adsabs.harvard.edu/abs/2010MNRAS.401..791S},
	doi = {10.1111/j.1365-2966.2009.15715.x},
	urldate = {2022-01-25},
	journal = {MNRAS},
	author = {Springel, Volker},
	month = jan,
	year = {2010},
	note = {ADS Bibcode: 2010MNRAS.401..791S},
	pages = {791--851},
}

@article{sijacki_illustris_2015,
	title = {The {Illustris} simulation: the evolving population of black holes across cosmic time},
	volume = {452},
	issn = {0035-8711, 1365-2966},
	shorttitle = {The {Illustris} simulation},
	url = {https://academic.oup.com/mnras/article-lookup/doi/10.1093/mnras/stv1340},
	doi = {10.1093/mnras/stv1340},
	language = {en},
	number = {1},
	urldate = {2022-01-25},
	journal = {MNRAS},
	author = {Sijacki, Debora and Vogelsberger, Mark and Genel, Shy and Springel, Volker and Torrey, Paul and Snyder, Gregory F. and Nelson, Dylan and Hernquist, Lars},
	month = sep,
	year = {2015},
	pages = {575--596},
}

@article{koudmani_little_2021,
	title = {A little {FABLE}: exploring {AGN} feedback in dwarf galaxies with cosmological simulations},
	volume = {503},
	issn = {0035-8711},
	shorttitle = {A little {FABLE}},
	url = {https://doi.org/10.1093/mnras/stab677},
	doi = {10.1093/mnras/stab677},
	number = {3},
	urldate = {2021-09-06},
	journal = {MNRAS},
	author = {Koudmani, Sophie and Henden, Nicholas A and Sijacki, Debora},
	month = may,
	year = {2021},
	pages = {3568--3591},
}

@article{johansson_evolution_2009,
	title = {{THE} {EVOLUTION} {OF} {BLACK} {HOLE} {SCALING} {RELATIONS} {IN} {GALAXY} {MERGERS}},
	volume = {707},
	issn = {0004-637X, 1538-4357},
	url = {https://iopscience.iop.org/article/10.1088/0004-637X/707/2/L184},
	doi = {10.1088/0004-637X/707/2/L184},
	language = {en},
	number = {2},
	urldate = {2022-01-25},
	journal = {ApJ},
	author = {Johansson, Peter H. and Burkert, Andreas and Naab, Thorsten},
	month = dec,
	year = {2009},
	pages = {L184--L189},
}

@article{reines_relations_2015,
	title = {{RELATIONS} {BETWEEN} {CENTRAL} {BLACK} {HOLE} {MASS} {AND} {TOTAL} {GALAXY} {STELLAR} {MASS} {IN} {THE} {LOCAL} {UNIVERSE}},
	volume = {813},
	issn = {1538-4357},
	url = {https://iopscience.iop.org/article/10.1088/0004-637X/813/2/82},
	doi = {10.1088/0004-637X/813/2/82},
	language = {en},
	number = {2},
	urldate = {2022-01-25},
	journal = {ApJ},
	author = {Reines, Amy E. and Volonteri, Marta},
	month = oct,
	year = {2015},
	pages = {82},
}

@article{planck_collaboration_planck_2016,
	title = {\textit{{Planck}} 2015 results: {XIII}. {Cosmological} parameters},
	volume = {594},
	issn = {0004-6361, 1432-0746},
	shorttitle = {\textit{{Planck}} 2015 results},
	url = {http://www.aanda.org/10.1051/0004-6361/201525830},
	doi = {10.1051/0004-6361/201525830},
	language = {en},
	urldate = {2022-01-25},
	journal = {A\&A},
	author = {{Planck Collaboration} and Ade, P. A. R. and Aghanim, N. and Arnaud, M. and Ashdown, M. and Aumont, J. and Baccigalupi, C. and Banday, A. J. and Barreiro, R. B. and Bartlett, J. G. and Bartolo, N. and Battaner, E. and Battye, R. and Benabed, K. and Benoît, A. and Benoit-Lévy, A. and Bernard, J.-P. and Bersanelli, M. and Bielewicz, P. and Bock, J. J. and Bonaldi, A. and Bonavera, L. and Bond, J. R. and Borrill, J. and Bouchet, F. R. and Boulanger, F. and Bucher, M. and Burigana, C. and Butler, R. C. and Calabrese, E. and Cardoso, J.-F. and Catalano, A. and Challinor, A. and Chamballu, A. and Chary, R.-R. and Chiang, H. C. and Chluba, J. and Christensen, P. R. and Church, S. and Clements, D. L. and Colombi, S. and Colombo, L. P. L. and Combet, C. and Coulais, A. and Crill, B. P. and Curto, A. and Cuttaia, F. and Danese, L. and Davies, R. D. and Davis, R. J. and de Bernardis, P. and de Rosa, A. and de Zotti, G. and Delabrouille, J. and Désert, F.-X. and Di Valentino, E. and Dickinson, C. and Diego, J. M. and Dolag, K. and Dole, H. and Donzelli, S. and Doré, O. and Douspis, M. and Ducout, A. and Dunkley, J. and Dupac, X. and Efstathiou, G. and Elsner, F. and Enßlin, T. A. and Eriksen, H. K. and Farhang, M. and Fergusson, J. and Finelli, F. and Forni, O. and Frailis, M. and Fraisse, A. A. and Franceschi, E. and Frejsel, A. and Galeotta, S. and Galli, S. and Ganga, K. and Gauthier, C. and Gerbino, M. and Ghosh, T. and Giard, M. and Giraud-Héraud, Y. and Giusarma, E. and Gjerløw, E. and González-Nuevo, J. and Górski, K. M. and Gratton, S. and Gregorio, A. and Gruppuso, A. and Gudmundsson, J. E. and Hamann, J. and Hansen, F. K. and Hanson, D. and Harrison, D. L. and Helou, G. and Henrot-Versillé, S. and Hernández-Monteagudo, C. and Herranz, D. and Hildebrandt, S. R. and Hivon, E. and Hobson, M. and Holmes, W. A. and Hornstrup, A. and Hovest, W. and Huang, Z. and Huffenberger, K. M. and Hurier, G. and Jaffe, A. H. and Jaffe, T. R. and Jones, W. C. and Juvela, M. and Keihänen, E. and Keskitalo, R. and Kisner, T. S. and Kneissl, R. and Knoche, J. and Knox, L. and Kunz, M. and Kurki-Suonio, H. and Lagache, G. and Lähteenmäki, A. and Lamarre, J.-M. and Lasenby, A. and Lattanzi, M. and Lawrence, C. R. and Leahy, J. P. and Leonardi, R. and Lesgourgues, J. and Levrier, F. and Lewis, A. and Liguori, M. and Lilje, P. B. and Linden-Vørnle, M. and López-Caniego, M. and Lubin, P. M. and Macías-Pérez, J. F. and Maggio, G. and Maino, D. and Mandolesi, N. and Mangilli, A. and Marchini, A. and Maris, M. and Martin, P. G. and Martinelli, M. and Martínez-González, E. and Masi, S. and Matarrese, S. and McGehee, P. and Meinhold, P. R. and Melchiorri, A. and Melin, J.-B. and Mendes, L. and Mennella, A. and Migliaccio, M. and Millea, M. and Mitra, S. and Miville-Deschênes, M.-A. and Moneti, A. and Montier, L. and Morgante, G. and Mortlock, D. and Moss, A. and Munshi, D. and Murphy, J. A. and Naselsky, P. and Nati, F. and Natoli, P. and Netterfield, C. B. and Nørgaard-Nielsen, H. U. and Noviello, F. and Novikov, D. and Novikov, I. and Oxborrow, C. A. and Paci, F. and Pagano, L. and Pajot, F. and Paladini, R. and Paoletti, D. and Partridge, B. and Pasian, F. and Patanchon, G. and Pearson, T. J. and Perdereau, O. and Perotto, L. and Perrotta, F. and Pettorino, V. and Piacentini, F. and Piat, M. and Pierpaoli, E. and Pietrobon, D. and Plaszczynski, S. and Pointecouteau, E. and Polenta, G. and Popa, L. and Pratt, G. W. and Prézeau, G. and Prunet, S. and Puget, J.-L. and Rachen, J. P. and Reach, W. T. and Rebolo, R. and Reinecke, M. and Remazeilles, M. and Renault, C. and Renzi, A. and Ristorcelli, I. and Rocha, G. and Rosset, C. and Rossetti, M. and Roudier, G. and Rouillé d’Orfeuil, B. and Rowan-Robinson, M. and Rubiño-Martín, J. A. and Rusholme, B. and Said, N. and Salvatelli, V. and Salvati, L. and Sandri, M. and Santos, D. and Savelainen, M. and Savini, G. and Scott, D. and Seiffert, M. D. and Serra, P. and Shellard, E. P. S. and Spencer, L. D. and Spinelli, M. and Stolyarov, V. and Stompor, R. and Sudiwala, R. and Sunyaev, R. and Sutton, D. and Suur-Uski, A.-S. and Sygnet, J.-F. and Tauber, J. A. and Terenzi, L. and Toffolatti, L. and Tomasi, M. and Tristram, M. and Trombetti, T. and Tucci, M. and Tuovinen, J. and Türler, M. and Umana, G. and Valenziano, L. and Valiviita, J. and Van Tent, F. and Vielva, P. and Villa, F. and Wade, L. A. and Wandelt, B. D. and Wehus, I. K. and White, M. and White, S. D. M. and Wilkinson, A. and Yvon, D. and Zacchei, A. and Zonca, A.},
	month = oct,
	year = {2016},
	pages = {A13},
}

@article{angles-alcazar_cosmological_2021,
	title = {Cosmological {Simulations} of {Quasar} {Fueling} to {Subparsec} {Scales} {Using} {Lagrangian} {Hyper}-refinement},
	volume = {917},
	issn = {0004-637X},
	url = {https://ui.adsabs.harvard.edu/abs/2021ApJ...917...53A},
	doi = {10.3847/1538-4357/ac09e8},
	urldate = {2022-05-26},
	journal = {ApJ},
	author = {Anglés-Alcázar, Daniel and Quataert, Eliot and Hopkins, Philip F. and Somerville, Rachel S. and Hayward, Christopher C. and Faucher-Giguère, Claude-André and Bryan, Greg L. and Kereš, Dušan and Hernquist, Lars and Stone, James M.},
	month = aug,
	year = {2021},
	note = {ADS Bibcode: 2021ApJ...917...53A},
	pages = {53},
}

@article{massonneau_how_2023,
	title = {How the super-{Eddington} regime affects black hole spin evolution in high-redshift galaxies},
	volume = {669},
	issn = {0004-6361},
	url = {https://ui.adsabs.harvard.edu/abs/2023A&A...669A.143M},
	doi = {10.1051/0004-6361/202244874},
	urldate = {2023-12-03},
	journal = {A\&A},
	author = {Massonneau, Warren and Dubois, Yohan and Volonteri, Marta and Beckmann, Ricarda S.},
	month = jan,
	year = {2023},
	note = {ADS Bibcode: 2023A\&A...669A.143M},
	pages = {A143},
}

@article{power_accretion_2011,
	title = {The accretion disc particle method for simulations of black hole feeding and feedback},
	volume = {412},
	issn = {0035-8711},
	url = {https://ui.adsabs.harvard.edu/abs/2011MNRAS.412..269P},
	doi = {10.1111/j.1365-2966.2010.17901.x},
	urldate = {2023-03-21},
	journal = {MNRAS},
	author = {Power, Chris and Nayakshin, Sergei and King, Andrew},
	month = mar,
	year = {2011},
	note = {ADS Bibcode: 2011MNRAS.412..269P},
	pages = {269--276},
}

@article{guo_toward_2023,
	title = {Toward {Horizon}-scale {Accretion} onto {Supermassive} {Black} {Holes} in {Elliptical} {Galaxies}},
	volume = {946},
	issn = {0004-637X},
	url = {https://ui.adsabs.harvard.edu/abs/2023ApJ...946...26G},
	doi = {10.3847/1538-4357/acb81e},
	urldate = {2023-12-03},
	journal = {ApJ},
	author = {Guo, Minghao and Stone, James M. and Kim, Chang-Goo and Quataert, Eliot},
	month = mar,
	year = {2023},
	note = {ADS Bibcode: 2023ApJ...946...26G},
	pages = {26},
}

@article{bustamante_spin_2019,
	title = {Spin evolution and feedback of supermassive black holes in cosmological simulations},
	volume = {490},
	issn = {0035-8711},
	url = {https://ui.adsabs.harvard.edu/abs/2019MNRAS.490.4133B},
	doi = {10.1093/mnras/stz2836},
	urldate = {2023-03-21},
	journal = {MNRAS},
	author = {Bustamante, Sebastian and Springel, Volker},
	month = dec,
	year = {2019},
	note = {ADS Bibcode: 2019MNRAS.490.4133B},
	pages = {4133--4153},
}

@article{dubois_black_2014,
	title = {Black hole evolution - {II}. {Spinning} black holes in a supernova-driven turbulent interstellar medium},
	volume = {440},
	issn = {0035-8711},
	url = {https://ui.adsabs.harvard.edu/abs/2014MNRAS.440.2333D},
	doi = {10.1093/mnras/stu425},
	urldate = {2023-03-21},
	journal = {MNRAS},
	author = {Dubois, Yohan and Volonteri, Marta and Silk, Joseph and Devriendt, Julien and Slyz, Adrianne},
	month = may,
	year = {2014},
	note = {ADS Bibcode: 2014MNRAS.440.2333D},
	pages = {2333--2346},
}

@article{berti_cosmological_2008,
	title = {Cosmological {Black} {Hole} {Spin} {Evolution} by {Mergers} and {Accretion}},
	volume = {684},
	issn = {0004-637X},
	url = {https://ui.adsabs.harvard.edu/abs/2008ApJ...684..822B},
	doi = {10.1086/590379},
	urldate = {2023-10-18},
	journal = {ApJ},
	author = {Berti, Emanuele and Volonteri, Marta},
	month = sep,
	year = {2008},
	note = {ADS Bibcode: 2008ApJ...684..822B},
	pages = {822--828},
}

@article{rosas-guevara_impact_2015,
	title = {The impact of angular momentum on black hole accretion rates in simulations of galaxy formation},
	volume = {454},
	issn = {0035-8711},
	url = {https://ui.adsabs.harvard.edu/abs/2015MNRAS.454.1038R},
	doi = {10.1093/mnras/stv2056},
	urldate = {2023-10-18},
	journal = {MNRAS},
	author = {Rosas-Guevara, Y. M. and Bower, R. G. and Schaye, J. and Furlong, M. and Frenk, C. S. and Booth, C. M. and Crain, R. A. and Dalla Vecchia, C. and Schaller, M. and Theuns, T.},
	month = nov,
	year = {2015},
	note = {ADS Bibcode: 2015MNRAS.454.1038R},
	pages = {1038--1057},
}

@article{beckmann_dense_2019,
	title = {Dense gas formation and destruction in a simulated {Perseus}-like galaxy cluster with spin-driven black hole feedback},
	volume = {631},
	issn = {0004-6361},
	url = {https://ui.adsabs.harvard.edu/abs/2019A&A...631A..60B},
	doi = {10.1051/0004-6361/201936188},
	urldate = {2023-10-18},
	journal = {A\&A},
	author = {Beckmann, R. S. and Dubois, Y. and Guillard, P. and Salome, P. and Olivares, V. and Polles, F. and Cadiou, C. and Combes, F. and Hamer, S. and Lehnert, M. D. and Pineau des Forets, G.},
	month = nov,
	year = {2019},
	note = {ADS Bibcode: 2019A\&A...631A..60B},
	pages = {A60},
}

@article{power_inner_2003,
	title = {The inner structure of Λ{CDM} haloes - {I}. {A} numerical convergence study},
	volume = {338},
	issn = {0035-8711},
	url = {https://ui.adsabs.harvard.edu/abs/2003MNRAS.338...14P},
	doi = {10.1046/j.1365-8711.2003.05925.x},
	urldate = {2023-07-26},
	journal = {MNRAS},
	author = {Power, C. and Navarro, J. F. and Jenkins, A. and Frenk, C. S. and White, S. D. M. and Springel, V. and Stadel, J. and Quinn, T.},
	month = jan,
	year = {2003},
	note = {ADS Bibcode: 2003MNRAS.338...14P},
	pages = {14--34},
}

@article{koudmani_two_2022,
	title = {Two can play at that game: constraining the role of supernova and {AGN} feedback in dwarf galaxies with cosmological zoom-in simulations},
	volume = {516},
	issn = {0035-8711},
	shorttitle = {Two can play at that game},
	url = {https://ui.adsabs.harvard.edu/abs/2022MNRAS.516.2112K},
	doi = {10.1093/mnras/stac2252},
	urldate = {2022-10-16},
	journal = {MNRAS},
	author = {Koudmani, Sophie and Sijacki, Debora and Smith, Matthew C.},
	month = oct,
	year = {2022},
	note = {ADS Bibcode: 2022MNRAS.516.2112K},
	pages = {2112--2141},
}

@article{bourne_resolution_2015,
	title = {The resolution bias: low-resolution feedback simulations are better at destroying galaxies},
	volume = {453},
	issn = {0035-8711},
	shorttitle = {The resolution bias},
	url = {https://ui.adsabs.harvard.edu/abs/2015MNRAS.453.1829B},
	doi = {10.1093/mnras/stv1730},
	urldate = {2022-07-27},
	journal = {MNRAS},
	author = {Bourne, Martin A. and Zubovas, Kastytis and Nayakshin, Sergei},
	month = oct,
	year = {2015},
	note = {ADS Bibcode: 2015MNRAS.453.1829B},
	pages = {1829--1842},
}

@article{angles-alcazar_gravitational_2017,
	title = {Gravitational torque-driven black hole growth and feedback in cosmological simulations},
	volume = {464},
	issn = {0035-8711},
	url = {https://ui.adsabs.harvard.edu/abs/2017MNRAS.464.2840A},
	doi = {10.1093/mnras/stw2565},
	urldate = {2022-07-27},
	journal = {MNRAS},
	author = {Anglés-Alcázar, Daniel and Davé, Romeel and Faucher-Giguère, Claude-André and Özel, Feryal and Hopkins, Philip F.},
	month = jan,
	year = {2017},
	note = {ADS Bibcode: 2017MNRAS.464.2840A},
	pages = {2840--2853},
}

@article{hopkins_analytic_2011,
	title = {An analytic model of angular momentum transport by gravitational torques: from galaxies to massive black holes},
	volume = {415},
	issn = {0035-8711},
	shorttitle = {An analytic model of angular momentum transport by gravitational torques},
	url = {https://ui.adsabs.harvard.edu/abs/2011MNRAS.415.1027H},
	doi = {10.1111/j.1365-2966.2011.18542.x},
	urldate = {2022-07-27},
	journal = {MNRAS},
	author = {Hopkins, Philip F. and Quataert, Eliot},
	month = aug,
	year = {2011},
	note = {ADS Bibcode: 2011MNRAS.415.1027H},
	pages = {1027--1050},
}

@article{ma_seeds_2021,
	title = {Seeds don't sink: even massive black hole 'seeds' cannot migrate to galaxy centres efficiently},
	volume = {508},
	issn = {0035-8711},
	shorttitle = {Seeds don't sink},
	url = {https://ui.adsabs.harvard.edu/abs/2021MNRAS.508.1973M},
	doi = {10.1093/mnras/stab2713},
	urldate = {2022-07-26},
	journal = {MNRAS},
	author = {Ma, Linhao and Hopkins, Philip F. and Ma, Xiangcheng and Anglés-Alcázar, Daniel and Faucher-Giguère, Claude-André and Kelley, Luke Zoltan},
	month = dec,
	year = {2021},
	note = {ADS Bibcode: 2021MNRAS.508.1973M},
	pages = {1973--1985},
}

@article{pfister_erratic_2019,
	title = {The erratic dynamical life of black hole seeds in high-redshift galaxies},
	volume = {486},
	issn = {0035-8711},
	url = {https://ui.adsabs.harvard.edu/abs/2019MNRAS.486..101P},
	doi = {10.1093/mnras/stz822},
	urldate = {2022-07-26},
	journal = {MNRAS},
	author = {Pfister, Hugo and Volonteri, Marta and Dubois, Yohan and Dotti, Massimo and Colpi, Monica},
	month = jun,
	year = {2019},
	note = {ADS Bibcode: 2019MNRAS.486..101P},
	pages = {101--111},
}

@article{springel_cosmological_2003,
	title = {Cosmological smoothed particle hydrodynamics simulations: a hybrid multiphase model for star formation},
	volume = {339},
	issn = {0035-8711},
	shorttitle = {Cosmological smoothed particle hydrodynamics simulations},
	url = {https://ui.adsabs.harvard.edu/abs/2003MNRAS.339..289S},
	doi = {10.1046/j.1365-8711.2003.06206.x},
	urldate = {2022-06-16},
	journal = {MNRAS},
	author = {Springel, Volker and Hernquist, Lars},
	month = feb,
	year = {2003},
	note = {ADS Bibcode: 2003MNRAS.339..289S},
	pages = {289--311},
}

@article{curtis_resolving_2015,
	title = {Resolving flows around black holes: numerical technique and applications},
	volume = {454},
	issn = {0035-8711},
	shorttitle = {Resolving flows around black holes},
	url = {https://ui.adsabs.harvard.edu/abs/2015MNRAS.454.3445C},
	doi = {10.1093/mnras/stv2246},
	urldate = {2022-05-26},
	journal = {MNRAS},
	author = {Curtis, Michael and Sijacki, Debora},
	month = dec,
	year = {2015},
	note = {ADS Bibcode: 2015MNRAS.454.3445C},
	pages = {3445--3463},
}

@article{fiacconi_galactic_2018,
	title = {Galactic nuclei evolution with spinning black holes: method and implementation},
	volume = {477},
	issn = {0035-8711},
	shorttitle = {Galactic nuclei evolution with spinning black holes},
	url = {https://ui.adsabs.harvard.edu/abs/2018MNRAS.477.3807F},
	doi = {10.1093/mnras/sty893},
	urldate = {2022-05-26},
	journal = {MNRAS},
	author = {Fiacconi, Davide and Sijacki, Debora and Pringle, J. E.},
	month = jul,
	year = {2018},
	note = {ADS Bibcode: 2018MNRAS.477.3807F},
	pages = {3807--3835},
}

@article{dubois_dancing_2014,
	title = {Dancing in the dark: galactic properties trace spin swings along the cosmic web},
	volume = {444},
	issn = {0035-8711},
	shorttitle = {Dancing in the dark},
	url = {https://ui.adsabs.harvard.edu/abs/2014MNRAS.444.1453D},
	doi = {10.1093/mnras/stu1227},
	urldate = {2022-05-24},
	journal = {MNRAS},
	author = {Dubois, Y. and Pichon, C. and Welker, C. and Le Borgne, D. and Devriendt, J. and Laigle, C. and Codis, S. and Pogosyan, D. and Arnouts, S. and Benabed, K. and Bertin, E. and Blaizot, J. and Bouchet, F. and Cardoso, J. -F. and Colombi, S. and de Lapparent, V. and Desjacques, V. and Gavazzi, R. and Kassin, S. and Kimm, T. and McCracken, H. and Milliard, B. and Peirani, S. and Prunet, S. and Rouberol, S. and Silk, J. and Slyz, A. and Sousbie, T. and Teyssier, R. and Tresse, L. and Treyer, M. and Vibert, D. and Volonteri, M.},
	month = oct,
	year = {2014},
	note = {ADS Bibcode: 2014MNRAS.444.1453D},
	pages = {1453--1468},
}

@article{volonteri_black_2020,
	title = {Black hole mergers from dwarf to massive galaxies with the {NewHorizon} and {Horizon}-{AGN} simulations},
	volume = {498},
	issn = {0035-8711},
	url = {https://ui.adsabs.harvard.edu/abs/2020MNRAS.498.2219V},
	doi = {10.1093/mnras/staa2384},
	urldate = {2022-04-29},
	journal = {MNRAS},
	author = {Volonteri, Marta and Pfister, Hugo and Beckmann, Ricarda S. and Dubois, Yohan and Colpi, Monica and Conselice, Christopher J. and Dotti, Massimo and Martin, Garreth and Jackson, Ryan and Kraljic, Katarina and Pichon, Christophe and Trebitsch, Maxime and Yi, Sukyoung K. and Devriendt, Julien and Peirani, Sébastien},
	month = oct,
	year = {2020},
	note = {ADS Bibcode: 2020MNRAS.498.2219V},
	pages = {2219--2238},
}

@article{bellovary_origins_2021,
	title = {The origins of off-centre massive black holes in dwarf galaxies},
	volume = {505},
	issn = {0035-8711},
	url = {https://ui.adsabs.harvard.edu/abs/2021MNRAS.505.5129B},
	doi = {10.1093/mnras/stab1665},
	urldate = {2022-04-29},
	journal = {MNRAS},
	author = {Bellovary, Jillian M. and Hayoune, Sarra and Chafla, Katheryn and Vincent, Donovan and Brooks, Alyson and Christensen, Charlotte R. and Munshi, Ferah D. and Tremmel, Michael and Quinn, Thomas R. and Van Nest, Jordan and Sligh, Serena K. and Luzuriaga, Michelle},
	month = aug,
	year = {2021},
	note = {ADS Bibcode: 2021MNRAS.505.5129B},
	pages = {5129--5141},
}

@article{reines_dwarf_2013,
	title = {Dwarf {Galaxies} with {Optical} {Signatures} of {Active} {Massive} {Black} {Holes}},
	volume = {775},
	issn = {0004-637X},
	url = {https://ui.adsabs.harvard.edu/abs/2013ApJ...775..116R},
	doi = {10.1088/0004-637X/775/2/116},
	urldate = {2022-04-28},
	journal = {ApJ},
	author = {Reines, Amy E. and Greene, Jenny E. and Geha, Marla},
	month = oct,
	year = {2013},
	note = {ADS Bibcode: 2013ApJ...775..116R},
	pages = {116},
}

@article{greene_active_2004,
	title = {Active {Galactic} {Nuclei} with {Candidate} {Intermediate}-{Mass} {Black} {Holes}},
	volume = {610},
	issn = {0004-637X},
	url = {https://ui.adsabs.harvard.edu/abs/2004ApJ...610..722G},
	doi = {10.1086/421719},
	urldate = {2022-04-28},
	journal = {ApJ},
	author = {Greene, Jenny E. and Ho, Luis C.},
	month = aug,
	year = {2004},
	note = {ADS Bibcode: 2004ApJ...610..722G},
	pages = {722--736},
}

@article{greene_new_2007,
	title = {A {New} {Sample} of {Low}-{Mass} {Black} {Holes} in {Active} {Galaxies}},
	volume = {670},
	issn = {0004-637X},
	url = {https://ui.adsabs.harvard.edu/abs/2007ApJ...670...92G},
	doi = {10.1086/522082},
	urldate = {2022-04-28},
	journal = {ApJ},
	author = {Greene, Jenny E. and Ho, Luis C.},
	month = nov,
	year = {2007},
	note = {ADS Bibcode: 2007ApJ...670...92G},
	pages = {92--104},
}

@article{shen_bolometric_2020,
	title = {The bolometric quasar luminosity function at z = 0-7},
	volume = {495},
	issn = {0035-8711},
	url = {https://ui.adsabs.harvard.edu/abs/2020MNRAS.495.3252S},
	doi = {10.1093/mnras/staa1381},
	urldate = {2022-03-28},
	journal = {MNRAS},
	author = {Shen, Xuejian and Hopkins, Philip F. and Faucher-Giguère, Claude-André and Alexander, D. M. and Richards, Gordon T. and Ross, Nicholas P. and Hickox, R. C.},
	month = jan,
	year = {2020},
	note = {ADS Bibcode: 2020MNRAS.495.3252S},
	pages = {3252--3275},
}

@article{xiao_exploring_2011,
	title = {Exploring the {Low}-mass {End} of the {M} {BH}-σ* {Relation} with {Active} {Galaxies}},
	volume = {739},
	issn = {0004-637X},
	url = {https://ui.adsabs.harvard.edu/abs/2011ApJ...739...28X},
	doi = {10.1088/0004-637X/739/1/28},
	urldate = {2022-03-23},
	journal = {ApJ},
	author = {Xiao, Ting and Barth, Aaron J. and Greene, Jenny E. and Ho, Luis C. and Bentz, Misty C. and Ludwig, Randi R. and Jiang, Yanfei},
	month = sep,
	year = {2011},
	note = {ADS Bibcode: 2011ApJ...739...28X},
	pages = {28},
}

@article{kormendy_coevolution_2013,
	title = {Coevolution ({Or} {Not}) of {Supermassive} {Black} {Holes} and {Host} {Galaxies}},
	volume = {51},
	issn = {0066-4146},
	url = {https://ui.adsabs.harvard.edu/abs/2013ARA%26A..51..511K/abstract},
	doi = {10.1146/annurev-astro-082708-101811},
	language = {en},
	number = {1},
	urldate = {2022-03-23},
	journal = {ARA\&A},
	author = {Kormendy, John and Ho, Luis C.},
	month = aug,
	year = {2013},
	pages = {511},
}

@article{henden_redshift_2019,
	title = {The redshift evolution of {X}-ray and {Sunyaev}-{Zel}'dovich scaling relations in the {FABLE} simulations},
	volume = {489},
	issn = {0035-8711},
	url = {https://ui.adsabs.harvard.edu/abs/2019MNRAS.489.2439H},
	doi = {10.1093/mnras/stz2301},
	urldate = {2025-09-03},
	journal = {MNRAS},
	author = {Henden, Nicholas A. and Puchwein, Ewald and Sijacki, Debora},
	month = oct,
	year = {2019},
	note = {ADS Bibcode: 2019MNRAS.489.2439H},
	pages = {2439--2470},
}

@article{henden_baryon_2020,
	title = {The baryon content of groups and clusters of galaxies in the {FABLE} simulations},
	volume = {498},
	issn = {0035-8711},
	url = {https://ui.adsabs.harvard.edu/abs/2020MNRAS.498.2114H},
	doi = {10.1093/mnras/staa2235},
	urldate = {2025-09-03},
	journal = {MNRAS},
	author = {Henden, Nicholas A. and Puchwein, Ewald and Sijacki, Debora},
	month = oct,
	year = {2020},
	note = {ADS Bibcode: 2020MNRAS.498.2114H},
	pages = {2114--2137},
}

@article{katz_cosmological_1996,
	title = {Cosmological {Simulations} with {TreeSPH}},
	volume = {105},
	issn = {0067-0049},
	url = {https://ui.adsabs.harvard.edu/abs/1996ApJS..105...19K},
	doi = {10.1086/192305},
	urldate = {2025-09-03},
	journal = {ApJS},
	author = {Katz, Neal and Weinberg, David H. and Hernquist, Lars},
	month = jul,
	year = {1996},
	note = {ADS Bibcode: 1996ApJS..105...19K},
	pages = {19},
}

@article{wiersma_effect_2009,
	title = {The effect of photoionization on the cooling rates of enriched, astrophysical plasmas},
	volume = {393},
	issn = {0035-8711},
	url = {https://ui.adsabs.harvard.edu/abs/2009MNRAS.393...99W},
	doi = {10.1111/j.1365-2966.2008.14191.x},
	urldate = {2025-09-03},
	journal = {MNRAS},
	author = {Wiersma, Robert P. C. and Schaye, Joop and Smith, Britton D.},
	month = feb,
	year = {2009},
	note = {ADS Bibcode: 2009MNRAS.393...99W},
	pages = {99--107},
}

@article{wiersma_chemical_2009,
	title = {Chemical enrichment in cosmological, smoothed particle hydrodynamics simulations},
	volume = {399},
	issn = {0035-8711},
	url = {https://ui.adsabs.harvard.edu/abs/2009MNRAS.399..574W},
	doi = {10.1111/j.1365-2966.2009.15331.x},
	urldate = {2025-09-03},
	journal = {MNRAS},
	author = {Wiersma, Robert P. C. and Schaye, Joop and Theuns, Tom and Dalla Vecchia, Claudio and Tornatore, Luca},
	month = oct,
	year = {2009},
	note = {ADS Bibcode: 2009MNRAS.399..574W},
	pages = {574--600},
}

@article{gaspari_chaotic_2013,
	title = {Chaotic cold accretion on to black holes},
	volume = {432},
	issn = {0035-8711},
	url = {https://ui.adsabs.harvard.edu/abs/2013MNRAS.432.3401G},
	doi = {10.1093/mnras/stt692},
	urldate = {2025-09-24},
	journal = {MNRAS},
	author = {Gaspari, M. and Ruszkowski, M. and Oh, S. Peng},
	month = jul,
	year = {2013},
	note = {ADS Bibcode: 2013MNRAS.432.3401G},
	pages = {3401--3422},
}

@article{marinacci_formation_2014,
	title = {The formation of disc galaxies in high-resolution moving-mesh cosmological simulations},
	volume = {437},
	issn = {0035-8711},
	url = {https://ui.adsabs.harvard.edu/abs/2014MNRAS.437.1750M},
	doi = {10.1093/mnras/stt2003},
	urldate = {2025-09-24},
	journal = {MNRAS},
	author = {Marinacci, Federico and Pakmor, Rüdiger and Springel, Volker},
	month = jan,
	year = {2014},
	note = {ADS Bibcode: 2014MNRAS.437.1750M},
	pages = {1750--1775},
}

@article{sijacki_unified_2007,
	title = {A unified model for {AGN} feedback in cosmological simulations of structure formation},
	volume = {380},
	issn = {0035-8711},
	url = {https://ui.adsabs.harvard.edu/abs/2007MNRAS.380..877S},
	doi = {10.1111/j.1365-2966.2007.12153.x},
	urldate = {2025-09-26},
	journal = {MNRAS},
	author = {Sijacki, Debora and Springel, Volker and Di Matteo, Tiziana and Hernquist, Lars},
	month = sep,
	year = {2007},
	note = {ADS Bibcode: 2007MNRAS.380..877S},
	pages = {877--900},
}

@article{maiolino_jades_2024,
	title = {{JADES}: {The} diverse population of infant black holes at 4 {\textless} z {\textless} 11: {Merging}, tiny, poor, but mighty},
	volume = {691},
	issn = {0004-6361},
	shorttitle = {{JADES}},
	url = {https://ui.adsabs.harvard.edu/abs/2024A&A...691A.145M},
	doi = {10.1051/0004-6361/202347640},
	urldate = {2025-09-26},
	journal = {A\&A},
	publisher = {EDP},
	author = {Maiolino, Roberto and Scholtz, Jan and Curtis-Lake, Emma and Carniani, Stefano and Baker, William and de Graaff, Anna and Tacchella, Sandro and Übler, Hannah and D'Eugenio, Francesco and Witstok, Joris and Curti, Mirko and Arribas, Santiago and Bunker, Andrew J. and Charlot, Stéphane and Chevallard, Jacopo and Eisenstein, Daniel J. and Egami, Eiichi and Ji, Zhiyuan and Jones, Gareth C. and Lyu, Jianwei and Rawle, Tim and Robertson, Brant and Rujopakarn, Wiphu and Perna, Michele and Sun, Fengwu and Venturi, Giacomo and Williams, Christina C. and Willott, Chris},
	month = nov,
	year = {2024},
	note = {ADS Bibcode: 2024A\&A...691A.145M},
	pages = {A145},
}

@article{di_matteo_energy_2005,
	title = {Energy input from quasars regulates the growth and activity of black holes and their host galaxies},
	volume = {433},
	issn = {0028-0836},
	url = {https://ui.adsabs.harvard.edu/abs/2005Natur.433..604D},
	doi = {10.1038/nature03335},
	urldate = {2025-09-26},
	journal = {Nature},
	author = {Di Matteo, Tiziana and Springel, Volker and Hernquist, Lars},
	month = feb,
	year = {2005},
	note = {ADS Bibcode: 2005Natur.433..604D},
	pages = {604--607},
}

@article{ciotti_feedback_2010,
	title = {Feedback from {Central} {Black} {Holes} in {Elliptical} {Galaxies}. {III}. {Models} with {Both} {Radiative} and {Mechanical} {Feedback}},
	volume = {717},
	issn = {0004-637X},
	url = {https://ui.adsabs.harvard.edu/abs/2010ApJ...717..708C},
	doi = {10.1088/0004-637X/717/2/708},
	urldate = {2025-09-26},
	journal = {ApJ},
	author = {Ciotti, Luca and Ostriker, Jeremiah P. and Proga, Daniel},
	month = jul,
	year = {2010},
	note = {ADS Bibcode: 2010ApJ...717..708C},
	pages = {708--723},
}

@article{torrey_instability_2017,
	title = {An instability of feedback-regulated star formation in galactic nuclei},
	volume = {467},
	issn = {0035-8711},
	url = {https://ui.adsabs.harvard.edu/abs/2017MNRAS.467.2301T},
	doi = {10.1093/mnras/stx254},
	urldate = {2025-09-26},
	journal = {MNRAS},
	author = {Torrey, Paul and Hopkins, Philip F. and Faucher-Giguère, Claude-André and Vogelsberger, Mark and Quataert, Eliot and Kereš, Dušan and Murray, Norman},
	month = may,
	year = {2017},
	note = {ADS Bibcode: 2017MNRAS.467.2301T},
	pages = {2301--2314},
}

@article{costa_quenching_2018,
	title = {Quenching star formation with quasar outflows launched by trapped {IR} radiation},
	volume = {479},
	issn = {0035-8711},
	url = {https://ui.adsabs.harvard.edu/abs/2018MNRAS.479.2079C},
	doi = {10.1093/mnras/sty1514},
	urldate = {2025-09-26},
	journal = {MNRAS},
	author = {Costa, Tiago and Rosdahl, Joakim and Sijacki, Debora and Haehnelt, Martin G.},
	month = sep,
	year = {2018},
	note = {ADS Bibcode: 2018MNRAS.479.2079C},
	pages = {2079--2111},
}

@article{mcnamara_heating_2007,
	title = {Heating {Hot} {Atmospheres} with {Active} {Galactic} {Nuclei}},
	volume = {45},
	issn = {0066-4146},
	url = {https://ui.adsabs.harvard.edu/abs/2007ARA&A..45..117M},
	doi = {10.1146/annurev.astro.45.051806.110625},
	urldate = {2025-09-26},
	journal = {ARA\&A},
	author = {McNamara, B. R. and Nulsen, P. E. J.},
	month = sep,
	year = {2007},
	note = {ADS Bibcode: 2007ARA\&A..45..117M},
	pages = {117--175},
}

@article{fabian_observational_2012,
	title = {Observational {Evidence} of {Active} {Galactic} {Nuclei} {Feedback}},
	volume = {50},
	issn = {0066-4146},
	url = {https://ui.adsabs.harvard.edu/abs/2012ARA&A..50..455F},
	doi = {10.1146/annurev-astro-081811-125521},
	urldate = {2025-09-26},
	journal = {ARA\&A},
	author = {Fabian, A. C.},
	month = sep,
	year = {2012},
	note = {ADS Bibcode: 2012ARA\&A..50..455F},
	pages = {455--489},
}

@article{davis_evolution_1985,
	title = {The evolution of large-scale structure in a universe dominated by cold dark matter},
	volume = {292},
	issn = {0004-637X},
	url = {https://ui.adsabs.harvard.edu/abs/1985ApJ...292..371D},
	doi = {10.1086/163168},
	urldate = {2025-09-26},
	journal = {ApJ},
	publisher = {IOP},
	author = {Davis, M. and Efstathiou, G. and Frenk, C. S. and White, S. D. M.},
	month = may,
	year = {1985},
	note = {ADS Bibcode: 1985ApJ...292..371D},
	pages = {371--394},
}

@article{springel_gadget_2001,
	title = {{GADGET}: a code for collisionless and gasdynamical cosmological simulations},
	volume = {6},
	issn = {1384-1076},
	shorttitle = {{GADGET}},
	url = {https://ui.adsabs.harvard.edu/abs/2001NewA....6...79S},
	doi = {10.1016/S1384-1076(01)00042-2},
	urldate = {2025-09-26},
	journal = {New Astronomy},
	author = {Springel, Volker and Yoshida, Naoki and White, Simon D. M.},
	month = apr,
	year = {2001},
	note = {ADS Bibcode: 2001NewA....6...79S},
	pages = {79--117},
}

@article{dolag_substructures_2009,
	title = {Substructures in hydrodynamical cluster simulations},
	volume = {399},
	issn = {0035-8711},
	url = {https://ui.adsabs.harvard.edu/abs/2009MNRAS.399..497D},
	doi = {10.1111/j.1365-2966.2009.15034.x},
	urldate = {2025-09-26},
	journal = {MNRAS},
	author = {Dolag, K. and Borgani, S. and Murante, G. and Springel, V.},
	month = oct,
	year = {2009},
	note = {ADS Bibcode: 2009MNRAS.399..497D},
	pages = {497--514},
}

@article{rodriguez-gomez_merger_2015,
	title = {The merger rate of galaxies in the {Illustris} simulation: a comparison with observations and semi-empirical models},
	volume = {449},
	issn = {0035-8711},
	shorttitle = {The merger rate of galaxies in the {Illustris} simulation},
	url = {https://ui.adsabs.harvard.edu/abs/2015MNRAS.449...49R},
	doi = {10.1093/mnras/stv264},
	urldate = {2025-09-26},
	journal = {MNRAS},
	author = {Rodriguez-Gomez, Vicente and Genel, Shy and Vogelsberger, Mark and Sijacki, Debora and Pillepich, Annalisa and Sales, Laura V. and Torrey, Paul and Snyder, Greg and Nelson, Dylan and Springel, Volker and Ma, Chung-Pei and Hernquist, Lars},
	month = may,
	year = {2015},
	note = {ADS Bibcode: 2015MNRAS.449...49R},
	pages = {49--64},
}

@article{lupi_size_2024,
	title = {Size matters: are we witnessing super-{Eddington} accretion in high-redshift black holes from {JWST}?},
	volume = {689},
	issn = {0004-6361},
	shorttitle = {Size matters},
	url = {https://ui.adsabs.harvard.edu/abs/2024A&A...689A.128L},
	doi = {10.1051/0004-6361/202451249},
	urldate = {2025-12-18},
	journal = {A\&A},
	publisher = {EDP},
	author = {Lupi, Alessandro and Trinca, Alessandro and Volonteri, Marta and Dotti, Massimo and Mazzucchelli, Chiara},
	month = sep,
	year = {2024},
	note = {ADS Bibcode: 2024A\&A...689A.128L},
	pages = {A128},
}

@article{rusakov_little_2026,
	title = {Little red dots as young supermassive black holes in dense ionized cocoons},
	volume = {649},
	issn = {0028-0836},
	url = {https://ui.adsabs.harvard.edu/abs/2026Natur.649..574R},
	doi = {10.1038/s41586-025-09900-4},
	urldate = {2026-01-26},
	journal = {Nature},
	author = {Rusakov, V. and Watson, D. and Nikopoulos, G. P. and Brammer, G. and Gottumukkala, R. and Harvey, T. and Heintz, K. E. and Damgaard, R. and Sim, S. A. and Sneppen, A. and Vijayan, A. P. and Adams, N. and Austin, D. and Conselice, C. J. and Goolsby, C. M. and Toft, S. and Witstok, J.},
	month = jan,
	year = {2026},
	note = {ADS Bibcode: 2026Natur.649..574R},
	pages = {574--579},
}

@article{greene_what_2026,
	title = {What {You} {See} {Is} {What} {You} {Get}: {Empirically} {Measured} {Bolometric} {Luminosities} of {Little} {Red} {Dots}},
	volume = {996},
	issn = {0004-637X},
	shorttitle = {What {You} {See} {Is} {What} {You} {Get}},
	url = {https://ui.adsabs.harvard.edu/abs/2026ApJ...996..129G},
	doi = {10.3847/1538-4357/ae1836},
	urldate = {2026-01-26},
	journal = {ApJ},
	publisher = {IOP},
	author = {Greene, Jenny E. and Setton, David J. and Furtak, Lukas J. and Naidu, Rohan P. and Volonteri, Marta and Dayal, Pratika and Labbe, Ivo and van Dokkum, Pieter and Bezanson, Rachel and Brammer, Gabriel and Cutler, Sam E. and Glazebrook, Karl and de Graaff, Anna and Hirschmann, Michaela and Hviding, Raphael E. and Kokorev, Vasily and Leja, Joel and Liu, Hanpu and Ma, Yilun and Matthee, Jorryt and Nanayakkara, Themiya and Oesch, Pascal A. and Pan, Richard and Price, Sedona H. and Spilker, Justin S. and Wang, Bingjie and Weaver, John R. and Whitaker, Katherine E. and Williams, Christina C. and Zitrin, Adi},
	month = jan,
	year = {2026},
	note = {ADS Bibcode: 2026ApJ...996..129G},
	pages = {129},
}

@article{matthee_little_2024,
	title = {Little {Red} {Dots}: {An} {Abundant} {Population} of {Faint} {Active} {Galactic} {Nuclei} at z ∼ 5 {Revealed} by the {EIGER} and {FRESCO} {JWST} {Surveys}},
	volume = {963},
	issn = {0004-637X},
	shorttitle = {Little {Red} {Dots}},
	url = {https://ui.adsabs.harvard.edu/abs/2024ApJ...963..129M},
	doi = {10.3847/1538-4357/ad2345},
	urldate = {2026-02-01},
	journal = {ApJ},
	publisher = {IOP},
	author = {Matthee, Jorryt and Naidu, Rohan P. and Brammer, Gabriel and Chisholm, John and Eilers, Anna-Christina and Goulding, Andy and Greene, Jenny and Kashino, Daichi and Labbe, Ivo and Lilly, Simon J. and Mackenzie, Ruari and Oesch, Pascal A. and Weibel, Andrea and Wuyts, Stijn and Xiao, Mengyuan and Bordoloi, Rongmon and Bouwens, Rychard and van Dokkum, Pieter and Illingworth, Garth and Kramarenko, Ivan and Maseda, Michael V. and Mason, Charlotte and Meyer, Romain A. and Nelson, Erica J. and Reddy, Naveen A. and Shivaei, Irene and Simcoe, Robert A. and Yue, Minghao},
	month = mar,
	year = {2024},
	note = {ADS Bibcode: 2024ApJ...963..129M},
	pages = {129},
}

@article{furtak_high_2024,
	title = {A high black-hole-to-host mass ratio in a lensed {AGN} in the early {Universe}},
	volume = {628},
	copyright = {2024 The Author(s), under exclusive licence to Springer Nature Limited},
	issn = {1476-4687},
	url = {https://www.nature.com/articles/s41586-024-07184-8},
	doi = {10.1038/s41586-024-07184-8},
	language = {en},
	number = {8006},
	urldate = {2026-02-01},
	journal = {Nature},
	publisher = {Nature Publishing Group},
	author = {Furtak, Lukas J. and Labbé, Ivo and Zitrin, Adi and Greene, Jenny E. and Dayal, Pratika and Chemerynska, Iryna and Kokorev, Vasily and Miller, Tim B. and Goulding, Andy D. and de Graaff, Anna and Bezanson, Rachel and Brammer, Gabriel B. and Cutler, Sam E. and Leja, Joel and Pan, Richard and Price, Sedona H. and Wang, Bingjie and Weaver, John R. and Whitaker, Katherine E. and Atek, Hakim and Bogdán, Ákos and Charlot, Stéphane and Curtis-Lake, Emma and van Dokkum, Pieter and Endsley, Ryan and Feldmann, Robert and Fudamoto, Yoshinobu and Fujimoto, Seiji and Glazebrook, Karl and Juneau, Stéphanie and Marchesini, Danilo and Maseda, Micheal V. and Nelson, Erica and Oesch, Pascal A. and Plat, Adèle and Setton, David J. and Stark, Daniel P. and Williams, Christina C.},
	month = apr,
	year = {2024},
	pages = {57--61},
}

@article{taylor_capers-lrd-z9_2025,
	title = {{CAPERS}-{LRD}-z9: {A} {Gas}-enshrouded {Little} {Red} {Dot} {Hosting} a {Broad}-line {Active} {Galactic} {Nucleus} at z = 9.288},
	volume = {989},
	issn = {0004-637X},
	shorttitle = {{CAPERS}-{LRD}-z9},
	url = {https://ui.adsabs.harvard.edu/abs/2025ApJ...989L...7T},
	doi = {10.3847/2041-8213/ade789},
	urldate = {2026-02-01},
	journal = {ApJ},
	publisher = {IOP},
	author = {Taylor, Anthony J. and Kokorev, Vasily and Kocevski, Dale D. and Akins, Hollis B. and Cullen, Fergus and Dickinson, Mark and Finkelstein, Steven L. and Arrabal Haro, Pablo and Bromm, Volker and Giavalisco, Mauro and Inayoshi, Kohei and Juneau, Stéphanie and Leung, Gene C. K. and Pérez-González, Pablo G. and Somerville, Rachel S. and Trump, Jonathan R. and Amorín, Ricardo O. and Barro, Guillermo and Burgarella, Denis and Brooks, Madisyn and Carnall, Adam C. and Casey, Caitlin M. and Cheng, Yingjie and Chisholm, John and Chworowsky, Katherine and Davis, Kelcey and Donnan, Callum T. and Dunlop, James S. and Ellis, Richard S. and Fernández, Vital and Fujimoto, Seiji and Grogin, Norman A. and Gupta, Ansh R. and Hathi, Nimish P. and Jung, Intae and Hirschmann, Michaela and Kartaltepe, Jeyhan S. and Koekemoer, Anton M. and Larson, Rebecca L. and Leung, Ho-Hin and Llerena, Mario and Lucas, Ray A. and McLeod, Derek J. and McLure, Ross and Napolitano, Lorenzo and Papovich, Casey and Stanton, Thomas M. and Tripodi, Roberta and Wang, Xin and Wilkins, Stephen M. and Yung, L. Y. Aaron and Zavala, Jorge A.},
	month = aug,
	year = {2025},
	note = {ADS Bibcode: 2025ApJ...989L...7T},
	pages = {L7},
}

@article{kovacs_candidate_2024,
	title = {A {Candidate} {Supermassive} {Black} {Hole} in a {Gravitationally} {Lensed} {Galaxy} at {Z} ≈ 10},
	volume = {965},
	issn = {0004-637X},
	url = {https://ui.adsabs.harvard.edu/abs/2024ApJ...965L..21K},
	doi = {10.3847/2041-8213/ad391f},
	urldate = {2026-02-01},
	journal = {ApJ},
	publisher = {IOP},
	author = {Kovács, Orsolya E. and Bogdán, Ákos and Natarajan, Priyamvada and Werner, Norbert and Azadi, Mojegan and Volonteri, Marta and Tremblay, Grant R. and Chadayammuri, Urmila and Forman, William R. and Jones, Christine and Kraft, Ralph P.},
	month = apr,
	year = {2024},
	note = {ADS Bibcode: 2024ApJ...965L..21K},
	pages = {L21},
}

@article{kennicutt_global_1998,
	title = {The {Global} {Schmidt} {Law} in {Star}-forming {Galaxies}},
	volume = {498},
	issn = {0004-637X},
	url = {https://ui.adsabs.harvard.edu/abs/1998ApJ...498..541K},
	doi = {10.1086/305588},
	urldate = {2026-02-02},
	journal = {ApJ},
	publisher = {IOP},
	author = {Kennicutt, Jr., Robert C.},
	month = may,
	year = {1998},
	note = {ADS Bibcode: 1998ApJ...498..541K},
	pages = {541--552},
}

@article{pacucci_jwst_2023,
	title = {{JWST} {CEERS} and {JADES} {Active} {Galaxies} at z = 4-7 {Violate} the {Local} {M} •-{M} ⋆ {Relation} at {\textgreater}3σ: {Implications} for {Low}-mass {Black} {Holes} and {Seeding} {Models}},
	volume = {957},
	issn = {0004-637X},
	shorttitle = {{JWST} {CEERS} and {JADES} {Active} {Galaxies} at z = 4-7 {Violate} the {Local} {M} •-{M} ⋆ {Relation} at {\textgreater}3σ},
	url = {https://ui.adsabs.harvard.edu/abs/2023ApJ...957L...3P},
	doi = {10.3847/2041-8213/ad0158},
	urldate = {2026-02-03},
	journal = {ApJ},
	publisher = {IOP},
	author = {Pacucci, Fabio and Nguyen, Bao and Carniani, Stefano and Maiolino, Roberto and Fan, Xiaohui},
	month = nov,
	year = {2023},
	note = {ADS Bibcode: 2023ApJ...957L...3P},
	pages = {L3},
}

@article{mezcua_overmassive_2023,
	title = {Overmassive {Black} {Holes} in {Dwarf} {Galaxies} {Out} to z 0.9 in the {VIPERS} {Survey}},
	volume = {943},
	issn = {0004-637X},
	url = {https://ui.adsabs.harvard.edu/abs/2023ApJ...943L...5M},
	doi = {10.3847/2041-8213/acae25},
	urldate = {2026-02-03},
	journal = {ApJ},
	publisher = {IOP},
	author = {Mezcua, Mar and Siudek, Malgorzata and Suh, Hyewon and Valiante, Rosa and Spinoso, Daniele and Bonoli, Silvia},
	month = jan,
	year = {2023},
	note = {ADS Bibcode: 2023ApJ...943L...5M},
	pages = {L5},
}

@article{suh_no_2020,
	title = {No {Significant} {Evolution} of {Relations} between {Black} {Hole} {Mass} and {Galaxy} {Total} {Stellar} {Mass} {Up} to z ∼ 2.5},
	volume = {889},
	issn = {0004-637X},
	url = {https://ui.adsabs.harvard.edu/abs/2020ApJ...889...32S},
	doi = {10.3847/1538-4357/ab5f5f},
	urldate = {2026-02-03},
	journal = {ApJ},
	publisher = {IOP},
	author = {Suh, Hyewon and Civano, Francesca and Trakhtenbrot, Benny and Shankar, Francesco and Hasinger, Günther and Sanders, David B. and Allevato, Viola},
	month = jan,
	year = {2020},
	note = {ADS Bibcode: 2020ApJ...889...32S},
	pages = {32},
}

@article{li_tip_2025,
	title = {Tip of the {Iceberg}: {Overmassive} {Black} {Holes} at 4 {\textless} z {\textless} 7 {Found} by {JWST} {Are} {Not} {Inconsistent} with the {Local} {Relation}},
	volume = {981},
	issn = {0004-637X},
	shorttitle = {Tip of the {Iceberg}},
	url = {https://ui.adsabs.harvard.edu/abs/2025ApJ...981...19L},
	doi = {10.3847/1538-4357/ada603},
	urldate = {2026-02-05},
	journal = {ApJ},
	publisher = {IOP},
	author = {Li, Junyao and Silverman, John D. and Shen, Yue and Volonteri, Marta and Jahnke, Knud and Zhuang, Ming-Yang and Scoggins, Matthew T. and Ding, Xuheng and Harikane, Yuichi and Onoue, Masafusa and Tanaka, Takumi S.},
	month = mar,
	year = {2025},
	note = {ADS Bibcode: 2025ApJ...981...19L},
	pages = {19},
}

@article{weller_discrepancies_2025,
	title = {Discrepancies between {JWST} {Observations} and {Simulations} of {Quenched} {Massive} {Galaxies} at z {\textgreater} 3: {A} {Comparative} {Study} with {IllustrisTNG} and {ASTRID}},
	volume = {979},
	issn = {0004-637X},
	shorttitle = {Discrepancies between {JWST} {Observations} and {Simulations} of {Quenched} {Massive} {Galaxies} at z {\textgreater} 3},
	url = {https://ui.adsabs.harvard.edu/abs/2025ApJ...979..181W},
	doi = {10.3847/1538-4357/ada360},
	urldate = {2026-02-05},
	journal = {ApJ},
	publisher = {IOP},
	author = {Weller, Emma Jane and Pacucci, Fabio and Ni, Yueying and Hernquist, Lars and Park, Minjung},
	month = feb,
	year = {2025},
	note = {ADS Bibcode: 2025ApJ...979..181W},
	pages = {181},
}

@article{carnall_surprising_2023,
	title = {A surprising abundance of massive quiescent galaxies at 3 {\textless} z {\textless} 5 in the first data from {JWST} {CEERS}},
	volume = {520},
	issn = {0035-8711},
	url = {https://doi.org/10.1093/mnras/stad369},
	doi = {10.1093/mnras/stad369},
	number = {3},
	urldate = {2026-02-05},
	journal = {MNRAS},
	author = {Carnall, A C and McLeod, D J and McLure, R J and Dunlop, J S and Begley, R and Cullen, F and Donnan, C T and Hamadouche, M L and Jewell, S M and Jones, E W and Pollock, C L and Wild, V},
	month = apr,
	year = {2023},
	pages = {3974--3985},
}

@article{long_efficient_2024,
	title = {Efficient {NIRCam} {Selection} of {Quiescent} {Galaxies} at 3 {\textless} z {\textless} 6 in {CEERS}},
	volume = {970},
	issn = {0004-637X},
	url = {https://doi.org/10.3847/1538-4357/ad4cea},
	doi = {10.3847/1538-4357/ad4cea},
	language = {en},
	number = {1},
	urldate = {2026-02-05},
	journal = {ApJ},
	publisher = {The American Astronomical Society},
	author = {Long, Arianna S. and Antwi-Danso, Jacqueline and Lambrides, Erini L. and Lovell, Christopher C. and de la Vega, Alexander and Valentino, Francesco and Zavala, Jorge A. and Casey, Caitlin M. and Wilkins, Stephen M. and Yung, L. Y. Aaron and Arrabal Haro, Pablo and Bagley, Micaela B. and Bisigello, Laura and Chworowsky, Katherine and Cooper, M. C. and Cooper, Olivia R. and Cooray, Asantha R. and Croton, Darren and Dickinson, Mark and Finkelstein, Steven L. and Franco, Maximilien and Gould, Katriona M. L. and Hirschmann, Michaela and Hutchison, Taylor A. and Kartaltepe, Jeyhan S. and Kocevski, Dale D. and Koekemoer, Anton M. and Lucas, Ray A. and McKinney, Jed and Nere, Rachel and Papovich, Casey and Pérez-González, Pablo G. and Pirzkal, Nor and Santini, Paola},
	month = jul,
	year = {2024},
	pages = {68},
}

@article{valentino_atlas_2023,
	title = {An {Atlas} of {Color}-selected {Quiescent} {Galaxies} at z {\textgreater} 3 in {Public} {JWST} {Fields}},
	volume = {947},
	issn = {0004-637X},
	url = {https://doi.org/10.3847/1538-4357/acbefa},
	doi = {10.3847/1538-4357/acbefa},
	language = {en},
	number = {1},
	urldate = {2026-02-05},
	journal = {ApJ},
	publisher = {The American Astronomical Society},
	author = {Valentino, Francesco and Brammer, Gabriel and Gould, Katriona M. L. and Kokorev, Vasily and Fujimoto, Seiji and Jespersen, Christian Kragh and Vijayan, Aswin P. and Weaver, John R. and Ito, Kei and Tanaka, Masayuki and Ilbert, Olivier and Magdis, Georgios E. and Whitaker, Katherine E. and Faisst, Andreas L. and Gallazzi, Anna and Gillman, Steven and Giménez-Arteaga, Clara and Gómez-Guijarro, Carlos and Kubo, Mariko and Heintz, Kasper E. and Hirschmann, Michaela and Oesch, Pascal and Onodera, Masato and Rizzo, Francesca and Lee, Minju and Strait, Victoria and Toft, Sune},
	month = apr,
	year = {2023},
	pages = {20},
}

@article{gould_cosmos2020_2023,
	title = {{COSMOS2020}: {Exploring} the {Dawn} of {Quenching} for {Massive} {Galaxies} at 3 {\textless} z {\textless} 5 with a {New} {Color}-selection {Method}},
	volume = {165},
	issn = {1538-3881},
	shorttitle = {{COSMOS2020}},
	url = {https://doi.org/10.3847/1538-3881/accadc},
	doi = {10.3847/1538-3881/accadc},
	language = {en},
	number = {6},
	urldate = {2026-02-05},
	journal = {AJ},
	publisher = {The American Astronomical Society},
	author = {Gould, Katriona M. L. and Brammer, Gabriel and Valentino, Francesco and Whitaker, Katherine E. and Weaver, John. R. and del P. Lagos, Claudia and Rizzo, Francesca and Franco, Maximilien and Hsieh, Bau-Ching and Ilbert, Olivier and Jin, Shuowen and Magdis, Georgios and J. McCracken, Henry and Mobasher, Bahram and Shuntov, Marko and Steinhardt, Charles L. and Strait, Victoria and Toft, Sune},
	month = may,
	year = {2023},
	pages = {248},
}

@article{de_graaff_efficient_2025,
	title = {Efficient formation of a massive quiescent galaxy at redshift 4.9},
	volume = {9},
	copyright = {2024 The Author(s)},
	issn = {2397-3366},
	url = {https://www.nature.com/articles/s41550-024-02424-3},
	doi = {10.1038/s41550-024-02424-3},
	language = {en},
	number = {2},
	urldate = {2026-02-05},
	journal = {Nat Astron},
	publisher = {Nature Publishing Group},
	author = {de Graaff, Anna and Setton, David J. and Brammer, Gabriel and Cutler, Sam and Suess, Katherine A. and Labbé, Ivo and Leja, Joel and Weibel, Andrea and Maseda, Michael V. and Whitaker, Katherine E. and Bezanson, Rachel and Boogaard, Leindert A. and Cleri, Nikko J. and De Lucia, Gabriella and Franx, Marijn and Greene, Jenny E. and Hirschmann, Michaela and Matthee, Jorryt and McConachie, Ian and Naidu, Rohan P. and Oesch, Pascal A. and Price, Sedona H. and Rix, Hans-Walter and Valentino, Francesco and Wang, Bingjie and Williams, Christina C.},
	month = feb,
	year = {2025},
	pages = {280--292},
}

@article{dome_mini-quenching_2024,
	title = {Mini-quenching of z = 4-8 galaxies by bursty star formation},
	volume = {527},
	issn = {0035-8711},
	url = {https://ui.adsabs.harvard.edu/abs/2024MNRAS.527.2139D},
	doi = {10.1093/mnras/stad3239},
	urldate = {2026-02-05},
	journal = {MNRAS},
	publisher = {OUP},
	author = {Dome, Tibor and Tacchella, Sandro and Fialkov, Anastasia and Ceverino, Daniel and Dekel, Avishai and Ginzburg, Omri and Lapiner, Sharon and Looser, Tobias J.},
	month = jan,
	year = {2024},
	note = {ADS Bibcode: 2024MNRAS.527.2139D},
	pages = {2139--2151},
}

@article{tripodi_extreme_2025,
	title = {Extreme properties of a compact and massive accreting black hole host in the first 500 {Myr}},
	volume = {16},
	issn = {2041-1723},
	url = {https://ui.adsabs.harvard.edu/abs/2025NatCo..16.9830T},
	doi = {10.1038/s41467-025-65070-x},
	urldate = {2026-02-05},
	journal = {Nat Comm},
	author = {Tripodi, Roberta and Martis, Nicholas and Markov, Vladan and Bradač, Maruša and Di Mascia, Fabio and Cammelli, Vieri and D'Eugenio, Francesco and Willott, Chris and Curti, Mirko and Bhatt, Maulik and Gallerani, Simona and Rihtaršič, Gregor and Singh, Jasbir and Gaspar, Gaia and Harshan, Anishya and Judež, Jon and Merida, Rosa M. and Desprez, Guillaume and Sawicki, Marcin and Goovaerts, Ilias and Muzzin, Adam and Noirot, Gaël and Sarrouh, Ghassan T. E. and Abraham, Roberto and Asada, Yoshihisa and Brammer, Gabriel and Estrada-Carpenter, Vicente and Felicioni, Giordano and Fujimoto, Seiji and Iyer, Kartheik and Mowla, Lamiya and Strait, Victoria},
	month = nov,
	year = {2025},
	note = {ADS Bibcode: 2025NatCo..16.9830T},
	pages = {9830},
}

@article{trefoloni_missing_2025,
	title = {The missing {Fe} {II} bump in faint {JWST} active galactic nuclei: {Possible} evidence of metal-poor broad-line regions at early cosmic times},
	volume = {700},
	issn = {0004-6361},
	shorttitle = {The missing {Fe} {II} bump in faint {JWST} active galactic nuclei},
	url = {https://ui.adsabs.harvard.edu/abs/2025A&A...700A.203T},
	doi = {10.1051/0004-6361/202452795},
	urldate = {2026-02-05},
	journal = {A\&A},
	publisher = {EDP},
	author = {Trefoloni, Bartolomeo and Ji, Xihan and Maiolino, Roberto and D'Eugenio, Francesco and Übler, Hannah and Scholtz, Jan and Marconi, Alessandro and Marconcini, Cosimo and Mazzolari, Giovanni},
	month = aug,
	year = {2025},
	note = {ADS Bibcode: 2025A\&A...700A.203T},
	pages = {A203},
}

@article{farcy_mistral_2025,
	title = {{MISTRAL}: a model for {AGN} winds from radiatively efficient accretion in cosmological simulations},
	volume = {543},
	issn = {0035-8711},
	shorttitle = {{MISTRAL}},
	url = {https://ui.adsabs.harvard.edu/abs/2025MNRAS.543..967F},
	doi = {10.1093/mnras/staf1464},
	urldate = {2026-02-05},
	journal = {MNRAS},
	publisher = {OUP},
	author = {Farcy, Marion and Hirschmann, Michaela and Somerville, Rachel S. and Choi, Ena and Koudmani, Sophie and Naab, Thorsten and Weinberger, Rainer and Bennett, Jake S. and Bhowmick, Aklant K. and Choi, Hyunseop and Hernquist, Lars and Hlavacek-Larrondo, Julie and Terrazas, Bryan A. and Valentino, Francesco},
	month = oct,
	year = {2025},
	note = {ADS Bibcode: 2025MNRAS.543..967F},
	pages = {967--993},
}

@article{bennert_local_2021,
	title = {A {Local} {Baseline} of the {Black} {Hole} {Mass} {Scaling} {Relations} for {Active} {Galaxies}. {IV}. {Correlations} {Between} {MBH} and {Host} {Galaxy} σ, {Stellar} {Mass}, and {Luminosity}},
	volume = {921},
	issn = {0004-637X},
	url = {https://ui.adsabs.harvard.edu/abs/2021ApJ...921...36B},
	doi = {10.3847/1538-4357/ac151a},
	urldate = {2026-02-05},
	journal = {ApJ},
	publisher = {IOP},
	author = {Bennert, Vardha N. and Treu, Tommaso and Ding, Xuheng and Stomberg, Isak and Birrer, Simon and Snyder, Tomas and Malkan, Matthew A. and Stephens, Andrew W. and Auger, Matthew W.},
	month = nov,
	year = {2021},
	note = {ADS Bibcode: 2021ApJ...921...36B},
	pages = {36},
}

@article{guo_search_2025,
	title = {A {Search} for z = 5 {Hα} and {Hβ} + [{O} {III}] {Dual}-line {Emitting} {Galaxies} in the {JWST} {CEERS} {Field}: {Implications} for the {Active} {Galactic} {Nucleus} {Abundance}},
	volume = {991},
	issn = {0004-637X},
	shorttitle = {A {Search} for z = 5 {Hα} and {Hβ} + [{O} {III}] {Dual}-line {Emitting} {Galaxies} in the {JWST} {CEERS} {Field}},
	url = {https://ui.adsabs.harvard.edu/abs/2025ApJ...991...74G},
	doi = {10.3847/1538-4357/adee0f},
	urldate = {2026-02-06},
	journal = {ApJ},
	publisher = {IOP},
	author = {Guo, Jingsong and Onoue, Masafusa and Inayoshi, Kohei and Kocevski, Dale D. and Finkelstein, Steven L. and Bagley, Micaela B. and McGrath, Elizabeth J.},
	month = sep,
	year = {2025},
	note = {ADS Bibcode: 2025ApJ...991...74G},
	pages = {74},
}

@article{kokorev_census_2024,
	title = {A {Census} of {Photometrically} {Selected} {Little} {Red} {Dots} at 4 {\textless} z {\textless} 9 in {JWST} {Blank} {Fields}},
	volume = {968},
	issn = {0004-637X},
	url = {https://ui.adsabs.harvard.edu/abs/2024ApJ...968...38K},
	doi = {10.3847/1538-4357/ad4265},
	urldate = {2026-02-06},
	journal = {ApJ},
	publisher = {IOP},
	author = {Kokorev, Vasily and Caputi, Karina I. and Greene, Jenny E. and Dayal, Pratika and Trebitsch, Maxime and Cutler, Sam E. and Fujimoto, Seiji and Labbé, Ivo and Miller, Tim B. and Iani, Edoardo and Navarro-Carrera, Rafael and Rinaldi, Pierluigi},
	month = jun,
	year = {2024},
	note = {ADS Bibcode: 2024ApJ...968...38K},
	pages = {38},
}

@article{akins_cosmos-web_2025,
	title = {{COSMOS}-{Web}: {The} {Overabundance} and {Physical} {Nature} of "{Little} {Red} {Dots}"—{Implications} for {Early} {Galaxy} and {SMBH} {Assembly}},
	volume = {991},
	issn = {0004-637X},
	shorttitle = {{COSMOS}-{Web}},
	url = {https://ui.adsabs.harvard.edu/abs/2025ApJ...991...37A},
	doi = {10.3847/1538-4357/ade984},
	urldate = {2026-02-06},
	journal = {ApJ},
	publisher = {IOP},
	author = {Akins, Hollis B. and Casey, Caitlin M. and Lambrides, Erini and Allen, Natalie and Andika, Irham T. and Brinch, Malte and Champagne, Jaclyn B. and Cooper, Olivia and Ding, Xuheng and Drakos, Nicole E. and Faisst, Andreas and Finkelstein, Steven L. and Franco, Maximilien and Fujimoto, Seiji and Gentile, Fabrizio and Gillman, Steven and Gozaliasl, Ghassem and Harish, Santosh and Hayward, Christopher C. and Hirschmann, Michaela and Ilbert, Olivier and Kartaltepe, Jeyhan S. and Kocevski, Dale D. and Koekemoer, Anton M. and Kokorev, Vasily and Liu, Daizhong and Long, Arianna S. and McCracken, Henry Joy and McKinney, Jed and Onoue, Masafusa and Paquereau, Louise and Renzini, Alvio and Rhodes, Jason and Robertson, Brant E. and Shuntov, Marko and Silverman, John D. and Tanaka, Takumi S. and Toft, Sune and Trakhtenbrot, Benny and Valentino, Francesco and Zavala, Jorge},
	month = sep,
	year = {2025},
	note = {ADS Bibcode: 2025ApJ...991...37A},
	pages = {37},
}

@article{richards_spectral_2006,
	title = {Spectral {Energy} {Distributions} and {Multiwavelength} {Selection} of {Type} 1 {Quasars}},
	volume = {166},
	issn = {0067-0049},
	url = {https://ui.adsabs.harvard.edu/abs/2006ApJS..166..470R},
	doi = {10.1086/506525},
	urldate = {2026-02-06},
	journal = {ApJS},
	publisher = {IOP},
	author = {Richards, Gordon T. and Lacy, Mark and Storrie-Lombardi, Lisa J. and Hall, Patrick B. and Gallagher, S. C. and Hines, Dean C. and Fan, Xiaohui and Papovich, Casey and Vanden Berk, Daniel E. and Trammell, George B. and Schneider, Donald P. and Vestergaard, Marianne and York, Donald G. and Jester, Sebastian and Anderson, Scott F. and Budavári, Tamás and Szalay, Alexander S.},
	month = oct,
	year = {2006},
	note = {ADS Bibcode: 2006ApJS..166..470R},
	pages = {470--497},
}

@article{habouzit_is_2025,
	title = {Is the {JWST} detecting too many {AGN} candidates?},
	volume = {537},
	issn = {0035-8711},
	url = {https://ui.adsabs.harvard.edu/abs/2025MNRAS.537.2323H},
	doi = {10.1093/mnras/staf167},
	urldate = {2026-02-06},
	journal = {MNRAS},
	publisher = {OUP},
	author = {Habouzit, Melanie},
	month = mar,
	year = {2025},
	note = {ADS Bibcode: 2025MNRAS.537.2323H},
	pages = {2323--2333},
}

@article{jones_m_rm_2025,
	title = {The \${M}\_\{{\textbackslash}rm {BH}\}-{M}\_\{*\}\$ {Relationship} at \$3},
	volume = {arXiv.2510.07376},
	url = {https://ui.adsabs.harvard.edu/abs/2025arXiv251007376J},
	doi = {10.48550/arXiv.2510.07376},
	urldate = {2026-02-06},
	journal = {arXiv e-prints},
	publisher = {arXiv},
	author = {Jones, Brenda L. and Kocevski, Dale D. and Pacucci, Fabio and Taylor, Anthony J. and Finkelstein, Steven L. and Buchner, Johannes and Trump, Jonathan R. and Somerville, Rachel S. and Hirschmann, Michaela and Yung, L. Y. Aaron and Barro, Guillermo and Bell, Eric F. and Bisigello, Laura and Calabro, Antonello and Cleri, Nikko J. and Dekel, Avishai and Dickinson, Mark and Gandolfi, Giovanni and Giavalisco, Mauro and Grogin, Norman A. and Inayoshi, Kohei and Kartaltepe, Jeyhan S. and Koekemoer, Anton M. and Napolitano, Lorenzo and Onoue, Masafusa and Ravindranath, Swara and Rodighiero, Giulia and Wilkins, Stephen M.},
	month = oct,
	year = {2025},
	note = {ADS Bibcode: 2025arXiv251007376J},
}

@article{taylor_broad-line_2025,
	title = {Broad-line {AGNs} at 3.5 {\textless} z {\textless} 6: {The} {Black} {Hole} {Mass} {Function} and a {Connection} with {Little} {Red} {Dots}},
	volume = {986},
	issn = {0004-637X},
	shorttitle = {Broad-line {AGNs} at 3.5 {\textless} z {\textless} 6},
	url = {https://ui.adsabs.harvard.edu/abs/2025ApJ...986..165T},
	doi = {10.3847/1538-4357/add15b},
	urldate = {2026-02-06},
	journal = {ApJ},
	publisher = {IOP},
	author = {Taylor, Anthony J. and Finkelstein, Steven L. and Kocevski, Dale D. and Jeon, Junehyoung and Bromm, Volker and Amorín, Ricardo O. and Arrabal Haro, Pablo and Backhaus, Bren E. and Bagley, Micaela B. and Banados, Eduardo and Bhatawdekar, Rachana and Brooks, Madisyn and Calabrò, Antonello and Chávez Ortiz, Óscar A. and Cheng, Yingjie and Cleri, Nikko J. and Cole, Justin W. and Davis, Kelcey and Dickinson, Mark and Donnan, Callum and Dunlop, James S. and Ellis, Richard S. and Fernández, Vital and Fontana, Adriano and Fujimoto, Seiji and Giavalisco, Mauro and Grazian, Andrea and Guo, Jingsong and Hathi, Nimish P. and Holwerda, Benne W. and Hirschmann, Michaela and Inayoshi, Kohei and Kartaltepe, Jeyhan S. and Khusanova, Yana and Koekemoer, Anton M. and Kokorev, Vasily and Larson, Rebecca L. and Leung, Gene C. K. and Lucas, Ray A. and McLeod, Derek J. and Napolitano, Lorenzo and Onoue, Masafusa and Pacucci, Fabio and Papovich, Casey and Pérez-González, Pablo G. and Pirzkal, Nor and Somerville, Rachel S. and Trump, Jonathan R. and Wilkins, Stephen M. and Yung, L. Y. Aaron and Zhang, Haowen},
	month = jun,
	year = {2025},
	note = {ADS Bibcode: 2025ApJ...986..165T},
	pages = {165},
}

@article{ortame_small_2026,
	title = {Small hosts, big appetites: unveiling rapid and early low-mass black hole growth in cosmological zoom-in simulations of dwarf galaxies},
	volume = {2603.00241},
	shorttitle = {Small hosts, big appetites},
	url = {https://arxiv.org/abs/2603.00241v1},
	language = {en},
	urldate = {2026-06-14},
	journal = {arXiv e-prints},
	author = {Ortame, Giulia and Bourne, Martin A. and Koudmani, Sophie and Sijacki, Debora and D'Eugenio, Francesco and Maiolino, Roberto},
	month = feb,
	year = {2026},
}

@article{geris_jades_2026,
	title = {{JADES} reveals a large population of low-mass black holes at high redshift},
	volume = {545},
	issn = {0035-8711},
	url = {https://doi.org/10.1093/mnras/staf1979},
	doi = {10.1093/mnras/staf1979},
	number = {1},
	urldate = {2026-06-15},
	journal = {MNRAS},
	author = {Geris, Sophia and Maiolino, Roberto and Isobe, Yuki and Scholtz, Jan and D’Eugenio, Francesco and Ji, Xihan and Juodžbalis, Ignas and Simmonds, Charlotte and Dayal, Pratika and Trinca, Alessandro and Schneider, Raffaella and Arribas, Santiago and Bhatawdekar, Rachana and Bunker, Andrew J and Carniani, Stefano and Charlot, Stéphane and Chevallard, Jacopo and Curtis-Lake, Emma and Johnson, Benjamin D and Parlanti, Eleonora and Rinaldi, Pierluigi and Robertson, Brant and Tacchella, Sandro and Übler, Hannah and Venturi, Giacomo and Williams, Christina C and Witstok, Joris},
	month = jan,
	year = {2026},
	pages = {staf1979},
}

@article{trinca_seeking_2023,
	title = {Seeking the growth of the first black hole seeds with {JWST}},
	volume = {519},
	issn = {0035-8711},
	url = {https://ui.adsabs.harvard.edu/abs/2023MNRAS.519.4753T},
	doi = {10.1093/mnras/stac3768},
	urldate = {2026-06-18},
	journal = {MNRAS},
	publisher = {OUP},
	author = {Trinca, Alessandro and Schneider, Raffaella and Maiolino, Roberto and Valiante, Rosa and Graziani, Luca and Volonteri, Marta},
	month = mar,
	year = {2023},
	note = {ADS Bibcode: 2023MNRAS.519.4753T},
	pages = {4753--4764},
}

@article{maiolino_black_2026,
	title = {A black hole in a near pristine galaxy 700 {Myr} after the big bang},
	volume = {548},
	issn = {0035-8711},
	url = {https://ui.adsabs.harvard.edu/abs/2026MNRAS.548f2109M},
	doi = {10.1093/mnras/staf2109},
	urldate = {2026-06-18},
	journal = {MNRAS},
	publisher = {OUP},
	author = {Maiolino, Roberto and Übler, Hannah and D'Eugenio, Francesco and Scholtz, Jan and Juodžbalis, Ignas and Ji, Xihan and Perna, Michele and Bromm, Volker and Dayal, Pratika and Koudmani, Sophie and Liu, Boyuan and Schneider, Raffaella and Sijacki, Debora and Valiante, Rosa and Trinca, Alessandro and Zhang, Saiyang and Volonteri, Marta and Inayoshi, Kohei and Carniani, Stefano and Nakajima, Kimihiko and Isobe, Yuki and Witstok, Joris and Jones, Gareth C. and Tacchella, Sandro and Arribas, Santiago and Bunker, Andrew and Cataldi, Elisa and Charlot, Stephane and Curti, Giovanni Cresci Mirko and Fabian, Andrew C. and Katz, Harley and Kumari, Nimisha and Laporte, Nicolas and Mazzolari, Giovanni and Robertson, Brant and Sun, Fengwu and Rodriguez Del Pino, Bruno and Venturi, Giacomo},
	month = may,
	year = {2026},
	note = {ADS Bibcode: 2026MNRAS.548f2109M},
	pages = {staf2109},
}

@article{zhang_primordial_2026,
	title = {Primordial {Black} {Holes} as {Seeds} for {Extremely} {Overmassive} {Active} {Galactic} {Nuclei} {Observed} by {JWST}},
	volume = {1000},
	issn = {0004-637X},
	url = {https://ui.adsabs.harvard.edu/abs/2026ApJ..1000L..19Z},
	doi = {10.3847/2041-8213/ae4bd0},
	urldate = {2026-06-19},
	journal = {ApJ},
	publisher = {IOP},
	author = {Zhang, Saiyang and Liu, Boyuan and Bromm, Volker and Kühnel, Florian},
	month = mar,
	year = {2026},
	note = {ADS Bibcode: 2026ApJ..1000L..19Z},
	pages = {L19},
}

@article{dayal_light_2026,
	title = {Light, heavy, primordial: exploring the diversity of black hole seeding and growth mechanisms in the {JWST} era},
	volume = {arXiv:2604.20966},
	shorttitle = {Light, heavy, primordial},
	url = {https://ui.adsabs.harvard.edu/abs/2026arXiv260420966D},
	doi = {10.48550/arXiv.2604.20966},
	urldate = {2026-06-19},
	journal = {arXiv e-prints},
	publisher = {arXiv},
	author = {Dayal, Pratika},
	month = apr,
	year = {2026},
	note = {ADS Bibcode: 2026arXiv260420966D},
}

@article{dayal_properties_2026,
	title = {The properties of primordially-seeded black holes and their hosts in the first billion years: implications for {JWST}},
	volume = {706},
	issn = {0004-6361},
	shorttitle = {The properties of primordially-seeded black holes and their hosts in the first billion years},
	url = {https://ui.adsabs.harvard.edu/abs/2026A&A...706A..72D},
	doi = {10.1051/0004-6361/202555959},
	urldate = {2026-06-19},
	journal = {A\&A},
	publisher = {EDP},
	author = {Dayal, Pratika and Maiolino, Roberto},
	month = feb,
	year = {2026},
	note = {ADS Bibcode: 2026A\&A...706A..72D},
	pages = {A72},
}

@article{dayal_exploring_2024,
	title = {Exploring a primordial solution for early black holes detected with {JWST}},
	volume = {690},
	issn = {0004-6361},
	url = {https://ui.adsabs.harvard.edu/abs/2024A&A...690A.182D},
	doi = {10.1051/0004-6361/202451481},
	urldate = {2026-06-19},
	journal = {A\&A},
	publisher = {EDP},
	author = {Dayal, Pratika},
	month = oct,
	year = {2024},
	note = {ADS Bibcode: 2024A\&A...690A.182D},
	pages = {A182},
}

@article{prole_primordial_2025,
	title = {Primordial black holes in cosmological simulations: growth prospects for supermassive black holes},
	volume = {8},
	issn = {2565-6120},
	shorttitle = {Primordial black holes in cosmological simulations},
	url = {https://ui.adsabs.harvard.edu/abs/2025OJAp....8E.126P},
	doi = {10.33232/001c.143643},
	urldate = {2026-06-19},
	journal = {OJAp},
	author = {Prole, Lewis R and Regan, John A and Mehta, Daxal and Coles, Peter and Dayal, Pratika},
	month = aug,
	year = {2025},
	note = {ADS Bibcode: 2025OJAp....8E.126P},
	pages = {126},
}

@article{kho_learning_2026,
	title = {Learning the {Universe} at {High} {Redshifts}: {Impact} of {Accretion} {Modeling} on {Early} {Black} {Hole} {Growth}},
	volume = {arXiv.2606.10036},
	shorttitle = {Learning the {Universe} at {High} {Redshifts}},
	url = {https://ui.adsabs.harvard.edu/abs/2026arXiv260610036K},
	doi = {10.48550/arXiv.2606.10036},
	urldate = {2026-06-25},
	journal = {arXiv e-prints},
	publisher = {arXiv},
	author = {Kho, Jonathan and Bhowmick, Aklant K. and Weinberger, Rainer and Torrey, Paul and Blecha, Laura and Hernquist, Lars and Bryan, Greg L. and Garcia, Alex M. and Ahvazi, Niusha and Saravia, Alejandro and Oh, Boon Kiat},
	month = jun,
	year = {2026},
	note = {ADS Bibcode: 2026arXiv260610036K},
}

@article{bhowmick_growth_2024,
	title = {Growth of high-redshift supermassive black holes from heavy seeds in the {BRAHMA} cosmological simulations: implications of overmassive black holes},
	volume = {533},
	issn = {0035-8711},
	shorttitle = {Growth of high-redshift supermassive black holes from heavy seeds in the {BRAHMA} cosmological simulations},
	url = {https://ui.adsabs.harvard.edu/abs/2024MNRAS.533.1907B},
	doi = {10.1093/mnras/stae1819},
	urldate = {2026-06-25},
	journal = {MNRAS},
	publisher = {OUP},
	author = {Bhowmick, Aklant K. and Blecha, Laura and Torrey, Paul and Somerville, Rachel S. and Kelley, Luke Zoltan and Vogelsberger, Mark and Weinberger, Rainer and Hernquist, Lars and Sivasankaran, Aneesh},
	month = sep,
	year = {2024},
	note = {ADS Bibcode: 2024MNRAS.533.1907B},
	pages = {1907--1926},
}

@article{gravity_collaboration_spatially_2026,
	title = {Spatially resolved broad-line region in a quasar at z = 4: {Dynamical} black hole mass and prominent outflow},
	volume = {706},
	issn = {0004-6361},
	shorttitle = {Spatially resolved broad-line region in a quasar at z = 4},
	url = {https://ui.adsabs.harvard.edu/abs/2026A&A...706A..99G},
	doi = {10.1051/0004-6361/202557285},
	urldate = {2026-06-25},
	journal = {A\&A},
	publisher = {EDP},
	author = {{Gravity+ Collaboration} and Abd El Dayem, K. and Aimar, N. and Berdeu, A. and Berger, J.-P. and Bourdarot, G. and Bourget, P. and Brandner, W. and Cao, Y. and Correia, C. and Cuevas Cardona, S. and Davies, R. and Defrère, D. and Drescher, A. and Eckart, A. and Eisenhauer, F. and Fabricius, M. and Farah, A. and Feuchtgruber, H. and Förster Schreiber, N. M. and Foschi, A. and Garcia, P. and Garcia Lopez, R. and Genzel, R. and Gillessen, S. and Gomes, T. and Gonté, F. and Gopinath, V. and Graf, J. and Hartl, M. and Haubois, X. and Haußmann, F. and Ho, L. C. and Hönig, S. and Houllé, M. and Joharle, S. and Keiman, C. and Kervella, P. and Kolb, J. and Kreidberg, L. and Labdon, A. and Lacour, S. and Lai, O. and Lai, S. and Laugier, R. and Le Bouquin, J.-B. and Leftley, J. and Li, R. and Lopez, B. and Lutz, D. and Mang, F. and Mérand, A. and Millour, F. and Montargès, M. and More, N. and Morujão, N. and Nowacki, H. and Nowak, M. and Oberti, S. and Onken, C. and Osorno, J. and Ott, T. and Paumard, T. and Perraut, K. and Perrin, G. and Petrov, R. and Petrucci, P.-O. and Pourré, N. and Rabien, S. and Rau, C. and Ribeiro, D. C. and Robbe-Dubois, S. and Sadun Bordoni, M. and Salman, M. and Sanchez-Bermudez, J. and Santos, D. and Sauter, J. and Scialpi, M. and Scigliuto, J. and Shangguan, J. and Shchekaturov, P. and Shimizu, T. and Soulez, F. and Straubmeier, C. and Sturm, E. and Subroweit, M. and Sykes, C. and Tacconi, L. J. and Übler, H. and Ulbricht, G. and Vincent, F. and Webster, R. and Wieprecht, E. and Woillez, J. and Wolf, C.},
	month = feb,
	year = {2026},
	note = {ADS Bibcode: 2026A\&A...706A..99G},
	pages = {A99},
}

@article{stevenson_primer_2026,
	title = {{PRIMER} and {JADES} reveal an abundance of massive quiescent galaxies at 2 {\textless} z {\textless} 5},
	volume = {545},
	issn = {0035-8711},
	url = {https://ui.adsabs.harvard.edu/abs/2026MNRAS.545f2087S},
	doi = {10.1093/mnras/staf2087},
	urldate = {2026-06-25},
	journal = {MNRAS},
	publisher = {OUP},
	author = {Stevenson, Struan D. and Carnall, Adam C. and Leung, Ho-Hin and Taylor, Elizabeth and Cullen, Fergus and Dunlop, James S. and McLeod, Derek J. and McLure, Ross J. and Begley, Ryan and Arellano-Córdova, Karla Z. and Barrufet, Laia and Bondestam, Cecilia and Donnan, Callum T. and Ellis, Richard S. and Grogin, Norman A. and Koekemoer, Anton M. and Liu, Feng-Yuan and Pérez-González, Pablo G. and Rowlands, Kate and Sanders, Ryan L. and Scholte, Dirk and Shapley, Alice E. and Skarbinski, Maya and Stanton, Thomas M. and Wild, Vivienne},
	month = jan,
	year = {2026},
	note = {ADS Bibcode: 2026MNRAS.545f2087S},
	pages = {staf2087},
}

@article{mehta_growth_2026,
	title = {The growth of light seed black holes in the early {Universe}},
	volume = {10},
	issn = {2397-3366},
	url = {https://ui.adsabs.harvard.edu/abs/2026NatAs..10..583M},
	doi = {10.1038/s41550-025-02767-5},
	urldate = {2026-06-25},
	journal = {Nature Astronomy},
	author = {Mehta, Daxal H. and Regan, John A. and Prole, Lewis},
	month = apr,
	year = {2026},
	note = {ADS Bibcode: 2026NatAs..10..583M},
	pages = {583--592},
}

@article{ma_undermassive_2026,
	title = {Undermassive {Hosts} of \$z = 4-6 \$ {AGN} from {JWST}/{NIRCam} {Image} {Decomposition} with {CONGRESS}, {FRESCO}, and {JADES}},
	volume = {arXiv.2601.15962},
	url = {http://arxiv.org/abs/2601.15962},
	doi = {10.48550/arXiv.2601.15962},
	urldate = {2026-06-25},
	journal = {arXiv e-prints},
	publisher = {arXiv},
	author = {Ma, Zheng and Egami, Eichi and Zhu, Yongda and Sun, Fengwu and Lyu, Jianwei and Zhang, Junyu and Willmer, Christopher N. A. and Bunker, Andrew J. and Carniani, Stefano and Curtis-Lake, Emma and Hausen, Ryan and Ji, Xihan and Ji, Zhiyuan and Juodžbalis, Ignas and Maiolino, Roberto and Rieke, George H. and Rinaldi, Pierluigi and Sun, Yang and Tacchella, Sandro and Übler, Hannah and Williams, Christina C.},
	month = jan,
	year = {2026},
	note = {arXiv:2601.15962 [astro-ph.GA]},
}

@article{marszewski_little_2026,
	title = {Little {Red} {Dots} on {FIRE}: {The} {Ability} of {Bursty} {Galaxies} to {Host} an {Abundant} {Population} of {High}-redshift {AGN}},
	volume = {1002},
	issn = {0004-637X},
	shorttitle = {Little {Red} {Dots} on {FIRE}},
	url = {https://ui.adsabs.harvard.edu/abs/2026ApJ..1002L..13M},
	doi = {10.3847/2041-8213/ae5d45},
	urldate = {2026-06-25},
	journal = {ApJ},
	publisher = {IOP},
	author = {Marszewski, Andrew and Faucher-Giguère, Claude-André and Sun, Guochao and Anglés-Alcázar, Daniel and Feldmann, Robert and Su, Kung-Yi and Miller, Tim B. and Roy, Niranjan Chandra},
	month = may,
	year = {2026},
	note = {ADS Bibcode: 2026ApJ..1002L..13M},
	pages = {L13},
}

@article{weinberger_accretion_2025,
	title = {Accretion onto supermassive and intermediate-mass black holes in cosmological simulations},
	volume = {700},
	issn = {0004-6361},
	url = {https://ui.adsabs.harvard.edu/abs/2025A&A...700A..52W},
	doi = {10.1051/0004-6361/202554174},
	urldate = {2026-06-25},
	journal = {A\&A},
	publisher = {EDP},
	author = {Weinberger, R. and Bhowmick, A. and Blecha, L. and Bryan, G. and Buchner, J. and Hernquist, L. and Hlavacek-Larrondo, J. and Springel, V.},
	month = aug,
	year = {2025},
	note = {ADS Bibcode: 2025A\&A...700A..52W},
	pages = {A52},
}

@article{schneider_are_2023,
	title = {Are we surprised to find {SMBHs} with {JWST} at z ≥ 9?},
	volume = {526},
	issn = {0035-8711},
	url = {https://ui.adsabs.harvard.edu/abs/2023MNRAS.526.3250S},
	doi = {10.1093/mnras/stad2503},
	urldate = {2026-06-25},
	journal = {MNRAS},
	publisher = {OUP},
	author = {Schneider, Raffaella and Valiante, Rosa and Trinca, Alessandro and Graziani, Luca and Volonteri, Marta and Maiolino, Roberto},
	month = dec,
	year = {2023},
	note = {ADS Bibcode: 2023MNRAS.526.3250S},
	pages = {3250--3261},
}

@article{keitaanranta_rapid_2026,
	title = {Rapid sinking and efficient mergers of supermassive black holes in compact high-redshift galaxies},
	volume = {549},
	issn = {0035-8711},
	url = {https://ui.adsabs.harvard.edu/abs/2026MNRAS.549ag756K},
	doi = {10.1093/mnras/stag756},
	urldate = {2026-06-25},
	journal = {MNRAS},
	publisher = {OUP},
	author = {Keitaanranta, Atte and Johansson, Peter H. and Rawlings, Alexander and Tuominen, Toni and Rantala, Antti and Naab, Thorsten and Liao, Shihong and Reinoso, Bastián},
	month = jun,
	year = {2026},
	note = {ADS Bibcode: 2026MNRAS.549ag756K},
	pages = {stag756},
}

@article{prole_seedz_2026,
	title = {The {SEEDZ} {Simulations}: {Methodology} and {First} {Results} on {Massive} {Black} {Hole} {Seeding} and {Early} {Galaxy} {Growth}},
	volume = {9},
	issn = {2565-6120},
	shorttitle = {The {SEEDZ} {Simulations}},
	url = {https://ui.adsabs.harvard.edu/abs/2026OJAp....959236P},
	doi = {10.33232/001c.159236},
	urldate = {2026-06-26},
	journal = {OJAp},
	author = {Prole, Lewis Robert and Regan, John A. and Mehta, Daxal and Pakmor, Rudiger and Koudmani, Sophie and Bourne, Martin A. and Glover, Simon C. O. and Wise, John H. and Klessen, Ralf S. and Tremmel, Michael and Sijacki, Debora and Beckmann, Ricarda S. and Haehnelt, Martin G. and Brennan, John and van de Bor, Pelle and Clark, Paul C.},
	month = mar,
	year = {2026},
	note = {ADS Bibcode: 2026OJAp....959236P},
	pages = {59236},
}

@article{habouzit_supermassive_2021,
	title = {Supermassive black holes in cosmological simulations {I}: {MBH} - {M}⋆ relation and black hole mass function},
	volume = {503},
	issn = {0035-8711},
	shorttitle = {Supermassive black holes in cosmological simulations {I}},
	url = {https://ui.adsabs.harvard.edu/abs/2021MNRAS.503.1940H},
	doi = {10.1093/mnras/stab496},
	urldate = {2026-06-26},
	journal = {MNRAS},
	publisher = {OUP},
	author = {Habouzit, Mélanie and Li, Yuan and Somerville, Rachel S. and Genel, Shy and Pillepich, Annalisa and Volonteri, Marta and Davé, Romeel and Rosas-Guevara, Yetli and McAlpine, Stuart and Peirani, Sébastien and Hernquist, Lars and Anglés-Alcázar, Daniel and Reines, Amy and Bower, Richard and Dubois, Yohan and Nelson, Dylan and Pichon, Christophe and Vogelsberger, Mark},
	month = may,
	year = {2021},
	note = {ADS Bibcode: 2021MNRAS.503.1940H},
	pages = {1940--1975},
}

@article{spinoso_multiflavour_2023,
	title = {Multiflavour {SMBH} seeding and evolution in cosmological environments},
	volume = {518},
	issn = {0035-8711},
	url = {https://ui.adsabs.harvard.edu/abs/2023MNRAS.518.4672S},
	doi = {10.1093/mnras/stac3169},
	urldate = {2026-06-26},
	journal = {MNRAS},
	publisher = {OUP},
	author = {Spinoso, D. and Bonoli, S. and Valiante, R. and Schneider, R. and Izquierdo-Villalba, D.},
	month = jan,
	year = {2023},
	note = {ADS Bibcode: 2023MNRAS.518.4672S},
	pages = {4672--4692},
}

@article{dave_simba_2019,
	title = {{SIMBA}: {Cosmological} simulations with black hole growth and feedback},
	volume = {486},
	issn = {0035-8711},
	shorttitle = {{SIMBA}},
	url = {https://ui.adsabs.harvard.edu/abs/2019MNRAS.486.2827D},
	doi = {10.1093/mnras/stz937},
	urldate = {2026-06-26},
	journal = {MNRAS},
	publisher = {OUP},
	author = {Davé, Romeel and Anglés-Alcázar, Daniel and Narayanan, Desika and Li, Qi and Rafieferantsoa, Mika H. and Appleby, Sarah},
	month = jun,
	year = {2019},
	note = {ADS Bibcode: 2019MNRAS.486.2827D},
	pages = {2827--2849},
}

@article{husko_effects_2025,
	title = {The effects of super-{Eddington} accretion and feedback on the growth of early supermassive black holes and galaxies},
	volume = {537},
	issn = {0035-8711},
	url = {https://ui.adsabs.harvard.edu/abs/2025MNRAS.537.2559H},
	doi = {10.1093/mnras/staf146},
	urldate = {2026-06-26},
	journal = {MNRAS},
	publisher = {OUP},
	author = {Huško, Filip and Lacey, Cedric G. and Roper, William J. and Schaye, Joop and Briggs, Jemima Mae and Schaller, Matthieu},
	month = mar,
	year = {2025},
	note = {ADS Bibcode: 2025MNRAS.537.2559H},
	pages = {2559--2578},
}

@article{lupi_sustained_2024,
	title = {Sustained super-{Eddington} accretion in high-redshift quasars},
	volume = {686},
	issn = {0004-6361},
	url = {https://ui.adsabs.harvard.edu/abs/2024A&A...686A.256L},
	doi = {10.1051/0004-6361/202348788},
	urldate = {2026-06-26},
	journal = {A\&A},
	publisher = {EDP},
	author = {Lupi, Alessandro and Quadri, Giada and Volonteri, Marta and Colpi, Monica and Regan, John A.},
	month = jun,
	year = {2024},
	note = {ADS Bibcode: 2024A\&A...686A.256L},
	pages = {A256},
}

@article{quadri_super-eddington_2025,
	title = {Super-{Eddington} accretion in high-redshift quasar hosts: {Black}-hole-driven outflows, galaxy quenching, and the nature of little red dots},
	volume = {704},
	issn = {0004-6361},
	shorttitle = {Super-{Eddington} accretion in high-redshift quasar hosts},
	url = {https://ui.adsabs.harvard.edu/abs/2025A&A...704A.248Q},
	doi = {10.1051/0004-6361/202555423},
	urldate = {2026-06-26},
	journal = {A\&A},
	publisher = {EDP},
	author = {Quadri, Giada and Trinca, Alessandro and Lupi, Alessandro and Colpi, Monica and Volonteri, Marta},
	month = dec,
	year = {2025},
	note = {ADS Bibcode: 2025A\&A...704A.248Q},
	pages = {A248},
}

@article{volonteri_exploring_2025,
	title = {Exploring active galactic nuclei and little red dots with the {Obelisk} simulation},
	volume = {695},
	issn = {0004-6361},
	url = {https://ui.adsabs.harvard.edu/abs/2025A&A...695A..33V},
	doi = {10.1051/0004-6361/202451963},
	urldate = {2026-06-26},
	journal = {A\&A},
	publisher = {EDP},
	author = {Volonteri, M. and Trebitsch, M. and Greene, J. E. and Dubois, Y. and Dong-Paez, C.-A. and Habouzit, M. and Lupi, A. and Ma, Y. and Beckmann, R. S. and Dayal, P. and Schneider, R.},
	month = mar,
	year = {2025},
	note = {ADS Bibcode: 2025A\&A...695A..33V},
	pages = {A33},
}

@article{bhowmick_dynamics_2025,
	title = {Dynamics of {Low}-mass {Black} {Hole} {Seeds} in the {BRAHMA} {Simulations} {Using} {Subgrid} {Dynamical} {Friction}: {Impact} on {Merger}-driven {Black} {Hole} {Growth} in the {High}-redshift {Universe}},
	volume = {991},
	issn = {0004-637X},
	shorttitle = {Dynamics of {Low}-mass {Black} {Hole} {Seeds} in the {BRAHMA} {Simulations} {Using} {Subgrid} {Dynamical} {Friction}},
	url = {https://doi.org/10.3847/1538-4357/adf96b},
	doi = {10.3847/1538-4357/adf96b},
	language = {en},
	number = {1},
	urldate = {2026-06-26},
	journal = {ApJ},
	publisher = {The American Astronomical Society},
	author = {Bhowmick, Aklant K. and Blecha, Laura and Kelley, Luke Z. and Sivasankaran, Aneesh and Torrey, Paul and Weinberger, Rainer and Chen, Nianyi and Vogelsberger, Mark and Hernquist, Lars and Natarajan, Priyamvada and Di Matteo, Tiziana},
	month = sep,
	year = {2025},
	pages = {81},
}

@article{greene_uncover_2024,
	title = {{UNCOVER} {Spectroscopy} {Confirms} the {Surprising} {Ubiquity} of {Active} {Galactic} {Nuclei} in {Red} {Sources} at z {\textgreater} 5},
	volume = {964},
	issn = {0004-637X},
	url = {https://ui.adsabs.harvard.edu/abs/2024ApJ...964...39G},
	doi = {10.3847/1538-4357/ad1e5f},
	urldate = {2026-06-26},
	journal = {ApJ},
	publisher = {IOP},
	author = {Greene, Jenny E. and Labbe, Ivo and Goulding, Andy D. and Furtak, Lukas J. and Chemerynska, Iryna and Kokorev, Vasily and Dayal, Pratika and Volonteri, Marta and Williams, Christina C. and Wang, Bingjie and Setton, David J. and Burgasser, Adam J. and Bezanson, Rachel and Atek, Hakim and Brammer, Gabriel and Cutler, Sam E. and Feldmann, Robert and Fujimoto, Seiji and Glazebrook, Karl and de Graaff, Anna and Khullar, Gourav and Leja, Joel and Marchesini, Danilo and Maseda, Michael V. and Matthee, Jorryt and Miller, Tim B. and Naidu, Rohan P. and Nanayakkara, Themiya and Oesch, Pascal A. and Pan, Richard and Papovich, Casey and Price, Sedona H. and van Dokkum, Pieter and Weaver, John R. and Whitaker, Katherine E. and Zitrin, Adi},
	month = mar,
	year = {2024},
	note = {ADS Bibcode: 2024ApJ...964...39G},
	pages = {39},
}

@article{omukai_can_2008,
	title = {Can {Supermassive} {Black} {Holes} {Form} in {Metal}-enriched {High}-{Redshift} {Protogalaxies}?},
	volume = {686},
	issn = {0004-637X},
	url = {https://ui.adsabs.harvard.edu/abs/2008ApJ...686..801O},
	doi = {10.1086/591636},
	urldate = {2026-06-26},
	journal = {AJ},
	publisher = {IOP},
	author = {Omukai, K. and Schneider, R. and Haiman, Z.},
	month = oct,
	year = {2008},
	note = {ADS Bibcode: 2008ApJ...686..801O},
	pages = {801--814},
}

@article{madau_massive_2001,
	title = {Massive {Black} {Holes} as {Population} {III} {Remnants}},
	volume = {551},
	issn = {0004-637X},
	url = {https://ui.adsabs.harvard.edu/abs/2001ApJ...551L..27M},
	doi = {10.1086/319848},
	urldate = {2026-06-26},
	journal = {ApJ},
	publisher = {IOP},
	author = {Madau, Piero and Rees, Martin J.},
	month = apr,
	year = {2001},
	note = {ADS Bibcode: 2001ApJ...551L..27M},
	pages = {L27--L30},
}

@article{bromm_formation_2003,
	title = {Formation of the {First} {Supermassive} {Black} {Holes}},
	volume = {596},
	issn = {0004-637X},
	url = {https://ui.adsabs.harvard.edu/abs/2003ApJ...596...34B},
	doi = {10.1086/377529},
	urldate = {2026-06-26},
	journal = {ApJ},
	publisher = {IOP},
	author = {Bromm, Volker and Loeb, Abraham},
	month = oct,
	year = {2003},
	note = {ADS Bibcode: 2003ApJ...596...34B},
	pages = {34--46},
}

@article{lodato_supermassive_2006,
	title = {Supermassive black hole formation during the assembly of pre-galactic discs},
	volume = {371},
	issn = {0035-8711},
	url = {https://ui.adsabs.harvard.edu/abs/2006MNRAS.371.1813L},
	doi = {10.1111/j.1365-2966.2006.10801.x},
	urldate = {2026-06-26},
	journal = {MNRAS},
	publisher = {OUP},
	author = {Lodato, Giuseppe and Natarajan, Priyamvada},
	month = oct,
	year = {2006},
	note = {ADS Bibcode: 2006MNRAS.371.1813L},
	pages = {1813--1823},
}

@article{portegies_zwart_runaway_2002,
	title = {The {Runaway} {Growth} of {Intermediate}-{Mass} {Black} {Holes} in {Dense} {Star} {Clusters}},
	volume = {576},
	issn = {0004-637X},
	url = {https://ui.adsabs.harvard.edu/abs/2002ApJ...576..899P},
	doi = {10.1086/341798},
	urldate = {2026-06-26},
	journal = {ApJ},
	publisher = {IOP},
	author = {Portegies Zwart, Simon F. and McMillan, Stephen L. W.},
	month = sep,
	year = {2002},
	note = {ADS Bibcode: 2002ApJ...576..899P},
	pages = {899--907},
}

@article{devecchi_formation_2009,
	title = {Formation of the {First} {Nuclear} {Clusters} and {Massive} {Black} {Holes} at {High} {Redshift}},
	volume = {694},
	issn = {0004-637X},
	url = {https://ui.adsabs.harvard.edu/abs/2009ApJ...694..302D},
	doi = {10.1088/0004-637X/694/1/302},
	urldate = {2026-06-26},
	journal = {ApJ},
	publisher = {IOP},
	author = {Devecchi, B. and Volonteri, M.},
	month = mar,
	year = {2009},
	note = {ADS Bibcode: 2009ApJ...694..302D},
	pages = {302--313},
}

@article{brazzini_ruling_2025,
	title = {Ruling out dominant electron scattering in {Little} {Red} {Dots}' {Rosetta} {Stone} using multiple hydrogen lines},
	volume = {544},
	issn = {0035-8711},
	url = {https://ui.adsabs.harvard.edu/abs/2025MNRAS.544L.167B},
	doi = {10.1093/mnrasl/slaf116},
	urldate = {2026-06-26},
	journal = {MNRAS},
	publisher = {OUP},
	author = {Brazzini, Matilde and D'Eugenio, Francesco and Maiolino, Roberto and Juodžbalis, Ignas and Ji, Xihan and Scholtz, Jan and Chang, Seok-Jun},
	month = nov,
	year = {2025},
	note = {ADS Bibcode: 2025MNRAS.544L.167B},
	pages = {L167--L173},
}

@article{deugenio_irony_2025,
	title = {Irony at z=6.68: a bright {AGN} with forbidden {Fe} emission and multi-component {Balmer} absorption},
	volume = {arXiv.2510.00101},
	shorttitle = {Irony at z=6.68},
	url = {https://ui.adsabs.harvard.edu/abs/2025arXiv251000101D},
	doi = {10.48550/arXiv.2510.00101},
	urldate = {2026-06-26},
	journal = {arXiv e-prints},
	publisher = {arXiv},
	author = {D'Eugenio, Francesco and Nelson, Erica and Ji, Xihan and Baggen, Josephine and Greene, Jenny and Labbé, Ivo and Pezzulli, Gabriele and Brown, Vanessa and Maiolino, Roberto and Matthee, Jorryt and Terlevich, Elena and Terlevich, Roberto and Torralba, Alberto and Carniani, Stefano},
	month = sep,
	year = {2025},
	note = {ADS Bibcode: 2025arXiv251000101D},
}

@article{deugenio_blackthunder_2026,
	title = {{BlackTHUNDER} strikes twice: {Balmer}-line absorption in an overmassive {Little} {Red} {Dot} at z = 7.04},
	volume = {547},
	issn = {0035-8711},
	shorttitle = {{BlackTHUNDER} strikes twice},
	url = {https://ui.adsabs.harvard.edu/abs/2026MNRAS.547ag401D},
	doi = {10.1093/mnras/stag401},
	urldate = {2026-06-26},
	journal = {MNRAS},
	publisher = {OUP},
	author = {D'Eugenio, Francesco and Maiolino, Roberto and Perna, Michele and Übler, Hannah and Ji, Xihan and McClymont, William and Koudmani, Sophie and Sijacki, Debora and Juodžbalis, Ignas and Scholtz, Jan and Bennett, Jake S. and Bunker, Andrew J. and Carniani, Stefano and Charlot, Stéphane and Cresci, Giovanni and Curtis-Lake, Emma and Bontà, Elena Dalla and Inayoshi, Kohei and Jones, Gareth C. and Lyu, Jianwei and Marconi, Alessandro and Mazzolari, Giovanni and Nelson, Erica J. and Parlanti, Eleonora and Robertson, Brant E. and Schneider, Raffaella and Simmonds, Charlotte and Tacchella, Sandro and Venturi, Giacomo and Willott, Chris and Witstok, Joris and Witten, Callum},
	month = apr,
	year = {2026},
	note = {ADS Bibcode: 2026MNRAS.547ag401D},
	pages = {stag401},
}

@article{scholtz_little_2026,
	title = {Little {Red} and {Blue} {Dots}: simply stratified {Broad} {Line} {Regions}},
	volume = {arXiv.2603.22277},
	shorttitle = {Little {Red} and {Blue} {Dots}},
	url = {https://ui.adsabs.harvard.edu/abs/2026arXiv260322277S},
	doi = {10.48550/arXiv.2603.22277},
	urldate = {2026-06-26},
	journal = {arXiv e-prints},
	publisher = {arXiv},
	author = {Scholtz, J. and D'Eugenio, F. and Maiolino, R. and Brazzini, M. and Übler, H. and Ji, X. and Perna, M. and Sun, F. and Brocchi, G. and Carniani, S. and Cresci, G. and Ivey, L. R. and Juodžbalis, I. and Marconi, A. and Mazzolari, G. and Risaliti, G. and Trefoloni, B.},
	month = mar,
	year = {2026},
	note = {ADS Bibcode: 2026arXiv260322277S},
}

@article{laor_evidence_2006,
	title = {Evidence for {Line} {Broadening} by {Electron} {Scattering} in the {Broad}-{Line} {Region} of {NGC} 4395},
	volume = {643},
	issn = {0004-637X},
	url = {https://ui.adsabs.harvard.edu/abs/2006ApJ...643..112L},
	doi = {10.1086/502798},
	urldate = {2026-06-26},
	journal = {ApJ},
	publisher = {IOP},
	author = {Laor, Ari},
	month = may,
	year = {2006},
	note = {ADS Bibcode: 2006ApJ...643..112L},
	pages = {112--119},
}

@article{schaye_colibre_2026,
	title = {The {COLIBRE} project: cosmological hydrodynamical simulations of galaxy formation and evolution},
	volume = {548},
	copyright = {https://creativecommons.org/licenses/by/4.0/},
	issn = {0035-8711, 1365-2966},
	shorttitle = {The {COLIBRE} project},
	url = {https://academic.oup.com/mnras/article/doi/10.1093/mnras/stag375/8650959},
	doi = {10.1093/mnras/stag375},
	language = {en},
	number = {1},
	urldate = {2026-07-02},
	journal = {MNRAS},
	author = {Schaye, Joop and Chaikin, Evgenii and Schaller, Matthieu and Ploeckinger, Sylvia and Huško, Filip and McGibbon, Robert J and Trayford, James W and Benítez-Llambay, Alejandro and Correa, Camila and Frenk, Carlos S and Richings, Alexander J and Forouhar Moreno, Victor J and Bahé, Yannick M and Borrow, Josh and Durrant, Anna and Gebek, Andrea and Helly, John C and Jenkins, Adrian and Lacey, Cedric G and Ludlow, Aaron and Nobels, Folkert S J},
	month = apr,
	year = {2026},
	pages = {stag375},
}

@article{zana_super-eddington_2026,
	title = {Super-{Eddington} accretion in protogalactic cores},
	volume = {708},
	issn = {0004-6361},
	url = {https://ui.adsabs.harvard.edu/abs/2026A&A...708A...7Z},
	doi = {10.1051/0004-6361/202557006},
	urldate = {2026-07-02},
	journal = {A\&A},
	publisher = {EDP},
	author = {Zana, Tommaso and Capelo, Pedro R. and Boresta, Mairo and Schneider, Raffaella and Lupi, Alessandro and Trinca, Alessandro and Mayer, Lucio and Valiante, Rosa and Graziani, Luca},
	month = mar,
	year = {2026},
	note = {ADS Bibcode: 2026A\&A...708A...7Z},
	pages = {A7},
}

@article{mcclymont_overmassive_2026,
	title = {Overmassive black holes in the early {Universe} can be explained by gas-rich, dark matter-dominated galaxies},
	volume = {545},
	issn = {0035-8711},
	url = {https://ui.adsabs.harvard.edu/abs/2026MNRAS.545f2092M},
	doi = {10.1093/mnras/staf2092},
	urldate = {2026-07-02},
	journal = {MNRAS},
	publisher = {OUP},
	author = {McClymont, William and Tacchella, Sandro and Ji, Xihan and Kannan, Rahul and Maiolino, Roberto and Simmonds, Charlotte and Smith, Aaron and Puchwein, Ewald and Garaldi, Enrico and Vogelsberger, Mark and D'Eugenio, Francesco and Keating, Laura and Shen, Xuejian and Trefoloni, Bartolomeo and Zier, Oliver},
	month = jan,
	year = {2026},
	note = {ADS Bibcode: 2026MNRAS.545f2092M},
	pages = {staf2092},
}

@article{gordon_conditions_2025,
	title = {Conditions for super-{Eddington} accretion onto the first black holes},
	volume = {537},
	issn = {0035-8711},
	url = {https://ui.adsabs.harvard.edu/abs/2025MNRAS.537..674G},
	doi = {10.1093/mnras/staf054},
	urldate = {2026-07-02},
	journal = {MNRAS},
	publisher = {OUP},
	author = {Gordon, Simone T. and Smith, Britton D. and Khochfar, Sadegh and Beckmann, Ricarda S.},
	month = feb,
	year = {2025},
	note = {ADS Bibcode: 2025MNRAS.537..674G},
	pages = {674--690},
}

@article{shin_mandelzoom_2025,
	title = {The {MandelZoom} project {I}: modelling black hole accretion through an α-disc in dwarf galaxies with a resolved interstellar medium},
	volume = {544},
	issn = {0035-8711},
	shorttitle = {The {MandelZoom} project {I}},
	url = {https://ui.adsabs.harvard.edu/abs/2025MNRAS.544.2467S},
	doi = {10.1093/mnras/staf1786},
	urldate = {2026-07-02},
	journal = {MNRAS},
	publisher = {OUP},
	author = {Shin, Eun-jin and Sijacki, Debora and Smith, Matthew C. and Bourne, Martin A. and Koudmani, Sophie},
	month = dec,
	year = {2025},
	note = {ADS Bibcode: 2025MNRAS.544.2467S},
	pages = {2467--2492},
}

@article{partmann_importance_2025,
	title = {The importance of nuclear star clusters for massive black hole growth and nuclear star formation in simulated low-mass galaxies},
	volume = {537},
	issn = {0035-8711},
	url = {https://ui.adsabs.harvard.edu/abs/2025MNRAS.537..956P},
	doi = {10.1093/mnras/staf002},
	urldate = {2026-07-02},
	journal = {MNRAS},
	publisher = {OUP},
	author = {Partmann, Christian and Naab, Thorsten and Lahén, Natalia and Rantala, Antti and Hirschmann, Michaela and Hislop, Jessica M. and Petersson, Jonathan and Johansson, Peter H.},
	month = feb,
	year = {2025},
	note = {ADS Bibcode: 2025MNRAS.537..956P},
	pages = {956--977},
}

@article{petersson_noctua_2025,
	title = {{NOCTUA} suite of simulations: {The} difficulty of growing massive black holes in low-mass dwarf galaxies},
	volume = {704},
	issn = {0004-6361},
	shorttitle = {{NOCTUA} suite of simulations},
	url = {https://ui.adsabs.harvard.edu/abs/2025A&A...704A.177P},
	doi = {10.1051/0004-6361/202555130},
	urldate = {2026-07-02},
	journal = {A\&A},
	publisher = {EDP},
	author = {Petersson, Jonathan and Hirschmann, Michaela and Tress, Robin G. and Farcy, Marion and Glover, Simon C. O. and Klessen, Ralf S. and Naab, Thorsten and Partmann, Christian and Whitworth, David J.},
	month = dec,
	year = {2025},
	note = {ADS Bibcode: 2025A\&A...704A.177P},
	pages = {A177},
}

@article{kao_novel_2026,
	title = {A novel subgrid model for super-{Eddington} accretion of spinning black holes in galaxy-scale simulations},
	volume = {546},
	issn = {0035-8711},
	url = {https://ui.adsabs.harvard.edu/abs/2026MNRAS.546ag003K},
	doi = {10.1093/mnras/stag003},
	urldate = {2026-07-02},
	journal = {MNRAS},
	publisher = {OUP},
	author = {Kao, Wei-Bo and Capelo, Pedro R. and Cenci, Elia and Mayer, Lucio and Lupi, Alessandro and Sala, Luca},
	month = feb,
	year = {2026},
	note = {ADS Bibcode: 2026MNRAS.546ag003K},
	pages = {stag003},
}

@article{shin_mandelzoom_2026,
	title = {The {MandelZoom} project ─ {II}. {The} impact of stellar feedback on black hole accretion through an α-disc in dwarf galaxies with a resolved interstellar medium},
	volume = {548},
	issn = {0035-8711},
	url = {https://ui.adsabs.harvard.edu/abs/2026MNRAS.548ag580S},
	doi = {10.1093/mnras/stag580},
	urldate = {2026-07-02},
	journal = {MNRAS},
	publisher = {OUP},
	author = {Shin, Eun-jin and Smith, Matthew C. and Sijacki, Debora and Bourne, Martin A. and Koudmani, Sophie},
	month = may,
	year = {2026},
	note = {ADS Bibcode: 2026MNRAS.548ag580S},
	pages = {stag580},
}

@article{buttigieg_premature_2025,
	title = {Premature supermassive black hole mergers in cosmological simulations of structure formation},
	volume = {542},
	issn = {0035-8711},
	url = {https://ui.adsabs.harvard.edu/abs/2025MNRAS.542.2019B},
	doi = {10.1093/mnras/staf1336},
	urldate = {2026-07-02},
	journal = {MNRAS},
	publisher = {OUP},
	author = {Buttigieg, Stephanie and Sijacki, Debora and Moore, Christopher J. and Bourne, Martin A.},
	month = sep,
	year = {2025},
	note = {ADS Bibcode: 2025MNRAS.542.2019B},
	pages = {2019--2038},
}

@article{hirashima_asura-fdps-ml_2025,
	title = {{ASURA}-{FDPS}-{ML}: {Star}-by-star {Galaxy} {Simulations} {Accelerated} by {Surrogate} {Modeling} for {Supernova} {Feedback}},
	volume = {987},
	issn = {0004-637X},
	shorttitle = {{ASURA}-{FDPS}-{ML}},
	url = {https://ui.adsabs.harvard.edu/abs/2025ApJ...987...86H},
	doi = {10.3847/1538-4357/add689},
	urldate = {2026-07-02},
	journal = {ApJ},
	publisher = {IOP},
	author = {Hirashima, Keiya and Moriwaki, Kana and Fujii, Michiko S. and Hirai, Yutaka and Saitoh, Takayuki R. and Makino, Junichiro and Steinwandel, Ulrich P. and Ho, Shirley},
	month = jul,
	year = {2025},
	note = {ADS Bibcode: 2025ApJ...987...86H},
	pages = {86},
}

@article{piana_blackholeweather_2026,
	title = {{BlackHoleWeather} -- {Spin}-coupled chaotic cold accretion across the meso-scale: {Morphology} and thermodynamics},
	volume = {arXiv.2605.27502},
	shorttitle = {{BlackHoleWeather} -- {Spin}-coupled chaotic cold accretion across the meso-scale},
	url = {https://ui.adsabs.harvard.edu/abs/2026arXiv260527502P},
	doi = {10.48550/arXiv.2605.27502},
	urldate = {2026-07-02},
	journal = {arXiv e-prints},
	publisher = {arXiv},
	author = {Piana, Olmo and Gaspari, Massimo and Barbani, Filippo and Cammelli, Vieri and Stel, Giovanni and Brustio, Davide M. and Olivares, Valeria and Salvestrini, Francesco and Danehkar, Ashkbiz and Tombesi, Francesco and Temi, Pasquale and Maccagni, Filippo M. and Fournier, Martin},
	month = may,
	year = {2026},
	note = {ADS Bibcode: 2026arXiv260527502P},
}

@article{piana_blackholeweather_2026-1,
	title = {{BlackHoleWeather} -- {Spin}-coupled chaotic cold accretion across the meso scale: {Variability} and kinematics},
	volume = {arXiv.2605.27508},
	shorttitle = {{BlackHoleWeather} -- {Spin}-coupled chaotic cold accretion across the meso scale},
	url = {https://ui.adsabs.harvard.edu/abs/2026arXiv260527508P},
	doi = {10.48550/arXiv.2605.27508},
	urldate = {2026-07-02},
	journal = {arXiv e-prints},
	publisher = {arXiv},
	author = {Piana, Olmo and Gaspari, Massimo and Cammelli, Vieri and Barbani, Filippo and Brustio, Davide M. and Stel, Giovanni and Olivares, Valeria and Danehkar, Ashkbiz and Temi, Pasquale and Serafinelli, Roberto and Salvestrini, Francesco and Maccagni, Filippo M. and Tombesi, Francesco},
	month = may,
	year = {2026},
	note = {ADS Bibcode: 2026arXiv260527508P},
}

@article{cenci_little_2025,
	title = {Little {Red} {Dots} as direct-collapse black hole nurseries},
	volume = {542},
	issn = {0035-8711},
	url = {https://ui.adsabs.harvard.edu/abs/2025MNRAS.542.2597C},
	doi = {10.1093/mnras/staf1362},
	urldate = {2026-07-02},
	journal = {MNRAS},
	publisher = {OUP},
	author = {Cenci, Elia and Habouzit, Melanie},
	month = sep,
	year = {2025},
	note = {ADS Bibcode: 2025MNRAS.542.2597C},
	pages = {2597--2609},
}

@article{degraf_cosmological_2020,
	title = {Cosmological simulations of massive black hole seeds: predictions for next-generation electromagnetic and gravitational wave observations},
	volume = {491},
	issn = {0035-8711},
	shorttitle = {Cosmological simulations of massive black hole seeds},
	url = {https://ui.adsabs.harvard.edu/abs/2020MNRAS.491.4973D},
	doi = {10.1093/mnras/stz3309},
	urldate = {2026-07-02},
	journal = {MNRAS},
	publisher = {OUP},
	author = {DeGraf, C. and Sijacki, D.},
	month = feb,
	year = {2020},
	note = {ADS Bibcode: 2020MNRAS.491.4973D},
	pages = {4973--4992},
}

@article{tremmel_off_2015,
	title = {Off the beaten path: a new approach to realistically model the orbital decay of supermassive black holes in galaxy formation simulations},
	volume = {451},
	issn = {0035-8711},
	shorttitle = {Off the beaten path},
	url = {https://ui.adsabs.harvard.edu/abs/2015MNRAS.451.1868T},
	doi = {10.1093/mnras/stv1060},
	urldate = {2026-07-02},
	journal = {MNRAS},
	publisher = {OUP},
	author = {Tremmel, M. and Governato, F. and Volonteri, M. and Quinn, T. R.},
	month = aug,
	year = {2015},
	note = {ADS Bibcode: 2015MNRAS.451.1868T},
	pages = {1868--1874},
}

@inproceedings{rantala_ketju_2017,
	title = {{KETJU}: {Post}-{Newtonian}-{Accurate} {Supermassive} {Black} {Hole} {Dynamics} in {GADGET}-3},
	volume = {324},
	shorttitle = {{KETJU}},
	doi = {10.1017/S1743921316013004},
	urldate = {2026-07-02},
	author = {Rantala, Antti and Pihajoki, Pauli and Johansson, Peter H.},
	month = jan,
	year = {2017},
	note = {ADS Bibcode: 2017IAUS..324..342R},
	pages = {342--346},
}

@article{mannerkoski_signatures_2022,
	title = {Signatures of the {Many} {Supermassive} {Black} {Hole} {Mergers} in a {Cosmologically} {Forming} {Massive} {Early}-type {Galaxy}},
	volume = {929},
	issn = {0004-637X},
	url = {https://ui.adsabs.harvard.edu/abs/2022ApJ...929..167M},
	doi = {10.3847/1538-4357/ac5f0b},
	urldate = {2026-07-02},
	journal = {ApJ},
	publisher = {IOP},
	author = {Mannerkoski, Matias and Johansson, Peter H. and Rantala, Antti and Naab, Thorsten and Liao, Shihong and Rawlings, Alexander},
	month = apr,
	year = {2022},
	note = {ADS Bibcode: 2022ApJ...929..167M},
	pages = {167},
}

@article{hirschmann_emission-line_2023,
	title = {Emission-line properties of {IllustrisTNG} galaxies: from local diagnostic diagrams to high-redshift predictions for {JWST}},
	volume = {526},
	issn = {0035-8711},
	shorttitle = {Emission-line properties of {IllustrisTNG} galaxies},
	url = {https://ui.adsabs.harvard.edu/abs/2023MNRAS.526.3610H},
	doi = {10.1093/mnras/stad2955},
	urldate = {2026-07-02},
	journal = {MNRAS},
	publisher = {OUP},
	author = {Hirschmann, Michaela and Charlot, Stephane and Feltre, Anna and Curtis-Lake, Emma and Somerville, Rachel S. and Chevallard, Jacopo and Choi, Ena and Nelson, Dylan and Morisset, Christophe and Plat, Adele and Vidal-Garcia, Alba},
	month = dec,
	year = {2023},
	note = {ADS Bibcode: 2023MNRAS.526.3610H},
	pages = {3610--3636},
}

@article{wilkins_first_2025,
	title = {First {Light} and {Reionization} {Epoch} {Simulations} ({FLARES}) -- {XVIII}: the ionising emissivities and hydrogen recombination line properties of early {AGN}},
	volume = {arXiv.2505.05257},
	shorttitle = {First {Light} and {Reionization} {Epoch} {Simulations} ({FLARES}) -- {XVIII}},
	url = {https://ui.adsabs.harvard.edu/abs/2025arXiv250505257W},
	doi = {10.48550/arXiv.2505.05257},
	urldate = {2026-07-02},
	journal = {arXiv e-prints},
	publisher = {arXiv},
	author = {Wilkins, Stephen M. and Vijayan, Aswin P. and Hagen, Scott and Caruana, Joseph and Conselice, Christopher J. and Done, Chris and Hirschmann, Michaela and Irodotou, Dimitrios and Lovell, Christopher C. and Matthee, Jorryt and Plat, Adèle and Roper, William J. and Taylor, Anthony J.},
	month = may,
	year = {2025},
	note = {ADS Bibcode: 2025arXiv250505257W},
}

@article{roy_little_2026,
	title = {Little {Red} {Dots} on {FIRE}: {Exploring} the formation and observational signatures of ultra-compact early galaxies},
	volume = {arXiv.2606.23683},
	shorttitle = {Little {Red} {Dots} on {FIRE}},
	url = {https://ui.adsabs.harvard.edu/abs/2026arXiv260623683R},
	doi = {10.48550/arXiv.2606.23683},
	urldate = {2026-07-03},
	journal = {arXiv e-prints},
	publisher = {arXiv},
	author = {Roy, Niranjan Chandra and Anglés-Alcázar, Daniel and Cochrane, Rachel K. and Richings, Alexander J. and Mercedes-Feliz, Jonathan and Hayward, Christopher C. and Faucher-Giguère, Claude-André and Lambrides, Erini and Feldmann, Robert and Oh, Boon Kiat and Marszewski, Andrew and Sun, Guochao and Davis, Kelcey and McKinney, Jed and Casey, Caitlin M. and Díaz-Santos, Tanio and Brooks, Madisyn and Farrell, Grace},
	month = jun,
	year = {2026},
	note = {ADS Bibcode: 2026arXiv260623683R},
}

@article{stern_type_2012,
	title = {Type 1 {AGN} at low z - {II}. {The} relative strength of narrow lines and the nature of intermediate type {AGN}},
	volume = {426},
	issn = {0035-8711},
	url = {https://ui.adsabs.harvard.edu/abs/2012MNRAS.426.2703S},
	doi = {10.1111/j.1365-2966.2012.21772.x},
	urldate = {2026-07-03},
	journal = {MNRAS},
	publisher = {OUP},
	author = {Stern, Jonathan and Laor, Ari},
	month = nov,
	year = {2012},
	note = {ADS Bibcode: 2012MNRAS.426.2703S},
	pages = {2703--2718},
}

@article{mcclymont_thesan-zoom_2025,
	title = {The {THESAN}-{ZOOM} project: burst, quench, repeat ─ unveiling the evolution of high-redshift galaxies along the star-forming main sequence},
	volume = {544},
	issn = {0035-8711},
	shorttitle = {The {THESAN}-{ZOOM} project},
	url = {https://ui.adsabs.harvard.edu/abs/2025MNRAS.544..513M},
	doi = {10.1093/mnras/staf1660},
	urldate = {2026-07-04},
	journal = {MNRAS},
	publisher = {OUP},
	author = {McClymont, William and Tacchella, Sandro and Smith, Aaron and Kannan, Rahul and Puchwein, Ewald and Borrow, Josh and Garaldi, Enrico and Keating, Laura and Vogelsberger, Mark and Zier, Oliver and Shen, Xuejian and Popovic, Filip and Simmonds, Charlotte},
	month = nov,
	year = {2025},
	note = {ADS Bibcode: 2025MNRAS.544..513M},
	pages = {513--534},
}

@article{regan_massive_2024,
	title = {Massive {Black} {Hole} {Seeds}},
	volume = {7},
	issn = {2565-6120},
	url = {https://ui.adsabs.harvard.edu/abs/2024OJAp....7E..72R},
	doi = {10.33232/001c.123239},
	urldate = {2026-07-04},
	journal = {OJAp},
	author = {Regan, John and Volonteri, Marta},
	month = sep,
	year = {2024},
	note = {ADS Bibcode: 2024OJAp....7E..72R},
	pages = {72},
}

@article{song_evolution_2016,
	title = {The {Evolution} of the {Galaxy} {Stellar} {Mass} {Function} at z = 4-8: {A} {Steepening} {Low}-mass-end {Slope} with {Increasing} {Redshift}},
	volume = {825},
	issn = {0004-637X},
	shorttitle = {The {Evolution} of the {Galaxy} {Stellar} {Mass} {Function} at z = 4-8},
	url = {https://ui.adsabs.harvard.edu/abs/2016ApJ...825....5S},
	doi = {10.3847/0004-637X/825/1/5},
	urldate = {2026-07-16},
	journal = {ApJ},
	publisher = {IOP},
	author = {Song, Mimi and Finkelstein, Steven L. and Ashby, Matthew L. N. and Grazian, A. and Lu, Yu and Papovich, Casey and Salmon, Brett and Somerville, Rachel S. and Dickinson, Mark and Duncan, K. and Faber, Sandy M. and Fazio, Giovanni G. and Ferguson, Henry C. and Fontana, Adriano and Guo, Yicheng and Hathi, Nimish and Lee, Seong-Kook and Merlin, Emiliano and Willner, S. P.},
	month = jul,
	year = {2016},
	note = {ADS Bibcode: 2016ApJ...825....5S},
	pages = {5},
}

@article{navarro-carrera_constraints_2024,
	title = {Constraints on the {Faint} {End} of the {Galaxy} {Stellar} {Mass} {Function} at z ≃ 4─8 from {Deep} {JWST} {Data}},
	volume = {961},
	issn = {0004-637X},
	url = {https://ui.adsabs.harvard.edu/abs/2024ApJ...961..207N},
	doi = {10.3847/1538-4357/ad0df6},
	urldate = {2026-07-16},
	journal = {ApJ},
	publisher = {IOP},
	author = {Navarro-Carrera, Rafael and Rinaldi, Pierluigi and Caputi, Karina I. and Iani, Edoardo and Kokorev, Vasily and van Mierlo, Sophie E.},
	month = feb,
	year = {2024},
	note = {ADS Bibcode: 2024ApJ...961..207N},
	pages = {207},
}

@article{weibel_galaxy_2024,
	title = {Galaxy build-up in the first 1.5 {Gyr} of cosmic history: insights from the stellar mass function at z   4-9 from {JWST} {NIRCam} observations},
	volume = {533},
	issn = {0035-8711},
	shorttitle = {Galaxy build-up in the first 1.5 {Gyr} of cosmic history},
	url = {https://ui.adsabs.harvard.edu/abs/2024MNRAS.533.1808W},
	doi = {10.1093/mnras/stae1891},
	urldate = {2026-07-16},
	journal = {MNRAS},
	publisher = {OUP},
	author = {Weibel, Andrea and Oesch, Pascal A. and Barrufet, Laia and Gottumukkala, Rashmi and Ellis, Richard S. and Santini, Paola and Weaver, John R. and Allen, Natalie and Bouwens, Rychard and Bowler, Rebecca A. A. and Brammer, Gabe and Carnall, Adam C. and Cullen, Fergus and Dayal, Pratika and Dickinson, Mark and Donnan, Callum T. and Dunlop, James S. and Giavalisco, Mauro and Grogin, Norman A. and Illingworth, Garth D. and Koekemoer, Anton M. and Labbe, Ivo and Marchesini, Danilo and McLeod, Derek J. and McLure, Ross J. and Naidu, Rohan P. and Pérez-González, Pablo G. and Shuntov, Marko and Stefanon, Mauro and Toft, Sune and Xiao, Mengyuan},
	month = sep,
	year = {2024},
	note = {ADS Bibcode: 2024MNRAS.533.1808W},
	pages = {1808--1838},
}

@article{baldry_galaxy_2012,
	title = {Galaxy {And} {Mass} {Assembly} ({GAMA}): the galaxy stellar mass function at z {\textless} 0.06},
	volume = {421},
	issn = {0035-8711},
	shorttitle = {Galaxy {And} {Mass} {Assembly} ({GAMA})},
	url = {https://ui.adsabs.harvard.edu/abs/2012MNRAS.421..621B},
	doi = {10.1111/j.1365-2966.2012.20340.x},
	urldate = {2026-07-16},
	journal = {MNRAS},
	publisher = {OUP},
	author = {Baldry, I. K. and Driver, S. P. and Loveday, J. and Taylor, E. N. and Kelvin, L. S. and Liske, J. and Norberg, P. and Robotham, A. S. G. and Brough, S. and Hopkins, A. M. and Bamford, S. P. and Peacock, J. A. and Bland-Hawthorn, J. and Conselice, C. J. and Croom, S. M. and Jones, D. H. and Parkinson, H. R. and Popescu, C. C. and Prescott, M. and Sharp, R. G. and Tuffs, R. J.},
	month = mar,
	year = {2012},
	note = {ADS Bibcode: 2012MNRAS.421..621B},
	pages = {621--634},
}

@article{li_distribution_2009,
	title = {The distribution of stellar mass in the low-redshift {Universe}},
	volume = {398},
	issn = {0035-8711},
	url = {https://ui.adsabs.harvard.edu/abs/2009MNRAS.398.2177L},
	doi = {10.1111/j.1365-2966.2009.15268.x},
	urldate = {2026-07-16},
	journal = {MNRAS},
	publisher = {OUP},
	author = {Li, Cheng and White, Simon D. M.},
	month = oct,
	year = {2009},
	note = {ADS Bibcode: 2009MNRAS.398.2177L},
	pages = {2177--2187},
}

@article{dsouza_massive_2015,
	title = {The massive end of the stellar mass function},
	volume = {454},
	issn = {0035-8711},
	url = {https://ui.adsabs.harvard.edu/abs/2015MNRAS.454.4027D},
	doi = {10.1093/mnras/stv2234},
	urldate = {2026-07-16},
	journal = {MNRAS},
	publisher = {OUP},
	author = {D'Souza, Richard and Vegetti, Simona and Kauffmann, Guinevere},
	month = dec,
	year = {2015},
	note = {ADS Bibcode: 2015MNRAS.454.4027D},
	pages = {4027--4036},
}

@article{bernardi_massive_2013,
	title = {The massive end of the luminosity and stellar mass functions: dependence on the fit to the light profile},
	volume = {436},
	issn = {0035-8711},
	shorttitle = {The massive end of the luminosity and stellar mass functions},
	url = {https://ui.adsabs.harvard.edu/abs/2013MNRAS.436..697B},
	doi = {10.1093/mnras/stt1607},
	urldate = {2026-07-16},
	journal = {MNRAS},
	publisher = {OUP},
	author = {Bernardi, M. and Meert, A. and Sheth, R. K. and Vikram, V. and Huertas-Company, M. and Mei, S. and Shankar, F.},
	month = nov,
	year = {2013},
	note = {ADS Bibcode: 2013MNRAS.436..697B},
	pages = {697--704},
}

@article{somerville_cosmic_2004,
	title = {Cosmic {Variance} in the {Great} {Observatories} {Origins} {Deep} {Survey}},
	volume = {600},
	issn = {0004-637X},
	url = {https://ui.adsabs.harvard.edu/abs/2004ApJ...600L.171S},
	doi = {10.1086/378628},
	urldate = {2026-07-16},
	journal = {ApJ},
	publisher = {IOP},
	author = {Somerville, Rachel S. and Lee, Kyoungsoo and Ferguson, Henry C. and Gardner, Jonathan P. and Moustakas, Leonidas A. and Giavalisco, Mauro},
	month = jan,
	year = {2004},
	note = {ADS Bibcode: 2004ApJ...600L.171S},
	pages = {L171--L174},
}

@article{trenti_cosmic_2008,
	title = {Cosmic {Variance} and {Its} {Effect} on the {Luminosity} {Function} {Determination} in {Deep} {High}-z {Surveys}},
	volume = {676},
	issn = {0004-637X},
	url = {https://ui.adsabs.harvard.edu/abs/2008ApJ...676..767T},
	doi = {10.1086/528674},
	urldate = {2026-07-16},
	journal = {ApJ},
	publisher = {IOP},
	author = {Trenti, M. and Stiavelli, M.},
	month = apr,
	year = {2008},
	note = {ADS Bibcode: 2008ApJ...676..767T},
	pages = {767--780},
}

@article{tinker_toward_2008,
	title = {Toward a {Halo} {Mass} {Function} for {Precision} {Cosmology}: {The} {Limits} of {Universality}},
	volume = {688},
	issn = {0004-637X},
	shorttitle = {Toward a {Halo} {Mass} {Function} for {Precision} {Cosmology}},
	url = {https://ui.adsabs.harvard.edu/abs/2008ApJ...688..709T},
	doi = {10.1086/591439},
	urldate = {2026-07-16},
	journal = {ApJ},
	publisher = {IOP},
	author = {Tinker, Jeremy and Kravtsov, Andrey V. and Klypin, Anatoly and Abazajian, Kevork and Warren, Michael and Yepes, Gustavo and Gottlöber, Stefan and Holz, Daniel E.},
	month = dec,
	year = {2008},
	note = {ADS Bibcode: 2008ApJ...688..709T},
	pages = {709--728},
}

@article{moster_galactic_2013,
	title = {Galactic star formation and accretion histories from matching galaxies to dark matter haloes},
	volume = {428},
	issn = {0035-8711},
	url = {https://ui.adsabs.harvard.edu/abs/2013MNRAS.428.3121M},
	doi = {10.1093/mnras/sts261},
	urldate = {2026-07-16},
	journal = {MNRAS},
	publisher = {OUP},
	author = {Moster, Benjamin P. and Naab, Thorsten and White, Simon D. M.},
	month = feb,
	year = {2013},
	note = {ADS Bibcode: 2013MNRAS.428.3121M},
	pages = {3121--3138},
}

@article{bigwood_case_2025,
	title = {The case for large-scale {AGN} feedback in galaxy formation simulations: insights from {XFABLE}},
	volume = {542},
	issn = {0035-8711},
	shorttitle = {The case for large-scale {AGN} feedback in galaxy formation simulations},
	url = {https://ui.adsabs.harvard.edu/abs/2025MNRAS.542.3206B},
	doi = {10.1093/mnras/staf1435},
	urldate = {2026-07-16},
	journal = {MNRAS},
	publisher = {OUP},
	author = {Bigwood, Leah and Bourne, Martin A. and Iršič, Vid and Amon, Alexandra and Sijacki, Debora},
	month = sep,
	year = {2025},
	note = {ADS Bibcode: 2025MNRAS.542.3206B},
	pages = {3206--3230},
}

@article{gupta_rapid_2026,
	title = {A {Rapid} {Evolution} in the {Observed} {Mbh}/{M}* {Relation} at z {\textgreater} 3 {Revealed} via {Spectro}-photometric {SED}-{Modeling}},
	volume = {arXiv.2605.30414},
	url = {https://ui.adsabs.harvard.edu/abs/2026arXiv260530414G},
	doi = {10.48550/arXiv.2605.30414},
	urldate = {2026-07-16},
	journal = {arXiv e-prints},
	publisher = {arXiv},
	author = {Gupta, Ansh R. and Taylor, Anthony and Curtis-Lake, Emma and Silcock, Maddie and Chávez Ortiz, Óscar A. and Finkelstein, Steven L. and Akins, Hollis B. and Backhaus, Bren E. and Barro, Guillermo and Bisigello, Laura and Brooks, Madisyn and Casey, Caitlin M. and Charlot, Stephane and Chevallard, Jacopo and Feltre, Anna and Gandolfi, Giovanni and Giavalisco, Mauro and Grogin, Norman A. and Hirschmann, Michaela and Hsiao, Tiger Yu-Yang and Jeon, Junehyoung and Jogee, Shardha and Kartaltepe, Jeyhan S. and Kocevski, Dale D. and Koekemoer, Anton M. and Kokorev, Vasily and Leung, Gene C. K. and Lucas, Ray A. and Pacucci, Fabio and Pirzkal, Nor and Plat, Adele and Somerville, Rachel S. and Trump, Jonathan R. and Vidal-García, Alba and Wang, Xin and Yung, L. Y. Aaron},
	month = may,
	year = {2026},
	note = {ADS Bibcode: 2026arXiv260530414G},
}

@article{kocevski_rise_2025,
	title = {The {Rise} of {Faint}, {Red} {Active} {Galactic} {Nuclei} at z {\textgreater} 4: {A} {Sample} of {Little} {Red} {Dots} in the {JWST} {Extragalactic} {Legacy} {Fields}},
	volume = {986},
	issn = {0004-637X},
	shorttitle = {The {Rise} of {Faint}, {Red} {Active} {Galactic} {Nuclei} at z {\textgreater} 4},
	url = {https://ui.adsabs.harvard.edu/abs/2025ApJ...986..126K},
	doi = {10.3847/1538-4357/adbc7d},
	urldate = {2026-07-16},
	journal = {ApJ},
	publisher = {IOP},
	author = {Kocevski, Dale D. and Finkelstein, Steven L. and Barro, Guillermo and Taylor, Anthony J. and Calabrò, Antonello and Laloux, Brivael and Buchner, Johannes and Trump, Jonathan R. and Leung, Gene C. K. and Yang, Guang and Dickinson, Mark and Pérez-González, Pablo G. and Pacucci, Fabio and Inayoshi, Kohei and Somerville, Rachel S. and McGrath, Elizabeth J. and Akins, Hollis B. and Bagley, Micaela B. and Bowler, Rebecca A. A. and Bisigello, Laura and Carnall, Adam and Casey, Caitlin M. and Cheng, Yingjie and Cleri, Nikko J. and Costantin, Luca and Cullen, Fergus and Davis, Kelcey and Donnan, Callum T. and Dunlop, James S. and Ellis, Richard S. and Ferguson, Henry C. and Fujimoto, Seiji and Fontana, Adriano and Giavalisco, Mauro and Grazian, Andrea and Grogin, Norman A. and Hathi, Nimish P. and Hirschmann, Michaela and Huertas-Company, Marc and Holwerda, Benne W. and Illingworth, Garth and Juneau, Stéphanie and Kartaltepe, Jeyhan S. and Koekemoer, Anton M. and Li, Wenxiu and Lucas, Ray A. and Magee, Dan and Mason, Charlotte and McLeod, Derek J. and McLure, Ross J. and Napolitano, Lorenzo and Papovich, Casey and Pirzkal, Nor and Rodighiero, Giulia and Santini, Paola and Wilkins, Stephen M. and Yung, L. Y. Aaron},
	month = jun,
	year = {2025},
	note = {ADS Bibcode: 2025ApJ...986..126K},
	pages = {126},
}

@article{jeon_observability_2023,
	title = {Observability of low-luminosity {AGNs} in the early {Universe} with {JWST}},
	volume = {524},
	issn = {0035-8711},
	url = {https://ui.adsabs.harvard.edu/abs/2023MNRAS.524..176J},
	doi = {10.1093/mnras/stad1877},
	urldate = {2026-07-16},
	journal = {MNRAS},
	publisher = {OUP},
	author = {Jeon, Junehyoung and Liu, Boyuan and Bromm, Volker and Finkelstein, Steven L.},
	month = sep,
	year = {2023},
	note = {ADS Bibcode: 2023MNRAS.524..176J},
	pages = {176--187},
}

@inproceedings{pontoppidan_pandeia_2016,
	title = {Pandeia: a multi-mission exposure time calculator for {JWST} and {WFIRST}},
	volume = {9910},
	shorttitle = {Pandeia},
	doi = {10.1117/12.2231768},
	language = {en},
	urldate = {2026-07-20},
	booktitle = {Observatory {Operations}: {Strategies}, {Processes}, and {Systems} {VI}},
	publisher = {SPIE},
	author = {Pontoppidan, Klaus M. and Pickering, Timothy E. and Laidler, Victoria G. and Gilbert, Karoline and Sontag, Christopher D. and Slocum, Christine and Jr, Mark J. Sienkiewicz and Hanley, Christopher and Earl, Nicholas M. and Pueyo, Laurent and Ravindranath, Swara and Karakla, Diane M. and Robberto, Massimo and Noriega-Crespo, Alberto and Barker, Elizabeth A.},
	month = jul,
	year = {2016},
	pages = {381},
}

@article{zamora_ga-nifs_2025,
	title = {{GA}-{NIFS}: {Understanding} the ionization nature of {EGSY8p7}/{CEERS}-1019. {Evidence} for a star formation-driven outflow at z = 8.6},
	volume = {arXiv.2512.09022},
	shorttitle = {{GA}-{NIFS}},
	url = {https://ui.adsabs.harvard.edu/abs/2025arXiv251209022Z},
	doi = {10.48550/arXiv.2512.09022},
	urldate = {2026-07-20},
	journal = {arXiv e-prints},
	publisher = {arXiv},
	author = {Zamora, Sandra and Carniani, Stefano and Bertola, Elena and Parlanti, Eleonora and Pérez-González, Pablo G. and Arribas, Santiago and Böker, Torsten and Bunker, Andrew J. and D'Eugenio, Francesco and Maiolino, Roberto and Perna, Michele and Rodríguez Del Pino, Bruno and Übler, Hannah and Cresci, Giovanni and Jones, Gareth C. and Lamperti, Isabella and Scholtz, Jan and Trefoloni, Bartolomeo and Venturi, Giacomo},
	month = dec,
	year = {2025},
	note = {ADS Bibcode: 2025arXiv251209022Z},
}

@article{rodriguez_del_pino_ga-nifs_2026,
	title = {{GA}-{NIFS}: {High} prevalence of dusty and metal-enriched outflows in massive and luminous star-forming galaxies at z ∼ 3─9},
	volume = {711},
	issn = {0004-6361},
	shorttitle = {{GA}-{NIFS}},
	url = {https://ui.adsabs.harvard.edu/abs/2026A&A...711A..66R},
	doi = {10.1051/0004-6361/202558697},
	urldate = {2026-07-20},
	journal = {A\&A},
	publisher = {EDP},
	author = {Rodríguez Del Pino, B. and Arribas, S. and Perna, M. and Lamperti, I. and Bunker, A. and Carniani, S. and Charlot, S. and D'Eugenio, F. and Maiolino, R. and Übler, H. and Bertola, E. and Böker, T. and Cresci, G. and Jones, G. C. and Marconcini, C. and Parlanti, E. and Scholtz, J. and Venturi, G. and Zamora, S.},
	month = jul,
	year = {2026},
	note = {ADS Bibcode: 2026A\&A...711A..66R},
	pages = {A66},
}

@article{deugenio_jades_2026,
	title = {{JADES} {Dark} {Horse}: demonstrating high-multiplex observations with {JWST}/{NIRSpec} dense-shutter spectroscopy in the {JADES} {Origins} {Field}},
	volume = {549},
	issn = {0035-8711},
	shorttitle = {{JADES} {Dark} {Horse}},
	url = {https://ui.adsabs.harvard.edu/abs/2026MNRAS.549ag824D},
	doi = {10.1093/mnras/stag824},
	urldate = {2026-07-20},
	journal = {MNRAS},
	publisher = {OUP},
	author = {D'Eugenio, Francesco and Nelson, Erica J. and Eisenstein, Daniel J. and Maiolino, Roberto and Carniani, Stefano and Scholtz, Jan and Curti, Mirko and Willmer, Christopher N. A. and Bunker, Andrew J. and Helton, Jakob M. and Juodžbalis, Ignas and Sun, Fengwu and Tacchella, Sandro and Arribas, Santiago and Cameron, Alex J. and Charlot, Stéphane and Curtis-Lake, Emma and Hainline, Kevin and Johnson, Benjamin D. and Robertson, Brant and Williams, Christina C. and Willott, Chris and Baker, William M. and Chevallard, Jacopo and Danhaive, A. Lola and Isobe, Yuki and Ji, Xihan and Ji, Zhiyuan and Jones, Gareth C. and Kumari, Nimisha and Looser, Tobias J. and Lyu, Jianwei and Parlanti, Eleonora and Perna, Michele and Puskás, Dávid and Rinaldi, Pierluigi and Simmonds, Charlotte and Sun, Yang and Übler, Hannah and Venturi, Giacomo and Witstok, Joris and Wu, Zihao and Zhu, Yongda},
	month = jun,
	year = {2026},
	note = {ADS Bibcode: 2026MNRAS.549ag824D},
	pages = {stag824},
}

@article{gravity_collaboration_size-luminosity_2024,
	title = {The size-luminosity relation of local active galactic nuclei from interferometric observations of the broad-line region},
	volume = {684},
	copyright = {© The Authors 2024},
	issn = {0004-6361, 1432-0746},
	url = {https://www.aanda.org/articles/aa/abs/2024/04/aa48167-23/aa48167-23.html},
	doi = {10.1051/0004-6361/202348167},
	language = {en},
	urldate = {2026-07-20},
	journal = {A\&A},
	publisher = {EDP Sciences},
	author = {{Gravity Collaboration} and Amorim, A. and Cao, Y. and Clénet, Y. and Davies, R. and Zeeuw, P. T. de and Dexter, J. and Drescher, A. and Eckart, A. and Eisenhauer, F. and Fabricius, M. and Feuchtgruber, H. and Schreiber, N. M. Förster and Garcia, P. J. V. and Genzel, R. and Gillessen, S. and Gratadour, D. and Hönig, S. and Kishimoto, M. and Lacour, S. and Lutz, D. and Millour, F. and Netzer, H. and Ott, T. and Paumard, T. and Perraut, K. and Perrin, G. and Peterson, B. M. and Petrucci, P. O. and Pfuhl, O. and Prieto, M. A. and Rabien, S. and Rouan, D. and Santos, D. J. D. and Shangguan, J. and Shimizu, T. and Sternberg, A. and Straubmeier, C. and Sturm, E. and Tacconi, L. J. and Tristram, K. R. W. and Widmann, F. and Woillez, J.},
	month = apr,
	year = {2024},
	pages = {A167},
}

@article{taylor_seeding_2014,
	title = {Seeding black holes in cosmological simulations},
	volume = {442},
	issn = {0035-8711},
	url = {https://ui.adsabs.harvard.edu/abs/2014MNRAS.442.2751T},
	doi = {10.1093/mnras/stu983},
	urldate = {2026-07-20},
	journal = {MNRAS},
	publisher = {OUP},
	author = {Taylor, P. and Kobayashi, C.},
	month = aug,
	year = {2014},
	note = {ADS Bibcode: 2014MNRAS.442.2751T},
	pages = {2751--2767},
}

@article{taylor_time_2016,
	title = {Time evolution of galaxy scaling relations in cosmological simulations},
	volume = {463},
	issn = {0035-8711},
	url = {https://ui.adsabs.harvard.edu/abs/2016MNRAS.463.2465T},
	doi = {10.1093/mnras/stw2157},
	urldate = {2026-07-20},
	journal = {MNRAS},
	publisher = {OUP},
	author = {Taylor, Philip and Kobayashi, Chiaki},
	month = dec,
	year = {2016},
	note = {ADS Bibcode: 2016MNRAS.463.2465T},
	pages = {2465--2479},
}

@article{hviding_rubies_2025,
	title = {{RUBIES}: {A} spectroscopic census of little red dots: {All} point sources with v-shaped continua have broad lines},
	volume = {702},
	issn = {0004-6361},
	shorttitle = {{RUBIES}},
	url = {https://ui.adsabs.harvard.edu/abs/2025A&A...702A..57H},
	doi = {10.1051/0004-6361/202555816},
	urldate = {2026-07-20},
	journal = {A\&A},
	publisher = {EDP},
	author = {Hviding, Raphael E. and de Graaff, Anna and Miller, Tim B. and Setton, David J. and Greene, Jenny E. and Labbé, Ivo and Brammer, Gabriel and Bezanson, Rachel and Boogaard, Leindert A. and Cleri, Nikko J. and Leja, Joel and Maseda, Michael V. and McConachie, Ian and Matthee, Jorryt and Naidu, Rohan P. and Oesch, Pascal A. and Wang, Bingjie and Whitaker, Katherine E. and Williams, Christina C.},
	month = oct,
	year = {2025},
	note = {ADS Bibcode: 2025A\&A...702A..57H},
	pages = {A57},
}

@article{narayanan_outshining_2024,
	title = {Outshining by {Recent} {Star} {Formation} {Prevents} the {Accurate} {Measurement} of {High}-z {Galaxy} {Stellar} {Masses}},
	volume = {961},
	issn = {0004-637X},
	url = {https://doi.org/10.3847/1538-4357/ad0966},
	doi = {10.3847/1538-4357/ad0966},
	language = {en},
	number = {1},
	urldate = {2026-07-20},
	journal = {ApJ},
	publisher = {The American Astronomical Society},
	author = {Narayanan, Desika and Lower, Sidney and Torrey, Paul and Brammer, Gabriel and Cui, Weiguang and Davé, Romeel and Iyer, Kartheik G. and Li, Qi and Lovell, Christopher C. and Sales, Laura V. and Stark, Daniel P. and Marinacci, Federico and Vogelsberger, Mark},
	month = jan,
	year = {2024},
	pages = {73},
}

@article{bourne_recent_2023,
	title = {Recent {Progress} in {Modeling} the {Macro}- and {Micro}-{Physics} of {Radio} {Jet} {Feedback} in {Galaxy} {Clusters}},
	volume = {11},
	url = {https://ui.adsabs.harvard.edu/abs/2023Galax..11...73B},
	doi = {10.3390/galaxies11030073},
	urldate = {2026-07-20},
	journal = {Galaxies},
	publisher = {MDPI},
	author = {Bourne, Martin A. and Yang, Hsiang-Yi Karen},
	month = jun,
	year = {2023},
	note = {ADS Bibcode: 2023Galax..11...73B},
	pages = {73},
}

@article{madau_super-critical_2014,
	title = {Super-critical {Growth} of {Massive} {Black} {Holes} from {Stellar}-mass {Seeds}},
	volume = {784},
	issn = {0004-637X},
	url = {https://ui.adsabs.harvard.edu/abs/2014ApJ...784L..38M},
	doi = {10.1088/2041-8205/784/2/L38},
	urldate = {2026-07-21},
	journal = {ApJ},
	publisher = {IOP},
	author = {Madau, Piero and Haardt, Francesco and Dotti, Massimo},
	month = apr,
	year = {2014},
	note = {ADS Bibcode: 2014ApJ...784L..38M},
	pages = {L38},
}

@article{sadowski_numerical_2014,
	title = {Numerical simulations of super-critical black hole accretion flows in general relativity},
	volume = {439},
	issn = {0035-8711},
	url = {https://ui.adsabs.harvard.edu/abs/2014MNRAS.439..503S},
	doi = {10.1093/mnras/stt2479},
	urldate = {2026-07-21},
	journal = {MNRAS},
	publisher = {OUP},
	author = {Sądowski, Aleksander and Narayan, Ramesh and McKinney, Jonathan C. and Tchekhovskoy, Alexander},
	month = mar,
	year = {2014},
	note = {ADS Bibcode: 2014MNRAS.439..503S},
	pages = {503--520},
}

@article{scholtz_jades_2026,
	title = {{JADES} {Data} {Release} 4 ─ {Paper} {II}. {Data} reduction, analysis, and emission-line fluxes of the complete spectroscopic sample},
	volume = {549},
	issn = {0035-8711},
	url = {https://ui.adsabs.harvard.edu/abs/2026MNRAS.549ag939S},
	doi = {10.1093/mnras/stag939},
	urldate = {2026-07-23},
	journal = {MNRAS},
	publisher = {OUP},
	author = {Scholtz, J. and Carniani, S. and Parlanti, E. and D'Eugenio, F. and Curtis-Lake, E. and Jakobsen, P. and Bunker, A. J. and Cameron, A. J. and Arribas, S. and Baker, W. M. and Charlot, S. and Chevellard, J. and Circosta, C. and Curti, M. and Duan, Q. and Eisenstein, D. J. and Hainline, K. and Ji, Z. and Johnson, B. D. and Jones, G. C. and Kumari, N. and Maiolino, R. and Maseda, M. V. and Perna, M. and Pérez-González, P. G. and Rawle, T. and Rieke, M. and Rinaldi, P. and Robertson, B. and Saxena, A. and Shivaei, I. and Silcock, M. S. and Sun, Y. and Rodríguez Del Pino, B. and Tacchella, S. and Übler, H. and Venturi, G. and Williams, C. C. and Willmer, C. N. A. and Willott, C. and Witstok, J.},
	month = jul,
	year = {2026},
	note = {ADS Bibcode: 2026MNRAS.549ag939S},
	pages = {stag939},
}

@article{scholtz_jades_2025,
	title = {{JADES}: {A} large population of obscured, narrow-line active galactic nuclei at high redshift},
	volume = {697},
	issn = {0004-6361},
	shorttitle = {{JADES}},
	url = {https://ui.adsabs.harvard.edu/abs/2025A&A...697A.175S},
	doi = {10.1051/0004-6361/202348804},
	urldate = {2026-07-23},
	journal = {A\&A},
	publisher = {EDP},
	author = {Scholtz, Jan and Maiolino, Roberto and D'Eugenio, Francesco and Curtis-Lake, Emma and Carniani, Stefano and Charlot, Stephane and Curti, Mirko and Silcock, Maddie S. and Arribas, Santiago and Baker, William and Bhatawdekar, Rachana and Boyett, Kristan and Bunker, Andrew J. and Chevallard, Jacopo and Circosta, Chiara and Eisenstein, Daniel J. and Hainline, Kevin and Hausen, Ryan and Ji, Xihan and Ji, Zhiyuan and Johnson, Benjamin D. and Kumari, Nimisha and Looser, Tobias J. and Lyu, Jianwei and Maseda, Michael V. and Parlanti, Eleonora and Perna, Michele and Rieke, Marcia and Robertson, Brant and Del Pino, Bruno Rodríguez and Sun, Fengwu and Tacchella, Sandro and Übler, Hannah and Venturi, Giacomo and Williams, Christina C. and Willmer, Christopher N. A. and Willott, Chris and Witstok, Joris},
	month = may,
	year = {2025},
	note = {ADS Bibcode: 2025A\&A...697A.175S},
	pages = {A175},
}

@article{sun_bursty_2023,
	title = {Bursty {Star} {Formation} {Naturally} {Explains} the {Abundance} of {Bright} {Galaxies} at {Cosmic} {Dawn}},
	volume = {955},
	issn = {0004-637X},
	url = {https://ui.adsabs.harvard.edu/abs/2023ApJ...955L..35S},
	doi = {10.3847/2041-8213/acf85a},
	urldate = {2026-07-23},
	journal = {ApJ},
	publisher = {IOP},
	author = {Sun, Guochao and Faucher-Giguère, Claude-André and Hayward, Christopher C. and Shen, Xuejian and Wetzel, Andrew and Cochrane, Rachel K.},
	month = oct,
	year = {2023},
	note = {ADS Bibcode: 2023ApJ...955L..35S},
	pages = {L35},
}

@article{bhagwat_spice_2024,
	title = {{SPICE}: the connection between cosmic reionization and stellar feedback in the first galaxies},
	volume = {531},
	issn = {0035-8711},
	shorttitle = {{SPICE}},
	url = {https://ui.adsabs.harvard.edu/abs/2024MNRAS.531.3406B},
	doi = {10.1093/mnras/stae1125},
	urldate = {2026-07-23},
	journal = {MNRAS},
	publisher = {OUP},
	author = {Bhagwat, Aniket and Costa, Tiago and Ciardi, Benedetta and Pakmor, Rüdiger and Garaldi, Enrico},
	month = jul,
	year = {2024},
	note = {ADS Bibcode: 2024MNRAS.531.3406B},
	pages = {3406--3430},
}

@article{scholtz_gn-z11_2024,
	title = {{GN}-z11: {The} environment of an active galactic nucleus at z = 10.603. {New} insights into the most distant {Lyα} detection},
	volume = {687},
	issn = {0004-6361},
	shorttitle = {{GN}-z11},
	url = {https://ui.adsabs.harvard.edu/abs/2024A&A...687A.283S},
	doi = {10.1051/0004-6361/202347187},
	urldate = {2026-07-23},
	journal = {A\&A},
	publisher = {EDP},
	author = {Scholtz, Jan and Witten, Callum and Laporte, Nicolas and Übler, Hannah and Perna, Michele and Maiolino, Roberto and Arribas, Santiago and Baker, William M. and Bennett, Jake S. and D'Eugenio, Francesco and Simmonds, Charlotte and Tacchella, Sandro and Witstok, Joris and Bunker, Andrew J. and Carniani, Stefano and Charlot, Stéphane and Cresci, Giovanni and Curtis-Lake, Emma and Eisenstein, Daniel J. and Kumari, Nimisha and Robertson, Brant and Rodríguez Del Pino, Bruno and Smit, Renske and Venturi, Giacomo and Williams, Christina C. and Willmer, Christopher N. A.},
	month = jul,
	year = {2024},
	note = {ADS Bibcode: 2024A\&A...687A.283S},
	pages = {A283},
}

@article{marshall_ga-nifs_2025,
	title = {{GA}-{NIFS} and {EIGER}: {A} merging quasar host at z = 7 with an overmassive black hole},
	volume = {702},
	issn = {0004-6361},
	shorttitle = {{GA}-{NIFS} and {EIGER}},
	url = {https://ui.adsabs.harvard.edu/abs/2025A&A...702A..50M},
	doi = {10.1051/0004-6361/202452650},
	urldate = {2026-07-23},
	journal = {A\&A},
	publisher = {EDP},
	author = {Marshall, Madeline A. and Yue, Minghao and Eilers, Anna-Christina and Scholtz, Jan and Perna, Michele and Willott, Chris J. and Maiolino, Roberto and Übler, Hannah and Arribas, Santiago and Bunker, Andrew J. and Charlot, Stephane and Rodríguez Del Pino, Bruno and Böker, Torsten and Carniani, Stefano and Circosta, Chiara and Cresci, Giovanni and D'Eugenio, Francesco and Jones, Gareth C. and Venturi, Giacomo and Bordoloi, Rongmon and Kashino, Daichi and Mackenzie, Ruari and Matthee, Jorryt and Naidu, Rohan and Simcoe, Robert A.},
	month = oct,
	year = {2025},
	note = {ADS Bibcode: 2025A\&A...702A..50M},
	pages = {A50},
}

@article{perna_ga-nifs_2023,
	title = {{GA}-{NIFS}: {The} ultra-dense, interacting environment of a dual {AGN} at z ∼ 3.3 revealed by {JWST}/{NIRSpec} {IFS}},
	volume = {679},
	issn = {0004-6361},
	shorttitle = {{GA}-{NIFS}},
	url = {https://ui.adsabs.harvard.edu/abs/2023A&A...679A..89P},
	doi = {10.1051/0004-6361/202346649},
	urldate = {2026-07-23},
	journal = {A\&A},
	publisher = {EDP},
	author = {Perna, M. and Arribas, S. and Marshall, M. and D'Eugenio, F. and Übler, H. and Bunker, A. and Charlot, S. and Carniani, S. and Jakobsen, P. and Maiolino, R. and Rodríguez Del Pino, B. and Willott, C. J. and Böker, T. and Circosta, C. and Cresci, G. and Curti, M. and Husemann, B. and Kumari, N. and Lamperti, I. and Pérez-González, P. G. and Scholtz, J.},
	month = nov,
	year = {2023},
	note = {ADS Bibcode: 2023A\&A...679A..89P},
	pages = {A89},
}

@article{parlanti_ga-nifs_2024,
	title = {{GA}-{NIFS}: {Early}-stage feedback in a heavily obscured active galactic nucleus at z = 4.76},
	volume = {684},
	issn = {0004-6361},
	shorttitle = {{GA}-{NIFS}},
	url = {https://ui.adsabs.harvard.edu/abs/2024A&A...684A..24P},
	doi = {10.1051/0004-6361/202347914},
	urldate = {2026-07-23},
	journal = {A\&A},
	publisher = {EDP},
	author = {Parlanti, Eleonora and Carniani, Stefano and Übler, Hannah and Venturi, Giacomo and Circosta, Chiara and D'Eugenio, Francesco and Arribas, Santiago and Bunker, Andrew J. and Charlot, Stéphane and Lützgendorf, Nora and Maiolino, Roberto and Perna, Michele and Rodríguez Del Pino, Bruno and Willott, Chris J. and Böker, Torsten and Cameron, Alex J. and Chevallard, Jacopo and Cresci, Giovanni and Jones, Gareth C. and Kumari, Nimisha and Lamperti, Isabella and Scholtz, Jan},
	month = apr,
	year = {2024},
	note = {ADS Bibcode: 2024A\&A...684A..24P},
	pages = {A24},
}

@article{deugenio_jades_2026-1,
	title = {{JADES} and {BlackTHUNDER}: rest-frame {Balmer}-line absorption and the local environment in a {Little} {Red} {Dot} at z = 5},
	volume = {545},
	issn = {0035-8711},
	shorttitle = {{JADES} and {BlackTHUNDER}},
	url = {https://ui.adsabs.harvard.edu/abs/2026MNRAS.545f2117D},
	doi = {10.1093/mnras/staf2117},
	urldate = {2026-07-23},
	journal = {MNRAS},
	publisher = {OUP},
	author = {D'Eugenio, Francesco and Juodžbalis, Ignas and Ji, Xihan and Scholtz, Jan and Maiolino, Roberto and Carniani, Stefano and Perna, Michele and Mazzolari, Giovanni and Übler, Hannah and Arribas, Santiago and Bhatawdekar, Rachana and Bunker, Andrew J. and Cresci, Giovanni and Curtis-Lake, Emma and Hainline, Kevin and Inayoshi, Kohei and Isobe, Yuki and Ji, Zhiyuan and Johnson, Benjamin D. and Jones, Gareth C. and Looser, Tobias J. and Nelson, Erica J. and Parlanti, Eleonora and Puskás, Dávid and Rinaldi, Pierluigi and Robertson, Brant and Rodríguez Del Pino, Bruno and Shivaei, Irene and Sun, Fengwu and Tacchella, Sandro and Venturi, Giacomo and Volonteri, Marta and Williams, Christina C. and Willmer, Christopher N. A. and Willott, Chris and Witstok, Joris},
	month = jan,
	year = {2026},
	note = {ADS Bibcode: 2026MNRAS.545f2117D},
	pages = {staf2117},
}

@article{jones_blackthunder_2026,
	title = {{BlackTHUNDER}: {Shedding} light on a dormant and extreme little red dot at z = 8.50},
	volume = {546},
	issn = {0035-8711},
	shorttitle = {{BlackTHUNDER}},
	url = {https://ui.adsabs.harvard.edu/abs/2026MNRAS.546ag115J},
	doi = {10.1093/mnras/stag115},
	urldate = {2026-07-23},
	journal = {MNRAS},
	publisher = {OUP},
	author = {Jones, Gareth C. and Übler, Hannah and Maiolino, Roberto and Ji, Xihan and Marconi, Alessandro and D'Eugenio, Francesco and Arribas, Santiago and Bunker, Andrew J. and Carniani, Stefano and Charlot, Stéphane and Cresci, Giovanni and Inayoshi, Kohei and Isobe, Yuki and Juodžbalis, Ignas and Mazzolari, Giovanni and Pérez-González, Pablo G. and Perna, Michele and Schneider, Raffaella and Scholtz, Jan and Tacchella, Sandro},
	month = mar,
	year = {2026},
	note = {ADS Bibcode: 2026MNRAS.546ag115J},
	pages = {stag115},
}

@article{madau_little_2026,
	title = {Little red dots as obscured little blue dots: relative abundances, luminosities, and black-hole masses},
	volume = {arXiv.2605.05074},
	shorttitle = {Little red dots as obscured little blue dots},
	url = {https://ui.adsabs.harvard.edu/abs/2026arXiv260505074M},
	doi = {10.48550/arXiv.2605.05074},
	urldate = {2026-07-26},
	journal = {arXiv e-prints},
	publisher = {arXiv},
	author = {Madau, Piero and Maiolino, Roberto},
	month = may,
	year = {2026},
	note = {ADS Bibcode: 2026arXiv260505074M},
}

@article{hainline_investigation_2025,
	title = {An {Investigation} into the {Selection} and {Colors} of {Little} {Red} {Dots} and {Active} {Galactic} {Nuclei}},
	volume = {979},
	issn = {0004-637X},
	url = {https://ui.adsabs.harvard.edu/abs/2025ApJ...979..138H},
	doi = {10.3847/1538-4357/ad9920},
	urldate = {2026-07-26},
	journal = {ApJ},
	publisher = {IOP},
	author = {Hainline, Kevin N. and Maiolino, Roberto and Juodžbalis, Ignas and Scholtz, Jan and Übler, Hannah and D'Eugenio, Francesco and Helton, Jakob M. and Sun, Yang and Sun, Fengwu and Robertson, Brant and Tacchella, Sandro and Bunker, Andrew J. and Carniani, Stefano and Charlot, Stephane and Curtis-Lake, Emma and Egami, Eiichi and Johnson, Benjamin D. and Lin, Xiaojing and Lyu, Jianwei and Pérez-González, Pablo G. and Rinaldi, Pierluigi and Silcock, Maddie S. and Venturi, Giacomo and Williams, Christina C. and Willmer, Christopher N. A. and Willott, Chris and Zhang, Junyu and Zhu, Yongda},
	month = feb,
	year = {2025},
	note = {ADS Bibcode: 2025ApJ...979..138H},
	pages = {138},
}

@article{valentini_impact_2020,
	title = {Impact of {AGN} feedback on galaxies and their multiphase {ISM} across cosmic time},
	volume = {491},
	issn = {0035-8711},
	url = {https://ui.adsabs.harvard.edu/abs/2020MNRAS.491.2779V},
	doi = {10.1093/mnras/stz3131},
	urldate = {2026-07-27},
	journal = {MNRAS},
	publisher = {OUP},
	author = {Valentini, Milena and Murante, Giuseppe and Borgani, Stefano and Granato, Gian Luigi and Monaco, Pierluigi and Brighenti, Fabrizio and Tornatore, Luca and Bressan, Alessandro and Lapi, Andrea},
	month = jan,
	year = {2020},
	note = {ADS Bibcode: 2020MNRAS.491.2779V},
	pages = {2779--2807},
}

@article{sivasankaran_agn_2025,
	title = {{AGN} feedback in isolated galaxies with a {SMUGGLE} multiphase {ISM}},
	volume = {537},
	issn = {0035-8711},
	url = {https://ui.adsabs.harvard.edu/abs/2025MNRAS.537..817S},
	doi = {10.1093/mnras/staf062},
	urldate = {2026-07-27},
	journal = {MNRAS},
	publisher = {OUP},
	author = {Sivasankaran, Aneesh and Blecha, Laura and Torrey, Paul and Kelley, Luke Zoltan and Bhowmick, Aklant and Vogelsberger, Mark and Hernquist, Lars and Marinacci, Federico and Sales, Laura V.},
	month = feb,
	year = {2025},
	note = {ADS Bibcode: 2025MNRAS.537..817S},
	pages = {817--830},
}

@article{li_ramcoal_2025,
	title = {{RAMCOAL}: {Tracking} on-the-fly massive black hole binary evolution and coalescence in galaxy simulations},
	volume = {701},
	issn = {0004-6361},
	shorttitle = {{RAMCOAL}},
	url = {https://ui.adsabs.harvard.edu/abs/2025A&A...701A.232L},
	doi = {10.1051/0004-6361/202452562},
	urldate = {2026-07-27},
	journal = {A\&A},
	publisher = {EDP},
	author = {Li, Kunyang and Volonteri, Marta and Dubois, Yohan and Beckmann, Ricarda and Trebitsch, Maxime},
	month = sep,
	year = {2025},
	note = {ADS Bibcode: 2025A\&A...701A.232L},
	pages = {A232},
}

@article{smith_lyman_2015,
	title = {The {Lyman} α signature of the first galaxies},
	volume = {449},
	issn = {0035-8711},
	url = {https://ui.adsabs.harvard.edu/abs/2015MNRAS.449.4336S},
	doi = {10.1093/mnras/stv565},
	urldate = {2026-07-27},
	journal = {MNRAS},
	publisher = {OUP},
	author = {Smith, Aaron and Safranek-Shrader, Chalence and Bromm, Volker and Milosavljević, Miloš},
	month = jun,
	year = {2015},
	note = {ADS Bibcode: 2015MNRAS.449.4336S},
	pages = {4336--4362},
}

@article{ji_blackthunder_2025,
	title = {{BlackTHUNDER} ─ {A} non-stellar {Balmer} break in a black hole-dominated little red dot at z = 7.04},
	volume = {544},
	issn = {0035-8711},
	url = {https://ui.adsabs.harvard.edu/abs/2025MNRAS.544.3900J},
	doi = {10.1093/mnras/staf1867},
	urldate = {2026-07-27},
	journal = {MNRAS},
	publisher = {OUP},
	author = {Ji, Xihan and Maiolino, Roberto and Übler, Hannah and Scholtz, Jan and D'Eugenio, Francesco and Sun, Fengwu and Perna, Michele and Turner, Hannah and Carniani, Stefano and Arribas, Santiago and Bennett, Jake S. and Bunker, Andrew and Charlot, Stéphane and Cresci, Giovanni and Curti, Mirko and Egami, Eiichi and Fabian, Andy and Inayoshi, Kohei and Isobe, Yuki and Jones, Gareth and Juodžbalis, Ignas and Kumari, Nimisha and Lyu, Jianwei and Mazzolari, Giovanni and Parlanti, Eleonora and Robertson, Brant and Rodríguez Del Pino, Bruno and Schneider, Raffaella and Sijacki, Debora and Tacchella, Sandro and Trinca, Alessandro and Valiante, Rosa and Venturi, Giacomo and Volonteri, Marta and Willott, Chris and Witten, Callum and Witstok, Joris},
	month = dec,
	year = {2025},
	note = {ADS Bibcode: 2025MNRAS.544.3900J},
	pages = {3900--3935},
}

@article{woo_new_2026,
	title = {New {Black} {Hole} {Mass} {Calibrations} and the {Fundamental} {Plane} of the {Broad}-line {Region} {Size}, {Luminosity}, and {Velocity}},
	volume = {1003},
	issn = {0004-637X},
	url = {https://ui.adsabs.harvard.edu/abs/2026ApJ..1003..123W},
	doi = {10.3847/1538-4357/ae6109},
	urldate = {2026-07-27},
	journal = {AJ},
	publisher = {IOP},
	author = {Woo, Jong-Hak and Kim, Jimin and Cho, Hojin and Wang, Shu},
	month = jun,
	year = {2026},
	note = {ADS Bibcode: 2026ApJ..1003..123W},
	pages = {123},
}

@article{geris_little_2026,
	title = {Little {Red} and {Blue} {Dots}: {AGN}-excited narrow lines, {Lyman}-\$α\$ emission, and resemblance to standard quasars},
	volume = {arXiv.2606.21614},
	shorttitle = {Little {Red} and {Blue} {Dots}},
	url = {https://ui.adsabs.harvard.edu/abs/2026arXiv260621614G},
	doi = {10.48550/arXiv.2606.21614},
	urldate = {2026-07-27},
	journal = {arXiv e-prints},
	publisher = {arXiv},
	author = {Geris, Sophia and Maiolino, Roberto and Ji, Xihan and Risaliti, Guido and Lanzuisi, Giorgio and D'Eugenio, Francesco and Isobe, Yuki and Jones, Gareth and Harshan, Anishya and Brazzini, Matilde and Juodžbalis, Ignas and Scholtz, Jan and Rinaldi, Pierluigi and Übler, Hannah and Baker, William and Bunker, Andrew J. and Brusa, Marcella and Carniani, Stefano and Charlot, Stephane and Curti, Mirko and Comastri, Andrea and Lake, Emma Curtis and Gilli, Roberto and Hainline, Kevin and Madau, Piero and Marchesi, Stefano and Mazzolari, Giovanni and Napolitano, Lorenzo and Parlanti, Eleonora and Pentericci, Laura and Ramos Almeida, Cristina and Robertson, Brant and Silcock, Maddie S. and Tripodi, Roberta and Venturi, Giacomo and Vignali, Cristian and Vito, Fabio and Zhu, Yongda},
	month = jun,
	year = {2026},
	note = {ADS Bibcode: 2026arXiv260621614G},
}

@article{su_self-regulation_2025,
	title = {Self-regulation of high-redshift black hole accretion via jets: challenges for {SMBH} formation},
	volume = {538},
	issn = {0035-8711},
	shorttitle = {Self-regulation of high-redshift black hole accretion via jets},
	url = {https://ui.adsabs.harvard.edu/abs/2025MNRAS.538...11S},
	doi = {10.1093/mnras/staf228},
	urldate = {2026-07-27},
	journal = {MNRAS},
	publisher = {OUP},
	author = {Su, Kung-Yi and Bryan, Greg L. and Haiman, Zoltán},
	month = mar,
	year = {2025},
	note = {ADS Bibcode: 2025MNRAS.538...11S},
	pages = {11--30},
}

@article{habouzit_supermassive_2022,
	title = {Supermassive black holes in cosmological simulations - {II}: the {AGN} population and predictions for upcoming {X}-ray missions},
	volume = {509},
	issn = {0035-8711},
	shorttitle = {Supermassive black holes in cosmological simulations - {II}},
	url = {https://ui.adsabs.harvard.edu/abs/2022MNRAS.509.3015H},
	doi = {10.1093/mnras/stab3147},
	urldate = {2026-07-28},
	journal = {MNRAS},
	publisher = {OUP},
	author = {Habouzit, Mélanie and Somerville, Rachel S. and Li, Yuan and Genel, Shy and Aird, James and Anglés-Alcázar, Daniel and Davé, Romeel and Georgiev, Iskren Y. and McAlpine, Stuart and Rosas-Guevara, Yetli and Dubois, Yohan and Nelson, Dylan and Banados, Eduardo and Hernquist, Lars and Peirani, Sébastien and Vogelsberger, Mark},
	month = jan,
	year = {2022},
	note = {ADS Bibcode: 2022MNRAS.509.3015H},
	pages = {3015--3042},
}

@article{wu_redshift_2026,
	title = {Redshift {Evolution} of the {Ratio} of {Supermassive} {Black} {Hole} {Mass} to {Stellar} {Mass}},
	volume = {1004},
	issn = {0004-637X},
	url = {https://ui.adsabs.harvard.edu/abs/2026ApJ..1004...61W},
	doi = {10.3847/1538-4357/ae6916},
	urldate = {2026-07-28},
	journal = {ApJ},
	publisher = {IOP},
	author = {Wu, Ziyong and Cen, Renyue and Teyssier, Romain},
	month = jun,
	year = {2026},
	note = {ADS Bibcode: 2026ApJ..1004...61W},
	pages = {61},
}

@article{sanati_rapid_2025,
	title = {On the rapid growth of {SMBHs} in high-z galaxies: the aftermath of {Population} {III}.1 stars},
	volume = {544},
	issn = {0035-8711},
	shorttitle = {On the rapid growth of {SMBHs} in high-z galaxies},
	url = {https://ui.adsabs.harvard.edu/abs/2025MNRAS.544.4317S},
	doi = {10.1093/mnras/staf2000},
	urldate = {2026-07-28},
	journal = {MNRAS},
	publisher = {OUP},
	author = {Sanati, Mahsa and Devriendt, Julien and Martin-Alvarez, Sergio and Slyz, Adrianne and Tan, Jonathan C.},
	month = dec,
	year = {2025},
	note = {ADS Bibcode: 2025MNRAS.544.4317S},
	pages = {4317--4335},
}

@article{somerville_density-modulated_2025,
	title = {Density-modulated star formation efficiency: implications for the observed abundance of ultraviolet luminous galaxies at z {\textgreater} 10},
	volume = {544},
	issn = {0035-8711},
	shorttitle = {Density-modulated star formation efficiency},
	url = {https://ui.adsabs.harvard.edu/abs/2025MNRAS.544.3774S},
	doi = {10.1093/mnras/staf1824},
	urldate = {2026-07-28},
	journal = {MNRAS},
	publisher = {OUP},
	author = {Somerville, Rachel S. and Yung, L. Y. Aaron and Lancaster, Lachlan and Menon, Shyam and Sommovigo, Laura and Finkelstein, Steven L.},
	month = dec,
	year = {2025},
	note = {ADS Bibcode: 2025MNRAS.544.3774S},
	pages = {3774--3798},
}

@article{shuntov_stellar_2026,
	title = {The stellar mass function of quiescent and star-forming galaxies and its dependence on morphology in {COSMOS}-{Web}},
	volume = {707},
	issn = {0004-6361},
	url = {https://ui.adsabs.harvard.edu/abs/2026A&A...707A.391S},
	doi = {10.1051/0004-6361/202558022},
	urldate = {2026-07-28},
	journal = {A\&A},
	publisher = {EDP},
	author = {Shuntov, Marko and Ilbert, Olivier and Lagos, Claudia del P. and Toft, Sune and Valentino, Francesco and Mercier, Wilfried and Akins, Hollis B. and Binh, Nguyen and Brinch, Malte and Casey, Caitlin M. and Franco, Maximilien and Gentile, Fabrizio and Gozaliasl, Ghassem and Haghjoo, Aryana and Harish, Santosh and Hirschmann, Michaela and Huertas-Company, Marc and Jin, Shuowen and Kartaltepe, Jeyhan S. and Koekemoer, Anton M. and Laigle, Clotilde and Lewis, Joseph S. W. and Magdis, Georgios E. and Joy McCracken, Henry and Mobasher, Bahram and Moutard, Thibaud and Oesch, Pascal A. and Paquereau, Louise and Renzini, Alvio and Rich, Michael R. and Sanders, David B. and Toni, Greta and Tresse, Laurence and Weibel, Andrea and Weaver, John R. and Yang, Lilan},
	month = mar,
	year = {2026},
	note = {ADS Bibcode: 2026A\&A...707A.391S},
	pages = {A391},
}

@article{keller_can_2023,
	title = {Can {Cosmological} {Simulations} {Reproduce} the {Spectroscopically} {Confirmed} {Galaxies} {Seen} at z ≥ 10?},
	volume = {943},
	issn = {0004-637X},
	url = {https://ui.adsabs.harvard.edu/abs/2023ApJ...943L..28K},
	doi = {10.3847/2041-8213/acb148},
	urldate = {2026-09-22},
	journal = {ApJL},
	publisher = {IOP},
	author = {Keller, B. W. and Munshi, F. and Trebitsch, M. and Tremmel, M.},
	month = feb,
	year = {2023},
	note = {ADS Bibcode: 2023ApJ...943L..28K},
	pages = {L28},
}

@article{mccaffrey_no_2023,
	title = {No {Tension}: {JWST} {Galaxies} at z {\textgreater} 10 {Consistent} with {Cosmological} {Simulations}},
	volume = {6},
	issn = {2565-6120},
	shorttitle = {No {Tension}},
	url = {https://ui.adsabs.harvard.edu/abs/2023OJAp....6E..47M},
	doi = {10.21105/astro.2304.13755},
	urldate = {2026-09-22},
	journal = {OJAp},
	author = {McCaffrey, Joe and Hardin, Samantha and Wise, John H. and Regan, John A.},
	month = sep,
	year = {2023},
	note = {ADS Bibcode: 2023OJAp....6E..47M},
	pages = {47},
}

@article{prole_black_2026,
	title = {Black {Hole} {Feedback}, {Galaxy} {Quenching} and {Outflows} at {Cosmic} {Dawn}: {Analysis} of the {SEEDZ} {Simulations}},
	volume = {9},
	issn = {2565-6120},
	shorttitle = {Black {Hole} {Feedback}, {Galaxy} {Quenching} and {Outflows} at {Cosmic} {Dawn}},
	url = {https://ui.adsabs.harvard.edu/abs/2026OJAp....963549P},
	doi = {10.33232/001c.163549},
	urldate = {2026-09-22},
	journal = {OJAp},
	author = {Prole, Lewis Robert and Regan, John A. and Mehta, Daxal and Pakmor, Rüdiger and Koudmani, Sophie and Bourne, Martin A. and Glover, Simon C. O. and Wise, John H. and Klessen, Ralf S. and Tremmel, Michael and Sijacki, Debora and Beckmann, Ricarda S. and Haehnelt, Martin G. and Brennan, John and van de Bor, Pelle and Clark, Paul C.},
	month = jun,
	year = {2026},
	note = {ADS Bibcode: 2026OJAp....963549P},
	pages = {63549},
}

@article{prole_seedz_2026-1,
	title = {{SEEDZ}: {Rapid} {Galaxy} {Assembly} as a {Pathway} to {Supermassive} {Stars}, {Dense} {Stellar} {Environments} and {Massive} {Black} {Hole} {Seeds}},
	shorttitle = {{SEEDZ}},
	url = {https://ui.adsabs.harvard.edu/abs/2026arXiv260627427P},
	doi = {10.48550/arXiv.2606.27427},
	urldate = {2026-09-22},
	journal = {arXiv e-prints},
	publisher = {arXiv},
	author = {Prole, Lewis R. and Regan, John A. and Mehta, Daxal and Nandal, Devesh and Pakmor, Rüdiger and Beckmann, Ricarda S. and Tremmel, Michael and Haehnelt, Martin G. and Glover, Simon C. O. and Klessen, Ralf S. and Wise, John H. and Koudmani, Sophie and Bourne, Martin A. and Sijacki, Debora and Brennan, John and van de Bor, Pelle and Clark, Paul C.},
	month = jun,
	year = {2026},
	note = {ADS Bibcode: 2026arXiv260627427P},
}

@article{chen_dynamical_2022,
	title = {Dynamical friction modelling of massive black holes in cosmological simulations and effects on merger rate predictions},
	volume = {510},
	issn = {0035-8711},
	url = {https://ui.adsabs.harvard.edu/abs/2022MNRAS.510..531C},
	doi = {10.1093/mnras/stab3411},
	urldate = {2026-09-23},
	journal = {MNRAS},
	publisher = {OUP},
	author = {Chen, Nianyi and Ni, Yueying and Tremmel, Michael and Di Matteo, Tiziana and Bird, Simeon and DeGraf, Colin and Feng, Yu},
	month = feb,
	year = {2022},
	note = {ADS Bibcode: 2022MNRAS.510..531C},
	pages = {531--550},
}

@article{nomoto_nucleosynthesis_2013,
	title = {Nucleosynthesis in {Stars} and the {Chemical} {Enrichment} of {Galaxies}},
	volume = {51},
	issn = {0066-4146},
	url = {https://ui.adsabs.harvard.edu/abs/2013ARA&A..51..457N},
	doi = {10.1146/annurev-astro-082812-140956},
	urldate = {2026-09-23},
	journal = {ARA\&A},
	author = {Nomoto, Ken'ichi and Kobayashi, Chiaki and Tominaga, Nozomu},
	month = aug,
	year = {2013},
	note = {ADS Bibcode: 2013ARA\&A..51..457N},
	pages = {457--509},
}

@article{maiolino_re_2019,
	title = {De re metallica: the cosmic chemical evolution of galaxies},
	volume = {27},
	issn = {0935-4956},
	shorttitle = {De re metallica},
	url = {https://ui.adsabs.harvard.edu/abs/2019A&ARv..27....3M},
	doi = {10.1007/s00159-018-0112-2},
	urldate = {2026-09-23},
	journal = {ARA\&A},
	publisher = {Springer},
	author = {Maiolino, R. and Mannucci, F.},
	month = feb,
	year = {2019},
	note = {ADS Bibcode: 2019A\&ARv..27....3M},
	pages = {3},
}

@article{karlsson_pregalactic_2013,
	title = {Pregalactic metal enrichment: {The} chemical signatures of the first stars},
	volume = {85},
	issn = {0034-6861},
	shorttitle = {Pregalactic metal enrichment},
	url = {https://ui.adsabs.harvard.edu/abs/2013RvMP...85..809K},
	doi = {10.1103/RevModPhys.85.809},
	urldate = {2026-09-23},
	journal = {Reviews of Modern Physics},
	publisher = {APS},
	author = {Karlsson, Torgny and Bromm, Volker and Bland-Hawthorn, Joss},
	month = apr,
	year = {2013},
	note = {ADS Bibcode: 2013RvMP...85..809K},
	pages = {809--848},
}

@article{sesana_reconstructing_2011,
	title = {Reconstructing the massive black hole cosmic history through gravitational waves},
	volume = {83},
	issn = {1550-79980556-2821},
	url = {https://ui.adsabs.harvard.edu/abs/2011PhRvD..83d4036S},
	doi = {10.1103/PhysRevD.83.044036},
	urldate = {2026-09-23},
	journal = {PRD},
	publisher = {APS},
	author = {Sesana, Alberto and Gair, Jonathan and Berti, Emanuele and Volonteri, Marta},
	month = feb,
	year = {2011},
	note = {ADS Bibcode: 2011PhRvD..83d4036S},
	pages = {044036},
}

@article{klein_science_2016,
	title = {Science with the space-based interferometer {eLISA}: {Supermassive} black hole binaries},
	volume = {93},
	issn = {1550-79980556-2821},
	shorttitle = {Science with the space-based interferometer {eLISA}},
	url = {https://ui.adsabs.harvard.edu/abs/2016PhRvD..93b4003K},
	doi = {10.1103/PhysRevD.93.024003},
	urldate = {2026-09-23},
	journal = {PRD},
	publisher = {APS},
	author = {Klein, Antoine and Barausse, Enrico and Sesana, Alberto and Petiteau, Antoine and Berti, Emanuele and Babak, Stanislav and Gair, Jonathan and Aoudia, Sofiane and Hinder, Ian and Ohme, Frank and Wardell, Barry},
	month = jan,
	year = {2016},
	note = {ADS Bibcode: 2016PhRvD..93b4003K},
	pages = {024003},
}

@article{amaro-seoane_astrophysics_2023,
	title = {Astrophysics with the {Laser} {Interferometer} {Space} {Antenna}},
	volume = {26},
	url = {https://ui.adsabs.harvard.edu/abs/2023LRR....26....2A},
	doi = {10.1007/s41114-022-00041-y},
	urldate = {2026-09-23},
	journal = {Living Reviews in Relativity},
	publisher = {Springer},
	author = {Amaro-Seoane, Pau and Andrews, Jeff and Arca Sedda, Manuel and Askar, Abbas and Baghi, Quentin and Balasov, Razvan and Bartos, Imre and Bavera, Simone S. and Bellovary, Jillian and Berry, Christopher P. L. and Berti, Emanuele and Bianchi, Stefano and Blecha, Laura and Blondin, Stéphane and Bogdanović, Tamara and Boissier, Samuel and Bonetti, Matteo and Bonoli, Silvia and Bortolas, Elisa and Breivik, Katelyn and Capelo, Pedro R. and Caramete, Laurentiu and Cattorini, Federico and Charisi, Maria and Chaty, Sylvain and Chen, Xian and Chruślińska, Martyna and Chua, Alvin J. K. and Church, Ross and Colpi, Monica and D'Orazio, Daniel and Danielski, Camilla and Davies, Melvyn B. and Dayal, Pratika and De Rosa, Alessandra and Derdzinski, Andrea and Destounis, Kyriakos and Dotti, Massimo and Duţan, Ioana and Dvorkin, Irina and Fabj, Gaia and Foglizzo, Thierry and Ford, Saavik and Fouvry, Jean-Baptiste and Franchini, Alessia and Fragos, Tassos and Fryer, Chris and Gaspari, Massimo and Gerosa, Davide and Graziani, Luca and Groot, Paul and Habouzit, Melanie and Haggard, Daryl and Haiman, Zoltan and Han, Wen-Biao and Istrate, Alina and Johansson, Peter H. and Khan, Fazeel Mahmood and Kimpson, Tomas and Kokkotas, Kostas and Kong, Albert and Korol, Valeriya and Kremer, Kyle and Kupfer, Thomas and Lamberts, Astrid and Larson, Shane and Lau, Mike and Liu, Dongliang and Lloyd-Ronning, Nicole and Lodato, Giuseppe and Lupi, Alessandro and Ma, Chung-Pei and Maccarone, Tomas and Mandel, Ilya and Mangiagli, Alberto and Mapelli, Michela and Mathis, Stéphane and Mayer, Lucio and McGee, Sean and McKernan, Berry and Miller, M. Coleman and Mota, David F. and Mumpower, Matthew and Nasim, Syeda S. and Nelemans, Gijs and Noble, Scott and Pacucci, Fabio and Panessa, Francesca and Paschalidis, Vasileios and Pfister, Hugo and Porquet, Delphine and Quenby, John and Ricarte, Angelo and Röpke, Friedrich K. and Regan, John and Rosswog, Stephan and Ruiter, Ashley and Ruiz, Milton and Runnoe, Jessie and Schneider, Raffaella and Schnittman, Jeremy and Secunda, Amy and Sesana, Alberto and Seto, Naoki and Shao, Lijing and Shapiro, Stuart and Sopuerta, Carlos and Stone, Nicholas C. and Suvorov, Arthur and Tamanini, Nicola and Tamfal, Tomas and Tauris, Thomas and Temmink, Karel and Tomsick, John and Toonen, Silvia and Torres-Orjuela, Alejandro and Toscani, Martina and Tsokaros, Antonios and Unal, Caner and Vázquez-Aceves, Verónica and Valiante, Rosa and van Putten, Maurice and van Roestel, Jan and Vignali, Christian and Volonteri, Marta and Wu, Kinwah and Younsi, Ziri and Yu, Shenghua and Zane, Silvia and Zwick, Lorenz and Antonini, Fabio and Baibhav, Vishal and Barausse, Enrico and Bonilla Rivera, Alexander and Branchesi, Marica and Branduardi-Raymont, Graziella and Burdge, Kevin and Chakraborty, Srija and Cuadra, Jorge and Dage, Kristen and Davis, Benjamin and de Mink, Selma E. and Decarli, Roberto and Doneva, Daniela and Escoffier, Stephanie and Gandhi, Poshak and Haardt, Francesco and Lousto, Carlos O. and Nissanke, Samaya and Nordhaus, Jason and O'Shaughnessy, Richard and Portegies Zwart, Simon and Pound, Adam and Schussler, Fabian and Sergijenko, Olga and Spallicci, Alessandro and Vernieri, Daniele and Vigna-Gómez, Alejandro},
	month = dec,
	year = {2023},
	note = {ADS Bibcode: 2023LRR....26....2A},
	pages = {2},
}

@article{hirata_reducing_2010,
	title = {Reducing the weak lensing noise for the gravitational wave {Hubble} diagram using the non-{Gaussianity} of the magnification distribution},
	volume = {81},
	issn = {1550-79980556-2821},
	url = {https://ui.adsabs.harvard.edu/abs/2010PhRvD..81l4046H},
	doi = {10.1103/PhysRevD.81.124046},
	urldate = {2026-09-23},
	journal = {PRD},
	publisher = {APS},
	author = {Hirata, Christopher M. and Holz, Daniel E. and Cutler, Curt},
	month = jun,
	year = {2010},
	note = {ADS Bibcode: 2010PhRvD..81l4046H},
	pages = {124046},
}

@article{mangiagli_observing_2020,
	title = {Observing the inspiral of coalescing massive black hole binaries with {LISA} in the era of multimessenger astrophysics},
	volume = {102},
	issn = {1550-79980556-2821},
	url = {https://ui.adsabs.harvard.edu/abs/2020PhRvD.102h4056M},
	doi = {10.1103/PhysRevD.102.084056},
	urldate = {2026-09-23},
	journal = {PRD},
	publisher = {APS},
	author = {Mangiagli, Alberto and Klein, Antoine and Bonetti, Matteo and Katz, Michael L. and Sesana, Alberto and Volonteri, Marta and Colpi, Monica and Marsat, Sylvain and Babak, Stanislav},
	month = oct,
	year = {2020},
	note = {ADS Bibcode: 2020PhRvD.102h4056M},
	pages = {084056},
}

@article{ubler_blackthunder_2026,
	title = {{BlackTHUNDER}: {Evidence} of three massive black holes in a z ∼ 5 galaxy},
	volume = {712},
	issn = {0004-6361},
	shorttitle = {{BlackTHUNDER}},
	url = {https://ui.adsabs.harvard.edu/abs/2026A&A...712A..80U},
	doi = {10.1051/0004-6361/202557419},
	urldate = {2026-10-05},
	journal = {A\&A},
	publisher = {EDP},
	author = {Übler, Hannah and Mazzolari, Giovanni and Maiolino, Roberto and D'Eugenio, Francesco and Davari, Nazanin and Juodžbalis, Ignas and Perna, Michele and Matthee, Jorryt and Schneider, Raffaella and Valiante, Rosa and Arribas, Santiago and Bertola, Elena and Bunker, Andrew J. and Bromm, Volker and Carniani, Stefano and Charlot, Stéphane and Cresci, Giovanni and Curti, Mirko and Davies, Richard and Eisenhauer, Frank and Fabian, Andrew and Förster Schreiber, Natascha M. and Genzel, Reinhard and Inayoshi, Kohei and Ivey, Lucy R. and Jones, Gareth C. and Liu, Boyuan and Lutz, Dieter and Kashino, Daichi and Mackenzie, Ruari and Parlanti, Eleonora and Robertson, Brant and Rodríguez del Pino, Bruno and Shimizu, T. Taro and Sijacki, Debora and Sturm, Eckhard and Tacchella, Sandro and Tacconi, Linda and Tozzi, Giulia and Trinca, Alessandro and Venturi, Giacomo and Volonteri, Marta and Willott, Chris and Zhang, Saiyang},
	month = aug,
	year = {2026},
	note = {ADS Bibcode: 2026A\&A...712A..80U},
	pages = {A80},
}

@article{de_graaff_little_2026,
	title = {Little red dots host black hole stars: a unified family of gas-reddened {AGN} revealed by {JWST}/{NIRSpec} spectroscopy},
	volume = {551},
	issn = {0035-8711},
	shorttitle = {Little red dots host black hole stars},
	url = {https://ui.adsabs.harvard.edu/abs/2026MNRAS.551g1567D},
	doi = {10.1093/mnras/stag1567},
	urldate = {2026-10-05},
	journal = {MNRAS},
	publisher = {OUP},
	author = {de Graaff, Anna and Hviding, Raphael E. and Naidu, Rohan P. and Greene, Jenny E. and Miller, Tim B. and Leja, Joel and Matthee, Jorryt and Brammer, Gabriel and Katz, Harley and Bezanson, Rachel and Boogaard, Leindert A. and Bose, Sownak and Chisholm, John and Cleri, Nikko J. and Dayal, Pratika and Feldmann, Robert and Fudamoto, Yoshinobu and Fujimoto, Seiji and Furtak, Lukas J. and Glazebrook, Karl and Gottumukkala, Rashmi and Heintz, Kasper E. and Kokorev, Vasily and Labbe, Ivo and Maseda, Michael V. and McConachie, Ian and Nanayakkara, Themiya and Nelson, Erica and Nowaczyk, Przemysław and Oesch, Pascal A. and Rix, Hans-Walter and Setton, David J. and Torralba, Alberto and Walter, Fabian and Wang, Bingjie and Weibel, Andrea and van der Wel, Arjen},
	month = sep,
	year = {2026},
	note = {ADS Bibcode: 2026MNRAS.551g1567D},
	pages = {stag1567},
}

@article{umeda_black_2026,
	title = {A {Black} {Hole} {Envelope} {Interpretation} for {Cosmological} {Demographics} of {Little} {Red} {Dots}},
	volume = {999},
	issn = {0004-637X},
	url = {https://ui.adsabs.harvard.edu/abs/2026ApJ...999..183U},
	doi = {10.3847/1538-4357/ae4101},
	urldate = {2026-10-05},
	journal = {ApJ},
	publisher = {IOP},
	author = {Umeda, Hiroya and Inayoshi, Kohei and Harikane, Yuichi and Murase, Kohta},
	month = mar,
	year = {2026},
	note = {ADS Bibcode: 2026ApJ...999..183U},
	pages = {183},
}

@article{trinca_episodic_2026,
	title = {Episodic super-{Eddington} accretion as a clue to overmassive black holes in the early {Universe}},
	volume = {552},
	issn = {0035-8711},
	url = {https://ui.adsabs.harvard.edu/abs/2026MNRAS.552g1608T},
	doi = {10.1093/mnras/stag1608},
	urldate = {2026-10-05},
	journal = {MNRAS},
	publisher = {OUP},
	author = {Trinca, Alessandro and Valiante, Rosa and Schneider, Raffaella and Juodžbalis, Ignas and Maiolino, Roberto and Graziani, Luca and Lupi, Alessandro and Natarajan, Priyamvada and Volonteri, Marta and Zana, Tommaso},
	month = oct,
	year = {2026},
	note = {ADS Bibcode: 2026MNRAS.552g1608T},
	pages = {stag1608},
}

@article{sullivan_arkenstone_2026,
	title = {{ARKENSTONE} {BH}: a model for high specific energy black hole feedback in cosmological simulations},
	volume = {551},
	issn = {0035-8711},
	shorttitle = {{ARKENSTONE} {BH}},
	url = {https://ui.adsabs.harvard.edu/abs/2026MNRAS.551g1520S},
	doi = {10.1093/mnras/stag1520},
	urldate = {2026-10-05},
	journal = {MNRAS},
	publisher = {OUP},
	author = {Sullivan, James M. and Bryan, Greg L. and Smith, Matthew C. and Bennett, Jake S. and Fielding, Drummond B. and Terrazas, Bryan A. and Koudmani, Sophie and Somerville, Rachel S. and Hirschmann, Michaela},
	month = sep,
	year = {2026},
	note = {ADS Bibcode: 2026MNRAS.551g1520S},
	pages = {stag1520},
}

@article{kho_signatures_2025,
	title = {Signatures of {Black} {Hole} {Seeding} on the {M}∙─σ {Relation}: {Predictions} from the {BRAHMA} {Simulations}},
	volume = {994},
	issn = {0004-637X},
	shorttitle = {Signatures of {Black} {Hole} {Seeding} on the {M}∙─σ {Relation}},
	url = {https://ui.adsabs.harvard.edu/abs/2025ApJ...994..172K},
	doi = {10.3847/1538-4357/ae0e1e},
	urldate = {2026-10-05},
	journal = {ApJ},
	publisher = {IOP},
	author = {Kho, Jonathan and Bhowmick, Aklant Kumar and Torrey, Paul and Garcia, Alex M. and Ahvazi, Niusha and Blecha, Laura and Vogelsberger, Mark},
	month = dec,
	year = {2025},
	note = {ADS Bibcode: 2025ApJ...994..172K},
	pages = {172},
}

@article{juodzbalis_direct_2026,
	title = {A direct black-hole mass measurement in a little red dot at high redshift},
	volume = {653},
	issn = {0028-0836},
	url = {https://ui.adsabs.harvard.edu/abs/2026Natur.653.1017J},
	doi = {10.1038/s41586-026-10579-4},
	urldate = {2026-10-05},
	journal = {Nature},
	publisher = {Springer},
	author = {Juodžbalis, Ignas and Marconcini, Cosimo and D'Eugenio, Francesco and Maiolino, Roberto and Marconi, Alessandro and Übler, Hannah and Scholtz, Jan and Ji, Xihan and Jones, Gareth C. and Perna, Michele and Arribas, Santiago and Bennett, Jake S. and Bromm, Volker and Bunker, Andrew J. and Carniani, Stefano and Charlot, Stéphane and Cresci, Giovanni and Dayal, Pratika and Egami, Eiichi and Fabian, Andrew and Inayoshi, Kohei and Isobe, Yuki and Ivey, Lucy R. and Koudmani, Sophie and Laporte, Nicolas and Liu, Boyuan and Lyu, Jianwei and Mazzolari, Giovanni and Monty, Stephanie and Parlanti, Eleonora and Pérez-González, Pablo G. and Robertson, Brant and Schneider, Raffaella and Sijacki, Debora and Tacchella, Sandro and Trinca, Alessandro and Valiante, Rosa and Volonteri, Marta and Witstok, Joris and Zhang, Saiyang},
	month = may,
	year = {2026},
	note = {ADS Bibcode: 2026Natur.653.1017J},
	pages = {1017--1021},
}

@article{juodzbalis_jades_2026,
	title = {{JADES}: comprehensive census of broad-line {AGN} from reionization to cosmic noon revealed by {JWST}},
	volume = {546},
	issn = {0035-8711},
	shorttitle = {{JADES}},
	url = {https://ui.adsabs.harvard.edu/abs/2026MNRAS.546ag086J},
	doi = {10.1093/mnras/stag086},
	urldate = {2026-10-05},
	journal = {MNRAS},
	publisher = {OUP},
	author = {Juodžbalis, Ignas and Maiolino, Roberto and Baker, William M. and Lake, Emma Curtis and Scholtz, Jan and D'Eugenio, Francesco and Trefoloni, Bartolomeo and Isobe, Yuki and Tacchella, Sandro and Bunker, Andrew J. and Carniani, Stefano and Charlot, Stéphane and Jones, Gareth C. and Parlanti, Eleonora and Perna, Michele and Rinaldi, Pierluigi and Robertson, Brant and Übler, Hannah and Venturi, Giacomo and Willott, Chris},
	month = mar,
	year = {2026},
	note = {ADS Bibcode: 2026MNRAS.546ag086J},
	pages = {stag086},
}

@article{juodzbalis_dormant_2024,
	title = {A dormant overmassive black hole in the early {Universe}},
	volume = {636},
	issn = {0028-0836},
	url = {https://ui.adsabs.harvard.edu/abs/2024Natur.636..594J},
	doi = {10.1038/s41586-024-08210-5},
	urldate = {2026-10-05},
	journal = {Nature},
	author = {Juodžbalis, Ignas and Maiolino, Roberto and Baker, William M. and Tacchella, Sandro and Scholtz, Jan and D'Eugenio, Francesco and Witstok, Joris and Schneider, Raffaella and Trinca, Alessandro and Valiante, Rosa and DeCoursey, Christa and Curti, Mirko and Carniani, Stefano and Chevallard, Jacopo and de Graaff, Anna and Arribas, Santiago and Bennett, Jake S. and Bourne, Martin A. and Bunker, Andrew J. and Charlot, Stéphane and Jiang, Brian and Koudmani, Sophie and Perna, Michele and Robertson, Brant and Sijacki, Debora and Übler, Hannah and Williams, Christina C. and Willott, Chris},
	month = dec,
	year = {2024},
	note = {ADS Bibcode: 2024Natur.636..594J},
	pages = {594--597},
}

@article{ivey_exploring_2026,
	title = {Exploring spatially resolved metallicities, dynamics, and outflows in low-mass galaxies at z ∼ 7.6},
	volume = {546},
	issn = {0035-8711},
	url = {https://ui.adsabs.harvard.edu/abs/2026MNRAS.546ag094I},
	doi = {10.1093/mnras/stag094},
	urldate = {2026-10-05},
	journal = {MNRAS},
	publisher = {OUP},
	author = {Ivey, L. R. and Scholtz, J. and Danhaive, A. L. and Koudmani, S. and Jones, G. C. and Maiolino, R. and Curti, M. and D'Eugenio, F. and Tacchella, S. and Baker, W. M. and Arribas, S. and Charlot, S. and Eisenstein, D. and Ji, Z. and Koller, M. and Laporte, N. and Perna, M. and Puskás, D. and Robertson, B. and Sijacki, D. and Trussler, J. A. A. and Witten, C.},
	month = mar,
	year = {2026},
	note = {ADS Bibcode: 2026MNRAS.546ag094I},
	pages = {stag094},
}

@article{ivey_cliff_2026,
	title = {The {Cliff}: a metal-poor little red dot hosting an overmassive black hole at z = 3.55},
	volume = {550},
	issn = {0035-8711},
	shorttitle = {The {Cliff}},
	url = {https://ui.adsabs.harvard.edu/abs/2026MNRAS.550g1220I},
	doi = {10.1093/mnras/stag1220},
	urldate = {2026-10-05},
	journal = {MNRAS},
	publisher = {OUP},
	author = {Ivey, L. R. and D'Eugenio, F. and Maiolino, R. and Isobe, Y. and Juodžbalis, I. and Koudmani, S. and Perna, M. and Zhang, S. and Bromm, V. and Bunker, A. J. and Carniani, S. and Fabian, A. C. and Inayoshi, K. and Ji, X. and Jones, G. C. and Liu, B. and Pascalau, R. and Rinaldi, P. and Robertson, B. and Scholtz, J. and Tacchella, S.},
	month = aug,
	year = {2026},
	note = {ADS Bibcode: 2026MNRAS.550g1220I},
	pages = {stag1220},
}

@article{golubchik_venus_2026,
	title = {{VENUS}: {When} red meets blue: {A} multiply imaged little red dot with an apparent blue companion behind the galaxy cluster {Abell} 383},
	volume = {710},
	issn = {0004-6361},
	shorttitle = {{VENUS}},
	url = {https://ui.adsabs.harvard.edu/abs/2026A&A...710A.226G},
	doi = {10.1051/0004-6361/202558407},
	urldate = {2026-10-05},
	journal = {A\&A},
	publisher = {EDP},
	author = {Golubchik, Miriam and Furtak, Lukas J. and Allingham, Joseph F. V. and Zitrin, Adi and Akins, Hollis B. and Kokorev, Vasily and Fujimoto, Seiji and {Abdurro'uf} and Amorín, Ricardo O. and Bauer, Franz E. and Bezanson, Rachel and Bradač, Marusa and Bradley, Larry D. and Brammer, Gabriel B. and Chisholm, John and Coe, Dan and Conselice, Christopher J. and Dayal, Pratika and Dessauges-Zavadsky, Miroslava and Diego, Jose M. and Faisst, Andreas L. and Fei, Qinyue and Ferguson, Henry C. and Finkelstein, Steven L. and Frye, Brenda L. and González-Otero, Mauro and Greene, Jenny E. and Harikane, Yuichi and Hsiao, Tiger Yu-Yang and Inayoshi, Kohei and Jiménez-Teja, Yolanda and Knudsen, Kirsten and Koekemoer, Anton M. and Labbé, Ivo and Lucas, Ray A. and Magdis, Georgios E. and Matthee, Jorryt and Messa, Matteo and Naidu, Rohan P. and Nakane, Minami and Noirot, Gaël and Pan, Richard and Papovich, Casey and Richard, Johan and Ricotti, Massimo and Robbins, Luke and Stark, Daniel P. and Sun, Fengwu and Treu, Tommaso and Tripodi, Roberta and Vanzella, Eros and Willott, Chris and Windhorst, Rogier A.},
	month = jun,
	year = {2026},
	note = {ADS Bibcode: 2026A\&A...710A.226G},
	pages = {A226},
}

@article{genina_calibrated_2024,
	title = {A calibrated model for {N}-body dynamical friction acting on supermassive black holes},
	volume = {534},
	issn = {0035-8711},
	url = {https://ui.adsabs.harvard.edu/abs/2024MNRAS.534..957G},
	doi = {10.1093/mnras/stae2144},
	urldate = {2026-10-05},
	journal = {MNRAS},
	publisher = {OUP},
	author = {Genina, Anna and Springel, Volker and Rantala, Antti},
	month = oct,
	year = {2024},
	note = {ADS Bibcode: 2024MNRAS.534..957G},
	pages = {957--977},
}

@article{fei_glimpse_2026,
	title = {A {GLIMPSE} of {Intermediate} {Mass} {Black} {Holes} in the {Epoch} of {Reionization}: {Witnessing} the {Descendants} of {Direct} {Collapse}?},
	volume = {1003},
	issn = {0004-637X},
	shorttitle = {A {GLIMPSE} of {Intermediate} {Mass} {Black} {Holes} in the {Epoch} of {Reionization}},
	url = {https://ui.adsabs.harvard.edu/abs/2026ApJ..1003..244F},
	doi = {10.3847/1538-4357/ae6248},
	urldate = {2026-10-05},
	journal = {ApJ},
	publisher = {IOP},
	author = {Fei, Qinyue and Fujimoto, Seiji and Naidu, Rohan P. and Chisholm, John and Atek, Hakim and Brammer, Gabriel and Asada, Yoshihisa and Berg, Danielle A. and Bromm, Volker and Furtak, Lukas J. and Greene, Jenny E. and Hsiao, Tiger Yu-Yang and Jeon, Junehyoung and Kokorev, Vasily and Matthee, Jorryt and Natarajan, Priyamvada and Pan, Richard and Richard, Johan and Saldana-Lopez, Alberto and Schaerer, Daniel and Volonteri, Marta and Zitrin, Adi},
	month = jun,
	year = {2026},
	note = {ADS Bibcode: 2026ApJ..1003..244F},
	pages = {244},
}

@article{chon_overmassive_2026,
	title = {Overmassive black holes and little red dots naturally form in simulations},
	volume = {657},
	issn = {0028-0836},
	url = {https://ui.adsabs.harvard.edu/abs/2026Natur.657..621C},
	doi = {10.1038/s41586-026-10985-8},
	urldate = {2026-10-05},
	journal = {Nature},
	publisher = {Springer},
	author = {Chon, Sunmyon and Hirano, Shingo and Ishiyama, Tomoaki and Chang, Seok-Jun and Springel, Volker},
	month = sep,
	year = {2026},
	note = {ADS Bibcode: 2026Natur.657..621C},
	pages = {621--625},
}

@article{chaikin_importance_2026,
	title = {The importance of super-{Eddington} black hole accretion for the emergence of massive quiescent galaxies at high redshift},
	volume = {550},
	issn = {0035-8711},
	url = {https://ui.adsabs.harvard.edu/abs/2026MNRAS.550g1148C},
	doi = {10.1093/mnras/stag1148},
	urldate = {2026-10-05},
	journal = {MNRAS},
	publisher = {OUP},
	author = {Chaikin, Evgenii and Schaye, Joop and Huško, Filip and Lacey, Cedric G. and Ploeckinger, Sylvia and Schaller, Matthieu},
	month = aug,
	year = {2026},
	note = {ADS Bibcode: 2026MNRAS.550g1148C},
	pages = {stag1148},
}



\appendix

\section{Scaling relations} \label{appsec:scal_rel}

In this appendix section, we present several supplementary scaling relation plots that support the analysis from the main text.

\subsection{Black hole mass -- stellar velocity dispersion scaling relations} \label{appsec:stellarveldisprel}

\begin{figure}
    \centering
    \includegraphics[width=\columnwidth]{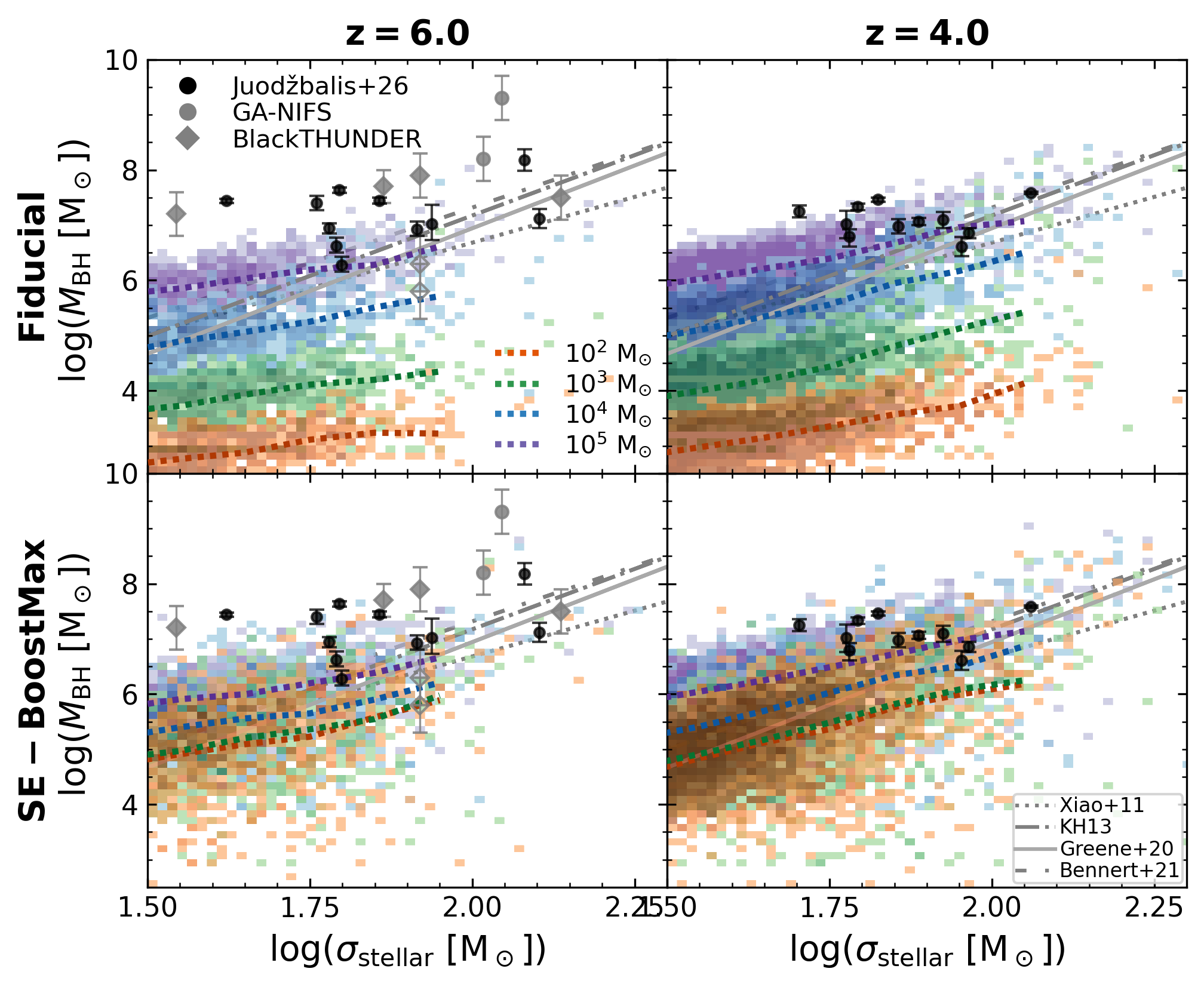}
    \caption{Black hole mass -- stellar velocity dispersion scaling relations at $z=6$ and $z=4$ for the \aesop~simulations including all four seed masses, as listed in the legend, and the `fiducial' (top row) as well as \textit{`SE-BoostMax'} accretion models (bottom row). We overplot the stellar velocity dispersions inferred for the host galaxies of high-z broad-line AGN from \citet{juodzbalis_jades_2026} as well as the high-redshift GA-NIFS AGN \citep{perna_ga-nifs_2023,ubler_ga-nifs_2023,parlanti_ga-nifs_2024,marshall_ga-nifs_2025} and BlackTHUNDER AGN \citep{deugenio_blackthunder_2026,deugenio_jades_2026-1,jones_blackthunder_2026,maiolino_black_2026,ubler_blackthunder_2026}. For reference, we also plot the observed local scaling relations from \citet{xiao_exploring_2011,kormendy_coevolution_2013,greene_intermediate-mass_2020} and \citet{bennert_local_2021}. With the fiducial model, only the heavy and intermediate seed runs overlap with the high-z AGN data, and the light seeds are undermassive. With the \textit{`SE-BoostMax'} model, all seed masses lead to broad agreement with the local relations.}
    \label{fig:scalrel_veldisp}
\end{figure}

In Section~\ref{subsec:results-scalrel-raw}, we presented the unprocessed `raw' black hole mass -- stellar mass scaling relations. Here we also show the corresponding black hole mass -- stellar velocity dispersion scaling relations, for reference, including the \textit{`fiducial'} (top row) and \textit{`SE-BoostMax'} accretion models (bottom row) as well as all four seed masses, as listed in the legend. We plot the mean relations as dotted lines and the full distribution as two-dimensional colour-coded histograms. We overplot the stellar velocity dispersions inferred for the host galaxies of high-z broad-line AGN from \citet{juodzbalis_jades_2026} as well as the high-redshift GA-NIFS AGN \citep{perna_ga-nifs_2023,ubler_ga-nifs_2023,parlanti_ga-nifs_2024,marshall_ga-nifs_2025} and BlackTHUNDER AGN \citep{deugenio_blackthunder_2026,deugenio_jades_2026-1,jones_blackthunder_2026,maiolino_black_2026,ubler_blackthunder_2026}. For reference, we also plot the observed local scaling relations from \citet{xiao_exploring_2011,kormendy_coevolution_2013,greene_intermediate-mass_2020} and \citet{bennert_local_2021}.

For each galaxy, the stellar velocity dispersion $\sigma_\mathrm{stellar}$ is computed from star particles belonging to the same friends-of-friends group as the host subhalo, using positions and velocities measured relative to the subhalo's centre and bulk velocity, respectively. Rather than computing the three-dimensional dispersion, we mimic an observational line-of-sight measurement by considering three orthogonal projections, corresponding to viewing the galaxy along each of the $x$, $y$, and $z$ axes in turn. For a given projection with line-of-sight axis $n$ and projected-plane axes $(l,m)$, we select star particles within a projected (cylindrical) radius of $2\,r_{1/2}$, where $r_{1/2}$ is the stellar half-mass radius, requiring at least 5 particles to proceed. For each valid projection we compute the mass-weighted standard deviation, $\sigma_n$, of the velocity component along the line of sight, $v_n$. The final velocity dispersion is then obtained by combining the $N_{\rm proj}\leq3$ valid projections in quadrature,
\begin{equation}
\sigma_\mathrm{stellar} = \sqrt{\frac{1}{N_{\rm proj}}\sum_{n} \sigma_n^2},
\end{equation}
providing an estimate that is more akin to observations whilst being more robust to line-of-sight orientation effects than a single projection would be. If none of the three orthogonal projections yields at least 5 particles within $2\,r_{1/2}$, $\sigma_\mathrm{stellar}$ is set to zero.

Focusing first on the `fiducial' accretion implementation, we can see that, as with the stellar mass scaling relation in Section~\ref{subsec:results-scalrel-raw}, the strong dependency of the Bondi accretion rate on the black hole mass leads to a clear separation between the mean relations for the runs with different seed masses. Only the distribution of the heavy and intermediate seed runs overlaps with observed high-z AGN. The mean relations for these runs stemming from the heavy and intermediate seeds are also in reasonable agreement with the observed local relations (deviations smaller than 1 dex). 

For the `\textit{SE-BoostMax}' implementation, however, we obtain convergence of all models onto the local relations as early as $z=6$. Due to the efficient accretion in the low-mass regime, the distributions for all seed masses overlap with the observed distribution and, indeed, when we apply the \textsc{balmersopica} filtering, the resulting synthetically `observable' AGN are consistent with the observations from JADES. This demonstrates that efficient early growth in our simulations leads to mean relations that have normalisations \textit{higher} than the local black hole mass -- stellar mass scaling relations, however, for the same set-ups agreement with the local relations for black hole mass -- stellar velocity dispersion is reached at high redshifts and only weakly evolves.

\subsection{Black hole mass -- dynamical mass scaling relations}

\begin{figure*}
    \centering
    \includegraphics[width=\columnwidth]{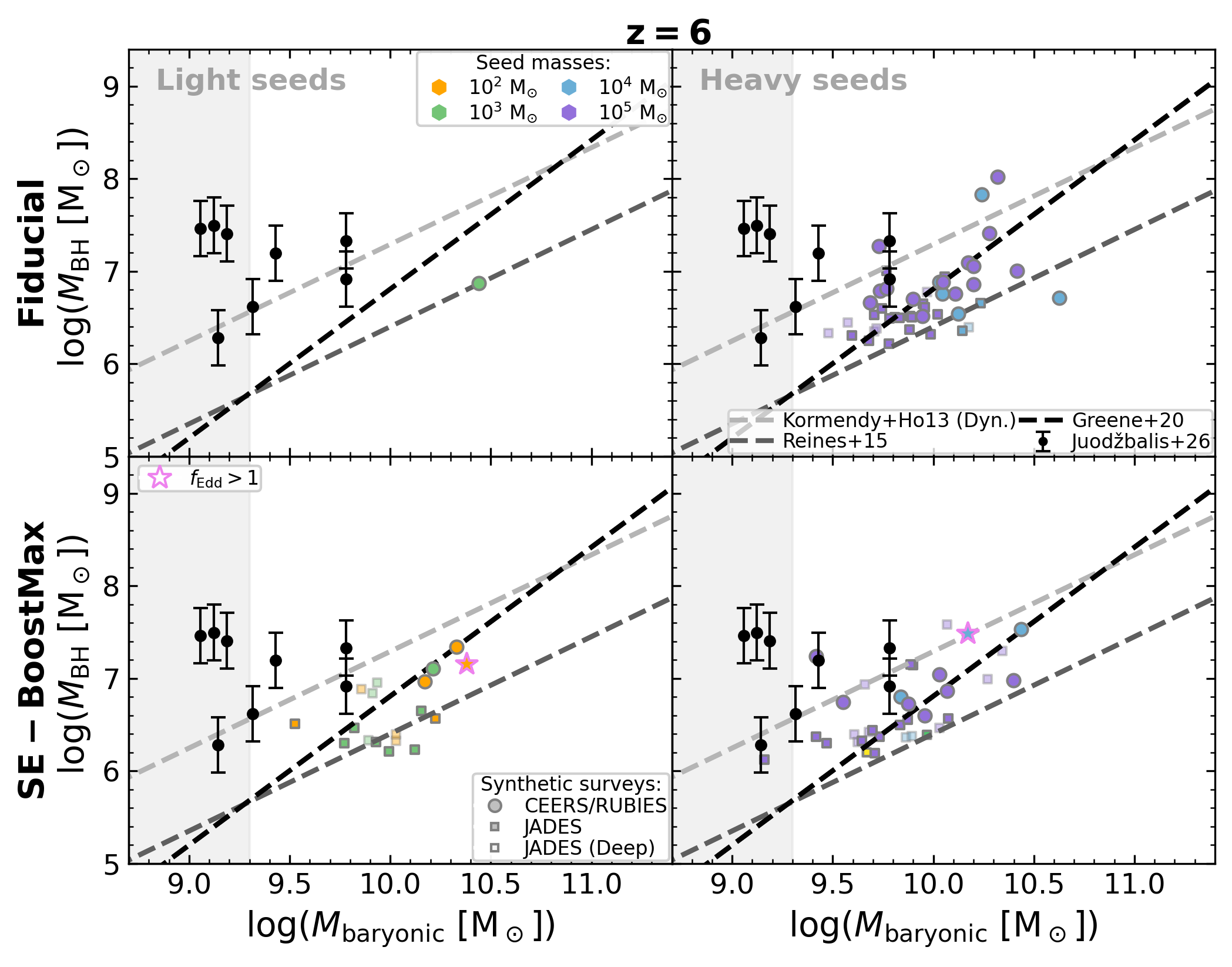}
    \includegraphics[width=\columnwidth]{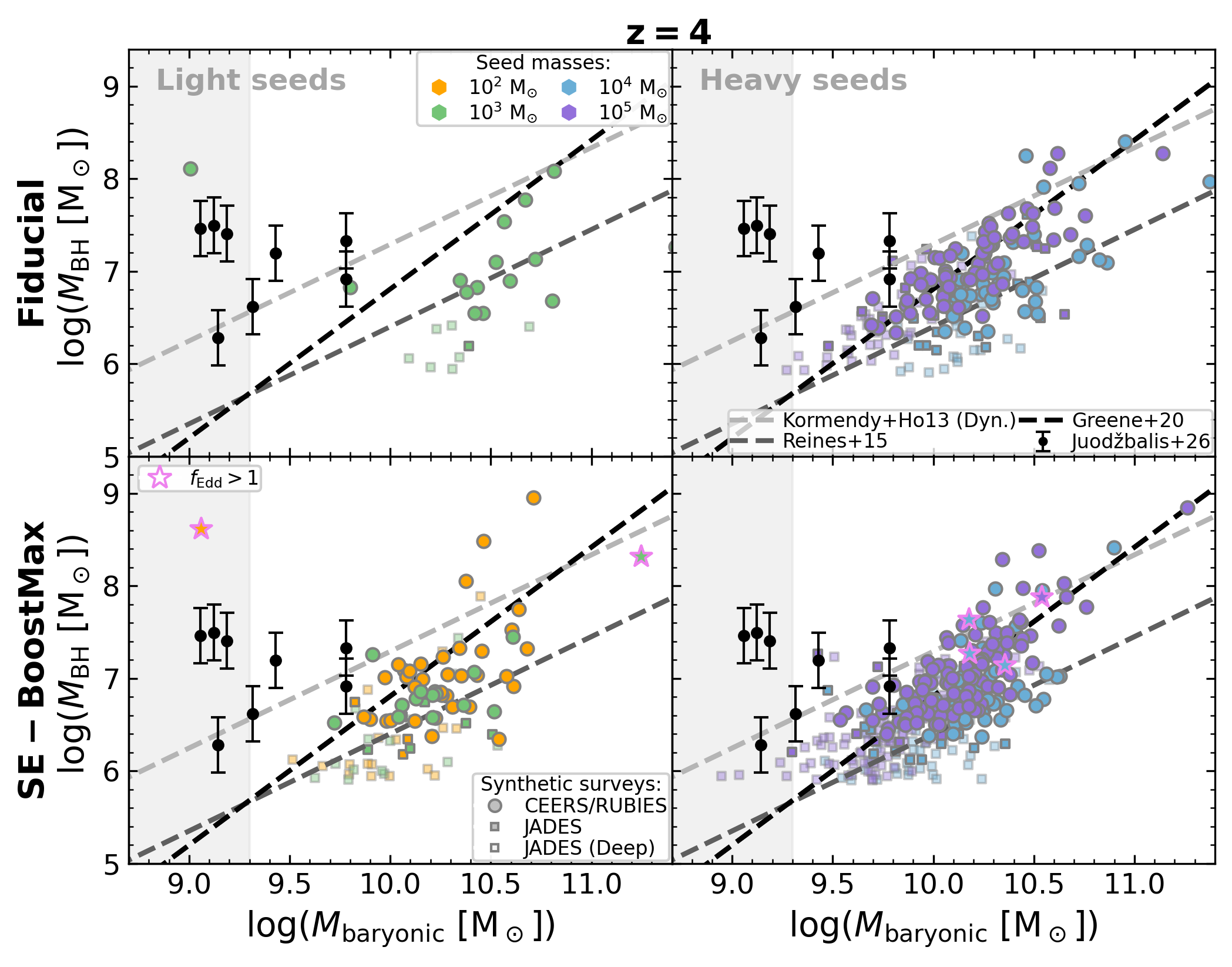}
    \caption{Black hole mass as a function of total baryonic mass (stellar mass plus gas mass within twice the stellar half-mass radius) from the \aesop~simulations for the fiducial (top row) and `\textit{SE-BoostMax}' accretion models (bottom row). Left and right panels show results at $z=6$ and $z=4$, respectively. The columns separate light seeds ($10^{2}$ and $10^{3}\,\Msun$, left) from heavy seeds ($10^{4}$ and $10^{5}\,\Msun$, right). We only include simulated active black holes that meet broad-line AGN selection criteria mimicking \jwst survey detection limits. Scatter symbols indicate detectability for different \jwst survey configurations: CEERS/RUBIES (circles), JADES Medium (opaque squares), and JADES Deep (transparent squares), assuming an R1000 grating and exposure times of 1\,hr, 2.6\,hr, and 7\,hr, respectively. Observed JADES AGN with dynamical mass estimates from \citet{juodzbalis_jades_2026} are overplotted for comparison. The grey shaded region indicates the resolution limit of the simulations and pink stars highlight the simulated black holes `observed' during super-Eddington accretion bursts. While the detectable \aesop~AGN appear overmassive relative to local black hole -- stellar mass relations, they are consistent with local scaling relations once the gas mass is included, indicating that sufficient baryonic mass is already present. This supports a `black hole first' scenario in which early black hole growth precedes the assembly of the stellar component.}
    \label{fig:scal_rel_filter_dyn}
\end{figure*}

\begin{figure*}
    \centering
    \includegraphics[width=0.49\textwidth]{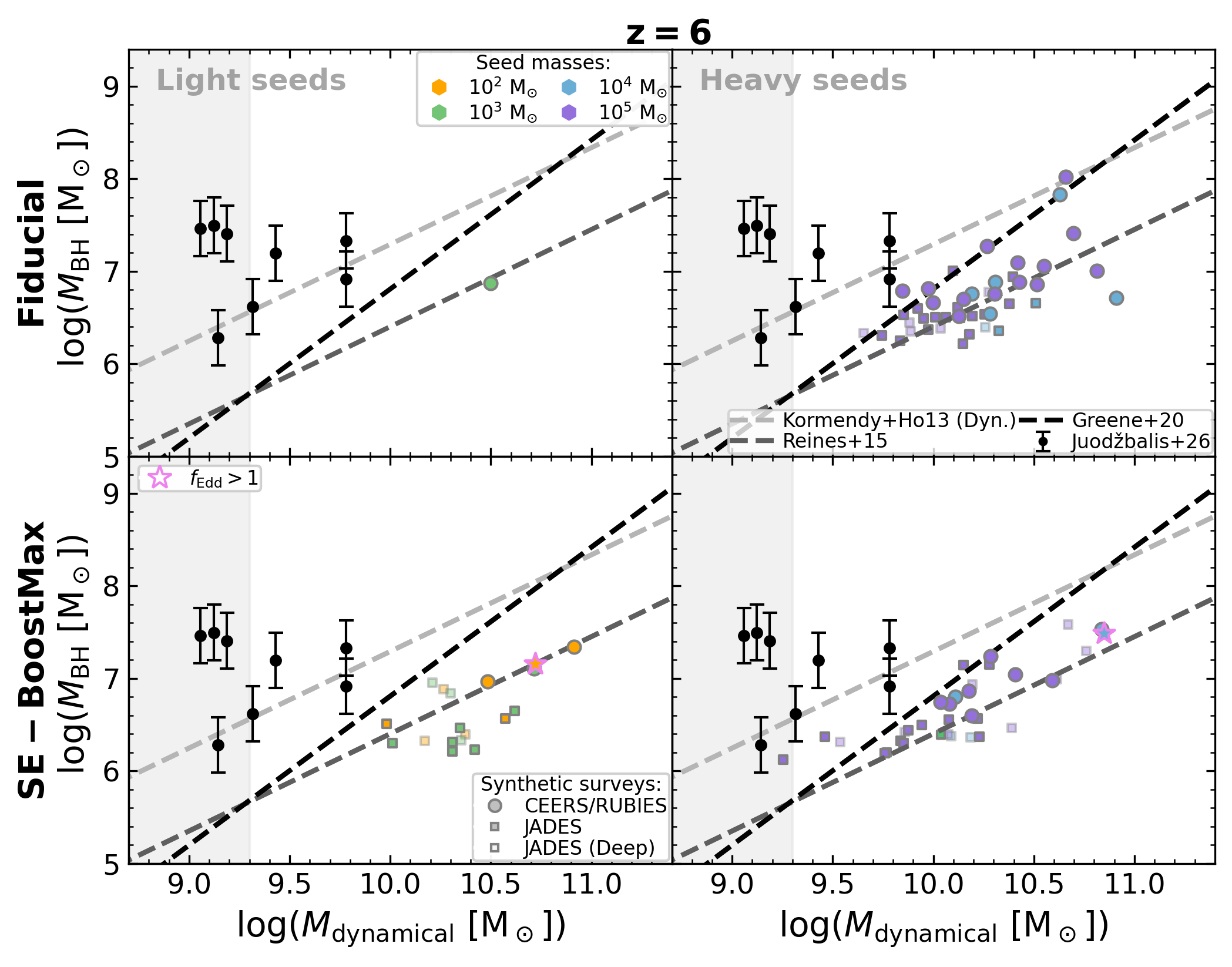}
    \includegraphics[width=0.49\textwidth]{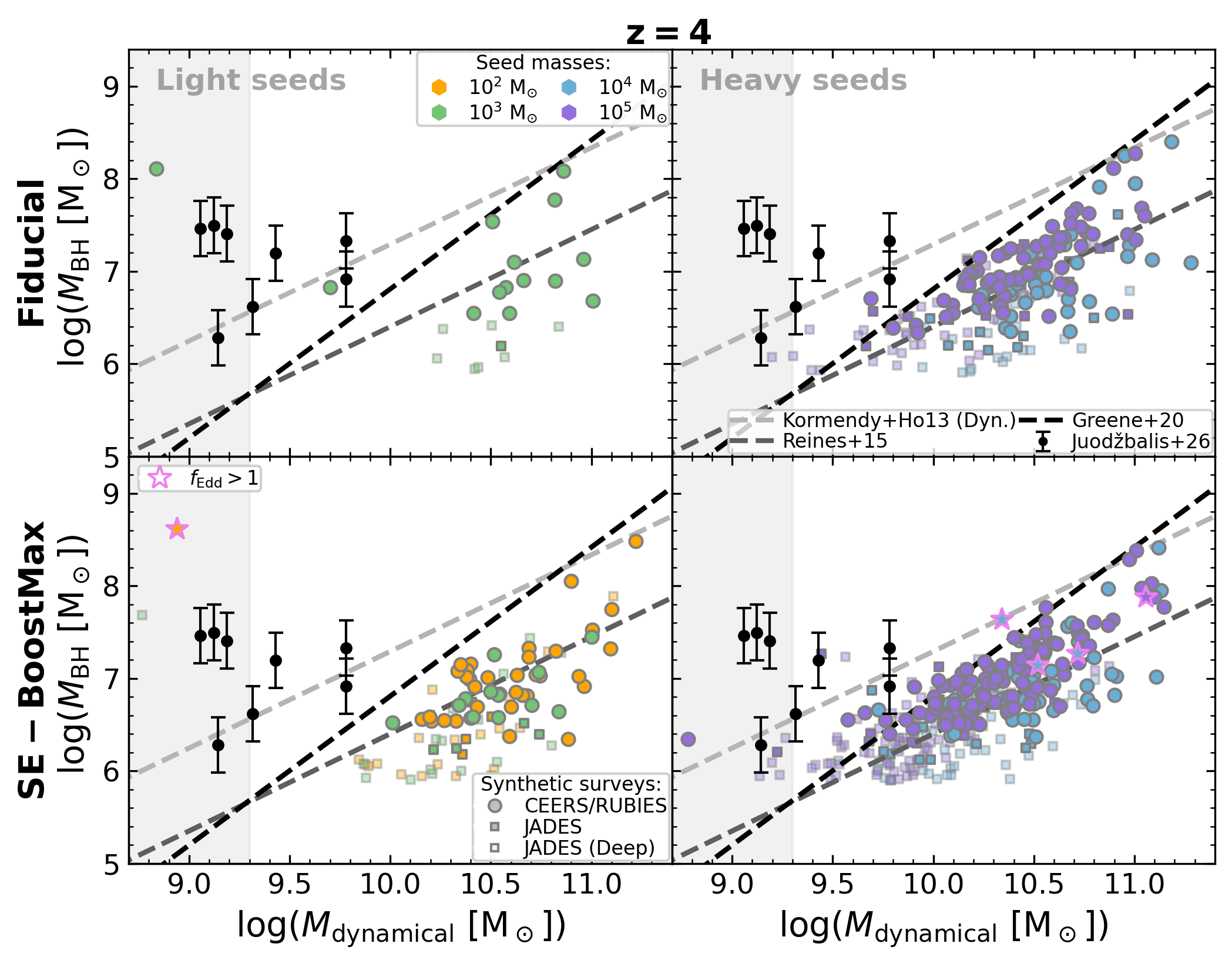}
    \caption{Black hole mass as a function of total dynamical mass (stellar mass plus gas mass plus dark matter mass within the stellar half-mass radius) from the \aesop~simulations for the fiducial (top row) and `\textit{SE-BoostMax}' accretion models (bottom row). Left and right panels show results at $z=6$ and $z=4$, respectively. The columns separate light seeds ($10^{2}$ and $10^{3}\,\Msun$, left) from heavy seeds ($10^{4}$ and $10^{5}\,\Msun$, right). We only include simulated active black holes that meet broad-line AGN selection criteria mimicking \jwst survey detection limits. Scatter symbols indicate detectability for different \jwst survey configurations: CEERS/RUBIES (circles), JADES Medium (opaque squares), and JADES Deep (transparent squares), assuming an R1000 grating and exposure times of 1\,hr, 2.6\,hr, and 7\,hr, respectively. Observed JADES AGN with dynamical mass estimates from \citet{juodzbalis_jades_2026} are overplotted for comparison. The grey shaded region indicates the resolution limit of the simulations and pink stars highlight the simulated black holes `observed' during super-Eddington accretion bursts. Despite the smaller aperture, including the dark matter mass worsens the agreement with the observed distribution (despite good agreement with stellar masses and stellar velocity dispersions) hinting at the complexities when transforming between velocity dispersions and dynamical masses at high redshift.}
    \label{fig:scalrel_dynmassdm}
\end{figure*}

In this section, we further discuss this aspect of the relation between the black hole and its host potential. In particular, we examine the relation between black hole mass and total baryonic mass as well as total dynamical mass. 

In Fig.~\ref{fig:scal_rel_filter}, we show that our simulated AGN that would be observable as per our synthetic selection tool \textsc{balmersopica} are overmassive with respect to the local black hole mass -- stellar mass scaling relations. Indeed, they occupy the same locus as the observational data. In Fig.~\ref{fig:scal_rel_filter_dyn}, we include the same synthetically selected AGN from our simulations, however, we plot their black hole masses against total baryonic mass, defined for our simulations as the sum of the stellar mass and gas mass within twice the stellar half-mass radius. For comparison, we also include the JADES AGN from \citet{juodzbalis_jades_2026} for which dynamical mass estimates of the host galaxies are available. Strikingly, while the `detectable' \aesop~AGN lie clearly above the local black hole mass--stellar mass scaling relations, they move much closer to the local relations once the gas mass is included in the host mass budget. This behaviour supports a `black hole first' scenario, in which the black hole assembles its mass early and the stellar component subsequently catches up. Importantly, this demonstrates that sufficient baryonic mass is already present in these systems to enable such evolution.

We note that the observed JADES AGN generally have lower inferred dynamical masses than those of the \aesop~hosts. This discrepancy is partly driven by the limited resolution of our simulations; the shaded region indicates our resolution limit, requiring at least 100 stellar particles and 100 gas cells for a galaxy to be considered resolved. However, substantial uncertainties also affect dynamical mass estimates at high redshift. These masses are typically inferred from gas velocity dispersions using relations calibrated in the local Universe, and may lead to both overestimates and underestimates of the true dynamical mass by up to an order of magnitude depending on the dynamical state of the system, in particular for dark-matter-dominated low-mass galaxies \citep[see][for details]{ortame_small_2026}.

Indeed, we find that when we include the dark matter mass in our dynamical mass estimate, the discrepancy between the observations and simulations becomes even more severe, with distributions offset by an order of magnitude. We note that we observe this significant offset despite choosing a smaller aperture of just the stellar half-mass radius for the total dynamical mass calculations in the simulations. This further hints at the complexity of translating between velocity dispersions and dynamical masses. Nevertheless, the analysis in the main body of the paper and this appendix clearly shows that co-evolution with stellar velocity dispersion, i.e. the potential, may be established first. This is in agreement with \citet{mcclymont_overmassive_2026} who demonstrate with the high-resolution THESAN-ZOOM simulations that gas-rich, dark-matter-dominated hosts may provide ideal growth conditions for overmassive black holes.

\section{Galaxy stellar mass function} \label{appsec:gsmf}

\begin{figure*}
    \centering
    \includegraphics[width=\textwidth]{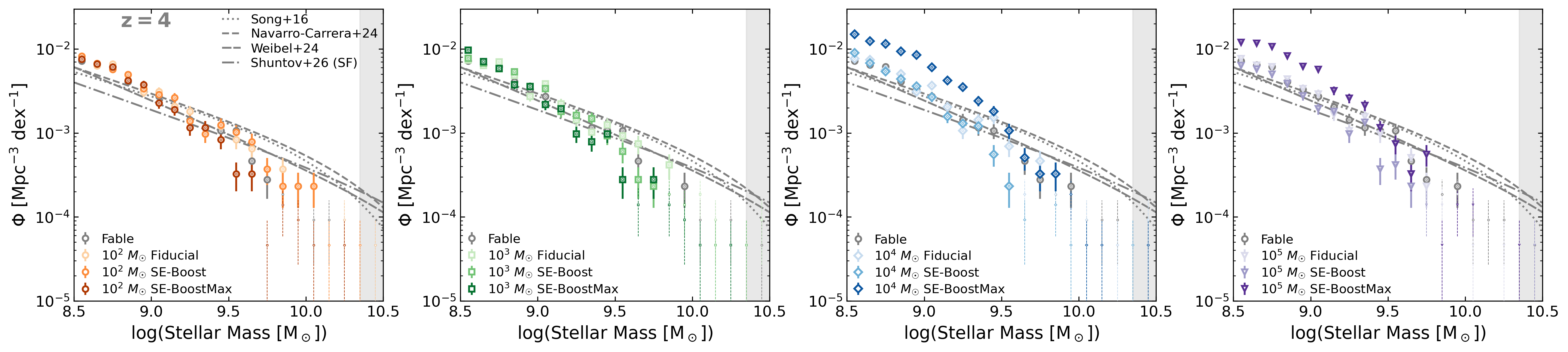}
    \includegraphics[width=\textwidth]{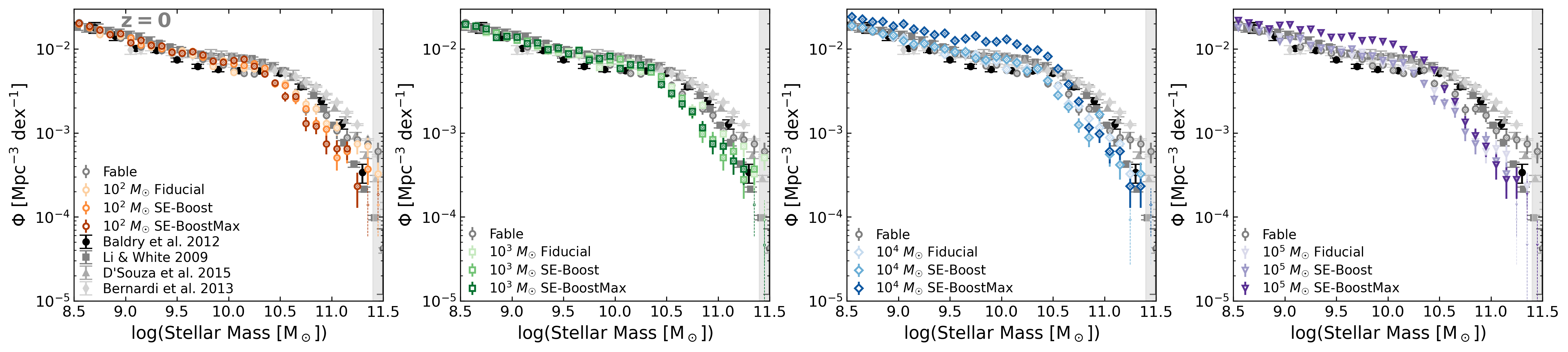}
    \caption{Stellar mass function at $z=4$ (top row) and $z=0$ (bottom row). The four columns show the simulated mass functions for seed masses $M_\mathrm{seed}=10^{2}~\Msun,~10^{3}~\Msun, 10^{4}~\Msun$ and $10^{5}~\Msun$, respectively. The line-shading indicates the accretion model employed, as indicated in the legend. Thin dashed lines indicate high statistical uncertainty due to low number counts in the simulations (fewer than five objects per bin). For comparison, we also show the galaxy stellar mass function from the original \fable~simulations in grey. The grey-shaded region indicates where our galaxy stellar mass function estimate would be expected to be significantly affected by cosmic variance due to the simulation box size ($L_\mathrm{box}= 60$~cMpc). Further, we indicate observed high-z galaxy stellar mass functions \citep{song_evolution_2016,navarro-carrera_constraints_2024,weibel_galaxy_2024,shuntov_stellar_2026} and observed local galaxy stellar mass functions \citep[][]{li_distribution_2009,baldry_galaxy_2012,bernardi_massive_2013,dsouza_massive_2015}. Overall, there is satisfactory agreement (variations smaller than 0.5 dex) for the majority of the models explored, however, we note that lowering the supernova feedback efficiency for the heavy seeds in the \textit{SE-BoostMax} model variation leads to an overabundance of low-mass galaxies compared to most observational constraints and the original \fable~simulations.}
    \label{fig:gsmf_z40}
\end{figure*}

In this paper, we have focused on analysing the evolution of active black holes in the early Universe both from the perspective of seeding and accretion modelling as well as in the context of \jwst~observations. It is, however, important to also ensure that we study this evolution within a realistic galaxy population. Whilst a full analysis of galaxy evolution and star formation in \aesop~is clearly beyond the scope of this paper, we compare the resulting galaxy stellar mass function from our different model variations both to the original \fable~simulations and to observational constraints at high redshift and in the local Universe.

Fig.~\ref{fig:gsmf_z40} shows the galaxy stellar mass function at $z=4$ (top row) and $z=0$ (bottom row). The four columns correspond to the four seed masses explored and the colour-coding indicates the accretion model employed, as indicated by the legend. For comparison, we also add the galaxy stellar mass functions from the original \fable~simulations. We calculate the simulated galaxy stellar mass function in bins of 0.1 dex. The error bars indicate the Poisson error for each bin. For all simulations, lines become dashed when there are fewer than 5 objects per bin. Furthermore, we indicate the uncertainty due to cosmic variance as a grey-shaded region. We estimate the stellar mass completeness limit of each box as the mass at which the expected halo count drops to $N=5$ based on the cumulative \citet{tinker_toward_2008} halo mass function and then convert this to a limiting stellar mass via abundance matching \citep{moster_galactic_2013}. At $N=5$, the $12~{\rm Mpc}$ (physical) box at $z=4$ is complete up to $\log(M_\mathrm{stellar} [M_\odot]) \approx 10.35$, while the $60~{\rm Mpc}$ box at $z=0$ is complete up to $\log(M_\mathrm{stellar}  [M_\odot]) \approx 11.40$. Above these masses, number densities are dominated by sample variance rather than counting statistics, also see discussion in \citet{somerville_cosmic_2004}, \citet{trenti_cosmic_2008} and \citet{bigwood_case_2025}. Furthermore, we plot high-z observational constraints (just showing the mean Schechter fits and omitting uncertainties for clarity) from \citet{song_evolution_2016,navarro-carrera_constraints_2024,weibel_galaxy_2024,shuntov_stellar_2026} in the top row as well as local galaxy stellar mass function constraints \citep[][]{li_distribution_2009,baldry_galaxy_2012,bernardi_massive_2013,dsouza_massive_2015} in the bottom row.

Most of the \aesop~models are in reasonable agreement (differences under 0.5 dex) with the observational constraints, given our simplified subgrid physics and limited resolution. The agreement is worse at high redshifts where the \aesop~simulations tend to underpredict the abundance of galaxies with $\log(M_\mathrm{stellar} \ [\Msun]) > 9.5$. Better agreement with the observations would require more star formation and therefore more dense gas in this regime which may then also lead to more black hole growth. This suggests that black hole growth in this mass regime starts from a difficult position: the same limited gas supply that suppresses star formation also restricts accretion. Nevertheless, we highlight that most of our analysis focuses on galaxies with $\log(M_\mathrm{stellar} \ [\Msun]) < 9.5$ as most of the observed \jwst AGN are hosted by low-mass galaxies.

Overall, we are therefore exploring high-redshift AGN evolution within a realistic galaxy framework as measured by the galaxy stellar mass abundances. Interestingly, despite the stark differences for the different \aesop~models with regard to the scaling relations and the black hole mass function, the models are in relatively close agreement with one another for the galaxy stellar mass function throughout cosmic time. The most significant discrepancy arises when reducing supernova feedback efficiency for intermediate and heavy seeds in the \textit{`SE-BoostMax'} variant. This produces too many low-mass galaxies relative to both observations and the original \fable~simulations and demonstrates that lowering stellar feedback activity throughout cosmic time may be too extreme a choice, at least without recalibrating the AGN feedback at the same time.


\bsp	
\label{lastpage}
\end{document}